\documentclass[preprint,12pt,eqsecnum,nofootinbib,amsmath,amssymb]{revtex4-2}
\usepackage{graphicx}  % Include figure files
\usepackage{dcolumn}  % Align table columns on decimal point
\usepackage{bm}             % Bold math: $\bm{\alpha}$
\usepackage{latexsym}  % Several additional symbols
\usepackage{fancyhdr}  % Fancy header package
\usepackage{wrapfig}
\usepackage{comment}
\usepackage{dsfont}
\usepackage{mathtools}
\usepackage{datetime}
\usepackage{mathrsfs}
\usepackage{amsbsy}
\usepackage{tikz}
\usepackage{bbold}
\usepackage{multirow}
\usepackage[export]{adjustbox}

\newcommand{\PROOF}{{\tiny PROOF}}

\newcommand{\ENDPROOF}{$\blacksquare$}

\newcommand{\smM}{{\scriptscriptstyle M}}

\newcommand{\smR}{{\scriptscriptstyle R}}

\newcommand{\smN}{{\scriptscriptstyle N}}
 
\newcommand{\bodyskip}{\baselineskip 18pt plus 1pt minus 1pt}
\newcommand{\bibskip}{\baselineskip16pt plus 1pt minus 1pt}
\newcommand{\tableofcontentsskip}{\baselineskip 14pt plus 1pt minus 1pt}
\newcommand{\footnoteskip}{\baselineskip 12pt plus 1pt minus 1pt}
\newcommand{\abstractskip}{\baselineskip 13pt plus 1pt minus 1pt}

\newcommand{\theoremskip}{\baselineskip 13pt plus 1pt minus 1pt}

\newtheorem{theorem}{Theorem}

\newtheorem{definition}[theorem]{Definition}

\begin{document}

\baselineskip 20pt plus 1pt minus 1pt

\bodyskip

%% title page 
\baselineskip 15pt plus 1pt minus 1pt
%\preprint{\preprintnumber}

% publication title page
\title{
  Shor's Factoring Algorithm  and \\ Modular Exponentiation Operators
}

\author{Robert L Singleton Jr$^\dagger$}

\affiliation{
Quantum Division\\
SavantX Research Center\\
Santa Fe,  New Mexico,  USA
}

%\date{6/15/2023}
\date{8/7/2023}

\begin{abstract}
\abstractskip
\vskip0.3cm 
\noindent
This manuscript provides a pedagogical presentation
of Shor's factoring 
algorithm,  which is a quantum algorithm for factoring
very large numbers (of order of hundreds to thousands 
of bits) in polynomial time.  In contrast,  all known 
classical algorithms for the factoring problem take 
an exponential time to factor such large numbers.  Shor's 
algorithm therefore has profound implication for 
public-key encryption such as RSA and Diffie-Hellman 
key exchange.  In these notes,  we assume no prior 
knowledge of Shor's algorithm beyond a basic familiarity 
with the circuit model of quantum computing.  Shor's 
algorithm contains a number of moving parts,  and can
 be rather daunting at first.  
The literature is replete with derivations and expositions of 
Shor's algorithm,  but most of them seem to be lacking 
in essential details,  and none of them provide a
pedagogical presentation.  They require a thicket of 
appendices and assume a knowledge of quantum 
algorithms and classical mathematics with which
the reader might not be familiar.  We therefore start 
with  first principle derivations of the quantum Fourier 
transform (QFT) and quantum phase estimation (QPE),  
which are the essential building blocks of Shor's algorithm.  
We then  go on to develop the theory of modular 
exponentiation (ME) operators,  one of the fundamental 
components of Shor's algorithm, and the place where 
most of the quantum resources are deployed.  We also 
delve into the number theory that establishes the link 
between factorization and the period of the modular 
exponential function. We then apply the QPE algorithm 
to obtain Shor's factoring algorithm.  We also discuss the
post-quantum processing and the method of continued 
fractions, which is used to extract the exact period of the
modular exponential function from the approximately
measured phase angles of the ME operator. The manuscript 
then moves on to a series of examples. We first verify 
the formalism by factoring $N=15$, the smallest number 
accessible to Shor's algorithm. We then proceed to factor 
larger integers,  developing a systematic procedure that 
will find the ME operators for any semi-prime $N =  p 
\times q$ (where $q$ and $p$ are prime). Finally, we 
factor the composite numbers $N=21, 33, 35, 143, 247$ using the 
Qiskit simulator.  It is observed that the ME operators 
are somewhat forgiving,  and truncated approximate 
forms are able to extract factors just as well as the 
exact operators.  This is because the method of continued
fractions only requires an approximate phase value 
for its input, which suggests that implementing Shor's 
algorithm might not be as difficult as first suspected.

\vfill
\noindent
$\dagger$
corresponding email: 
robert.singleton@savantx.com
or 
bobs1@comcast.net
\end{abstract}

\maketitle

\vfill
%\fi
%%

% to change page settings
%\thispagestyle{empty}
%\pagestyle{empty}
%\setcounter{page}{0}

\pagebreak
\tableofcontentsskip
\tableofcontents
%\thispagestyle{empty}

%\pagebreak
\newpage
\bodyskip

\pagebreak
\clearpage

\section{Introduction}

In this manuscript,  we present a pedagogical construction 
of Shor's factoring algorithm\,\cite{shor},  which can factor
exponentially large integers in polynomial time.  All known 
classical algorithms for factorization require an exponential 
time because they work by brute force,  essentially testing 
all (or most) numbers less than the number being factored.  
In contrast,  Shor's algorithm exploits the massive parallelism
inherent in quantum mechanics,  so that all numbers can be 
tested simultaneously rather than sequentially, thereby 
providing for a polynomial factorization process. Since 
Shor's algorithm can factor massively large numbers very 
quickly,  it has major implications for the security of 
encryption standards such as RSA\,\cite{rsa} and 
Diffie-Hellman\,\cite{df1, df2} key exchange,  rendering 
these encryption methods severely compromised. As 
most internet communication is based on such public 
key encryption schemes,  Shor's algorithm has profound 
implications for digital security.  

Shor's algorithm rests upon two fundamental quantum 
algorithms,  the quantum Fourier transform (QFT) and 
quantum phase estimation (QPE).  The QFT,  as the name 
suggests, implements the discrete Fourier transform on 
a gated quantum computer.  Like the classical Fourier 
transform,  it extracts 
frequency signals from an input source,  except that the 
QFT works by manipulating quantum bits or qubits (two 
state quantum systems) on a gated quantum computer.  
The QPE algorithm,  in contrast,  finds the complex phases
or the Eigenvalues of an arbitrary {\em unitary} linear operator.  
Shor's algorithm elegantly combines the QFT and QPE to 
construct a powerful quantum algorithm for factoring 
very large integers.   More precisely,  by employing a specific 
and well chosen unitary operator called the {\em modular 
exponentiation} (ME) operator,  quantum parallelism 
allows the QPE to extract the factors of exponentially 
large numbers in polynomial time.  One might say that 
the QPE algorithm is the workhorse of Shor's algorithm\,\cite{gen}, 
and we shall spend most of our time on the associated ME 
operators. 
The mathematics behind Shor's algorithm is based on a 
simple but profound
result from number theory,  which maps the factoring 
problem onto another mathematical problem that finds
the period of the {\em modular exponential function}.  The 
period of this function is directly related to the factors 
of the number in question,  and the QPE extracts this 
period using the method of continued fractions,  thereby 
providing the sought after factors. 

Historically, Shor's algorithm was motivated by Simon's 
algorithm\,\cite{simon}, but we shall not discuss these 
(interesting) details. We assume no prior knowledge 
of Shor's algorithm
beyond a basic familiarity with the circuit model of 
quantum computing.  Shor's algorithm has a number
of moving parts,  and it is rather complex.  It contains 
a pre-processing phase that happens on a classical 
computer,  then the QPE is conducted on a quantum 
computer (using the associated ME operators),  and 
finally there is a post-processing phase 
that takes place on a classical computer (employing 
the method of continued fractions). The literature 
is thick with expositions of Shor's algorithm,  but most 
of them seem to be lacking in some essential respect,  
and none of them provide a satisfying pedagogical 
presentation.  Consequently,  this manuscript is an attempt 
to derive Shor's algorithm from first principles in a 
self-contained manner assuming minimal familiarity 
with the requisite quantum computing machinery.  In 
Sections~\ref{sec_QFT} and \ref{sec_QPE},  we therefore
provide complete derivations the QFT and the QPE 
algorithms,   respectively.  As we have emphasized,  
these algorithms form the essential building blocks 
of Shor's algorithm.  For the classical post-processing 
stage,  one must utilize the theory of {\em continued fractions}, 
and in Section~\ref{sec_cont_frac} we provide a brief
introduction to the subject,  proving a number of 
fundamental theorems.  In Section~\ref{sec_factoring}, 
we are finally ready to address Shor's algorithm,  which 
involves a rigorous construction of the appropriate modular
exponentiation 
operator $U$. Shor's algorithm then follows by applying 
the QPE algorithm to this operator.  After this, we move on to discuss 
the post-quantum processing in more detail, where we 
apply the theory of continued fractions from 
Section~\ref{sec_cont_frac} to extract the 
{\em exact} period of the modular exponential function 
$f(x)$ from the {\em approximately} measured phase of the 
ME operator $U$. 

In Section~\ref{sec_examples} we apply the formalism 
to factor $N = 15$, the smallest number accessible to 
Shor's algorithm, and we use this section to develop an 
all purpose factoring script.  The difficulty in factoring a 
number with Shor's algorithm does not lie in the magnitude 
of the number itself, but in the size of the period $r$ of the 
modular exponential function $f(x)$\,\cite{pretend}. 
In Section~\ref{sec_further_ex}, we apply the formalism 
to the composites $N=21, 33, 35, 143, 247$,  which have 
periods ranging from $r = 2$ to $r = 36$. 
One might think that we have accomplished nothing, 
since knowing the exact ME operator is equivalent to 
knowing the period $r$ of the function $f(x)$, and
Shor's algorithm would therefore be unnecessary. However, 
it turns out that we do not require the {\em exact} ME 
operators! It is observed that the ME operators are 
somewhat 
forgiving, and truncated approximate forms are able 
to extract factors just as well as the exact operators.  
This is because the method of continued fractions only 
requires an approximate phase value for its input,  
which suggests that implementing Shor's algorithm 
might not be as difficult as first suspected. 
Finally, Section~\ref{sec_conclusions} provides some 
conclusions and closing remarks. 

\vfill
\pagebreak

\section{The Quantum Fourier Transform}
\label{sec_QFT}

\subsection{General Definitions}

In this section we formulate of the quantum Fourier transform 
(QFT),  where our primary references are Refs. \!\!\!\cite{qcqi} 
and \cite{des}.  Given an $M$-vector of complex numbers 
$\psi = (\psi_0,  \psi_1,  \cdots,  \psi_{\smM-1})$,   the {\em 
discrete Fourier transform} \hbox{$\tilde\psi = (\tilde \psi_0,  
\tilde \psi_1,  \cdots,  \tilde \psi_{\smM-1})$} is defined by 
\begin{eqnarray}
  \tilde \psi_\ell
  &=&
  \frac{1}{\sqrt{M}}\sum_{k=0}^{M-1} e^{2 \pi i \,  \ell k  /M}\, 
  \psi_k 
  \ ,
\label{eq_FT}
\end{eqnarray}
and the {\em discrete inverse Fourier transform} is therefore 
given by 
\begin{eqnarray}
  \psi_k 
  &=&
  \frac{1}{\sqrt{M}}\sum_{\ell=0}^{M-1} e^{-2 \pi i\, \ell  k /M}\, 
  \tilde \psi_\ell
  \ ,
\label{eq_FTinv}
\end{eqnarray}
where the indices $\ell, k \in \{0, 1,  \cdots, M -1\}$.   We wish 
to implement the Fourier transform using an \hbox{$m$-qubit} 
quantum system,  where $M = 2^m$ is the number of possible 
quantum states.  The corresponding {\em quantum Fourier 
transform} will be a linear unitary operator on the 
\hbox{$m$-qubit} Hilbert space,  denoted by $QFT$, whose 
action on the {\em computational basis} elements reproduces 
the classical transform~(\ref{eq_FT}), 
\begin{eqnarray}
  QFT\, \vert \ell \rangle
  &=&
  \frac{1}{\sqrt{M}}\sum_{k=0}^{M-1} e^{2 \pi i \,  \ell k  /M}\, 
  \vert k \rangle
  \ .
\label{eq_QFT}
\end{eqnarray}
Recall that a linear operator defined only on the basis states 
is sufficient to give the operator on any state in the Hilbert space. We can 
now express the $QFT$ operator in a very useful basis-dependent 
form,  
\begin{eqnarray}
  QFT
  =
  QFT \cdot \mathbb{1}
  =
  QFT \cdot
  \underbrace{
  \sum_{\ell = 0}^{M-1}  \vert \ell \rangle \langle \ell \vert
  }_{\mathbb{1}}
  = 
  \frac{1}{\sqrt{M}}\sum_{k = 0}^{M-1}\sum_{\ell=0}^{M-1} 
  e^{2\pi i\, k  \ell / M}\, \vert k \rangle \langle \ell \vert \ ,
\label{eq_QFT_basis}
\end{eqnarray}
where we have used the decomposition of unity $\mathbb{1}
= \sum_{\ell=0}^{\smM-1} \vert \ell \rangle \langle \ell \vert$.  
Since the quantum Fourier transform is unitary (an easy proof), 
that is to say $QFT \cdot QFT^\dagger = \mathbb{1}$,  then 
the {\em inverse quantum  Fourier transform} is given by 
\begin{eqnarray}
  QFT^{-1}
  =
  QFT^\dagger
  = 
  \frac{1}{\sqrt{M}}\sum_{\ell = 0}^{M-1}\sum_{k=0}^{M-1} 
  e^{-2\pi i\, k  \ell / M}\, \vert \ell \rangle \langle k \vert 
  \ .
\label{eq_QFTdagger}
\end{eqnarray}
Our aim in this Section is to construct a quantum circuit 
to implement the $QFT$ operator and its inverse.

%\vfill
\pagebreak
\subsection{Qubit Ordering and the QFT Circuit}
\label{sec_conventions}

\begin{figure}[b]
\begin{centering}
\includegraphics[width=\textwidth]{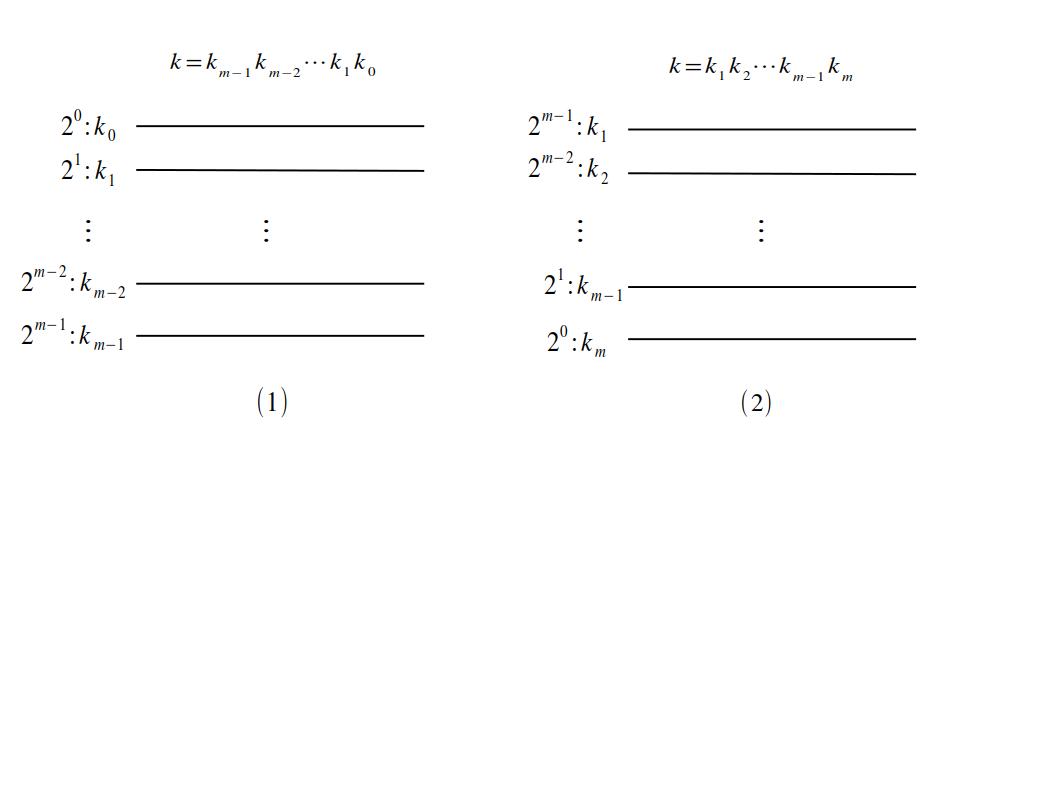} 
\par\end{centering}
\vskip-5.5cm 
\caption{\footnoteskip  
An $m$-bit binary integer $k$ can be encoded on a gated 
quantum computer in either of the conventions described 
in the text. Convention 1 is called the OpenQASM/Qiskit 
convention, while Convention 2 is called the Physics/Mathematics
convention.  Convention 1 labels the top qubit by 0,  and works 
its way down to the last qubit labeled by $m-1$.  Binary integers 
are expressed with the standard bit encoding $k = k_{m-1} 
\cdots k_1 k_0$,  which places the lowest order bit $k_0$ 
at the top of the circuit.   Convention 2 numbers the qubits 
from 1 to $m$ running from top to bottom,  and the bit-ordering 
of integers is flipped to $k = k_1 k_2 \cdots k_m$,  placing the 
lowest order bit $k_m$ at the bottom of the circuit. 
}
\label{fig_conventions}
\end{figure} 
A quantum circuit has four distinct qubit ordering 
conventions of which we must be cognizant, as they 
are all present in the literature. We can order the qubits 
on the quantum circuit in two ways,  and we can order 
the bits of a binary integer in two ways, thereby giving 
four possible conventions. We will be concerned with 
two such conventions. In computer science, one represents
an $m$-bit integer $k$ by $m$ binary digits $k_r \in \{0, 1\}$
using the standard notation $k =  k_{m-1} \cdots k_{1} k_{0}$.
The bit-ordering convention 
is that $k_0$ is the lowest-order bit, so that the value 
of the integer is given by $k = k_0\, 2^0 + k_1\, 2^1 
+ \cdots + k_{m-1}\,2^{m-1} $.  We must also label 
the qubits of the quantum circuit, which gives two 
more possible conventions. In 
OpenQASM/Qiskit\,\cite{OpenQASM_ref},  
the upper qubit of an $m$-qubit circuit is labeled
by~0, working its way down the circuit and ending 
with qubit $m-1$ at the bottom.   For a general 
\hbox{$m$-bit}
binary integer $k= k_{m-1} \cdots k_1, k_0$,  the least 
significant bit $k_0$ is therefore placed at the {\em top} 
of the circuit,
encoded by the computational basis state $\vert k_0 
\rangle$ of the upper qubit. The integers $k$
therefore correspond to the computational basis 
states $\vert k \rangle \equiv \vert k_{m-1}\cdots k_0 
\rangle \equiv \vert k_0 \rangle  \otimes \vert k_1 
\rangle \otimes \cdots \otimes \vert k_{m-2} \rangle
\otimes \vert k_{m-1} \rangle$. Note that the qubit 
ordering is opposite to the bit-string ordering. This 
will be one of our primary conventions. Alternatively, 
we can number the qubits on the circuit  from $1$ 
to $m$, and express binary numbers by $m$-bit strings 
of the form \hbox{$k = k_{1} k_{2} \cdots k_{m-1} 
k_{m}$}.  In this convention, the lowest order bit 
$k_m$ is placed at the {\em bottom} of the circuit, and 
the value of the index integer is $k = k_m\, 2^0 + k_{m-1}\, 
2^1 + \cdots + k_1\,2^{m-1} $. The 
qubit ordering is the same as the bit-string ordering, 
and the integers $k$ correspond to the computational 
basis states $\vert k \rangle \equiv \vert k_1 \cdots k_m 
\rangle \equiv \vert k_1 \rangle  \otimes \vert k_2 \rangle 
\otimes \cdots \otimes \vert k_{m-1} \rangle\otimes 
\vert k_{m} \rangle$.  This is the standard physics 
and mathematics convention. We shall use both 
conventions, which are illustrated in Fig.~\ref{fig_conventions}.  
Quantum circuits will be inverted horizontally 
between these two conventions,  so it is important 
to keep track of which convention is in use.

\subsubsection{Convention 2: Standard Physics/Mathematics }

We first work through the details of the QFT for Convention 2,  
the physics and mathematics convention.  We consider 
an $m$-qubit system in which the qubits are ordered from top
to bottom,  starting with qubit-$1$ in the upper position of 
the circuit and qubit-$m$ at the bottom of the circuit. The quantum system
has $M = 2^m$ distinct states that can be indexed by an
integer \hbox{$k \in \{0, 1,  \cdots, M-1\}$}. We can express 
this integer in the binary form, 
\begin{eqnarray}
  k &=& 
  k_1 k_2 \cdots k_{m-1} k_{m}
\nonumber
\\
  &=&
  2^{m-1} k_{1} +  2^{m-2} k_{2} +
  \cdots  + 2^1 \, k_{m-1} +   2^0 \, k_{m}
  \ ,
\end{eqnarray}
where $k_m$ is the least significant bit. The computational 
basis elements are then defined by
\begin{eqnarray}
   \vert k \rangle 
  &=&
  \vert k_1 k_2 \cdots k_{m-1} k_{m} \rangle
\nonumber\\
  &=&
  \vert k_1 \rangle  \otimes \vert k_2 \rangle  
  \otimes \cdots \otimes
 \vert k_{m-1} \rangle  \otimes \vert k_{m} \rangle 
  ~~\text{where}~~ k_r \in \{0,1\}
  \ . 
\end{eqnarray}
For example, the state labeled by $k=1$ is represented by 
\begin{eqnarray}
  \vert 1 \rangle
  = 
  \vert 0 \cdots 0 1 \rangle
  =
  \vert 0 \rangle \otimes \cdots \otimes\vert 0 \rangle
  \otimes \vert 1 \rangle
  \ ,
\label{eq_one_conv2b}
\end{eqnarray}
with $k_m = 1$ and all other bits $k_r = 0$.  This state will 
play a critical role in Shor's algorithm. Note that we are using
a slightly ambiguous notation in which $\vert 1 \rangle$
is used in different senses on the left- and right-hand sides
of equation (\ref{eq_one_conv2b}). However, the meaning 
of the state $\vert 1 \rangle$ will be clear from context, 
so this should cause no 
problems. Furthermore, this bit convention implies the useful relation 
\begin{eqnarray}
  \frac{k}{M} 
  =
  \frac{k_{1}}{2^1} +  \frac{k_{2}}{2^2} +
  \cdots + \frac{k_{m-1}}{2^{m-1}} +   \frac{k_{m}}{2^m}
  \ .
\end{eqnarray}

Note that a sum over the index $k \in \{0, 1,  \cdots,  M-1\}$ 
can be converted into $m$ sums over the binary components 
$k_r \in \{0, 1\}$ for $r \in \{1,  2,  \cdots,  m \}$,  
\begin{eqnarray}
 \sum_{k=0}^{M-1}
  ~=
  \sum_{k_m \in\{0,1\}} \cdots \sum_{k_2 \in\{0,1\}} 
  \sum_{k_1 \in\{0,1\}}
  \ .
\end{eqnarray}
This allows us to express the quantum Fourier transform
(\ref{eq_QFT}) in the form 
\begin{eqnarray}
  QFT\, \vert \ell \rangle
  &\equiv&
  \frac{1}{\sqrt{M}}\sum_{k=0}^{M-1} e^{2 \pi i \,  \ell k  /M}\,
  \vert k \rangle
\label{eq_qft_aa}
\\[5pt]
  &=& 
  \frac{1}{2^{m/2}}   \sum_{k_{m}} \cdots 
  \sum_{k_2} \sum_{k_1} e^{2 \pi i \,  \ell 
  \big( 
  2^{m-1} k_{1} +  2^{m-2} k_{2} +
  \cdots  + 2^1 \, k_{m-1} +   2^0 \, k_{m}
  \big) \big/ 2^m} \,
  \vert k_{1} k_{2}  \cdots k_{m-1} k_{m} \rangle
  ~~~~
\nonumber\\[5pt]
%\end{eqnarray}
%\begin{eqnarray}
  &=& 
  \nonumber
  \frac{1}{2^{m/2}}   
  \sum_{k_1=0, 1} e^{2\pi i \, \ell k_1/2^1} \vert k_1 \rangle
  \otimes
  \sum_{k_2=0, 1} e^{2\pi i \, \ell k_2 /2^2} \vert k_2 \rangle
  \otimes \cdots \otimes 
  \sum_{k_m=0, 1} e^{2\pi i \, \ell k_m/2^m} \vert k_m \rangle
  ~~~
\\[5pt]
  &=& 
  \frac{1}{2^{m/2}} 
  \Big(
  \vert 0 \rangle + e^{2\pi i \, \ell /2^1} \, \vert 1 \rangle 
  \Big)_{1}
  \otimes
  \Big(
  \vert 0 \rangle + e^{2\pi i \, \ell /2^2} \, \vert 1 \rangle 
  \Big)_{2}
  \otimes \cdots \otimes
\nonumber \\[5pt] && \hskip2.0cm
  \Big(
  \vert 0 \rangle + e^{2\pi i \, \ell /2^{m-1}} \, \vert 1 \rangle 
  \Big)_{m-1}
  \otimes 
  \Big(
  \vert 0 \rangle + e^{2\pi i \, \ell /2^m} \, \vert 1 \rangle 
  \Big)_{m}
  \ . ~~~
\label{eq_c2b_QFT_ell}
\end{eqnarray}
In our current convention,  the quantum state indexed 
by the integer $\ell$ is given by 
\begin{eqnarray}
  \vert \ell \rangle 
  &=&
  \vert \ell_{1} \ell_{2} \,\cdots\, \ell_{m-1} \ell_{m} \rangle
  ~~~\text{with}~~ 
  \ell_r \in \{0, 1\} 
  \nonumber\\[5pt]
   &=&
   \vert \ell_1 \rangle \otimes  \vert \ell_2 \rangle \otimes
   \cdots \otimes  \vert \ell_{m-1}   \rangle \otimes
   \vert \ell_{m}   \rangle 
  \ ,
\end{eqnarray}
where $\ell$ takes the binary form 
\begin{eqnarray}
  \ell &=& 
  \ell_1 \ell_2 \cdots \ell_{m-1} \ell_{m}
  =
  2^{m-1} \ell_{1} +  2^{m-2} \ell_{2} +
  \cdots  + 2^1 \, \ell_{m-1} +   2^0 \, \ell_{m}
  \ .
\end{eqnarray}
We shall also introduce the notion of {\em binary fractions}
corresponding to non-negative $m$-bit phase angles, 
\begin{eqnarray}
  \Omega
  \equiv
  0.\ell_1 \ell_2 \cdots \ell_{m-1} \ell_{m}
  \equiv
   \frac{\ell_{1}}{2^{1}} +  \frac{\ell_{2}}{2^{2} } +
  \cdots  + 
  \frac{\ell_{m-1} }{ 2^{m-1}}+    \frac{\ell_{m}}{2^{m} }
  \ .
\end{eqnarray}
Note that the phase $\Omega = 0.\ell_1 \ell_2 \cdots \ell_m$ 
and the corresponding integer index $\ell = \ell_1 \ell_2 \cdots 
\ell_m$ are related by 
\begin{eqnarray}
  \ell = M \Omega = 2^m \, \Omega  \ ,
\end{eqnarray}
an expression we shall employ throughout the sequel. We 
now rewrite the exponential terms in (\ref{eq_c2b_QFT_ell}) 
as follows,  working slowly through the algebra,  starting 
with qubit-1:
\begin{eqnarray}
  2\pi i \, \frac{\ell}{2^1}
  &=& 
  \frac{2\pi i }{2^1} \Big[ 
  \Big(
  2^{m-1} \ell_{1} +  2^{m-2} \ell_{2} +
  \cdots  + 2^1 \, \ell_{m-1} \Big) 
  +   2^0 \, \ell_{m}
  \Big]
\nonumber\\[5pt]
\label{eq_exp_ell_a}
  &=& 
  2\pi i  \Big[ 
  \Big(
  2^{m-2} \ell_{1} +  2^{m-3} \ell_{2} +
  \cdots  + 2^0 \, \ell_{m-1} \Big) 
  +   \Omega_1
  \Big]
\end{eqnarray}
\begin{eqnarray}
  \nonumber\\[10pt]
  2\pi i \, \frac{\ell}{2^2}
  &=& 
   \frac{2\pi i }{2^2} \Big[ 
  \Big(
  2^{m-1} \ell_{1} +  2^{m-2} \ell_{2} +
  \cdots  2^2 \ell_{m-2} \Big) 
  + 2^1 \, \ell_{m-1}   +   2^0 \, \ell_{m}
  \Big]
  ~~~~~~~~~~
\nonumber
\\[5pt]
  &=& 
   2\pi i  \Big[ 
  \Big(
  2^{m-3} \ell_{1} +  2^{m-4} \ell_{2} +
  \cdots  2^0 \ell_{m-2} \Big) 
  + \Omega_2
  \Big]
  ~~~~~~~~~~
\\[5pt]\nonumber
  & \cdots& 
\\[5pt] 
  2\pi i \, \frac{\ell}{2^{r}}
  &=& 
   \frac{2\pi i }{2^{r}} \Big[
  \Big(
  2^{m-1} \ell_{1} +  2^{m-2} \ell_{2} + \cdots  + 
  2^{r} \, \ell_{m-r} \Big) +  2^{r-1} \, \ell_{m-r+1} 
\nonumber\\ && \hskip6.0cm
  + \cdots +
  2^1 \, \ell_{m-1} +   2^0 \, \ell_{m}
  \Big] 
\nonumber\\
  &=& 
   2\pi i  \Big[  \Big(
  2^{m-r-1} \ell_{1} +  2^{m-r-2} \ell_{2} + \cdots  + 
  2^0 \, \ell_{m-r} \Big) +  
 \Omega_{r}
  \Big] 
\\\nonumber
  &\cdots&
%\end{eqnarray}
%\begin{eqnarray}
\\[5pt]
    2\pi i \, \frac{\ell}{2^{m-1}}
  &=& 
   \frac{2\pi i }{2^{m-1}}\Big[ 
  \big( 2^{m-1} \ell_{1} \big) 
  +  2^{m-2} \ell_{2} + \cdots  + 2^1 \, \ell_{m-1} +   2^0 \, \ell_{m}
  \Big]
\nonumber
\\[5pt]
  &=& 
  2\pi i \Big[ 
  \big( \ell_{1} \big) + \Omega_{m-1}
  \Big]
\\[10pt]
\label{eq_exp_ell_e}
    2\pi i \, \frac{\ell}{2^{m}}
  &=& 
   \frac{2\pi i }{2^m} \Big[ 
  2^{m-1} \ell_{1} +  2^{m-2} \ell_{2} +
  \cdots  + 2^1 \, \ell_{m-1} +   2^0 \, \ell_{m}
  \Big]
\nonumber
\\[5pt]
  &=&
   2\pi i \Big[ 
  \Omega_{m}
  \Big]
  \ ,
\end{eqnarray}
where the {\em partial-phase angles} are defined by 
\begin{eqnarray}
  \Omega_{1} &\equiv& 
  \frac{\ell_{m}}{2^1}
  =
  0.\ell_m
%\hskip7.3cm : ~ \vert \ell_{1} \rangle
\\[5pt]
  \Omega_{2} 
  &\equiv&
  \frac{\ell_{m-1}}{2^1} +  \frac{ \ell_{m}}{2^2}
  =
 0.\ell_{m-1} \ell_m
%\hskip5.1cm : ~ \vert \ell_{2} \rangle
\\[5pt]\nonumber
  & \cdots& 
\\[5pt]
  \Omega_{r}
  &\equiv&
  \frac{\ell_{m-r +1} }{2^1}
  +     \cdots + \frac{\ell_{m-1}}{2^{r-1}} +  \frac{\ell_{m}}{2^{r}}
  =
  0.\ell_{m-r+1} \cdots \ell_{m-1}\ell_m
%\hskip0.45cm : ~ \vert \ell_{r} \rangle
%\end{eqnarray}
%\begin{eqnarray}
\\\nonumber
  & \cdots& 
\\[5pt]
%\end{eqnarray}
%\begin{eqnarray}
  \Omega_{m-1} 
  &\equiv&
    \frac{\ell_{2}}{2^1} +     \frac{\ell_{3}}{2^2} +
  \cdots  + \frac{\ell_{m-2}}{2^{m-1}} +  \frac{\ell_{m}}{2^{m-1}}
  =
  0.\ell_2 \cdots \ell_{m-1}\ell_m
%\hskip0.8cm : ~ \vert \ell_{m-1} \rangle
\\[5pt]
  \Omega_{m}
  &\equiv&
    \frac{\ell_{1} }{2^1} +  \frac{\ell_{2} }{2^2}
  + \cdots  + \frac{\ell_{m-1}}{2^{m - 1}} +  \frac{\ell_{m}}{2^m}
  =
  0.\ell_1 \cdots \ell_{m-1}\ell_m
  \ .
%\hskip1.0cm : ~ \vert \ell_{m} \rangle
\end{eqnarray}
Note that integer multiples of $2\pi i$ in the parentheses 
of equations (\ref{eq_exp_ell_a})--(\ref{eq_exp_ell_e}) do not 
contribute,  as $e^{2\pi i \, n} = 1$ for any integer $n$,  thereby 
permitting us to express  the $QFT$ operation only in terms 
of the partial phases
\begin{eqnarray}
  \Omega_{r}
  &=&
  \sum_{k=1}^{r} \, \frac{\ell_{m -r + k}}{2^k}
  ~~~\text{for}~~ r \in \{1, 2,  \cdots,  m \} 
  \  .
\end{eqnarray}
The quantum Fourier transform (\ref{eq_c2b_QFT_ell}) 
of the $\ell$-state can therefore be written in any one 
of three useful forms: 
\begin{eqnarray}
  QFT\, \vert \ell \rangle
  &=&
  \frac{1}{2^{m/2}} 
  \Big(
  \vert 0 \rangle + e^{2\pi i \, \ell /2^1} \, \vert 1 \rangle 
  \Big)_{1}
  \otimes
  \Big(
  \vert 0 \rangle + e^{2\pi i \, \ell /2^2} \, \vert 1 \rangle 
  \Big)_{2}
  \otimes \cdots \otimes
 \\[5pt] && \hskip2.0cm
  \Big(
  \vert 0 \rangle + e^{2\pi i \, \ell /2^{m-1}} \, \vert 1 \rangle 
  \Big)_{m-1}
  \otimes 
  \Big(
  \vert 0 \rangle + e^{2\pi i \, \ell /2^m} \, \vert 1 \rangle 
  \Big)_{m}
  \nonumber
\nonumber\\[10pt]
  &=&
  \frac{1}{2^{m/2}} 
  \Big(
  \vert 0 \rangle + e^{2\pi i \, 0.\ell_m} \, \vert 1 \rangle 
  \Big)_{1}
  \otimes
  \Big(
  \vert 0 \rangle + e^{2\pi i \, 0.\ell_{m-1}\ell_m } \, \vert 1 \rangle 
  \Big)_{2}
  \otimes \cdots \otimes
 \\[5pt] && \hskip1.5cm
  \Big(
  \vert 0 \rangle + e^{2\pi i \,0.\ell_2 \cdots \ell_{m-1}\ell_m } \, 
  \vert 1 \rangle 
  \Big)_{m-1}
  \otimes
  \Big(
  \vert 0 \rangle + e^{2\pi i \, 0.\ell_1 \ell_2 \cdots\ell_{m-1}\ell_m} \, 
  \vert 1 \rangle 
  \Big)_{m}
\nonumber
%\end{eqnarray}
%\begin{eqnarray}
\nonumber\\[10pt]
  &=&
  \frac{1}{2^{m/2}} 
  \Big(
  \vert 0 \rangle + e^{2\pi i \, \Omega_{1}} \, \vert 1 \rangle 
  \Big)_{1}
  \otimes
  \Big(
  \vert 0 \rangle + e^{2\pi i \, \Omega_{2}} \, \vert 1 \rangle 
  \Big)_{2}
  \otimes \cdots \otimes
 \nonumber\\[5pt] && \hskip2.5cm
  \Big(
  \vert 0 \rangle + e^{2\pi i \, \Omega_{m-1}} \, \vert 1 \rangle 
  \Big)_{m-1}
  \otimes
  \Big(
  \vert 0 \rangle + e^{2\pi i \, \Omega_{m}} \, \vert 1 \rangle 
  \Big)_{m}
  \ .
\label{eq_QFTell_Omega}
\end{eqnarray}
We next reverse the order of the qubits with a string of SWAP
gates to form the state
\begin{eqnarray}
 \vert \psi_{\rm rev} \rangle
  &=&
  \frac{1}{2^{m/2}} 
  \Big(
  \vert 0 \rangle + e^{2\pi i \, \Omega_{m}} \, \vert 1 \rangle 
  \Big)_{1}
  \otimes
  \Big(
  \vert 0 \rangle + e^{2\pi i \, \Omega_{m-1}} \, \vert 1 \rangle 
  \Big)_{2}
  \otimes \cdots \otimes
\label{eq_convention1b_back}
 \\[5pt] && \hskip4.8cm
  \Big(
  \vert 0 \rangle + e^{2\pi i \, \Omega_{2}} \, \vert 1 \rangle 
  \Big)_{m-1}
  \otimes
  \Big(
  \vert 0 \rangle + e^{2\pi i \, \Omega_{1}} \, \vert 1 \rangle 
  \Big)_{m}
  \ .
\nonumber
\end{eqnarray}
The state $\vert \psi_{\rm rev} \rangle$ can be represented 
quite readily by a quantum circuit.  To see this,  let us start with 
qubit-$m$ of (\ref{eq_convention1b_back}).  Since $e^{2\pi 
i\,\Omega_{1}} = e^{2\pi i \, \ell_{m}/2} = (-1)^{\ell_{m}}$,  and 
since $\ell_{m}$ takes the binary values $0$ and $1$,  we 
have 
\begin{eqnarray}
  \left.
  \begin{matrix}
  \displaystyle{
  \frac{1}{\sqrt{2}}\Big(
  \vert 0 \rangle +e^{2\pi i \, \Omega_{1}(\ell_{m}=0)} \, \vert 1 \rangle 
  \Big)_{m}
  =
  \frac{\vert 0 \rangle_{m} + \vert 1 \rangle_{m}}{\sqrt{2}}
  = 
  H \vert 0 \rangle_{m}
  }
\\[15pt]
  \displaystyle{
  \frac{1}{\sqrt{2}} \Big( 
  \vert 0 \rangle + e^{2\pi i \, \Omega_{1}(\ell_{m}=1)}\, 
  \vert 1 \rangle  \Big)_{m}
  =
  \frac{\vert 0 \rangle_{m} - \vert 1 \rangle_{m}}{\sqrt{2}}
  = 
  H \vert 1 \rangle_{m} 
  }
  \end{matrix}
  ~\right\}
  =  H \vert \ell_m \rangle
 ~~~~~~~~
\label{eq_Omegaone}
\end{eqnarray}
for $\ell_m \in \{0,1\}$,  where the single-qubit Hadamard gate
is defined by
\begin{eqnarray}
  H =
  \frac{1}{\sqrt{2}}
  \left[
  \begin{matrix}
  1 & ~1\\
  1 & -1
  \end{matrix}
  \right]
  ~~~\text{with}~~
  \vert 0 \rangle
  =
  \left[
  \begin{matrix}
  ~1&\\
  ~0& 
  \end{matrix}
  \right]
  ~~\text{and}~~
  \vert 1 \rangle
  =
  \left[
  \begin{matrix}
  ~0&\\
  ~1& 
  \end{matrix}
  \right]
  \ .
\label{eq_H_def}
\end{eqnarray}
\begin{figure}[b!]
\begin{centering}
\includegraphics[width=\textwidth]{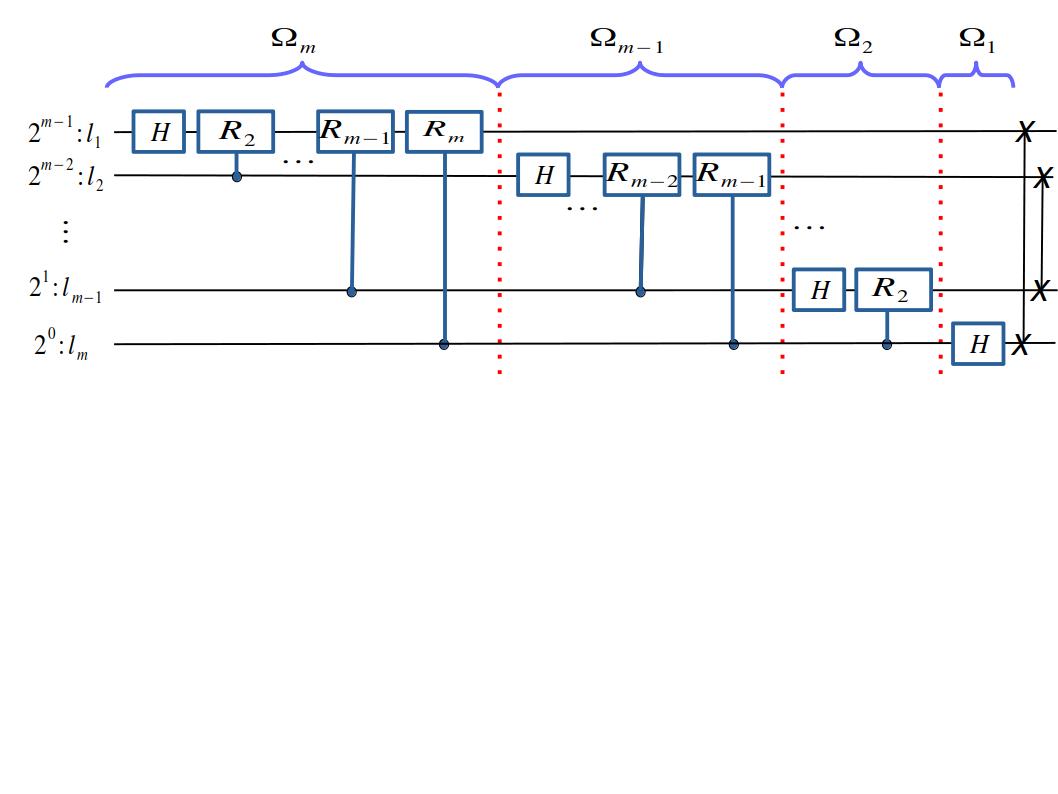} 
\par\end{centering}
\vskip-6.5cm 
\caption{\footnoteskip  
  Convention 2 for the $QFT$: The standard 
  physics/mathematics convention. 
}
\label{fig_QFT_conv2b}
\end{figure}

\noindent
We will often place subscripts on the single-qubit basis states
such as $\vert \ell_m \rangle_m$  to explicitly indicate their 
qubit position in the quantum circuit.  We will also denote the 
qubit upon which the Hadamard gate acts by a superscript,  
{\em e.g.} $H^m$ explicitly states that $H$ acts on the $m$-th
qubit.  Therefore,  we can express (\ref{eq_Omegaone}) in the 
form 
\begin{eqnarray}
  \frac{1}{\sqrt{2}}
  \Big(
  \vert 0 \rangle + e^{2\pi i \, \Omega_{1}} \, \vert 1 \rangle 
  \Big)_{m}
  &=&
  H^m \vert \ell_{m} \rangle_m
  ~~\text{with}~~ \ell_m \in \{0,1\} 
  \  .
\end{eqnarray}
Moving on to the next qubit,  $m - 1$,  we find 
\begin{eqnarray}
  \frac{1}{\sqrt{2}}
  \Big(
  \vert 0 \rangle + e^{2\pi i \, \Omega_{2}} \, \vert 1 \rangle 
  \Big)_{m-1}
  &=&
  \frac{1}{\sqrt{2}} \,
  \Big(
  \vert 0 \rangle + e^{2\pi i \, \ell_{m-1}/2^1} \cdot
  e^{2 \pi i \, \ell_{m}/2^2}\, \vert 1 \rangle 
  \Big)_{m-1}
  \ .
\end{eqnarray}
The first exponential $e^{2\pi i \, \ell_{m-1}/2}= e^{\pi i \, 
\ell_{m-1}} = (-1)^{\ell_{m-1}}$ gives a Hadamard gate acting 
on $\vert \ell_{m-1} \rangle_{m-1}$,  and the second exponential 
$e^{2 \pi i \, \ell_{m}/2^2}$  produces a controlled phase gate
with angle $\theta = 2\pi/2^2$ (call it $CR_2$) acting on the target 
qubit $\vert \ell_{m-1} \rangle_{m-1}$ with the control 
qubit $\vert \ell_{m} \rangle_m$,  so that
\begin{eqnarray}
  \frac{1}{\sqrt{2}}
  \Big(
  \vert 0 \rangle + e^{2\pi i \, \Omega_{2}} \, \vert 1 \rangle 
  \Big)_{m-1}
  &=&
  H^{m-1} \cdot C^{m} R_2^{m-1} \, 
  \vert \ell_{m-1}   \rangle_{m-1} 
  ~~\text{where}~~ \ell_{m-1} \in \{0,1\}
  \ . ~~~
\end{eqnarray}
We have used superscripts on the controlled-$R$ gate,
writing $C^c R_2^t$ to explicitly indicate the control qubit
$c$ and the  target qubit $t$.   Finally,  let us examine the 
$1$-st qubit,  where we find 
\begin{eqnarray}
  \frac{1}{\sqrt{2}}
  \Big(
  \vert 0 \rangle + e^{2\pi i \, \Omega_{m}} \, \vert 1 \rangle 
  \Big)_{1}
  &=&
  \frac{1}{\sqrt{2}}
    \Big(
  \vert 0 \rangle + e^{2\pi i \, \ell_{1}/2} \cdot
  e^{2\pi i \, \ell_{2}/2^2} \cdots e^{2 \pi i \, \ell_{m-1}/2^{m-1}}
  \cdot e^{2 \pi i \, \ell_{m}/2^{m}}
  \, \vert 1 \rangle 
  \Big)_{1} ~~~~~~~~~~~
\\[5pt]
  &=&
  H^1 \cdot C^{2}R_2^{1} \,\cdots\, C^{m-1}R_{m-1}^{1} 
  \cdot C^{m} R_m^{1} \, \vert \ell_{1} \rangle_1
  ~~~\text{with}~~ \ell_1 \in \{0,1\}
  \ ,
\end{eqnarray}
where the single-qubit phase operator is defined by 
\begin{eqnarray}
  R_n &=&
  \left[
  \begin{matrix}
  1 & 0\\[-5pt]
  0 & e^{2\pi i/ 2^n}
  \end{matrix}
  \right] 
  \ .
\end{eqnarray}
In terms of the standard phase gate $P(\theta)$,  
we can express the phase rotation by 
\begin{eqnarray}
  R_n 
  = 
  P(\theta_n) 
  =
  \left[
  \begin{matrix}
  1 & 0\\[-5pt]
  0 & e^{i \theta_n }
  \end{matrix}
  \right] 
  ~~~\text{where}~~~
  \theta_n = \frac{2\pi}{2^n} =  \frac{\pi}{2^{n-1}} 
  \ .
\end{eqnarray}
Figure~\ref{fig_QFT_conv2b}   reproduces the corresponding 
$QFT$ circuit.  Note that SWAP gates are required at the end
of the circuit to place the states back into their original order.  
The inverse $QFT^\dagger$ is given by reading the circuit 
backwards from right to left, starting with the SWAP gates,  
and inverting all phase angles (replacing $R_n$ by $R_n^\dagger$).  
The circuit uses $1 + 2 + \cdots + m = \frac{1}{2} m (m+1) 
= {\cal  O}(m^2)$ distinct gates,  plus  ${\cal O}(m/2)$ 
SWAP gates.

What would have happened if we had not used the SWAP 
gates to reverse the qubit order in state (\ref{eq_convention1b_back}),
but instead appealed directly to (\ref{eq_QFTell_Omega})?
We would have found terms like $\vert \ell_m \rangle_1$,  
and the index of $\ell$ would not have paired properly with 
the associated qubit.  By performing the SWAP operations,  
we only find states of the form $\vert \ell_k \rangle_k$.  It
should therefore cause no  confusion if we henceforth drop 
the subscript on the basis states  and simply write~$\vert 
\ell_k \rangle$.

%\vfill
%\clearpage
\subsubsection{Convention 1: OpenQASM/Qiskit}

We now look at the convention used by OpenQASM/Qiskit. 
Consider an $m$ qubit system in which  the circuit for
the computational basis states start with qubit $0$ in 
the upper position and qubit $m-1$ in the lower position.
There are $M = 2^m$ quantum states in the system,  and  
the index integer \hbox{$k \in \{0, 1,  \cdots,  M-1\}$} can 
be expressed in a binary form where $k_0$ is the least 
significant bit,  
\begin{eqnarray}
  k &=& 
  k_{m-1} k_{m-2} \cdots k_{1} k_{0}
\nonumber\\
  &=&
  2^{m-1} k_{m-1} +  2^{m-2} k_{m-2} +
  \cdots  + 2^1 \, k_{1} +   2^0 \, k_{0}
  \ .
\end{eqnarray}
The integer $k$ can then be used to label the 
computational basis states, 
\begin{eqnarray}
   \vert k \rangle 
  &=&
  \vert k_{m-1}  k_{m-2} \cdots k_1 k_0  \rangle
\nonumber\\
  &=&
  \vert k_0 \rangle  \otimes \vert k_1 \rangle  \otimes \cdots \otimes
 \vert k_{m-2} \rangle  \otimes \vert k_{m-1} \rangle 
  \ . 
\end{eqnarray}
In this convention, the state labeled by $k=1$ is given by 
\begin{eqnarray}
  \vert 1 \rangle
  = 
  \vert 0 \cdots 0 1 \rangle
  =
  \vert 1 \rangle \otimes \vert 0 \rangle \otimes
  \cdots \otimes\vert 0 \rangle
  \ ,
\label{eq_one_conv1}
\end{eqnarray}
with $k_0 = 1$ and all other bits $k_r = 0$. As previously 
mentioned, this state will play a critical role in Shor's
algorithm. We also record here the convenient relation
\begin{eqnarray}
  \frac{k}{M} 
  =
  \frac{k_{m-1}}{2^1} +  \frac{k_{m-2}}{2^2} +
  \cdots + \frac{k_{1}}{2^{m-1}} +   \frac{k_{0}}{2^m}
  \ .
\end{eqnarray}

As before, 
we can replace a sum over the index integer $k$ by 
$m$ sums over the binary components $k_r \in \{0, 1\}$ 
for  $r \in \{0,  1,  \cdots,  m-1\}$,  
\begin{eqnarray}
 \sum_{k=0}^{M-1}
  ~=
  \sum_{k_{m-1}\in\{0,1\}} \cdots \sum_{k_1\in\{0,1\}} 
  \sum_{k_0 \in\{0,1\}}
  \ ,
\end{eqnarray}
so that  expression (\ref{eq_QFT}) for the quantum 
Fourier transform becomes
\begin{eqnarray}
  QFT\, \vert \ell \rangle
  &\equiv&
  \frac{1}{\sqrt{M}}\sum_{k=0}^{M-1} e^{2 \pi i \,  \ell k  /M}\,
  \vert k \rangle
\\[5pt]
 &=&  % \hskip-2.0cm = 
  \frac{1}{2^{m/2}}  \sum_{k_{m-1}} \cdots 
  \sum_{k_1} \sum_{k_0} e^{2 \pi i \,  \ell 
  \big( 
  2^0 \, k_{0} + 2^1 \, k_{1} +  \cdots +
  2^{m-2} k_{m-2} + 2^{m-1} k_{m-1} 
  \big) \big/ 2^m} \,
  \vert k_{m-1}  k_{m-2} \cdots k_1 k_0   \rangle
\nonumber
\\[5pt]
%\end{eqnarray}
%\begin{eqnarray}
  &=&  
  \frac{1}{2^{m/2}}   
  \sum_{k_0=0, 1} e^{2\pi i \, \ell k_0/2^m} \vert k_0 \rangle
  \otimes
  \sum_{k_1=0, 1}  e^{2\pi i \, \ell k_1 /2^{m-1}} \vert k_1 \rangle
  \otimes \cdots \otimes \!\!
  \sum_{k_{m-1}=0, 1} e^{2\pi i \, \ell k_{m-1}/2^1} \vert k_{m-1} \rangle
\nonumber\\[5pt]
  &=&  
  \frac{1}{2^{m/2}} 
  \Big(
  \vert 0 \rangle + e^{2\pi i \, \ell /2^m} \, \vert 1 \rangle 
  \Big)_{0}
  \otimes
  \Big(
  \vert 0 \rangle + e^{2\pi i \, \ell /2^{m-1}} \, \vert 1 \rangle 
  \Big)_{1}
  \otimes \cdots \otimes
\nonumber \\[5pt] && \hskip2.0cm
  \Big(
  \vert 0 \rangle + e^{2\pi i \, \ell /2^{2}} \, \vert 1 \rangle 
  \Big)_{m-2}
  \otimes 
  \Big(
  \vert 0 \rangle + e^{2\pi i \, \ell /2^1} \, \vert 1 \rangle 
  \Big)_{m-1}
  \ . ~~~
\label{QFT_ell_conv1}
\end{eqnarray}
Next we use the form of the $\ell$-state given by our 
OpenQASM convention, 
\begin{eqnarray}
  \vert \ell \rangle 
  &=&
  \vert \ell_{m-1}  \ell_{m-2} \cdots \ell_1 \ell_0\, \rangle
  ~~~\text{with}~~ 
  \ell_r \in \{0, 1\} 
\nonumber\\[5pt]
   &=&
   \vert \ell_0 \rangle \otimes  \vert \ell_1 \rangle \otimes
   \cdots \otimes  \vert \ell_{m-2}   \rangle \otimes
   \vert \ell_{m-1}   \rangle 
  \ ,
\end{eqnarray}
where the $m$-bit integer $\ell$ takes the binary form
\begin{eqnarray}
  \ell &=& 
  \ell_{m-1} \ell_{m-2} \cdots \ell_{1} \ell_{0}
\nonumber\\
  &=&
  2^{m-1} \ell_{m-1} +  2^{m-2} \ell_{m-2} +
  \cdots  + 2^1 \, \ell_{1} +   2^0 \, \ell_{0}
  \ ,
\end{eqnarray}
while $m$-bit binary fractions can be expressed by
\begin{eqnarray}
  \Omega
  =
  0.\ell_{m-1} \ell_{m-2} \cdots \ell_{1} \ell_{0}
  \equiv
   \frac{\ell_{m-1}}{2^{1}} +  \frac{\ell_{m-2}}{2^{2} } +
  \cdots  + 
  \frac{\ell_{1} }{ 2^{m-1}}+    \frac{\ell_{0}}{2^{m} }
  \ .
\end{eqnarray}
Similarly to the previous case, the exponential terms 
in (\ref{QFT_ell_conv1}) can be written
\begin{eqnarray}
    2\pi i \, \frac{\ell}{2^{m}}
  &=&
   \frac{2\pi i }{2^m} \Big[ 
  2^{m-1} \ell_{m-1} +  2^{m-2} \ell_{m-2} +
  \cdots  + 2^1 \, \ell_{1} +   2^0 \, \ell_{0}
  \Big]
\nonumber
\\[5pt]
  &=& 
   2\pi i \Big[ 
  \Omega_{0}
  \Big]
\\[10pt]
    2\pi i \, \frac{\ell}{2^{m-1}}
  &=& 
   \frac{2\pi i }{2^{m-1}}\Big[ 
  \big( 2^{m-1} \ell_{m-1} \big) 
  +  2^{m-2} \ell_{m-2} + \cdots  + 2^1 \, \ell_{1} +   2^0 \, \ell_{0}
  \Big]
\nonumber
\\[5pt]
  &=& 
  2\pi i \Big[ 
  \big( \ell_{m-1} \big) + \Omega_{1}
  \Big]
\\[5pt]\nonumber
  &\cdots&
\\[5pt] 
  2\pi i \, \frac{\ell}{2^{m-r}}
  &=& 
   \frac{2\pi i }{2^{m-r}} \Big[
  \Big(
  2^{m-1} \ell_{m-1} +  2^{m-2} \ell_{m-2} +
  \cdots  + 2^{m-r}\,\ell_{m-r} \Big) + 
  2^{m-r-1}\,\ell_{m-r-1} 
\nonumber \\ && \hskip7.3cm 
  +
  \cdots +
  2^1 \, \ell_{1} +   2^0 \, \ell_{0}
  \Big] 
\nonumber
\\[5pt]
  &=& 
   2\pi i  \Big[
  \Big(
  2^{r-1} \ell_{m-1} +  2^{r-2} \ell_{m-2} +
  \cdots  + 2^{0}\,\ell_{m-r} \Big) +\Omega_{r}
  \Big] 
\\[5pt]\nonumber
  & \cdots& 
\end{eqnarray}
\begin{eqnarray}
%\\[5pt]
  2\pi i \, \frac{\ell}{2^2}
  &=& 
   \frac{2\pi i }{2^2} \Big[ 
  \Big(
  2^{m-1} \ell_{m-1} +  2^{m-2} \ell_{m-2} +
  \cdots  2^2 \ell_{2} \Big) 
  + 2^1 \, \ell_{1}   +   2^0 \, \ell_{0}
  \Big]
  ~~~~~~~~~~
\nonumber\\[5pt]
  &=& 
   2\pi i  \Big[ 
  \Big(
  2^{m-3} \ell_{m-1} +  2^{m-4} \ell_{m-2} +
  \cdots  2^0 \ell_{2} \Big) 
  + \Omega_{m-2}
  \Big]
  ~~~~~~~~~~
%\end{eqnarray}
%\begin{eqnarray}
\\[10pt]
  2\pi i \, \frac{\ell}{2^1}
  &=& 
  \frac{2\pi i }{2^1} \Big[ 
  \Big(
  2^{m-1} \ell_{m-1} +  2^{m-2} \ell_{m-2} +
  \cdots  + 2^1 \, \ell_{1} \Big) 
  +   2^0 \, \ell_{0}
  \Big]
\nonumber\\[5pt]
  &=& 
  2\pi i  \Big[ 
  \Big(
  2^{m-2} \ell_{m-1} +  2^{m-3} \ell_{m-2} +
  \cdots  + 2^0 \, \ell_{1} \Big) 
  +   \Omega_{m-1}
  \Big]
  \ ,
\end{eqnarray}
where the partial phases are now defined by

\begin{eqnarray}
  \Omega_{0}
  &\equiv& 
    \frac{\ell_{m-1} }{2^1} +  \frac{\ell_{m-2} }{2^2}
  + \cdots  + \frac{\ell_{1}}{2^{m - 1}} +  \frac{\ell_{0}}{2^m}
  =
  0.\ell_{m-1} \cdots \ell_1 \ell_0
%\hskip1.0cm : ~ \vert \ell_{0} \rangle
\\[5pt]
  \Omega_{1} 
  &\equiv& 
    \frac{\ell_{m-2}}{2^1} +     \frac{\ell_{m-3}}{2^2} +
  \cdots  + \frac{\ell_{1}}{2^{m-2}} +  \frac{\ell_{0}}{2^{m-1}}
  =
  0.\ell_{m-2} \cdots \ell_1 \ell_0
%\hskip0.65cm : ~ \vert \ell_{1} \rangle
\\[5pt]
\nonumber
  & \cdots& 
%\end{eqnarray}
%\begin{eqnarray}
\\[5pt]
  \Omega_{r}
  &\equiv& 
  \frac{\ell_{m-r-1} }{2^1} + \frac{\ell_{m-r-2} }{2^2}
  +     \cdots + \frac{\ell_{1}}{2^{m-r-1}} +  \frac{\ell_{0}}{2^{m-r}}
  =
  0.\ell_{m-r-1} \cdots \ell_1 \ell_0
%\hskip0.4cm : ~ \vert \ell_{r} \rangle~~~~
\\[5pt]\nonumber
  & \cdots& 
\\[5pt]
  \Omega_{m-2} 
  &\equiv& 
\frac{\ell_{1}}{2^1} +  \frac{ \ell_{0}}{2^2}
  =
  0.\ell_1 \ell_0
%\hskip6.9cm : ~ \vert \ell_{m-2} \rangle
\\[5pt]
  \Omega_{m-1} &\equiv& 
  \frac{\ell_{0}}{2^1}
  =
  0.\ell_0   \ .
%\hskip8.0cm : ~ \vert \ell_{m-1} \rangle
\end{eqnarray}
The general phase takes the form 
\begin{eqnarray}
  \Omega_{r}
  &=& 
  \sum_{k=1}^{m-r} \, \frac{\ell_{m-r-k}}{2^k}
  ~~~\text{for}~~ r \in \{0,  1,  \cdots,  m-1 \} 
  \  ,
\end{eqnarray}
and the quantum Fourier transform (\ref{QFT_ell_conv1})
can now be expressed in one of three equivalent ways:
\begin{eqnarray}
  QFT\, \vert \ell \rangle
  &=& 
  \frac{1}{2^{m/2}} 
  \Big(
  \vert 0 \rangle + e^{2\pi i \, \ell /2^m} \, \vert 1 \rangle 
  \Big)_{0}
  \otimes
  \Big(
  \vert 0 \rangle + e^{2\pi i \, \ell /2^{m-1}} \, \vert 1 \rangle 
  \Big)_{1}
  \otimes \cdots \otimes
\\ && \hskip2.0cm
  \Big(
  \vert 0 \rangle + e^{2\pi i \, \ell /2^{2}} \, \vert 1 \rangle 
  \Big)_{m-2}
  \otimes 
  \Big(
  \vert 0 \rangle + e^{2\pi i \, \ell /2^1} \, \vert 1 \rangle 
  \Big)_{m-1}
  \nonumber
\\[10pt]
  &=&
  \frac{1}{2^{m/2}} 
  \Big(
  \vert 0 \rangle + e^{2\pi i \, 0.\ell_0} \, 
  \vert 1 \rangle 
  \Big)_{0}
  \otimes
  \Big(
  \vert 0 \rangle +    e^{2\pi i \, 0.\ell_1 \ell_0} \,  
  \vert 1 \rangle 
  \Big)_{1}
  \otimes \cdots \otimes
 \\[5pt] && \hskip1.8cm
  \Big(
  \vert 0 \rangle +    e^{2\pi i \, 0.\ell_{m-2} \cdots \ell_{1} \ell_0 }  \, 
  \vert 1 \rangle 
  \Big)_{m-2}
  \otimes
  \Big(
  \vert 0 \rangle + e^{2\pi i \, 0.\ell_{m-1} \cdots \ell_{1} \ell_0 } \, 
  \vert 1 \rangle 
  \Big)_{m-1}
\nonumber
%\end{eqnarray}
%\begin{eqnarray}
\\[10pt]
  &=&
  \frac{1}{2^{m/2}} 
  \Big(
  \vert 0 \rangle + e^{2\pi i \, \Omega_{0}} \, \vert 1 \rangle 
  \Big)_{0}
  \otimes
  \Big(
  \vert 0 \rangle + e^{2\pi i \, \Omega_{1}} \, \vert 1 \rangle 
  \Big)_{1}
  \otimes \cdots \otimes
 \nonumber\\[5pt] && \hskip2.5cm
  \Big(
  \vert 0 \rangle + e^{2\pi i \, \Omega_{m-2}} \, \vert 1 \rangle 
  \Big)_{m-2}
  \otimes
  \Big(
  \vert 0 \rangle + e^{2\pi i \, \Omega_{m-1}} \, \vert 1 \rangle 
  \Big)_{m-1}
  \ .
\end{eqnarray}
As before, we  invert the qubits with SWAP gates to 
form  the state
\begin{eqnarray}
 \vert \psi_{\rm rev} \rangle
  &=&
  \frac{1}{2^{m/2}} 
  \Big(
  \vert 0 \rangle + e^{2\pi i \, \Omega_{m-1}} \, \vert 1 \rangle 
  \Big)_{0}
  \otimes
  \Big(
  \vert 0 \rangle + e^{2\pi i \, \Omega_{m-2}} \, \vert 1 \rangle 
  \Big)_{1}
  \otimes \cdots \otimes
\label{eq_convention2_back}
 \\[5pt] && \hskip5.0cm
  \Big(
  \vert 0 \rangle + e^{2\pi i \, \Omega_{1}} \, \vert 1 \rangle 
  \Big)_{m-2}
  \otimes
  \Big(
  \vert 0 \rangle + e^{2\pi i \, \Omega_{0}} \, \vert 1 \rangle 
  \Big)_{m-1}
  \ ,
\nonumber
\end{eqnarray}
which can be expressed in terms of basic gates to 
give the following circuit.   Starting with qubit-$0$ 
of (\ref{eq_convention2_back}),  and using $e^{2\pi i\,
\Omega_{m-1}} = e^{2\pi i \, \ell_{0}/2} = (-1)^{\ell_{0}}$,  
we find a state similar to the previous case, 
\begin{eqnarray}
  \frac{1}{\sqrt{2}}
  \Big(
  \vert 0 \rangle + e^{2\pi i \, \Omega_{m-1}} \, \vert 1 \rangle 
  \Big)_{0}
  &=&
  H^0 \vert \ell_{0} \rangle
  ~~~\text{where}~~ \ell_0 \in \{0,1\}
  \ .
\end{eqnarray}
%As usual, we have place a superscript over the $H$ 
%gate to explicitly indicate the target qubit.  
%Moving on to qubit-$1$,  we can write  
The other qubits give corresponding results, so we 
move on to the bottom of the circuit, expressing 
qubit $m-1$ by
%%%
%\begin{eqnarray}
%  \frac{1}{\sqrt{2}}
%  \Big(
%  \vert 0 \rangle + e^{2\pi i \, \Omega_{m-2}} \, \vert 1 \rangle 
%  \Big)_{1}
%  &=&
%  \frac{1}{\sqrt{2}}
%  \Big(
%  \vert 0 \rangle + e^{2\pi i \, \ell_{1}/2^1} \cdot
%  e^{2 \pi i \, \ell_{0}/2^2}\, \vert 1 \rangle 
%  \Big)_{1}
%\label{eq_mminusone}
%\\[5pt] 
%  &=&
%  H^1 \cdot C^{0} R_2^{1} \, \vert \ell_{1} \rangle 
%  ~~~\text{where}~~ \ell_1 \in \{0,1\}
%  \ .
%\end{eqnarray}
%%%
%%
%The first exponential $e^{2\pi i \, \ell_{1}/2}$ of 
%(\ref{eq_mminusone}) gives a Hadamard gate acting 
%on $\vert \ell_{1} \rangle$,  and the second exponential 
%$e^{2 \pi i \, \ell_{0}/2^2}$ gives a controlled phase 
%with angle $\theta = 2\pi/2^2$
%(again call it $CR_2$) on the target qubit $\vert \ell_{1} 
%\rangle $ with control qubit $\vert \ell_{0} \rangle$
%(hence the subscript notation $C^0 R^1_2$).
%As always,  we are using superscripts on the 
%controlled-$R$ gate $CR_2$ to explicitly indicate 
%the control and target qubits.  Finally,  the $m-1$ 
%qubit gives
%%
\begin{eqnarray}
  \frac{1}{\sqrt{2}}
  \Big(
  \vert 0 \rangle + e^{2\pi i \, \Omega_{0}} \, \vert 1 \rangle 
  \Big)_{m-1}
  &=&
  \frac{1}{\sqrt{2}}
  \Big(
  \vert 0 \rangle + e^{2\pi i \, \ell_{m-1}/2} \cdot
  e^{2\pi i \, \ell_{m-2}/2^2} \cdots e^{2 \pi i \, \ell_{1}/2^{m-1}}
  \cdot e^{2 \pi i \, \ell_{0}/2^{m}}
  \, \vert 1 \rangle 
  \Big)_{m-1} 
\nonumber
\\[5pt]
  &=&
  H^{m-1} \cdot C^{m-2}R_2^{m-1} \cdots 
  C^{1}R_{m-1}^{m-1} \cdot 
  C^{0} R_m^{m-1} \, \vert \ell_{m-1} \rangle
\end{eqnarray}
with $\ell_{m-1} \in \{0,1\}$,   where the phase operators 
are defined by 
\begin{eqnarray}
  R_n &=&
  \left[
  \begin{matrix}
  1 & 0\\[-5pt]
  0 & e^{2\pi i/ 2^n}
  \end{matrix}
  \right] 
  =
  P\big(\pi/2^{n-1}\big)
  \ .
\end{eqnarray}
%%

\begin{figure}[h!]
\begin{centering}
\includegraphics[width=\textwidth]{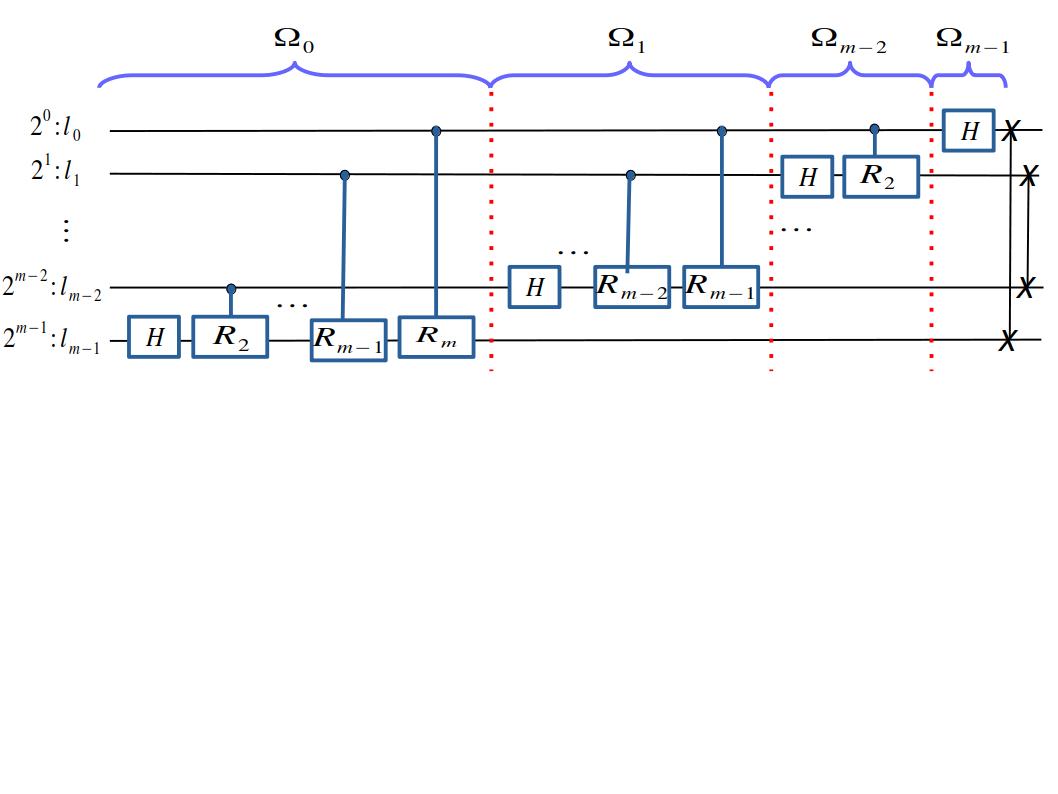} 
\par\end{centering}
\vskip-6.5cm 
\caption{\footnoteskip  
  Convention 1 for the $QFT$: OpenQASM/Qiskit.
}
\label{fig_conv1}
\end{figure}

%\vfill
%
\noindent
Figure~\ref{fig_conv1} reproduces the commensurate 
QFT circuit in the OpenQASM/Qiskit convention. The 
inverse $QFT^\dagger$ is given by reading the circuit 
in reverse order from right to left,  starting with the 
SWAP gates,  and inverting the sign of the phase gates.  
Note that the QFT circuits are reversed between the 
two conventions. 

\vfill
\pagebreak
\clearpage
\section{Quantum Phase Estimation}
\label{sec_QPE}

We now turn to  quantum phase estimation (QPE),  which is  
the workhorse of Shor's algorithm.   Our primary references 
for this section are Refs. \cite{qcqi}  and \cite{des}.  We shall  
work through the calculations in both Conventions~1 and 2 
as outlined in Section~\ref{sec_conventions}, corresponding
to the  OpenQASM and the standard physics conventions,  
respectively.

\subsection{Convention 2: Standard Physics/Mathematics}

\begin{figure}[t!]
\begin{centering}
\includegraphics[width=\textwidth]{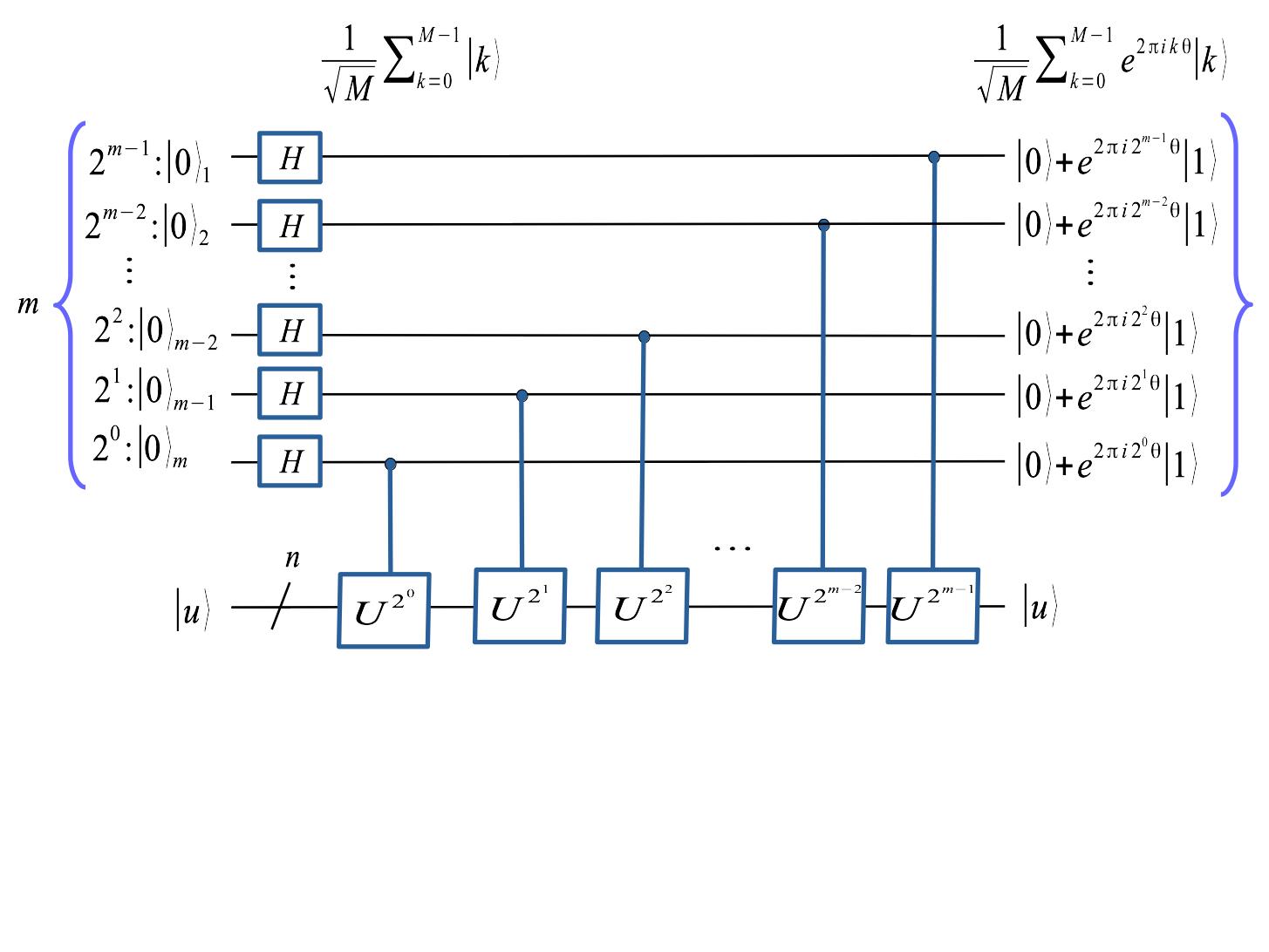} 
\par\end{centering}
\vskip-3.5cm 
\caption{\footnoteskip  
  QPE front-end: Convention 2 (physics and mathematics).
}
\label{fig_feQPE_conv2b}
\end{figure}

We first examine Convention 2,  the traditional physics and 
mathematics convention.  Consider a linear unitary operator $U$ 
with Eigenvalue $e^{2\pi i \theta}$ and Eigenstate $\vert u 
\rangle$,  where $\theta$ is a real phase such that $0 \le 
\theta < 1$:
\begin{eqnarray}
   U \vert u \rangle 
  =
  e^{2\pi i \theta} \, \vert u \rangle 
  \ .
\label{eq_U_eigen}
\end{eqnarray}
Since $U$ is unitary,  its Eigenvalues have a norm of unity.   
We wish to build a QPE circuit that will output an approximate 
($m$-bit binary) value for the phase angle $\theta$.  The circuit 
will consist of a {\em front-end} and a {\em back-end}.  There 
are two registers in the QPE front-end circuit: (i) a control 
register consisting of $m$ qubits and (ii) a work register 
containing $n$ qubits.
%(not to be confused with the index $\ell$ of the previous section).  
We store the Eigenstate 
$\vert u \rangle$ in the work register. To construct the 
front-end circuit, we first apply Hadamard gates to every 
control qubit, forming the state
\begin{eqnarray}
  \vert \psi_1 \rangle
  = H \vert 0 \rangle \otimes \cdots \otimes H \vert 0 \rangle
  \otimes \vert u \rangle
  =
  \frac{1}{\sqrt{M}} \sum_{k-0}^{M-1} \vert k \rangle
  \otimes \vert u \rangle
  \ ,
\end{eqnarray}
where $M = 2^m$ is the total number of computational
basis states. 
Given the unitary operator $U$, we assume that we are able 
to build a family of $m$ \hbox{controlled-$U^p$} gates for 
\hbox{$p \in \{ 2^0,  2^1,\cdots,  2^{m-1}\}$} that operate 
on the work register containing the state~$\vert u \rangle$.  
Note that the action of a $CU^p$ operator for a single 
control qubit takes the form 
%%
%\vskip-1.0cm
\begin{eqnarray}
  CU^p H \vert 0 \rangle \otimes \vert u \rangle
  &=&
  \frac{1}{\sqrt{2}} \, CU^p  \,\Big(\vert 0 \rangle 
  + \vert 1 \rangle\Big) \otimes \vert u \rangle
  =
  \frac{1}{\sqrt{2}}\, \vert 0 \rangle \otimes \vert u \rangle 
  +
   \frac{1}{\sqrt{2}}\,\vert 1 \rangle \otimes U^p \vert u \rangle 
\\[5pt]
  &=&
  \frac{1}{\sqrt{2}} \Big(\vert 0 \rangle \otimes \vert u \rangle 
  +
   \vert 1 \rangle \otimes e^{2\pi i p\, \theta} \vert u \rangle \Big) 
  =
  \frac{1}{\sqrt{2}} \Big(\vert 0 \rangle 
  +
    e^{2\pi i  p\, \theta} \,  \vert 1 \rangle \Big) \otimes \vert u \rangle 
  \ . ~~~~
\end{eqnarray}
This is an example of {\em phase kickback},  in which the
phase operation in the target register makes its way back
into the control register.  We now string these gates together 
to form the front-end of the circuit composed of the gates 
$C^n U^{2^n}$ for $ n \in \{0, 1, \cdots, m-1\}$, as illustrated 
in Fig.~\ref{fig_feQPE_conv2b},  with the least significant 
power $p = 2^0$ attached to the least significant $m$-th target qubit (as 
Convention~2 dictates). We see that the output state of
the front-end becomes
\begin{eqnarray}
  \vert \psi_{2} \rangle 
  &=&
  \frac{1}{2^{m/2}} \Big(\vert 0 \rangle 
  +
  e^{2\pi i\, 2^{m-1} \theta} \,  \vert 1 \rangle \Big)_1
  \otimes
  \Big(\vert 0 \rangle 
  +
  e^{2\pi i\, 2^{m-2} \theta} \,  \vert 1 \rangle \Big)_2
  \otimes \cdots \otimes  
\label{eq_theaout_a}
%\\ && \hskip7.5cm 
\\[3pt] && \hskip5.0cm 
  \Big(\vert 0 \rangle 
  +
  e^{2\pi i\, 2^{1} \theta} \,  \vert 1 \rangle \Big)_{m-1}
  \otimes
  \Big(\vert 0 \rangle 
  +
  e^{2\pi i\, 2^{0} \theta} \,  \vert 1 \rangle \Big)_m
  \otimes \vert u \rangle
\nonumber
\\[5pt]
  &=& 
  \frac{1}{2^{m/2}}   
  \sum_{k_1=0, 1} e^{2\pi i \,  \theta\,2^{m-1} k_1} \vert k_1 \rangle
  \otimes
  \sum_{k_2=0, 1} e^{2\pi i \,  \theta\, 2^{m-2} k_2}  \vert k_2 \rangle
  \otimes \cdots \otimes 
  \sum_{k_m=0, 1} e^{2\pi i \,  \theta\, 2^0 k_m}  \vert k_m \rangle
  \otimes \vert u \rangle
  ~~~
\nonumber\\[5pt]
  &=& 
  \frac{1}{2^{m/2}}   
  \sum_{k_1} \sum_{k_2} \cdots \sum_{k_m}
  e^{2\pi i \, \theta \big(2^{m-1} k_1 +  2^{m-2} k_2 +
  \cdots + 2^0 k_m    \big)}
  \vert k_1 \rangle \otimes  \vert k_2 \rangle 
  \otimes \dots \otimes  \vert k_m \rangle 
 \otimes \vert u \rangle
\nonumber\\[5pt]
  &=&
  \frac{1}{\sqrt{M}} \sum_{k=0}^{M-1} e^{2\pi i k \, \theta} \,
 \vert k \rangle \otimes \vert u \rangle
  \ ,
\label{eq_theta_out_d}
\end{eqnarray}
where $M = 2^m$ and $k = 2^{m-1} k_1 + 2^{m-2} k_2 + 
\cdots 2^1 k_{m-1} + 2^0 k_m$,  with $k_r \in \{0, 1\}$.   
This result is valid for a general phase angle $\theta$.  

\begin{figure}[t!]
\begin{centering}
\includegraphics[width=\textwidth]{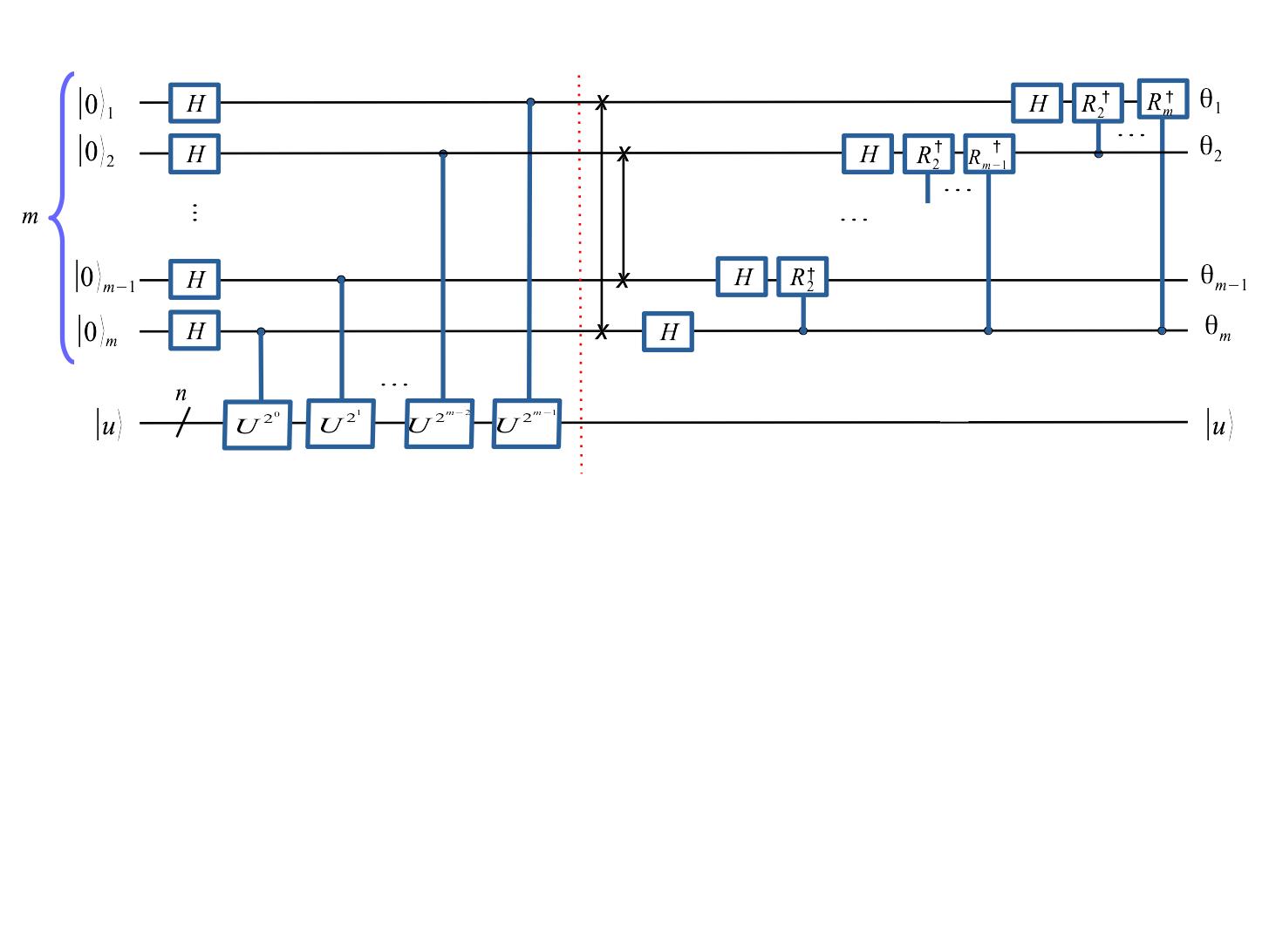} % n
\par\end{centering}
\vskip-6.5cm
\caption{\footnoteskip
  The $QFT^\dagger$ circuit has been expanded to
  illustrate the full complexity of the QPE circuit. 
}
\label{fig_QPE_conv2b_b}
\end{figure}

For simplicity, let us first suppose that the binary form of
the phase angle terminates after exactly $m$ bits,  so that 
\begin{eqnarray}
 \theta 
  &=& 
  0.\theta_1 \, \theta_2  \, \cdots \theta_{m-1} \theta_m
  ~~~\text{where}~~   \theta_r \in \{0, 1\}
\label{eq_theta_one}
\\[5pt]
  &=&
  \frac{\theta_1}{2} + \frac{\theta_2}{2^2} +  
  \cdots + \frac{\theta_{m-1}}{2^{m-1}} + \frac{\theta_m}{2^m} 
  \ .
\label{eq_theta_m}
\end{eqnarray}
We will shortly extend the argument to general phase angles 
that do not not terminate. We see that (\ref{eq_theta_m})
implies that $M \theta$ can be written as a binary integer, 
\begin{eqnarray}
  \ell_\theta
  \equiv
  M  \theta
   =
 2^m \theta 
  &=&
  2^{m-1}\,\theta_1 +   2^{m-2}\,\theta_2 + \cdots + 2^0 \, \theta_m
\\
  &=&
  \theta_1 \, \theta_2 \, \cdots \, \theta_m 
  \in \{0, 1, \cdots, M-1 \}
  \ .
\label{eq_2mtheta_m}
\end{eqnarray}
\begin{figure}[b] 
\vskip-0.5cm
\begin{centering}
\includegraphics[width=\textwidth]{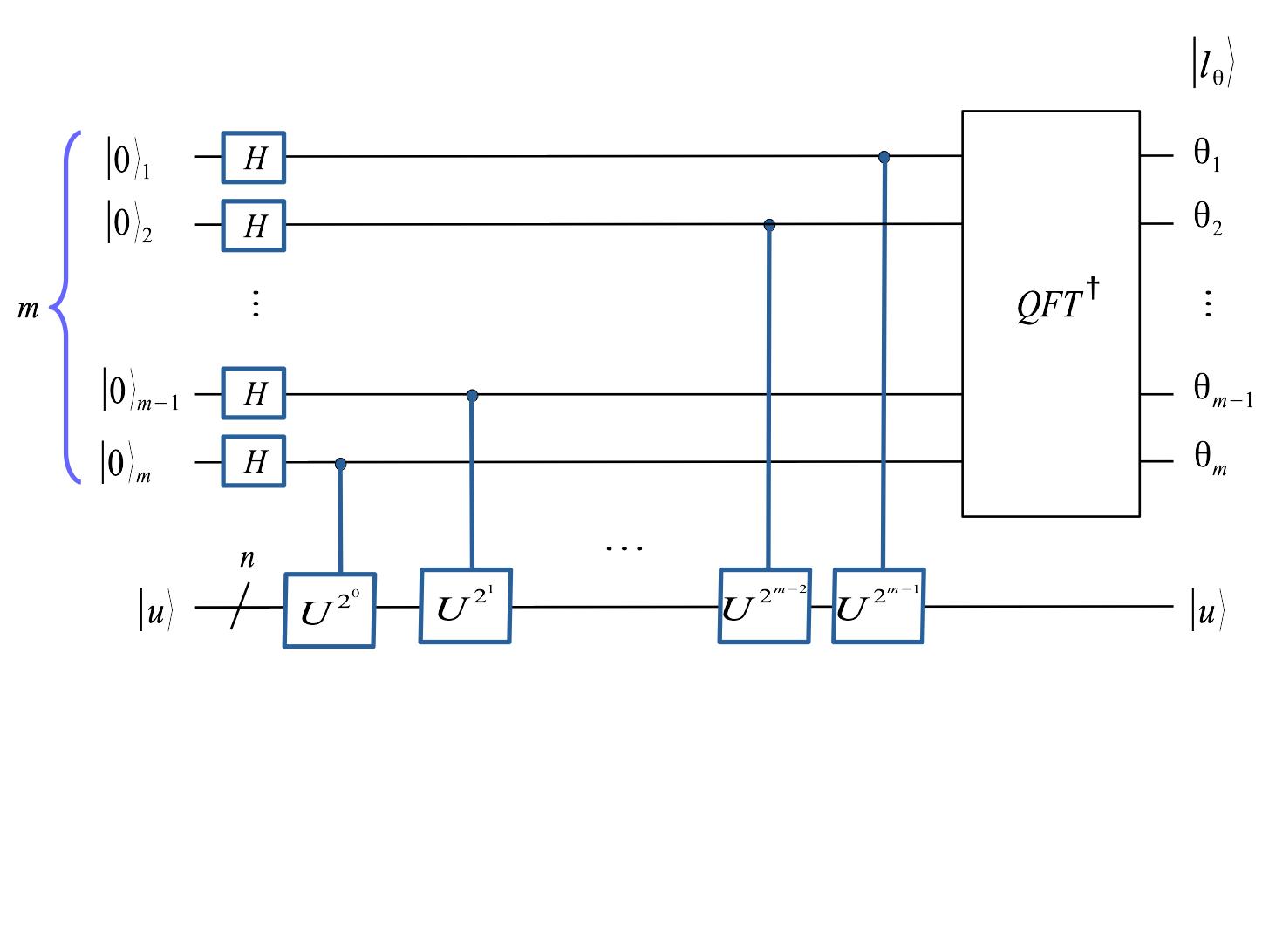} % n
\par\end{centering}
\vskip-3.5cm
\caption{\footnoteskip
  QPE: Convention 2.  The QPE  circuit with the inverse 
  quantum Fourier transform $QFT^\dagger$ represented 
  abstractly by a box.  
}
\label{fig_QPE_conv2b_a}
\end{figure}
\noindent
Upon using the relation $\theta = \ell_\theta/M$, we can now
express the output state of the front-end as 
\begin{eqnarray}
  \vert \psi_{2} \rangle 
  &=&
    \frac{1}{\sqrt{M}} \sum_{k=0}^{M-1} 
  e^{2\pi i \, k  \ell_\theta/M}
  \vert k \rangle \otimes \vert u \rangle
\\[5pt]
  &=&
  QFT \, \vert \ell_\theta \rangle \otimes \vert u \rangle
  \ ,
\end{eqnarray}
where we have used the definition of the $QFT$ operator
(\ref{eq_QFT}).   Therefore,  the back-end of the QPE circuit 
will consist of an inverse $QFT$ acting on the control register,  
as illustrated in Fig.~\ref{fig_QPE_conv2b_a}. In 
Fig.~\ref{fig_QPE_conv2b_b} we expand the $QFT^\dagger$ 
circuit explicitly.  In either case, the final output state is given 
by 
\begin{eqnarray}
  \vert \psi_{3} \rangle
  =
  QFT^\dagger \, \vert \psi_2 \rangle
  =
 \vert \ell_\theta \rangle
  \otimes \vert u \rangle
% \equiv
%  \vert 2^m \theta \rangle 
%  \otimes \vert u \rangle
  \ ,
\end{eqnarray}
where $\ell_\theta \equiv 2^m \theta \in \{0, 1, \cdots, M-1\}$. 
Some authors denote the state $\vert\ell_\theta \rangle$ 
by $\vert 2^m\theta \rangle$.  Upon measuring the control 
register, we will find $\ell_\theta = \theta_1 \cdots \theta_m$ 
for $\theta_r \in \{0, 1\}$, and the corresponding phase is 
then given {\em exactly} by $\theta = \ell_\theta/2^m = 
0.\theta_1 \cdots \theta_m$,  in agreement with (\ref{eq_theta_one}).
We see that the QPE circuit in Fig.~\ref{fig_QPE_conv2b_a}
does indeed extract the correct phase.

\subsection{Convention 1: OpenQASM/Qiskit}

\begin{figure}[h!]
\begin{centering}
\includegraphics[width=\textwidth]{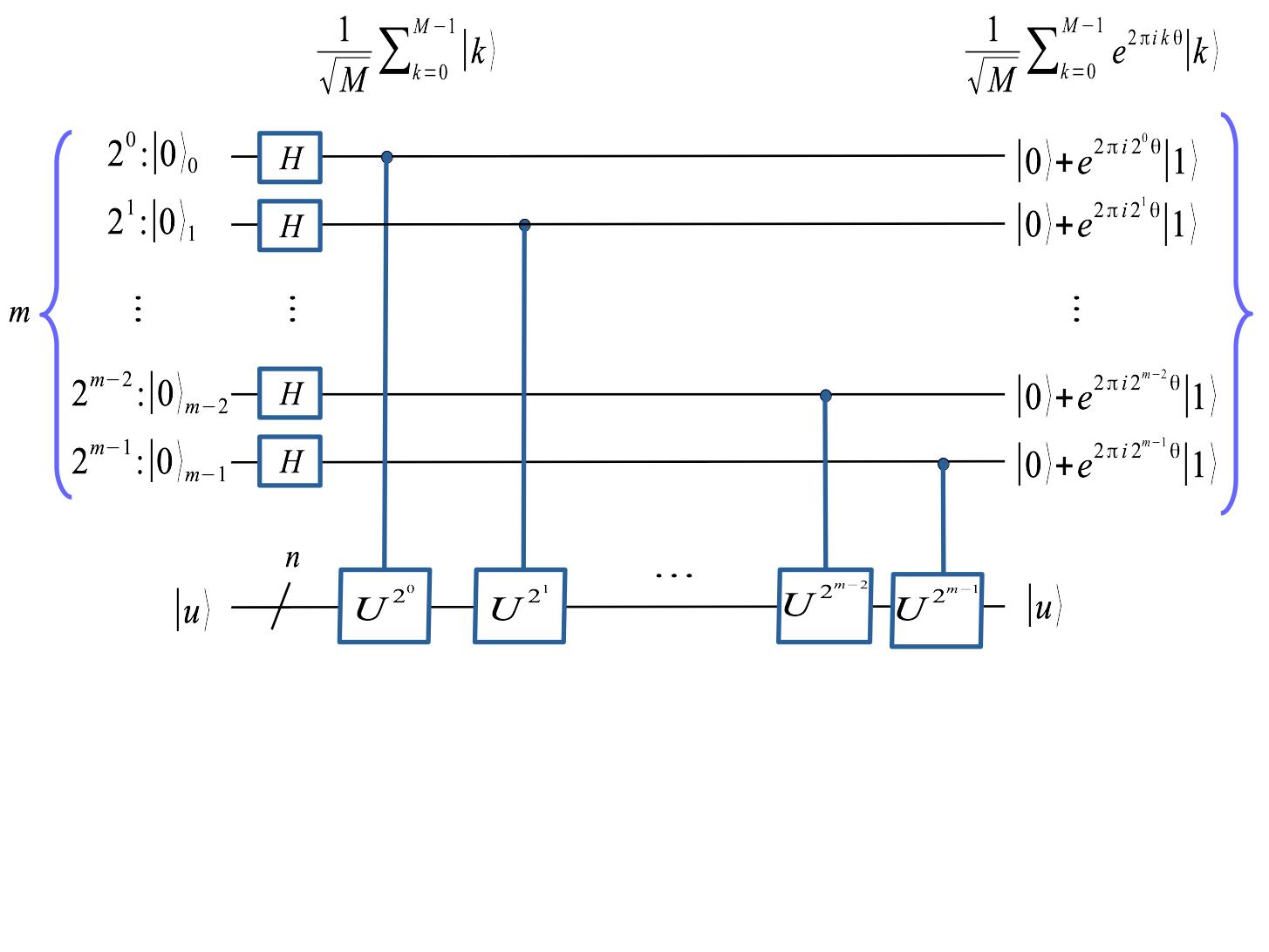} 
\par\end{centering}
\vskip-4.0cm 
\caption{\footnoteskip  
  QPE front-end: Convention 1 (OpenQASM/Qiskit).
}
\label{fig_qpe_1_conv1}
\end{figure}

\begin{figure}[h!]
\vskip-0.4cm
\begin{centering}
\includegraphics[width=\textwidth]{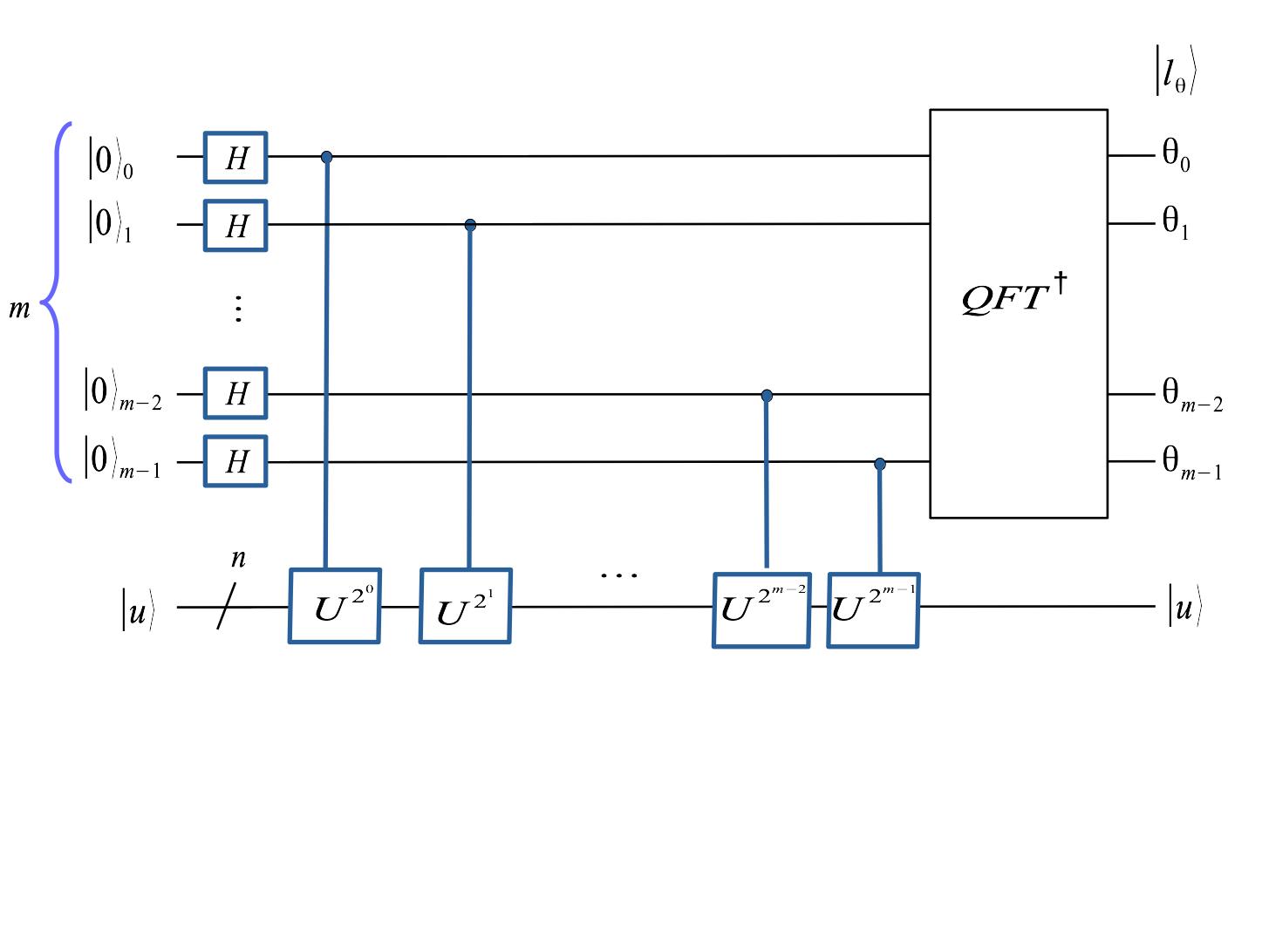}
\par\end{centering}
%\includegraphics[scale=0.38]{04_qpe_2_conv1_b.jpg} 
%\caption{Image B}
\vskip-4.0cm
\caption{\footnoteskip
  QPE front- and back-end: Convention 1 (OpenQASM/Qiskit)
\label{fig_QPE_conv1}
}
\end{figure}

\begin{figure}[h!]
%\vskip-1.0cm
\begin{centering}
\includegraphics[width=\textwidth]{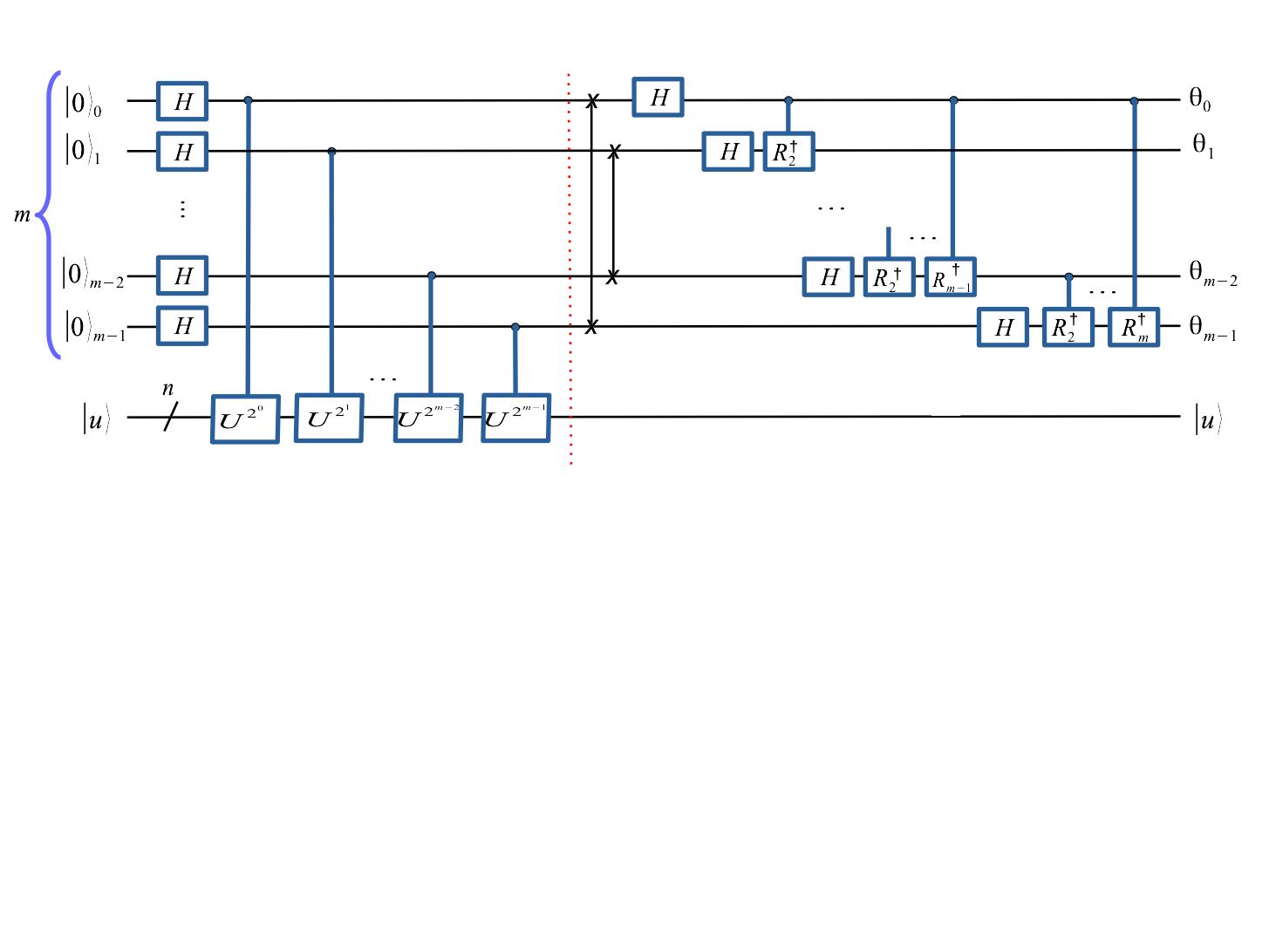}  % n
\par\end{centering}
\vskip-6.5cm
\caption{\footnoteskip
  QPE: Convention 1.  Expanded $QFT^\dagger$. 
}
\label{fig_QPE_3_conv1}
\end{figure}

We now examine the OpenQASM/Qiskit convention.
As before,  we string the $CU^p$ operators together 
to form the front-end of the circuit, with the smallest 
power $p$ being attached to the $0$-th qubit (the 
upper and least significant qubit of the circuit). The 
corresponding front-end is illustrated in 
Fig.~\ref{fig_qpe_1_conv1}. 
\vfill
\pagebreak
\noindent
Similarly to (\ref{eq_theta_out_d}),  the output of the
front-end can be expressed by
\begin{eqnarray}
  \vert \psi_{2} \rangle 
  &=&
  \frac{1}{2^{m/2}} \Big(\vert 0 \rangle 
  +
  e^{2\pi i\, 2^{0} \theta} \,  \vert 1 \rangle \Big)_0
  \otimes
  \Big(\vert 0 \rangle 
  +
  e^{2\pi i\, 2^{1} \theta} \,  \vert 1 \rangle \Big)_1
  \otimes \cdots \otimes  
\\[3pt] && \hskip0.9cm 
\Big(\vert 0 \rangle 
  +
  e^{2\pi i\, 2^{m-2} \theta} \,  \vert 1 \rangle \Big)_{m-2}
  \otimes
  \Big(\vert 0 \rangle 
  +
  e^{2\pi i\, 2^{m-1} \theta} \,  \vert 1 \rangle \Big)_{m-1}
  \otimes \vert u \rangle
\nonumber
\\[5pt]
  &=&
    \frac{1}{\sqrt{M}} \sum_{k=0}^{M-1} e^{2\pi i  k\,  \theta} \,
  \vert k \rangle \otimes \vert u \rangle
  \ ,
\label{eq_fe_out}
\end{eqnarray}
where $M = 2^m$ and $k = 2^{m-1} k_{m-1} + 2^{m-2} k_{m-2} 
+ \cdots 2^1 k_{1} + 2^0 k_0$,  with $k_r \in \{0, 1\}$.  
We first consider the simple case of an $m$-bit phase angle,
\begin{eqnarray}
 \theta 
  &=& 
  0.\theta_{m-1} \, \theta_{m-2} \, \cdots \,
  \theta_1 \, \theta_0
  ~~~\text{where}~~   \theta_r \in \{0, 1\}
\nonumber\\[5pt]
  &=&
  \frac{\theta_{m-1}}{2^1} + \frac{\theta_{m-2}}{2^2} 
  +   \cdots + \frac{\theta_1}{2^{m-1}} + \frac{\theta_0}{2^m} 
  \ .
\label{eq_theta_two_m}
\end{eqnarray}
As before, this expression implies that $M \theta$ is a binary 
integer between $0$ and $M-1$:
\begin{eqnarray}
  \ell_\theta
  \equiv
  M\theta
  =
 2^m \theta 
  &=&
  2^{m-1}\,\theta_{m-1} +   2^{m-2}\,\theta_{m-2}  
  + \cdots + 
  2^1 \, \theta_{1} + 2^0 \, \theta_{0}
\label{eq_theta_def}
\\
  &=&
  \theta_{m-1} \, \theta_{m-2} \, \cdots \, \theta_1 \, \theta_0
  \in \{0, 1, \cdots, M-1\}
  \ .
\end{eqnarray}
Again, the relation $\theta = \ell_\theta/M$ and definition 
(\ref{eq_QFT}) of  the QFT imply that the output of the front-end
becomes
\begin{eqnarray}
 \vert \psi_{2} \rangle 
  &=&
  \frac{1}{\sqrt{M}} \sum_{k=0}^{M-1} e^{2\pi i  k\,  \theta}
  \vert k \rangle \otimes \vert u \rangle
  =
  \frac{1}{\sqrt{M}} \sum_{k=0}^{M-1} e^{2\pi i \,  k \ell_\theta/ M}
  \vert k \rangle \otimes \vert u \rangle
\nonumber\\[5pt]
  &=&
  QFT \, \vert \ell_\theta \rangle 
  \otimes \vert u \rangle
  \ .
\end{eqnarray}
Upon taking the inverse Fourier transform,  the final 
state of the QPE circuit is
\begin{eqnarray}
  \vert \psi_{3} \rangle
  =
  QFT^\dagger \, \vert \psi_2 \rangle
  =
 \vert \ell_\theta \rangle
  \otimes \vert u \rangle
  \ .
\label{eq_theta_out}
\end{eqnarray}
The back-end of the QPE circuit therefore consists of an 
inverse $QFT$ operator,  as illustrated in Fig.~\ref{fig_QPE_conv1}.  
We also give the full circuit for the inverse $QFT$ in 
Fig.~\ref{fig_QPE_3_conv1}.  We see that a measurement
of the control register gives the integer $\ell_\theta
=\theta_{m-1}\cdots \theta_0$, which in turn provides
the correct measured phase $\theta = \ell_\theta/2^m = 
0.\theta_{m-1}\cdots \theta_0$, in agreement with
(\ref{eq_theta_two_m}).


\vfill
\pagebreak
\subsection{General Phase Angles}
\label{eq_gen_phase}

\begin{figure}[h!]
\includegraphics[scale=0.60]{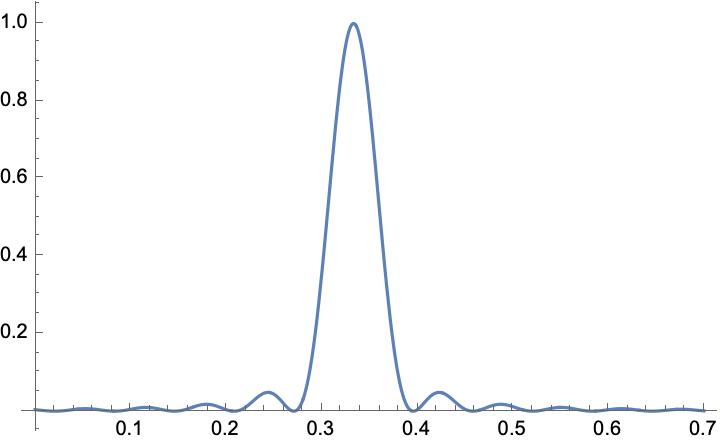} 
%\vskip-0.8cm 
\caption{\footnoteskip  
  The probability $P_\ell(\theta) = \vert A_\ell(\theta) \vert^2$ 
  as a function of  the measured phase angle $\theta_\ell \equiv 
  \ell/M$ for $M=2^4 = 16$,  where $\theta = 1/3$. Note that 
  $A_\ell(\theta) \approx 1$ for $\theta_\ell \approx \theta$.
}
\label{fig_pell}
\end{figure}

We now turn to the case in which the phase angle 
$\theta$ in (\ref{eq_U_eigen}) is a general real 
number between 0 and~1. The analysis here applies 
for both Conventions.  Recall that  the  output of 
the front-end in Fig.~\ref{fig_feQPE_conv2b} or 
Fig.~\ref{fig_qpe_1_conv1} is
\begin{eqnarray}
 \vert \psi_{2} \rangle 
  &=&
  \frac{1}{\sqrt{M}} \sum_{k=0}^{M-1}  
  e^{2\pi i k \, \theta } 
  \, \vert k \rangle  \otimes \vert u \rangle
  \ ,
\label{eq_psi2_gen}
\end{eqnarray}
where this result holds for general phase angles $\theta$. 
The output of the QPE circuit is then given by $\vert  \psi_3 \rangle 
= QFT^\dagger \,\vert \psi_{2} \rangle$, where the inverse 
of the $QFT$ operator takes the form 
\begin{eqnarray}
  QFT^\dagger
  = 
  \frac{1}{\sqrt{M}}
  \sum_{\ell = 0}^{M-1}\sum_{k=0}^{M-1} 
  e^{-2\pi i\, k  \ell / M}\, \vert \ell \rangle \langle k \vert 
  \ .
\label{eq_QFTdagger_a}
\end{eqnarray}
We therefore obtain the final state
\begin{eqnarray}
  \vert  \psi_{3} \rangle
    &=&
  QFT^\dagger \vert \psi_{2} \rangle
  =
  \frac{1}{M} 
  \sum_{\ell = 0}^{M-1}\sum_{k=0}^{M-1} 
  e^{-2\pi i\, k  \ell / M}\,   e^{2\pi i k \, \theta } \, 
  \vert \ell \rangle  \otimes \vert u \rangle
\\[5pt]
  &=& 
  \frac{1}{M} 
  \sum_{\ell = 0}^{M-1}\sum_{k=0}^{M-1}
  \big[e^{2\pi i\, ( \theta - \ell/M)}\big]^k \, \vert \ell \rangle 
 \otimes \vert u \rangle
 =
  \frac{1}{M} 
  \sum_{\ell = 0}^{M-1} \frac{1 - e^{2\pi i\, ( \theta - \ell/M)M}}
  {1-e^{2\pi i\,  ( \theta - \ell/M)}}
 \, \vert \ell \rangle  \otimes \vert u \rangle
  \ , ~~~~~~
\label{eq_geom_sum}
\end{eqnarray}
where we have performed an exact finite geometric sum.
Since this is such an important result, we summarize it below:
\begin{eqnarray}
  \vert  \psi_{3} \rangle
 =
  \sum_{\ell = 0}^{M-1} A_\ell(\theta)\, 
  \vert \ell \rangle  \otimes \vert u \rangle
\label{eq_psi3_def}
\end{eqnarray}
with amplitudes
\begin{eqnarray}
  A_\ell(\theta) 
  &\equiv&
  \frac{1}{M}\,\frac{1 - e^{2\pi i\, ( \theta - \ell/M)M}}
  {1-e^{2\pi i\,  ( \theta - \ell/M)}}
\label{eq_Aktheta_def}
\\[5pt]
  &=&
  \frac{1}{M}\,\frac{1 - e^{2\pi i\, ( \theta - \theta_\ell)M}}
  {1-e^{2\pi i\,  ( \theta - \theta_\ell)}}
  =
  \frac{1}{M}\,\frac{1 - e^{2\pi i\, ( \ell_\theta - \ell)}}
  {1-e^{2\pi i\,  ( \ell_\theta - \ell)/M}}
  \ .
\label{eq_Aktheta_def_x}
\end{eqnarray}
In (\ref{eq_Aktheta_def}) we have expressed the amplitude in
terms of the fundamental quantities $\theta$ and $\ell$,  while
(\ref{eq_Aktheta_def_x}) expresses the amplitude  in terms of 
the $m$-bit measured phase $\theta_\ell \equiv \ell/M$ and the 
``mode number'' $\ell_\theta \equiv \theta M$ (which might or might 
not be an integer). We now find that the probability of measuring 
the $\ell$-th state is
\begin{eqnarray}
  P_\ell(\theta)
  =
  \big\vert A_\ell(\theta) \big\vert^2
  &=&
  \frac{1}{M^2}\, \frac{\sin^2\left[ \pi \left(\theta - 
  \displaystyle\frac{\ell}{M} \right) M\right]}
  {\sin^2\left[ \pi \left(\theta - \displaystyle\frac{\ell}{M}  
  \right) \right]}
\\[5pt]
  &=&
  \frac{1}{M^2}\, \frac{\sin^2\left[ \pi \big(\theta - 
  \theta_\ell\big) M\right]}
  {\sin^2\left[ \pi \big(\theta - \theta_\ell
  \big) \right]}
  =
  \frac{1}{M^2}\, \frac{\sin^2\left[ \pi \big(\ell_\theta - \ell\big)\right]}
  {\sin^2\left[ \pi \big(\ell_\theta - \ell\big)/M\right]}
  \ . ~~~~~
\end{eqnarray}
The probability $P_\ell(\theta)$ is maximum for the state $\ell \in
\{0, 1, \cdots, M-1\}$ for which  $\delta = \theta - \ell/M = \theta - 
\theta_\ell$ is minimum.  For large values of $M = 2^m$,  the probability 
$P_\ell(\theta)$ is sharply peaked about~$\theta$,  as illustrated
in Fig.~\ref{fig_pell}.  We note that the value of $\ell_\theta \equiv 
M\theta$ need not be an $m$-bit integer (nor indeed, an integer 
at all).  The point to be emphasized is that when $\ell_\theta$ is 
not an integer, then the QPE circuit becomes probabilistic in nature. 
This will turn out to be a key feature of Shor's factoring circuit.  
%%
%\vskip-0.5cm

In fact, this analysis implicitly assumes that $\ell_\theta = M 
\theta$ in amplitude (\ref{eq_Aktheta_def}) is not an integer, 
as the expression for $A_\ell(\theta)$ then requires a more 
delicate treatment. To see this, let us write the amplitude 
in terms of the quantity $\ell_\theta$, so that 
\begin{eqnarray}
  A_{\ell, \ell_\theta}
  \equiv
   A_\ell(\ell_\theta/M)
   =
  \frac{1}{M}\,\frac{1 - e^{2\pi i\, ( \ell_\theta- \ell)}}
  {1-e^{2\pi i\,  ( \ell_\theta - \ell)/M}}
  \ .
\label{eq_A_int_def}
\end{eqnarray}
We have already considered the case in which $\ell_\theta$ 
is a non-negative integer in  the previous section, and 
consistency demands that amplitude (\ref{eq_A_int_def}),
or equivalently (\ref{eq_Aktheta_def}), must collapses to 
the Kronecker-delta form
\begin{eqnarray}
  A_\ell(\theta) 
  =
  \delta_{\ell, \ell_\theta}  
  %~~~\text{for}~~ \ell_\theta = M\theta \in \{0, 1, \cdots, M-1\}
  %~~\text{is a non-negative integer}.
\label{eq_A_int}
\end{eqnarray}
when $\ell_\theta$ becomes an integer. 
We will now show how the expression for $A_\ell(\theta)$
in (\ref{eq_A_int_def}) reduces to this simpler form. 
Note that the numerator in (\ref{eq_A_int_def}) vanishes 
for every integer $\ell_\theta$ (and therefore for every 
value of $\ell_\theta -\ell$, as $\ell$  is itself an integer). 
Consequently, the only way one can obtain a non-zero 
probability is when the denominator also vanishes, so 
that we have the indeterminate form $0/0$.  However,
the denominator vanishes only for $\ell_\theta - \ell$ 
such that 
\begin{eqnarray}
 (\ell_ \theta - \ell)/M = n 
  ~~\text{for any}~ n \in \mathbb{Z}
  \ ,
\end{eqnarray}
or equivalently, 
\begin{eqnarray}
  \ell = \ell_\theta - n M \ .
\end{eqnarray}
The index $\ell \in \{0, 1, \cdots, M-1\}$ is consequently out 
of range for every value of $n \in \mathbb{Z}$ except $n = 0$; therefore, 
$A_\ell(\theta)= 0$ except when 
\begin{eqnarray}
  \ell = \ell_\theta \equiv M \theta \ .
\end{eqnarray}
The value of the amplitude for $\ell = \ell_\theta$ must of 
course be unity (up to an arbitrary phase), and indeed it is. 
As previously mentioned, when $\ell = \ell_\theta$ in 
(\ref{eq_A_int_def}), we obtain the indeterminate form 
$0/0$. We therefore replace $\ell_\theta - \ell$ by a 
small displacement $\varepsilon$ and the take the
limit $\varepsilon \to 0$, in which case the associated 
amplitude becomes
\begin{eqnarray}
  A = 
  \lim_{\varepsilon \to 0}
  \frac{1}{M}\,\frac{1 - e^{2\pi i\, \varepsilon}}
  {1-e^{2\pi i \, \varepsilon/M}}
  =
  1
  \ ,
\end{eqnarray}
as expected. This  establishes expression (\ref{eq_A_int}).
In this case, the final state (\ref{eq_psi3_def}) becomes the
Eigenstate
\begin{eqnarray}
  \vert  \psi_{3} \rangle
 =
 \vert \ell_\theta \rangle  \otimes \vert u \rangle \ , 
 ~~\text{where}~ \ell_\theta = M \theta \in \{0, 1, \cdots, M-1 \}
  \ ,
\label{eq_psi3_eigen_def}
\end{eqnarray}
as we obtained in the previous section.

%\pagebreak
\subsection{Generalized Input States}
\label{sec_gen_input}

Suppose now that we choose the work state $\vert u 
\rangle$ to be a linear sum over all Eigenstates of $U$.
Then the output $\vert \psi_{2} \rangle$ in equation 
(\ref{eq_psi2_gen}) will become a corresponding sum 
over these Eigenstates. That is to say, for the Eigenstates
\begin{eqnarray}
  U \vert u_s \rangle 
  =
  e^{2\pi i \, \theta_s} \,  \vert u_s \rangle \ ,
\label{eq_U_gen}
\end{eqnarray}
let us populate the work register with the linear 
combination of states
\begin{eqnarray}
  \vert u \rangle
  = 
  {\sum}_{s}  a_s \,  \vert u_s \rangle 
  \ .
\label{eq_gen_u}
\end{eqnarray}
We now see that the output of the front-end (\ref{eq_psi2_gen}) becomes
\begin{eqnarray}
 \vert \psi_{2} \rangle
  &=&
  \frac{1}{\sqrt{M}} \, 
  \sum_{k=0}^{M-1} {\sum}_s \, a_s \, e^{2\pi i k\, \theta_s } 
  \, \vert k \rangle  \otimes \vert u_s \rangle
  \ ,
\label{eq_psi2_sums}
\end{eqnarray}
and that the control and work registers are now entangled.
The output of QPE circuit is given by $\vert \psi_3 \rangle 
= QFT^\dagger \vert \psi_2 \rangle$, and using the form 
of $QFT^\dagger$ in (\ref{eq_QFTdagger_a}), we find 
\begin{eqnarray}
  \vert  \psi_{3} \rangle
  &=&
  QFT^\dagger \vert \psi_{2} \rangle
  =
  \sum_{\ell = 0}^{M-1} {\sum}_s  A_\ell(\theta_s)\,
 \vert \ell \rangle  \otimes \vert u_s \rangle
  \ ,
\label{eq_psi3_sums}
\end{eqnarray}
where the amplitudes $A_\ell(\theta_s)$ are defined by 
\begin{eqnarray}
  A_\ell(\theta_s) 
  &\equiv&
  \frac{a_s}{M}\,\frac{1 - e^{2\pi i\, ( \theta_s - \ell/M)M}}
  {1-e^{2\pi i\,  ( \theta_s - \ell/M)}} 
\label{eq_Aktheta_def_as}
\\[5pt]
  &=&
  \frac{a_s}{M}\,\frac{1 - e^{2\pi i\, ( \theta_s - \theta_\ell)M}}
  {1-e^{2\pi i\,  ( \theta_s - \theta_\ell)}} 
  =
  \frac{a_s}{M}\,\frac{1 - e^{2\pi i\, ( \ell_s - \ell)}}{1-e^{2\pi i\,  
  ( \ell_s - \ell)/M}} 
\label{eq_Aktheta_def_as_x}
\end{eqnarray}
when $\ell_s \equiv M \theta_s$ is not an integer.  
In contrast, for integer values $\ell_s \in \{0, 1, \cdots, M-1\}$, 
then expression (\ref{eq_Aktheta_def_as}) reduces to
\begin{eqnarray}
  A_\ell(\theta_s)
  \equiv
  a_s \, \delta_{\ell, \ell_s} \ .
\label{eq_Aktheta_def_as_int}
\end{eqnarray}
For general values of $\theta_s$, we therefore find that 
the probability of measuring a specific $\ell$-$s$ state 
is given by
\begin{eqnarray}
  P_{\ell}(\theta_s)
  =
  \big\vert A_\ell(\theta_s) \big\vert^2
  &=&
  \frac{\vert a_s \vert^2 }{M^2} ~
  \frac{\sin^2\left[ \pi \left(\theta_s  - 
  \displaystyle\frac{\ell}{M} \right) M\right]}
  {\sin^2\left[ \pi \left(\theta_s  - \displaystyle\frac{\ell}{M}  
  \right) \right]}
\\[5pt]
  &=&
  \frac{\vert a_s \vert^2 }{M^2} ~
  \frac{\sin^2\left[ \pi \big(\theta_s  - \theta_\ell \big) M\right]}
  {\sin^2\left[ \pi \big(\theta_s  - \theta_\ell
  \big) \right]}
  =
  \frac{\vert a_s \vert^2 }{M^2} ~
  \frac{\sin^2\left[ \pi \big(\ell_s  -\ell \big) \right]}
  {\sin^2\left[ \pi \big(\ell_s  - \ell\big)/M \right]}
  \ .   ~~~~~
\end{eqnarray}
As before, for large values of $M = 2^m$, the probability 
is sharply peaked about the control states for which $\ell 
\approx \ell_s \equiv M \theta_s$, while the amplitudes 
$a_s$ determine the most likely values of~$\theta_s$. 

When all  of the angles $\theta_s$ become $m$-bit rational
numbers, this analysis simplifies considerably. In this
case,  the parameters
\begin{eqnarray}
  \ell_s \equiv M \theta_s = 2^m \theta_s 
\end{eqnarray}
are all  integers,  and the output of the front-end
 (\ref{eq_psi2_sums}) can be written 
\begin{eqnarray}
 \vert \psi_2 \rangle
  &=&
  \frac{1}{\sqrt{M}} \, 
  \sum_{k=0}^{M-1} {\sum}_s \, a_s \, e^{2\pi i\, k \theta_s  } 
  \, \vert k \rangle  \otimes \vert u_s \rangle
  \\
  &=&
   {\sum}_s \, a_s \, \frac{1}{\sqrt{M}} \, 
  \sum_{k=0}^{M-1} \, e^{2\pi i\, k \ell_s / M } 
  \, \vert k \rangle  \otimes \vert u_s \rangle
  \\[5pt]
  &=&
   {\sum}_s \, a_s \,  QFT \,
  \, \vert \ell_s \rangle  \otimes \vert u_s \rangle
   \ ,
\end{eqnarray}
and therefore
\begin{eqnarray}
  \vert \psi_3 \rangle
  =
  QFT^\dagger \,   \vert \psi_2 \rangle
  =
  {\sum}_s a_s \,  \vert \ell_s \rangle \otimes \vert u_s \rangle 
  \ .
\label{eq_psi_simple}
\end{eqnarray}
A measurement of the system now yields one of the states 
$\vert \ell_s \rangle \otimes \vert u_s \rangle$ with 
probability \hbox{$P_ s = \vert a_s \vert^2$}.  All other states have 
vanishing probability! This result can also be obtained directly 
from the general final state (\ref{eq_psi3_sums}). In this
case, when all phase angles $\theta_s$ becomes $m$-bit
fractions, then the amplitudes (\ref{eq_Aktheta_def_as}) 
reduce to $  A_\ell(\theta_s) = a_s \, \delta_{\ell, \ell_s}$,
so that only the terms for which $\ell = \ell_s \equiv M
\theta_s$ for some value of $s$ will contribute, and the 
general expression (\ref{eq_psi3_sums}) collapses to 
(\ref{eq_psi_simple}).

\vfill
\pagebreak
\clearpage
\section{Continued Fractions}
\label{sec_cont_frac}

In an effort to render this manuscript self-contained,  
in this section we take a mathematical digression 
to briefly introduce the theory of {\em continued 
fractions},  a topic with which many readers might 
not be  entirely familiar.   In Section~\ref{sec_extract_phase},   
we will employ continued fractions in the post-quantum 
processing stage of Shor's algorithm to extract the 
{\em exact} phase $\phi_s = s/r$ of the modular 
exponentiation operator $U_{a, \smN}$ from the 
measured (and {\em approximate}) $m$-bit phase 
value $\tilde \phi$.  As we have emphasized,  this 
will provide the requisite period $r$ from which the
factors of $N$ can be inferred.  Continued fractions,  however, 
are interesting in their own right,  and they provide a number 
of fascinating connections between the integers and 
the real numbers.  In classical mathematics,  for example,  
continued fractions were employed to find rational 
approximations to many irrational numbers of interest 
(something quite useful before the advent of modern 
computers and calculators).  We introduce the subject 
by considering the following infinite continued fraction  
\begin{eqnarray}
  x
  = 
  1 + \displaystyle\frac{1}{1 +  \displaystyle\frac{1}
  {1 + \displaystyle\frac{1}{1 + \displaystyle\frac{1}
  {1 + \,\ddots}}}}
  \ ,
\label{eq_phi_cf}
\end{eqnarray}
where the fraction $x$ telescopes downward without 
end.   Note that the denominator of the fractional piece after 
the initial 1 takes the same form as the continued fraction 
itself,  so that we can express  (\ref{eq_phi_cf}) by the
equation
\begin{eqnarray}
  x = 1 + \frac{1}{x}
 \ .
\label{eq_varphi}
\end{eqnarray}
At the risk of introducing a spurious (negative) solution,  
we multiply (\ref{eq_varphi}) by $x$ to obtain the quadratic 
equation $x^2 = x + 1$.  The positive solution to this equation 
is \hbox{$x = (1 + \sqrt{5})/2$},  which we recognize as the 
golden mean! Continued fractions are interesting indeed.
In fact,  any real number can be expressed as a continued 
fraction consisting of a sequence of well-chosen integers
using a simple and easily executed algorithm. 

We shall concentrate on continued fractions that terminate
after a finite number of iterations,  thereby producing a 
rational number.   We define a {\em finite continued fraction} 
as a number of the form 
\begin{eqnarray}
  x
  =
  a_0 + \displaystyle\frac{1}{a_1 +  \displaystyle\frac{1}
  {a_2 +   \displaystyle\frac{1}{a_3 + \ddots + \frac{1}
  {a_n}}}}
  \ , 
\label{eq_a0dotsaN_x}
\end{eqnarray}
where $a_0$ is an integer (positive or negative) and $a_1,  
a_2,  \cdots,  a_n$ are all positive integers.   We will denote 
a continued fraction by enumerating its integer coefficients 
in square brackets,  so that $x = [a_0; a_1,  a_2,  \cdots,  
a_n]$.   It is traditional to offset the first integer $a_0$ 
with a semicolon.   For our purposes,  the most important 
attributes of continued fractions are their so-called 
{\em convergents},  whose definition is formalized below. 

\begin{definition}
\theoremskip
Suppose $x = [a_0;  a_1,  a_2,  \cdots,  a_n]$ is a
continued fraction.  Any continued fraction of the 
form $[a_0;  a_1,  a_2,\cdots,  a_m]$ for $m \le n$ 
is called a {\em convergent} of the original continued 
fraction for $x$. 
\label{def_convergent}
\end{definition}
For example,  
\begin{eqnarray}
  [a_0;  a_1,  a_2,  a_3]
  =
  a_0 + \displaystyle\frac{1}{a_1 +  \displaystyle\frac{1}{a_2
 + \displaystyle\frac{1}{a_3} }}
\end{eqnarray}
is a convergent of the continued fraction $x = [a_0; 
a_1,  a_2,  a_3,  \cdots,  a_n]$.  In fact,  we can regard the 
convergents of a continued fraction of $x$ as 
systematically improving rational approximations 
to $x$.  As we have emphasized,  any finite continued 
fraction gives a rational number.  It turns out that the 
converse is also true,  namely,   that any rational number 
can be represented as a {\em finite} continued fraction 
with integer coefficients.  There exists a simple and 
efficient (polynomial time) algorithm for determining 
the associated continued fraction of a rational number.  
The algorithm is best explained through an example,  
so let us consider the following rational approximation 
to $\pi$:
\begin{eqnarray}
  3.1415
  &=&
  \frac{31415}{10000}
  =
  \frac{6283}{2000}
  =
  3 + \frac{283}{2000}
  =
  3 + \displaystyle\frac{1}{\displaystyle\frac{~2000~}{~283~}}
\label{eq_pi_a}
\\[5pt]
  &=&
  3 + \displaystyle\frac{1}{~7 + \displaystyle\frac{19}{283}~}
  =
  3 +  \displaystyle\frac{1}{~7 + \displaystyle\frac{1}
  {\displaystyle\frac{283}{19}}~ }
  =
  3 +  \displaystyle\frac{1}{~7 + \displaystyle\frac{1}
  {14 + \displaystyle\frac{17}{19}}~}
\label{eq_pi_b}
\\[5pt]
  &=&
   3 +  \displaystyle\frac{1}{~7 + \displaystyle\frac{1}
  {14 + \displaystyle\frac{1}{\displaystyle\frac{19}{17}}}~}
  =
  3 +  \displaystyle\frac{1}{~7 + \displaystyle\frac{1}
  {14 + \displaystyle\frac{1}{ 1 + \displaystyle\frac{2}{17}}}~}
\label{eq_pi_c}
\\[5pt]
  &=&
  3 +  \displaystyle\frac{1}{~7 + \displaystyle\frac{1}
  {14 + \displaystyle\frac{1}{ 1 + \displaystyle\frac{1}
  {\displaystyle\frac{17}{2}}}}~}
  = 
  3 +  \displaystyle\frac{1}{~7 + \displaystyle\frac{1}
  {14 + \displaystyle\frac{1}{ 1 + \displaystyle\frac{1}
  {8 + \displaystyle\frac{1}{2}}}}~}  ~~~ ,
\label{eq_pi_d}
\end{eqnarray}
and therefore $3.1415 = [3; 7, 14, 1, 8, 2]$.  This algorithm
consists of successive inversions of rational numbers,
followed by splitting the inverted form into an integer 
plus a rational piece,  and continuing this process again.
For example,  in line (\ref{eq_pi_a}) the rational contribution 
$283/2000$ is inverted to form the equivalent number 
$1/(2000/283)$,  and in the first line of (\ref{eq_pi_b}),   
the denominator $2000/283 > 1$ is then split into its 
equivalent form $7 + 19/283$.  We then invert the 
rational piece to give $1/(283/19)$,  and we split 
$283/19 > 1$ into an equivalent form $14 + 17/19$.  
We continue in this fashion until the procedure 
terminates. 

In a certain sense,  continued fractions are a more 
natural representation of real numbers than their 
decimal counterparts.  This is because the continued 
fraction representation of rational numbers always 
terminates after a finite number of iterations,  {\em i.e.} 
$x \in \mathbb{Q}$ iff $x = [a_0; a_1,  \cdots,  a_n]$ 
for some finite sequence of integer coefficients $a_\ell$.  
In contrast,  decimal representations of rational numbers 
need not be finite,  {\em e.g.} the rational number 
\hbox{$2/3 = 0.666 \cdots$} has an infinite number 
of digits,  whereas the continued fraction expansion 
$2/3 = [0; 1, 2]$ has only two non-zero coefficients.

Note that we have required the coefficients of 
continued fractions to be integers.  This is because 
the infinite continued fraction expansion of an irrational 
number is then unique.  Furthermore,  the continued 
fraction expansion of a rational number is 
almost unique.   It turns out that there are only {\em 
two} possible continued fraction expansions for 
any given rational number,  provided the coefficients 
$a_\ell$ are {\em integers} (we call this {\em 
semi-uniqueness}).  To see this,  suppose that $x 
=  [a_0;  a_1,  \cdots,  a_n]$ with $a_n > 1$ is a 
continued fraction 
expansion with integer coefficients.  We can rewrite
the last coefficient as $a_n = (a_n-1) + 1/1$,  and 
consequently we can also express the rational
number by the continued fraction $x = [a_0;  a_1,  
\cdots,  a_n-1,  1]$.   This means that,  without loss 
of generality,   
we may take a continued fraction representation of 
a  rational number to have either an even or an odd 
number of terms,  providing the coefficients are integers 
(and we shall use this fact in proving Theorem~\ref{thm_c} 
below).  We will,  however,  sometimes find it convenient 
to generalize the notion of a continued fraction to 
allow for rational (or even real) coefficients $a_\ell$.  
But we pay a price for doing so,  as we can no longer 
be assured of the semi-uniqueness of the associated 
continued fraction.   

We now prove three essential 
theorems concerning continued fractions.  The first
two establish that the convergents of a continued
fraction take the form $p_n/q_n$,  where $p_n$
and $q_n$ are special sequences of relatively prime
integers. The third theorem can be used to 
relate the period $r$ to the sequence of denominators
$q_n$,  and as we shall see in the next section,  it will 
be essential in extracting the period from the measured 
phase.  Our primary references  for this section are 
Refs. \cite{qcqi} and \cite{des}.

\vbox{ 
\begin{theorem}
\theoremskip
Let $[a_0; a_1,  \cdots,  a_m]$ be a continued fraction,
where the coefficients $a_\ell$ can be either rational 
numbers or integers. The convergents $x_n \equiv [a_0; 
a_1,  \cdots,  a_n]$ for $n \le m$ are equal to the ratio 
\hbox{$x_n = p_n/q_n$},  where $p_n$ and $q_n$ are 
defined through the sequence
\begin{eqnarray}
  p_n &=& a_n p_{n-1} + p_{n-2}
\nonumber\\
  q_n &=& a_n q_{n-1} + q_{n-2}
\label{eq_pm_qm}
\end{eqnarray}
for $2 \le n \le m$,  with the seed values
\begin{eqnarray}
  p_0 &=& a_0 \hskip1.5cm q_0 = 1
\label{eq_0pq}
\\
  p_1 &=& a_1 a_0 + 1 ~~~ q_1 = a_1
  \ .
\label{eq_1pq}
\end{eqnarray}
Furthermore,  if the coefficients $a_\ell$ are positive 
integers,  then $p_n$ and $q_n$ are also positive
integers (and they strictly increase in magnitude). 
\label{thm_a}
\end{theorem}
} % vbox
\vskip-0.3cm 
\PROOF.  
The proof will be through induction on $n$.  For $n=0$,  the
convergent is just $x_0 \equiv [a_0]$,  which corresponds to 
$x_0 = p_0/q_0$ for $p_0 = a_0$ and $q_0 = 1$,  thereby
validating (\ref{eq_0pq}).  For $n=1$,  we have the convergent
\begin{eqnarray}
  x_1 
  \equiv 
  [a_0 ; a_1] 
  = 
  a_0 + \frac{1}{a_1}
  =
  \frac{a_1 a_0 + 1}{a_1}
  \ ,
\end{eqnarray}
so that $x_1 = p_1/q_1$ with $p_1 = a_1 a_0 + 1$ and 
$q_1 = a_1$,  thereby validating (\ref{eq_1pq}).  This takes 
care of the initial seeding.  To provide a bit of intuition,  
let us explicitly verify the $n=2$ case for $x_2 \equiv
[a_0;  a_1,  a_2]$.   From (\ref{eq_pm_qm}) we have
\begin{eqnarray}
  p_2 
  &=&
   a_2 p_1 + p_0
%  =
%  a_2 ( a_1 a_0 + 1) + a_0
  =
  a_2 a_1 a_0 + a_2 + a_0
\label{eq_2pq}
\\
  q_2
  &=&
  a_2 q_1 + q_0
  =
  a_2 a_1 + 1
  \ ,
\label{eq_3pq}
\end{eqnarray}
and we can therefore express the convergent $x_2$ 
in the form:
\begin{eqnarray}
  x_2 
  &\equiv&
  [a_0;  a_1,  a_2]
  =
  a_0 + \displaystyle\frac{1}{a_1 +  \displaystyle\frac{1}{a_2}}
\label{eq_xtwo_a}
\\[10pt]
  &=&
  a_0 + \frac{a_2}{a_2 a_1 + 1}
  =
  \frac{a_2 a_1 a_0 + a_0 + a_2}{a_2 a_1 + 1}
  =
  \frac{p_2}{q_2}
  \ ,
\label{eq_xtwo_b}
\end{eqnarray}
thereby validating the theorem for $n=2$.  We now assume 
(\ref{eq_pm_qm}) holds for some $n \ge 3$ with $x_n = 
[a_0;  a_1,  a_2,  \cdots,  a_n]$,  and we wish to prove that  it 
continues to hold for $n + 1$.   Note that any convergent 
$x_{n+1} = [a_0; a_1,    \cdots,  a_{n-1},  a_n,  a_{n+1}]$
(which has $n+1$ coefficients) may be expressed in the  
alternative form
\begin{eqnarray}
  x_{n+1}
  =
  \underbrace{
  [a_0; a_1,   \cdots,  a_{n-1},  a_{n} + 1/a_{n+1}]
  }_{n ~\text{coefficients}}
  \ ,
\label{eq_xnpone}
\end{eqnarray}
which contains only $n$ coefficients,  albeit rational
coefficients.   We can therefore apply the induction 
hypothesis to (\ref{eq_xnpone}).  
To this end,  let $\tilde p_\ell/ \tilde q_\ell$ be the 
sequence of convergents associated with the second 
form of the continued fraction for $x_{n+1}$.  The
induction hypothesis now gives
\begin{eqnarray}
  x_{n+1}
  =
  [a_0; a_1,   \cdots,  a_{n-1}
  \,,\,
  \underbrace{a_{n} + 1/a_{n+1}}_{\tilde a_n}]
  =
  \frac{\tilde p_{n}}{\tilde q_{n}}
  \ , 
\end{eqnarray}
where
\begin{eqnarray}
  \frac{\tilde p_{n}}{\tilde q_{n}}
  =
  \frac{\tilde a_n  \, \tilde p_{n-1} + \tilde p_{n-2}  }
  {\tilde a_n \, \tilde q_{n-1} + \tilde q_{n-2} }
  \ .
\end{eqnarray}
It is clear that $\tilde p_{n-2} = p_{n-2}$,  $\tilde 
p_{n-1} = p_{n-1}$ and $\tilde q_{n-2} = q_{n-2}$,  
$\tilde q_{n-1} = q_{n-1}$,  and we therefore find 
\begin{eqnarray}
  x_{n+1}
  &=&
  \frac{\displaystyle{\Big(a_n + \frac{1}{a_{n+1}}\Big)} 
   p_{n-1} + p_{n-2}  }
  {\displaystyle{\Big(a_n + \frac{1}{a_{n+1}}\Big)} 
  q_{n-1} + q_{n-2} }
  =
  \frac{(a_n a_{n+1} + 1) p_{n-1} + p_{n-2}a_{n+1}}
  {(a_n a_{n+1} + 1) p_{n-1} + p_{n-2}a_{n+1}}
\\[5pt]
  &=&
  \frac{a_{n+1} (a_n p_{n-1} + p_{n-2}) + p_{n-1}}
  {a_{n+1} (a_n q_{n-1} + q_{n-2}) + q_{n-1}}
  ~~~\Leftarrow~~
  \text{use}~(\ref{eq_pm_qm})
\\[5pt]
  &=&
  \frac{a_{n+1} p_n + p_{n-1}}
  {a_{n+1} q_n + q_{n-1}}
  \equiv
  \frac{p_{n+1}}{q_{n+1}}
  \ .
\end{eqnarray}
Thus,  the theorem is true for $n+1$.  It is obvious that 
if the coefficients $a_\ell$ are positive integers,  then 
$p_n$ and $q_n$ are as well.  This completes the proof.
\ENDPROOF

\begin{theorem}
\theoremskip
If the coefficients of the continued fraction $[a_0; a_1,  a_2, 
 \cdots,  a_m]$ are integers,  then the integers $p_n$ and 
$q_n$ of Theorem~\ref{thm_a} are {\em relatively prime},
and satisfy the relation
\begin{eqnarray}
  q_n p_{n-1} - p_n q_{n-1}  = (-1)^n 
\label{eq_pnqnmone}
\end{eqnarray}
for $n \ge 1$. 
\label{thm_b}
\end{theorem}
\vskip-0.3cm 
\PROOF.   We prove (\ref{eq_pnqnmone}) by induction on 
$n$.  From   (\ref{eq_0pq}) and (\ref{eq_1pq}) we find
\begin{eqnarray}
  n=1 : ~~
  q_1 p_{0} - p_1 q_{0}  
  &=&
  a_1 a_0 - (a_1 a_0 + 1) \cdot 1
  = 
  - 1 
  =
  (-1)^1
  \ ,
\end{eqnarray}
so that (\ref{eq_pnqnmone}) holds for $n = 1$.  
Similarly,  expressions (\ref{eq_2pq}) and (\ref{eq_3pq}) 
imply that 
\begin{eqnarray}
  n=2 : ~~
  q_2 p_{1} - p_2 q_{1}  
  &=&
 (a_2 a_1 + 1)(a_1 a_0 + 1) - (a_2 a_1 a_0 + 
  a_2 + a_0) a_1
  = 
  1 
  = 
  (-1)^2 
  \ ,
  ~~~~~
\end{eqnarray}
so that (\ref{eq_pnqnmone}) also holds for $n=2$.  Let us now 
assume that (\ref{eq_pnqnmone})  holds for some $n \ge 3$,  
and let us prove that it continues to hold for $n+1$.  Taking
$n \to n+1$ in (\ref{eq_pm_qm}),  the induction hypothesis 
gives
\begin{eqnarray}
  n+1: ~~
  q_{n+1}  p_{n} - p_{n+1} q_{n}  
  &=&
  (a_{n+1} q_n + q_{n-1}) p_n - (a_{n+1} p_n + p_{n-1}) q_n
%\nonumber\\
%  &=&
%  \big(a_{n+1} q_n p_n + q_{n-1} p_n \big)
% - 
%  \big(a_{n+1} p_n q_n + p_{n-1} q_n \big)
% ~~~~~~
\nonumber\\
  &=&
  -\big(q_n p_{n-1} - p_n q_{n-1}   \big)
  =
  - (-1)^n 
  =
  (-1)^{n+1}
  \ .
\end{eqnarray}
Therefore,  (\ref{eq_pm_qm}) holds for $n+1$,  and 
this completes the proof of the first part of the theorem. 

We must now show that $p_n$ and $q_n$ have 
no common factors other than unity.  Let us therefore
assume that $k_n \ge 1$ is a common factor of 
$p_n$ and $q_n$,   so that 
\begin{eqnarray}
  p_n = k_n \tilde p_n 
  ~~\text{and}~~ 
  q_n = k_n \tilde q_n 
  ~~\text{with}~~ 
  \frac{p_n}{q_n} =  \frac{\tilde p_n}{\tilde q_n}
  \ .
\end{eqnarray}
Let us also assume that $k_{n-1} \ge 1$ is a common factor
of $p_{n-1}$ and $q_{n-1}$,  so that 
\begin{eqnarray}
  p_{n-1} = k_{n-1} \tilde p_{n-1} 
 ~~\text{and}~~ 
  q_{n-1} = k_{n-1} \tilde q_{n-1} 
  ~~\text{with}~~ 
  \frac{p_{n-1} }{q_{n-1} } =  \frac{\tilde p_{n-1} }{\tilde q_{n-1} }
  \ .
\end{eqnarray}
Then (\ref{eq_pnqnmone}) now takes the form 
\begin{eqnarray}
  k_n k_{n-1}
  \big( \tilde q_n \tilde p_{n-1} - \tilde p_n \tilde q_{n-1} 
  \big) 
  = (-1)^n 
  \ .
\label{eq_pnqnmone_k}
\end{eqnarray}
It is obvious that $\tilde p_n/\tilde q_n$ has the same
continued fraction representation as $p_n/q_n$ for all $n \ge 1$,  so
that $\tilde q_n \tilde p_{n-1} - \tilde p_n \tilde q_{n-1} 
  = (-1)^n $,  from which it follows that
\begin{eqnarray}
  k_n  k_{n-1}
  =
  1
  \ .
\label{eq_p_kk_one}
\end{eqnarray}
For $n=1$,  we have $k_1 k_0 =1$.   Since $1 = q_0 = k_0 
\,\tilde q_0$,   and since $k_0$ and $q_0$ are both integers,  
we must have $k_0 = 1$ (and $\tilde q_0 = 1$).   Therefore, 
$k_1 = 1$.    For $n=2$,  expression (\ref{eq_p_kk_one}) 
becomes $k_2 k_1 = 1$,  and therefore $k_2=1$.  Continuing 
in this fashion,  we find $k_n = 1$ for all $n \ge 1$,  and 
hence $p_n$ and $q_n$ are relatively prime. 
\ENDPROOF

\vskip0.3cm
\noindent
We move on to our primary result,  the theorem that
will allow us to extract the desired period from an
approximately measured phase angle.

\begin{theorem}
\theoremskip
 Let $x$ be a rational number.  If two relatively prime 
integers $p$ and $q$ satisfy
\begin{eqnarray}
  \left\vert    \frac{p}{q} - x \right\vert
  \le 
  \frac{1}{2q^2}
  \ ,
\label{eq_qpx }
\end{eqnarray}
then $p/q$ is necessarily a convergent of $x$. 
\label{thm_c}
\end{theorem}
\vskip-0.0cm 
\PROOF.  
Let $p/q = [a_0;  a_1,  \cdots,  a_n]$ be the continued 
fraction representation for $p/q$,  and define the
convergents $p_\ell/q_\ell$ for $\ell = 0,  1,  \cdots,
n$ as in Theorem~\ref{thm_a},  so that $p_n/q_n 
= p/q$.  The object of the proof is to construct the 
continued fraction representation for $x$,  and this 
will explicitly show that $p/q$ is one of its convergents. 

Let us first define the error $\delta$ for $p/q = p_n/q_n$ 
by
\begin{eqnarray}
  x - \frac{p_n}{q_n} 
  \equiv
  \frac{\delta}{2 q_n^2}
  \ .
\label{eq_x_delta}
\end{eqnarray}
Note that inequality (\ref{eq_qpx }) gives $0 \le \delta 
\le 1$.  We now define the parameter
\begin{eqnarray}
  \lambda
  \equiv
  2\, \frac{q_n p_{n-1} - p_n q_{n-1}}{\delta} 
  -
   \frac{q_{n-1}}{q_n}
  \ ,
\label{eq_lam}
\end{eqnarray}
and with some algebra  we can show that equations 
(\ref{eq_x_delta}) and (\ref{eq_lam}) imply
\begin{eqnarray}
  x 
  =
  \frac{\lambda p_n + p_{n-1}}{\lambda q_n + q_{n-1}}
  \ .
\label{eq_x_fin}
\end{eqnarray}
To see this,  note that (\ref{eq_lam}) allows us to write
\begin{eqnarray}
  &&
  \frac{2}{\delta} \,  \big(q_n p_{n-1} - p_n q_{n-1}\big)
  =
  \lambda +    \frac{q_{n-1}}{q_n}
  =
  \frac{\lambda q_n + q_{n-1}}{q_n}
  ~~~\Rightarrow~~
\\[5pt]
  &&
  \frac{\delta}{2}
  =
  \frac{q_n }{\lambda q_n + q_{n-1}}  
  \big( q_n p_{n-1} - p_n q_{n-1} \big)
  \ .
\end{eqnarray}
Using this result in equation (\ref{eq_x_delta}) gives
\begin{eqnarray}
  x
  &=&
  \frac{p_n}{q_n}  
  +
  \frac{\delta}{2}\,\frac{1}{q_n^2} 
  =
   \frac{p_n}{q_n}  
  + 
  \frac{q_n p_{n-1} - p_n q_{n-1}}
  {q_n \big(\lambda q_n + q_{n-1}\big)}  
\\[5pt]
  &=&
   \frac{
  p_n\big(\lambda q_n + q_{n-1}\big)
  + 
  \big( q_n p_{n-1} - p_n q_{n-1} \big)
  }{q_n\big(\lambda q_n + q_{n-1})}  
\\[5pt]
  &=&
   \frac{
  \big(\lambda p_n + p_n q_{n-1}/q_n \big)
  + 
  \big(p_{n-1} - p_n q_{n-1}/q_n \big)
  }{\lambda q_n + q_{n-1}}  
  =
  \frac{
  \lambda p_n + p_{n-1}}{\lambda q_n + q_{n-1}}  
  \ .
\end{eqnarray}
Now that we have established (\ref{eq_x_fin}),  
Theorem~\ref{thm_a} implies $x$ is a continued
fraction with coefficients  $a_0,  a_1,  \cdots,  
a_n,  a_{n+1}$,  where $a_{n+1} = \lambda$,
{\em i.e.}
\begin{eqnarray}
  x 
  =
  [a_0; a_1, \cdots,  a_n, a_{n+1}] 
  =
 [a_0; a_1, \cdots,  a_n, \lambda] \ . 
\end{eqnarray}
Without loss of generality,  we can assume that 
$n$ is even,  so that Theorem~\ref{thm_b} gives
\begin{eqnarray}
  \lambda
  =
  \frac{2}{\delta} - \frac{q_{n-1}}{q_n}
  >
  2 - \frac{q_{n-1}}{q_n}  
  > 
  2 - 1 > 1
  \ .
\end{eqnarray}
Thus,  $\lambda$ is a rational number greater one,  and 
it therefore has a continued fraction expansion of the 
form $\lambda = [b_0;  b_1,  \cdots,  b_m]$ (since 
$\lambda > 0$ we must have $b_0 > 0$,  and we
henceforth drop the semicolon after the initial 
coefficient $b_0$).  Therefore,  we find
\begin{eqnarray}
  x = [a_0; a_1, \cdots, a_n,  b_0,  \cdots,  b_m] 
  \ ,
\end{eqnarray}
which shows that $x$ is a finite continued fraction 
with $p/q = p_n/q_n$ as one of its 
convergents.~\ENDPROOF

\vfill

\clearpage
\section{Factoring with Shor's Algorithm}
\label{sec_factoring}

\subsection{Basic Observation}

We now address an essential observation on our way 
to building Shor's algorithm.   Let $N$ be the positive 
integer we wish to factor.  We 
assume that $N$ is not even,  and not a power of a prime 
number (otherwise we can find a factor quickly). We say 
that  two integers $a,  b \in \mathbb{Z}$ are {\em congruent 
modulo} $N$  provided the difference $a - b$ is divisible 
by $N$,  and we express this by writing 
\begin{eqnarray}
  a = b ~({\rm mod}~ N) \ .
\label{eq_ab_eqiv_modN}
\end{eqnarray}
That is to say,   $a$ and $b$ are congruent modulo $N$ 
provided there exists another integer $m \in \mathbb{Z}$ 
such that 
\begin{eqnarray}
  a - b = m N \ .
\label{eq_ab_modN}
\end{eqnarray}
Shor's factoring algorithm relies on the following observation. 
Suppose we can find a non-trivial or proper square root of unity 
modulo $N$.  In other words,  suppose that we have found an 
integer~$b$ such that
\begin{eqnarray}
  b^2 = 1 ~({\rm mod}~ N) \ .
\label{eq_b2_modN}
\end{eqnarray}
Then $b$ is a modular square root of unity. 
By a {\em proper} square root,  we mean that  
\begin{eqnarray}
  b \ne \pm 1 ~({\rm mod}~ N) \ .
\label{eq_b2_nontriv}
\end{eqnarray}
Equation (\ref{eq_b2_modN}) implies that there
exists an integer $m \in \mathbb{Z}$ such that
\begin{eqnarray}
   b^2 -1 = m N \ .
\label{b21mN}
\end{eqnarray}
The latter equation is the key to the factoring algorithm,
as it can be expressed as 
\begin{eqnarray}
  (b + 1) (b - 1) = m N \ ,
\label{eq_factor_N}
\end{eqnarray}
and therefore,  we see that the greatest common divisors 
$d_\pm \equiv {\rm gcd}( b \pm1,  N )$ are factors of~$N$.
We note that finding the greatest common divisor of two 
integers can be performed very quickly (in polynomial 
time) on a classical computer.  It can be shown that 
(\ref{eq_b2_modN}) and (\ref{eq_b2_nontriv}) indeed 
lead to non-trivial or proper factors,  in that $d_\pm \ne 
1,   N$.  This is formalized in the following theorem. 

\vfill
\pagebreak

\begin{theorem}
\theoremskip
  Let $b \in \mathbb{Z}$ be a proper square root of unity
  modulo $N$.  That is to say,  let \hbox{
  $b^2 = 1 ~({\rm mod}~ N)$} and \hbox{$b \ne \pm 1 
  ~({\rm mod}~ N)$}.  Then  \hbox{${\rm gcd}( b +1,  N )$} 
  and \hbox{${\rm gcd}( b - 1,  N )$} are {\em proper} factors 
  of $N$.
\label{thm_nontrivial}
\end{theorem}

\PROOF.  
First consider $d = {\rm gcd}(b-1,  N )$,   which is indeed
a factor of $N$. We will show that $d \ne 1$ and $d \ne N$.  
The proof will be by contradiction. 

\vskip0.2cm
\noindent
- First assume $d={\rm gcd}(b-1,  N) = N$.  Then $N$ divides
$b-1$,  so that $b - 1 = m N$ for some $m \in \mathbb{Z}$, 
or equivalently,  $b = 1 ~({\rm mod}~ N)$.  This contradicts
the fact that $b$ is a proper root of unity. 

\vskip0.2cm
\noindent
- Now assume $d={\rm gcd}(b-1,  N) = 1$.   Since $b-1$ and $N$
are relatively prime,  there exists integers $u,  v \in \mathbb{Z}$ 
such that
\begin{eqnarray}
  (b - 1) u + N v = 1 \ .
\label{b1N}
\end{eqnarray}
Multiplying both sides by $b+1$ gives the expression
\begin{eqnarray}
  b + 1 
  = 
  (b ^2- 1) u + (b + 1) N v \ .
\label{eq_bpsone}
\end{eqnarray}
Let us divide both sides of this equation by $N$,  and 
employ equation (\ref{b21mN}) for the first term 
$(b^2-1)u$ on the right-hand side of (\ref{eq_bpsone}):
\begin{eqnarray}
  \frac{b+1}{N} 
  = 
  \underbrace{~\frac{b ^2- 1}{N}~}_{\text{integer}
  ~\equiv~ m} \! \cdot \, u + (b + 1) \cdot v
  =
  m  \cdot u + (b + 1) \cdot v   \in \mathbb{Z} 
  \ ,
\end{eqnarray}
where $u,  v,  m,  b$ are all integers. Thus,  $N$ divides 
$b+1$,  so that $b = -1~({\rm mod}~N)$. Again,  this 
contradicts the fact that $b$ is a proper root of unity.
Similar reasoning holds for $d = {\rm gcd}(b + 1,  N)$.  
\ENDPROOF

%%
%\pagebreak
\subsection{Period Finding and Factorization}

Consider two integers $a$ and $N$ such that $a < N$.  
The integer $N$ is the number we wish to factor,  while 
$a$ is an initial ``guess" for one of the factors.  We will
usually refer to $a$ as the {\em base}.  In fact,  we can 
randomly choose the base from $\{2,  3,  \cdots,  N-1\}$,  
provided that $a$ and $N$ are relatively prime,  so that 
${\rm gcd}(a, N) = 1$ (otherwise we have found a 
non-trivial factor of $N$).   Let us now define the {\em 
order} of $a$ modulo $N$ as the {\em least} positive 
integer $r$ such that
\begin{eqnarray}
  a^r = 1 ~({\rm mod}~ N)
  \ .
\end{eqnarray}
Therefore,  if $r$ is even,  then $b = a^{r/2}$ is an integer,
and it is 
a square root of unity. If it is also non-trivial,  we can 
perform the factoring algorithm outlined above based 
on this value of $b$.  If the order $r$ is odd and $a$ 
is not a perfect square,  then we must try a new base 
$a$.  On the other hand,  if $r$ is odd and $a$ is a 
perfect square,  then $b = a^{r/2}$ is still an integer
square root of unity, and it can be used in the 
algorithm\,\cite{twoone}. 
For these cases of $r$ and $a$, we see that $b = 
a^{r/2}$ is a square root of unity, and provided 
that it is non-trivial, then the factors of $N$ are 
given by \hbox{${\rm gcd}(a^{r/2} \pm 1,  N)$}.

\begin{figure}[h!]
\vskip-0.4cm
\begin{minipage}[c]{0.5\linewidth}
\includegraphics[scale=0.50]{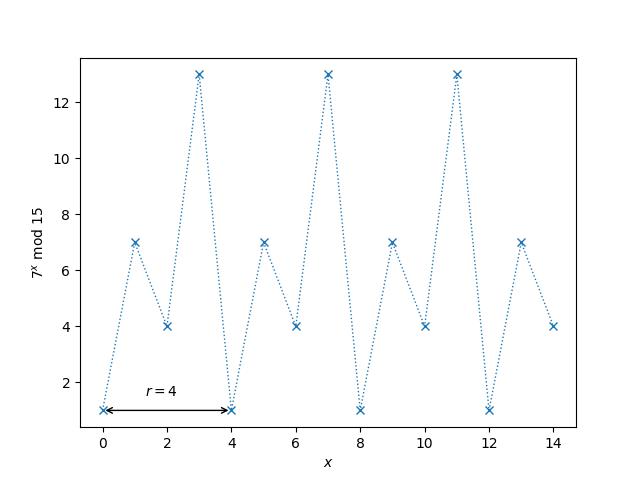} 
\end{minipage}
\begin{minipage}[c]{0.4\linewidth}
\includegraphics[scale=0.50]{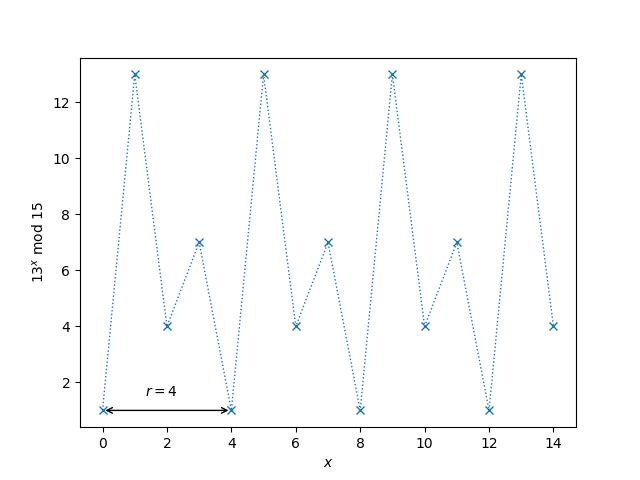} 
\end{minipage}
\caption{\footnoteskip
The function $f(x) = a^x ~({\rm mod}~ N)$ for 
$N =15$ and the bases  $a \in \{7,13\}$.   
}
\label{fig_ax_modN_b}
\end{figure}

There is an equivalent way of looking at this that 
employs the periodic {\em modular exponential 
function} defined by 
\begin{eqnarray}
  f_{a \, \smN}(x) = a^x ~({\rm mod}~ N)
  \ ,
\label{fx_ax}
\end{eqnarray}
\begin{figure}[b!]
\begin{minipage}[c]{0.4\linewidth}
\includegraphics[scale=0.50]{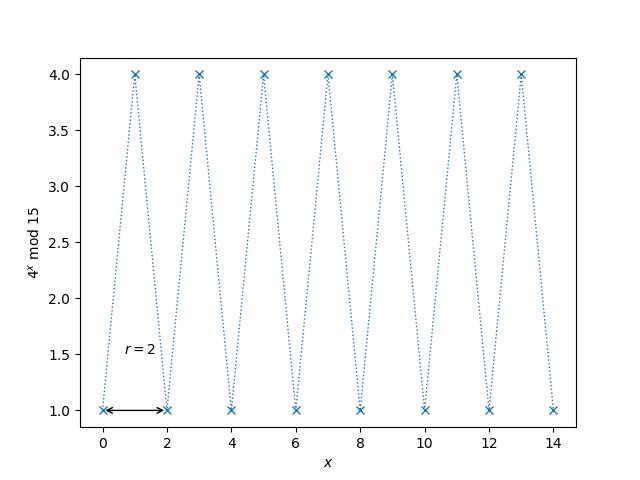} 
\end{minipage}
%\hfill
\hskip1.3cm
\begin{minipage}[c]{0.4\linewidth}
\includegraphics[scale=0.50 ]{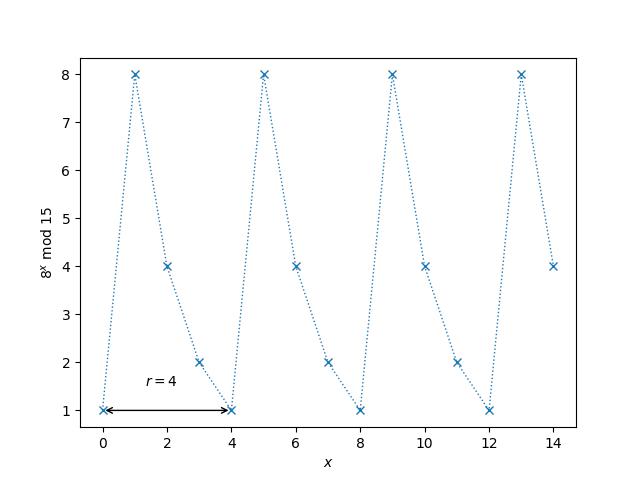} 
\end{minipage}
\begin{minipage}[c]{0.4\linewidth}
\includegraphics[scale=0.50]{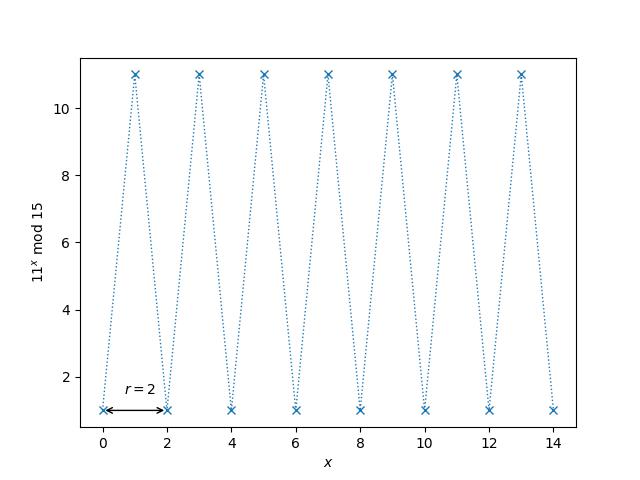} 
\end{minipage}
%\hfill
\hskip1.3cm
\begin{minipage}[c]{0.4\linewidth}
\includegraphics[scale=0.50]{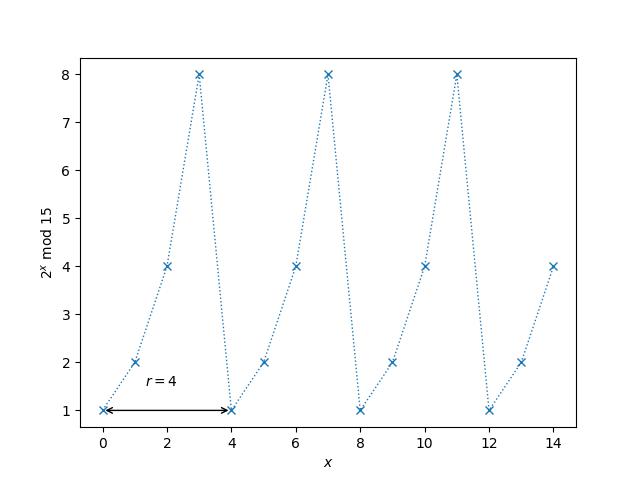} 
\end{minipage}
%\vskip-2.0cm
\caption{\footnoteskip
The function $f(x) = a^x ~({\rm mod}~ N)$ for 
$N =15$ and the bases $a \in \{4, 8,11, 2\}$. 
}
\label{fig_ax_modN_a}
\end{figure}

\noindent
where $a,  x$, and $N$ are non-negative integers. We
will usually drop the subscripts and simply write $f(x)$.  
The modular order $r$ is nothing more than the period 
of $f(x)$. To see this,  note that $f(0)=1$,  and since $r$ 
is the smallest integer such that \hbox{$f(r) = a^r 
~({\rm mod}~N) = 1$},  we see that $f(r)=1$.   In fact,  
for any argument $x$,  we have $f(x + r) = f(x)$,  and 
thus the order $r$ is the period of $f(x)$.  As an example,  
let us take $N = 15$. The base $a$ and the number $N$ 
cannot have any non-trivial common factors, and the 
base must satisfy $1 < a < N$,  which limits the allowed 
values to $a \in \{ 2,4,7,8,  11,  13, 14 \}$. Figures~\ref{fig_ax_modN_a}  
and \ref{fig_ax_modN_b} illustrate the functions $f(x)$ 
for $N=15$ for these values of $a$  (note that $a = 14$ 
gives a trivial square root with no factors, so we do not 
bother to provide a plot).
We summarize below the factorization algorithm from the
last section based on the periods $r$ for the cases specified 
in the Figures.   For the bases $a$ in Fig.~\ref{fig_ax_modN_a}  
we find:

\vskip0.1cm
\noindent
$\bullet~ a=4 \Rightarrow r = 2$:
\hskip4.8cm 
$\bullet~ a=2 \Rightarrow r = 4$:\\
- $a^{r/2} - 1 = 4^1 -1 = 3  ~\Rightarrow~ {\rm gcd}(3,15)=3$
\hskip0.9cm 
- $a^{r/2} - 1 = 2^2 -1 = 4 - 1 =3 ~\Rightarrow~ {\rm gcd}(3,15)=3$
\\[-3pt]
- $a^{r/2} + 1 = 4^1 +1 = 5  ~\Rightarrow~ {\rm gcd}(5,15)=5$
\hskip0.9cm
- $a^{r/2} + 1 = 2^2 +1 = 4 + 1 =5~\Rightarrow~ {\rm gcd}(5,15)=5$

\vskip0.2cm
\noindent
$\bullet~ a=11 \Rightarrow r = 2$:
\\
\vbox{
\noindent
- $a^{r/2} - 1 = 11^1 -1 = 10  ~\Rightarrow~ {\rm gcd}(10,15)=5$\\[-3pt]
- $a^{r/2} + 1 = 11^1 +1 = 12  ~\Rightarrow~ {\rm gcd}(12,15)=3$
}

\vskip0.3cm
\noindent
$\bullet~ a=8 \Rightarrow r = 4$:
\hskip8.0cm $\bullet~ a=14 \Rightarrow r = 2$:\\
- $a^{r/2} - 1 = 8^2 -1 = 64 - 1 =63 ~\Rightarrow~ {\rm gcd}(63,15)=3$
\hskip1.5cm - $a^{r/2} -1 =13$ \\[-3pt]
- $a^{r/2} + 1 = 8^2 +1 = 64 + 1 =65~\Rightarrow~ {\rm gcd}(65,15)=5$
\hskip1.5cm - $a^{r/2} + 1 =15 \Leftarrow$ trivial

%% ax_modN.py

%\vskip-0.3cm
\noindent
And for the bases $a$ in  Fig.~\ref{fig_ax_modN_b}  we
find:  %$^{\displaystyle\text{\hskip4.2cm  does not give factors}}$
\\
%\vskip0.3cm
\noindent
$\bullet~ a=7 \Rightarrow r = 4$:\\
- $a^{r/2} - 1 = 7^2 -1 = 49 - 1 =48 ~\Rightarrow~ {\rm gcd}(48,15)=3$\\[-3pt]
- $a^{r/2} + 1 = 7^2 +1 = 49+ 1 =50~\Rightarrow~ {\rm gcd}(50,15)=5$ 

%\vskip-0.1cm
\noindent
$\bullet~ a=13 \Rightarrow r = 4$:\\
- $a^{r/2} - 1 = 13^2 -1 = 169 - 1 =168 ~\Rightarrow~ {\rm gcd}(168,15)=3$\\[-3pt]
- $a^{r/2} + 1 = 13^2 +1 = 169 + 1 =170~\Rightarrow~ {\rm gcd}(170,15)=5$

\vfill
\clearpage
\subsection{The Factorization Circuit}
\label{sec_factor_circuit}

Our primary references for this section are 
Refs.~\cite{qcqi,des,ibm_shor}.  We will show that one 
can find the period $r$ of the modular exponential function 
$f_{a \, \smN} (x) = a^x ~({\rm mod}~N)$ by exploiting 
the quantum phase estimation (QPE) algorithm developed 
in Section~\ref{sec_QPE}.  The crucial step in this procedure
is to define a unitary operator $U_{a \, \smN}$ whose 
Eigen-phases contain information about the period $r$.
We then employ the QPE algorithm to find the phases 
of $U_{a \, \smN}$,  thereby permitting us to determine 
the exact value of $r$.  
%(with the help of the theory of 
%continued fractions to the measured a values of the phase). 
Recall that the QPE algorithm consists of two quantum registers,  
a control register and a work register.   The control register 
contains $m$ qubits that dictate the resolution of the 
measured output phase of $U_{a \, \smN}$,  while the work 
register encodes information about the number $N$,  and 
we therefore take the number of work qubits to be 
$n = \lceil \log_2 N \rceil$ (the binary length of $N$). 
We shall define a linear unitary operator $U_{a \, \smN}$ 
by its action on the computational basis states of the 
work register,
\begin{eqnarray}
  U_{a \, \smN}\, \vert w \rangle 
  =
  \vert a \cdot w ~({\rm mod}~ N) \rangle 
  \ .
\label{eq_UaN_def}
\end{eqnarray}
We will usually drop the $N$ and $a$ subscripts and write 
$U = U_{a \, \smN}$ for simplicity.  We shall refer to $U$ 
as either the {\em phase} operator or the {\em modular 
exponentiation} (ME) operator.  
To continue, let us now solve the Eigenvalue problem for the 
ME operator $U$.  This is accomplished by the simple 
observation that 
\begin{eqnarray}
  U \, \vert a^x ~({\rm mod}~ N)  \rangle 
  &=&
  \vert a^{x+1} ~({\rm mod}~ N) \rangle
\label{eq_Uk}
\end{eqnarray}
for any non-negative integer $x$.  
Let us now define the $r$ states 
\begin{eqnarray}
  \vert u_s \rangle
  =
  \frac{1}{\sqrt{r}}
  \sum_{k=0}^{r-1} \, e^{-2\pi i k\, s / r} \, 
  \vert a^k ~({\rm mod}~ N) \rangle 
 ~~~\text{for}~~
 s \in \{0,  1,  \cdots,  r-1\}
  \ ,
\label{eq_us}
\end{eqnarray}
from which equation (\ref{eq_Uk}) gives
\begin{eqnarray}
  U \vert u_s \rangle = e^{2\pi i \, \phi_s} \, \vert u_s \rangle 
  ~~~\text{with}~~ \phi_s = \frac{s}{r}
  \ .
\label{eq_U_phi_s}
\end{eqnarray}
The possible phases of the ME operator $U$ are therefore 
$\phi_s = s/r$ for $s \in \{0,  1,  \cdots,  r-1\}$,  where $r$
is the period of the function $f(x) = a^x ~({\rm mod}~N)$.
To prove this result, it is instructive to expand the states in
(\ref{eq_us}) term-by-term, 
\begin{eqnarray}
  \vert u_s \rangle
  &=&
  \frac{1}{\sqrt{r}} \, \Big[\, 
  \vert a^0 ~({\rm mod}~ N) \rangle  
  +
  e^{-2\pi i \, s/r} \, \vert a^1 ~({\rm mod}~ N) \rangle 
  +
  e^{-2\pi i \, 2s/r}\, \vert a^2 ~({\rm mod}~ N) \rangle   
  + \cdots +
  ~~~
\nonumber\\ && \hskip1.0cm
  e^{-2\pi i \, (r-2)s/r} \, \vert a^{r-2} ~({\rm mod}~ N) \rangle  
  +
    e^{-2\pi i \, (r-1)s/r} \, \vert a^{r-1} ~({\rm mod}~ N) \rangle  
  \Big]
  \ .
\label{eq_us_expand}
\end{eqnarray}
The series terminates after $r$ terms because 
$a^r ~({\rm mod}~N) = 1$,  which leads us back 
to the first term $\vert 1 \rangle = \vert a^0 ~({\rm 
mod}~N) \rangle$.   From relation (\ref{eq_Uk}), 
we can now easily prove expression (\ref{eq_U_phi_s}):
\begin{eqnarray}
  U \vert u_s \rangle
  &=&
  \frac{1}{\sqrt{r}} \Big[\, 
  \vert a^1 ~({\rm mod}~ N) \rangle  
  +
  e^{-2\pi i \, s/r} \vert \,a^2 ~({\rm mod}~ N) \rangle 
  +
  e^{-2\pi i \, 2s/r} \vert \,a^3 ~({\rm mod}~ N) \rangle   
  + \cdots +
\nonumber\\[5pt] && \hskip1.0cm
  e^{-2\pi i \, (r-2)s/r} \vert \,a^{r-1} ~({\rm mod}~ N) \rangle  
  +
    e^{-2\pi i \, (r-1)s/r} \vert \, a^{r} ~({\rm mod}~ N) \rangle  
  \Big]
\\[10pt] 
%\end{eqnarray}
%\begin{eqnarray}
  &=& %\hskip-1.5cm =
  e^{2\pi i \, s/r} \, 
  \frac{1}{\sqrt{r}} \Big[\, 
  e^{-2\pi i \, s/r} \,
  \vert a^1 ~({\rm mod}~ N) \rangle  
  +
  e^{-2\pi i \, 2s/r} \,\vert a^2 ~({\rm mod}~ N) \rangle 
  +
  e^{-2\pi i \, 3s/r} \,\vert a^3 ~({\rm mod}~ N) \rangle   
\nonumber
\nonumber\\[5pt] && \hskip1.0cm
  + \cdots +
  e^{-2\pi i \, (r-1)s/r} \vert \, a^{r-1} ~({\rm mod}~ N) \rangle  
  \,+\,
  \underbrace{\,~e^{-2\pi i \, s}~\,}_{1} \,
  \vert \, a^{0} ~({\rm mod}~ N) \rangle  
  \Big]
\\
  &=&
  e^{2\pi i \, s/r} \, \vert u_s \rangle
  \ . 
\end{eqnarray}
Also note that the phases $e^{2\pi i k \, s/r}$ sum to zero 
over $s$ for any value of the non-zero integers $r$ and $k$. This is 
easy to prove,  as the sum is geometric and can be performed 
exactly:
\begin{eqnarray}
  \sum_{s=0}^{r-1} e^{2\pi i\, k s/r}
  &=&
  \sum_{s=0}^{r-1} \big[e^{2\pi i \, k /r}\big]^s
  = 
  \frac{1 - \big[e^{2\pi i \, k/r}\big]^r}{1 - e^{2\pi i \,k/r}}
  =
  \frac{1 - e^{2\pi i\, k}}{1 - e^{2\pi i\, k/r}}
  =
  0 
\end{eqnarray}
for any integer $k \ne 0$.   % \in \{0, 1, \cdots, M-1\}$.  
The sum over $s$ in (\ref{eq_us_expand}) therefore gives 
\begin{eqnarray}
  \frac{1}{\sqrt{r}}\sum_{s=0}^{r-1} \vert u_s \rangle 
  = \vert 1 \rangle
  \ ,
\label{eq_sum_us}
\end{eqnarray}
which will prove to be a key ingredient for Shor's algorithm.   
Note that only the first term \hbox{$\vert a^0 ~({\rm mod}~N) \rangle = 
\vert 1 \rangle$} contributes to the sum,  as all other terms 
have phases that add to zero.  In fact, if we multiply 
(\ref{eq_us}) or (\ref{eq_us_expand}) by $e^{2\pi i\, k s/r}$,
we remove the phase of the term $\vert a^k ~({\rm mod}~N) 
\rangle$, and upon summing over $s$ we find  the generalized 
result
\begin{eqnarray}
  \frac{1}{\sqrt{r}}\sum_{s=0}^{r-1} 
  e^{2\pi i k\, \phi_s} \, \vert u_s \rangle 
  =
  \vert a^k ~({\rm mod}~N) \rangle 
  =
  \vert f(k) \rangle
~~~\text{where}~~ \phi_s = s/r
  \ .
\label{eq_sum_kus}
\end{eqnarray}
\begin{figure}[t!]
\begin{centering}
\includegraphics[width=\textwidth]{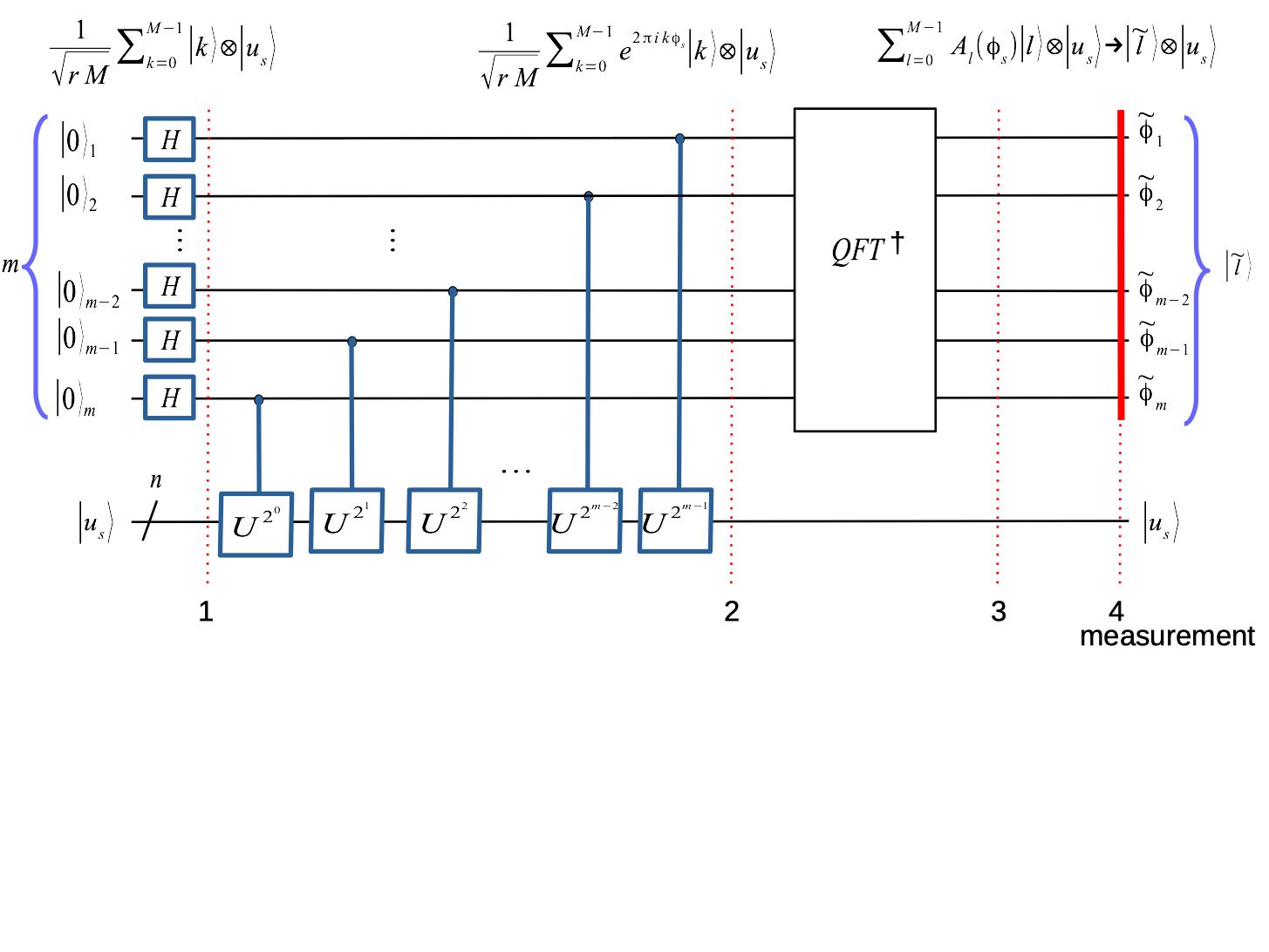} % n
\par\end{centering}
\vskip-4.0cm 
\caption{\footnoteskip  
First attempt at Shor's factoring circuit for the physics
and mathematics Convention~2. The control register 
has $m$ qubits, and the work register has $n = \lceil 
{\rm log}_2  N \rceil$ qubits.
We populate the work register with one of the Eigenstates 
$\vert u_s \rangle$,  where $U \vert u_s \rangle = e^{2\pi i 
\phi_s} \vert u_s \rangle$ with phase $\phi_s = s/r$ for some 
$s \in \{0, 1, \cdots, r-1\}$.  Measuring the control register 
projects the wavefunction into an Eigenstate $\vert \tilde 
\ell \, \rangle \otimes \vert u_s \rangle$, where the state 
is indexed by the $m$-bit integer $\tilde \ell = \tilde\phi_1
\cdots \tilde\phi_m$, from which we obtain the measured 
phase $\tilde \phi_\ell = \tilde\ell/2^m = 0.\tilde\phi_1 \dots 
\tilde\phi_m$ to $m$ bits of accuracy. We expect $\tilde\phi_\ell
\approx \phi_s$,  
thereby allowing us to determine $r$.  The only problem 
with this reasoning is that we do not know the state 
$\vert u_s \rangle$ in advance since we do not 
{\em a priori} know the period $r$.
}
\label{fig_shor_1}
\end{figure}

The central observation here is that every phase 
$\phi_s = s/r$, except $\phi_0 = 0$,  contains the period~$r$.  
This is the basis of Shor's algorithm: by measuring a phase
$\phi_s = s/r$ of the ME operator $U$ for which $s$ and $r$ 
are relatively prime,  we can infer the period $r$ of the function 
$f(x)$. And from the corresponding factoring procedure outlined 
above, we can then find non-trivial factors of~$N$.  

We can now address the quantum circuit for Shor's algorithm,
the first attempt of which is illustrated in Fig.~\ref{fig_shor_1}. 
We employ a QPE algorithm with the phase operator $U$ as 
defined in (\ref{eq_UaN_def}).  We emphasize that the QPE 
circuit consists of a {\em control} register of $m$ qubits and 
a {\em work} register of $n$ qubits.  As stated above,  the 
work register stores information about the number $N$. 
Suppose that the work register is populated by a specific 
Eigenstate $\vert u_s \rangle$ with phase $\phi_s = s/r$,  
so that $n =\lceil {\rm log}_2  N \rceil$. Note  that there 
are only $r$ phase states 
$\vert u_s \rangle$ out  of a possible $2^n$ work states,  
so that the states $\vert  u_s \rangle$ are very sparse indeed.
Upon measuring the final state of the control register, the 
wavefunction collapses into an Eigenstate $\vert \tilde\ell 
\, \rangle \otimes \vert u_s \rangle$ for $\tilde\ell \in \{0, 1, 
\cdots, M-1\}$.  The control register index takes the measured
value $\tilde \ell = \tilde\phi_1\cdots \tilde\phi_m$,  where 
each $\tilde\phi_k\in \{0, 1\}$ is the measured outcome 
of qubit $k \in \{1, 2, \cdots, m\}$ of the control register. 
From this we can readily infer the measured phase to be 
$\tilde \phi_\ell = \tilde\ell/2^m = 0.\tilde\phi_1\dots 
\tilde\phi_m$. From here on, we will place a tilde over 
measured quantities. If the $m$-bit resolution is sufficiently 
large,  then we expect $\tilde\phi_\ell \approx \phi_s$, 
from which we can extract the value of $r$ (provided 
$s$ and $r$ are relatively prime). In fact,  as we will 
establish in the next section, we should choose $m = 
2  n + 1$ to ensure sufficient phase resolution.

By populating the work register with $\vert u_s \rangle$,
Fig.~\ref{fig_shor_1} uses the QPE algorithm to find the 
phase $\phi_s =s/r$.  The problem with this method, however,   
is that we do not know the state $\vert u_s \rangle$ in 
advance,  as it depends upon the unknown period $r$.  
In fact,  if we knew the Eigenstates $\vert u_s \rangle$,
then we should also know the value of the period $r$,
and we would have no need for Shor's algorithm. 
We can circumvent this difficulty by employing 
(\ref{eq_sum_us}),  which we repeat here for emphasis:
\begin{eqnarray}
  \vert 1 \rangle 
  =
  \frac{1}{\sqrt{r}}\sum_{s=0}^{r-1} \vert u_s \rangle 
  \ .
\end{eqnarray}
\begin{figure}[t!]
\begin{centering}
\includegraphics[width=\textwidth]{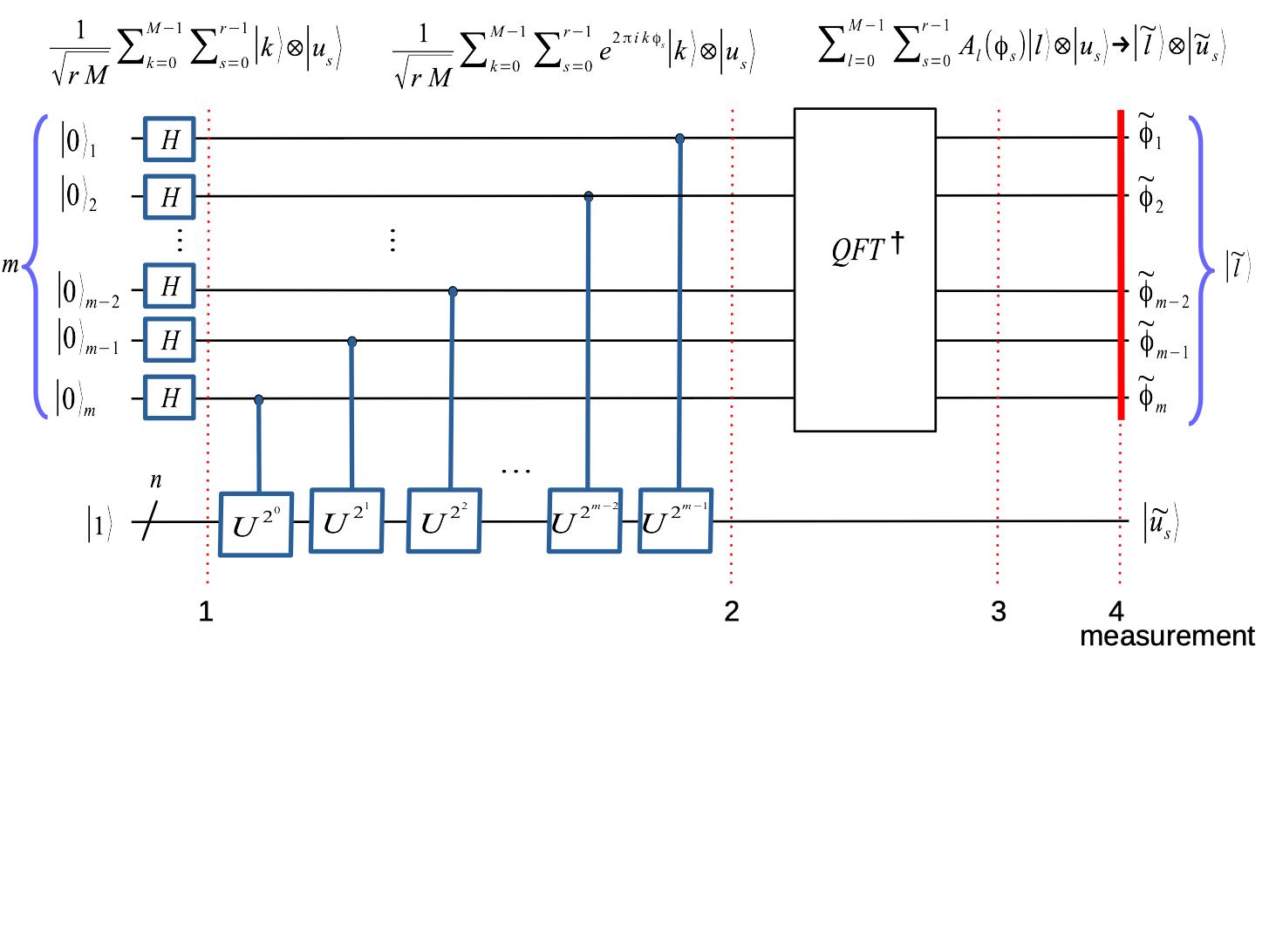} 
\par\end{centering}
\vskip-4.0cm 
\caption{\footnoteskip  
Shor's factoring circuit.   Convention 2.   We populate 
the work register with the state~$\vert 1 \rangle$,  which
is just a uniform linear combination of the Eigenstates
$\vert u_s \rangle$.  This requires no prior knowledge 
of these Eigenstates.  Through phase kickback,  the control 
register becomes a linear combination of terms involving 
the states $\vert u_s \rangle$ and their phases~$\phi_s$.  
Measuring the control register thereby projects the system 
into a state $\vert \tilde \ell \, \rangle \otimes \vert  \tilde u_s
\rangle$ for some $\tilde\ell \in \{0, 1, \cdots, M-1\}$ and
$s \in \{0,  1,  \cdots,  r-1\}$.  The control 
state is indexed by the $m$-bit integer $\tilde \ell = 
\tilde\phi_1\cdots \tilde\phi_m$,  where each $\tilde\phi_k
\in \{0, 1\}$ is the measured outcome of qubit $k \in \{1, 2, 
\cdots, m\}$ of the control register.  We then evaluate the 
$m$-bit measured phase $\tilde \phi_\ell = \tilde\ell /2^m 
= 0.\tilde\phi_1 \cdots \tilde\phi_m$, from which we can 
extract the exact phase $\phi_s = s/r$ using the method 
of continued fractions (provided $s$ and $r$ are relatively 
prime). 
}
\label{fig_shor_2}
\end{figure}
This suggests that we populate the work register with 
the easily prepared state \hbox{$\vert 1 \rangle = \vert 0 \cdots
0 1 \rangle$},  which is just 
a uniform linear combination of the phase Eigenstates 
$\vert u_s \rangle$.  This is achieved by initializing the 
lowest order qubit in the work register to the 1-state,  
and all other work qubits to 0-states.   This new circuit
is illustrated in Fig.~\ref{fig_shor_2}. Populating the 
work register with a linear superposition of the states 
$\vert u_s \rangle$ has the effect of rendering the 
state of the control register (through phase kickback) 
as a linear combination of the Eigenstates $\vert u_s 
\rangle$ and their corresponding phases $\phi_s$. 
More specifically, the state of the quantum system 
right after the front-end of the circuit (position 2 in 
the Figure) will be given by 
\begin{eqnarray}
  \vert \psi_2 \rangle
  =
  \frac{1}{\sqrt{r  M}}
  \sum_{k=0}^{M-1} \sum_{s=0}^{r-1} 
  e^{2\pi i k\,  \phi_s} \, 
  \vert k \rangle \otimes  \vert u_s \rangle
  \ .
\label{eq_theta_a}
\end{eqnarray}
Finally, the $QFT^\dagger$ operation (position 3 in the
Figure) transforms this state into
\begin{eqnarray}
  \vert \psi_3 \rangle
  =
  \sum_{\ell=0}^{M-1} \sum_{s=0}^{r-1} A_{\ell}(\phi_s) \,
  \vert \ell \rangle \otimes \vert u_s \rangle
   ~~~\text{with}~~ \phi_s = s/r
  \ ,
\label{eq_theta_b}
\end{eqnarray}
where the amplitudes are defined by 
\begin{eqnarray}
  A_\ell(\phi_s)
  &\equiv&
  \frac{1}{\sqrt{r} M}\,\frac{1 - e^{2\pi i\, ( \phi_s - \ell/M)M}}
  {1-e^{2\pi i\,  ( \phi_s - \ell/M)}}
\label{eq_Aktheta_def_a}
\\[5pt]
  &=&
  \frac{1}{\sqrt{r} M}\,\frac{1 - e^{2\pi i\, ( \phi_s - \phi_\ell)M}}
  {1-e^{2\pi i\,  ( \phi_s - \phi_\ell)}}
  =
  \frac{1}{\sqrt{r} M}\,\frac{1 - e^{2\pi i\, ( \ell_s - \ell)}}{1-e^{2\pi i\,  
  ( \ell_s - \ell)}/M}
  \ ,
\label{eq_Aktheta_def_a_x}
\end{eqnarray}
where $\phi_\ell \equiv \ell/M$ and $\ell_s \equiv  M \phi_s$.
The control register has therefore obtained knowledge 
of the phases $\phi_s = s/r$ and their corresponding 
Eigenstates~$\vert u_s  \rangle$.

Finally,  we must measure the control register (position 
4 in the Figure with the bold red line),  which collapses
the quantum state $\vert \psi_3 \rangle$ of (\ref{eq_theta_b}) 
into a definite Eigenstate state
\begin{eqnarray}
 \vert \tilde \ell \, \rangle \otimes \vert u_{\tilde s} \rangle 
  \ .
\label{eq_es_ks}
\end{eqnarray}
In other words, 
the control register collapses to a state $\vert \tilde\ell \,\rangle$  
labeled by the $m$-bit integer index $\tilde\ell = \tilde\phi_1 \,
\tilde\phi_2 \cdots \tilde\phi_m\in \{0, 1, \cdots, M-1\}$,
where the measured phase is then given by $\tilde\phi_\ell 
\equiv \tilde \ell/M = 0.\tilde\phi_1\, \tilde\phi_2 \cdots 
\tilde \phi_m$. Likewise, the work register is projected into a 
state $\vert u_{\tilde s} \rangle$ determined by a random 
choice of the phase integer \hbox{$\tilde s \in \{0, 1, \cdots, r-1\}$}.   
The ME phase angle will be written $\tilde\phi_s = \tilde s/r$,  
and we define the corresponding mode $\tilde\ell_s = M 
\tilde\phi_s$.  The probability of measuring the Eigenstate 
state (\ref{eq_es_ks}) is therefore
\begin{eqnarray}
  P_{\tilde\ell,  \tilde s}
  &=&
   \big\vert A_{\tilde\ell}(\tilde \phi_s) \big\vert^2
  =
  \frac{1 }{r M^2} ~
  \frac{\sin^2\left[ \pi \left(\tilde\phi_s  - 
  \displaystyle\frac{\tilde\ell}{M} \right) M\right]}
  {\sin^2\left[ \pi \left(\tilde\phi_s  - \displaystyle\frac{\tilde\ell}{M}  
  \right) \right]}
\\[5pt]
  &=&
  \frac{1 }{r M^2} ~
  \frac{\sin^2\left[ \pi \big(\tilde\phi_s  -  \tilde\phi_\ell \big) M \right]}
  {\sin^2\left[ \pi \big(\tilde\phi_s  - \tilde\phi_\ell \big) \right]}
  =
  \frac{1 }{r M^2} ~
  \frac{\sin^2\left[ \pi \big(\tilde\ell_s  -  \tilde\ell \,\big) \right]}
  {\sin^2\left[ \pi \big(\tilde\ell_s  - \tilde\ell \,\big)/M \right]}
  \ . ~~~~~~
\end{eqnarray}
In applying Shor's algorithm it will be critical that $\tilde\phi_{\ell}$ 
is a rational number,  and that $\tilde s$ and $r$  are relatively 
prime.  We will usually denote the state $\vert u_{\tilde s} \rangle$ 
by the simpler form $\vert \tilde u_s \rangle$, in which case we will 
drop the tilde from the ME phase and write $\phi_s = s/r$.  Note
that the control and work registers are entangled,  and that the 
most likely control-output states $\tilde\ell$ are those for which 
$\tilde\phi_\ell \approx \phi_s$ for some $s \in\{0, 1, \cdots, r-1\}$. 
The rub is then to extract the {\em exact} phase $\phi_s$ 
from the {\em approximately} 
measured $m$-bit phase $\tilde \phi_\ell$,  and in particular 
we must pull out the integer period $r$ from the rational 
number $\tilde \phi_\ell$.  This requires the mathematical 
technique of {\em continued fractions},  which we shall 
apply in the next section.  If we obtain the zero-phase 
state $\phi_0 = 0$,  or if $s$ and $r$ have non-trivial
common factors,  then the measurement will fail,  and 
we must try again.  Once we have a potential value
for the period $r$, we must explicitly check that $r$ 
is even,  and that $a^r = 1 ~({\rm mod}~N)$. We must 
also ensure that  $a^{r/2} \ne \pm 1 ~({\rm mod}~N)$.   
If these criteria are not met, then we must try again.
Thus,  Shor's algorithm is probabilistic in nature, but
 the chance of a successful run is quite high, and usually 
 requires at most two or three attempts.  

Let us now work through the circuit in Fig.~\ref{fig_shor_2}
in more detail.  First,  we must prepare the zero-state
\begin{eqnarray}
   \vert 0 \rangle^{\otimes m } \otimes \vert 0 
  \rangle^{\otimes n}
  \ ,
\end{eqnarray}
and then apply an $X$ operator to the lowest order qubit 
of the work register,  thereby producing the initial state 
\begin{eqnarray}
  \vert \psi_0 \rangle
  &=&
   \vert 0 \rangle^{\otimes m } \otimes \vert 1 \rangle
  \ .
\end{eqnarray}
We are using a short-hand notation for the states in 
the $n$-qubit work register,   where we denote the 
computational basis states by their corresponding 
integer index, 
%in the range $0,  1,  \cdots,  2^\ell-1$,  
rather than breaking them out into their respective 
tensor products of single-particle qubit states.  In 
the physics Convention 2,  the explicit form of the 
work register would be 
\begin{eqnarray}
   \vert 1 \rangle
  =
  \vert 0 \rangle_{1}  
  \otimes \cdots \otimes 
  \vert 0 \rangle_{n-1} \otimes  \vert 1 \rangle_n
  \ ,
\end{eqnarray}
and in the Qiskit Convention 1 we would have
\begin{eqnarray}
   \vert 1 \rangle
  =
  \vert 1 \rangle_{0}  \otimes \vert 0 \rangle_{1}   
  \otimes \cdots \otimes \vert 0 \rangle_{n -1}
%  =
%  \vert 0 \cdots 0 1 \rangle
  \ .
\end{eqnarray}
We should also note that the phase operator $U$ is distributed 
over the work register,  and it takes different forms between 
Conventions 1 and 2.  

Hadamard gates $H^{\otimes m}$ are then applied to the 
control register,  which splits the state $\vert 0 \rangle^{\otimes m}$ 
into a linear superposition of all possible control states $\vert 
k \rangle$ for $k \in \{0, 1,  \cdots,  M -1\}$ with $M = 2^m$.   
Upon using (\ref{eq_sum_us}),  we can therefore express the 
resulting state in either of two forms,  
\begin{eqnarray}
  \vert \psi_1 \rangle
  &=&
  \frac{1}{\sqrt{M}}
  \sum_{k=0}^{M-1} \vert k \rangle \otimes \vert 1 \rangle
\label{eq_psi1_one}
\\[5pt]
  &=&
  \frac{1}{\sqrt{r M}} 
 \sum_{k=0}^{M-1} \sum_{s=0}^{r-1}
  \, \vert k \rangle \otimes \vert u_s \rangle
  \ ,
\label{eq_psi1_us}
\end{eqnarray}
each of which reveals something essential about the 
quantum system.   For example,  (\ref{eq_psi1_us}) tells 
us that the state $\vert \psi_1 \rangle$
 is in fact just a uniform linear superposition 
involving the Eigenstates~$\vert u_s \rangle$,   which
was crucial to the above measurement analysis.  Next,  
the circuit operates on the state $\vert \psi_1 \rangle$ 
with a sequence of $m$ controlled-phase operators $CU^p$ for 
\hbox{$p\in \{2^0,  2^1,  \cdots,  2^{m-1}\}$},  giving 
the state
\begin{eqnarray}
  \vert \psi_2 \rangle
  &=&
   \frac{1}{\sqrt{ r\, M}}
   \sum_{k=0}^{M-1}  \sum_{s=0}^{r-1}
  e^{2\pi i k \, \phi_s} \, \vert k \rangle
  \otimes\,  \vert u_s \rangle 
  ~~~\text{for}~~ \phi_s = \frac{s}{r}
  \ .
\label{eq_psi2_us}
\end{eqnarray}
By employing (\ref{eq_sum_kus}),  we can express this
state in the alternative form 
\begin{eqnarray}
  \vert \psi_2 \rangle
  &=&
   \frac{1}{\sqrt{M}}
  \sum_{k=0}^{M-1}    \, 
  \vert k \rangle \otimes\,  \left(
  \frac{1}{\sqrt{r}}\sum_{s=0}^{r-1} e^{2\pi i\, k \phi_s} 
  \vert u_s \rangle 
  \right)
  \\
  &=&
    \frac{1}{\sqrt{M}}
   \sum_{k=0}^{M - 1} \vert k \rangle \otimes 
   \vert a^k ~({\rm mod}~N) \rangle
  =
  \frac{1}{\sqrt{M}}
   \sum_{k=0}^{M - 1} \vert k \rangle \otimes 
   \vert f(k) \rangle
  \ .
\label{eq_psi2_akmodN}
\end{eqnarray}
Recall that the period of the function 
$f(x) = a^x ~({\rm mod}~N)$ is $r$,  and we therefore 
find the following sequence 
of operations on the state $\vert 1 \rangle$:
\begin{eqnarray}
  U^1 \vert 1 \rangle 
  &=&
  \vert a  ~({\rm mod}~ N) \rangle 
\nonumber\\
  U^2 \vert 1 \rangle 
  &=&
  U   \vert a  ~({\rm mod}~ N) \rangle 
 =
  \vert a^2 ~({\rm mod}~ N) \rangle
\nonumber\\
  &\cdots&
\nonumber\\
  U^k \vert 1 \rangle 
  &=&
  U   \vert a^{k-1}  ~({\rm mod}~ N) \rangle 
 =
  \vert a^k ~({\rm mod}~ N) \rangle
\\
  &\cdots&
\nonumber
\\
%\end{eqnarray}
%\vfill
%\pagebreak
%\noindent
%\begin{eqnarray}
  U^{r-1} \vert 1 \rangle 
  &=&
  U   \vert a^{r-2}  ~({\rm mod}~ N) \rangle 
 =
  \vert a^{r-1} ~({\rm mod}~ N) \rangle
\nonumber\\[5pt]
  U^r \vert 1 \rangle 
  &=&
  U   \vert a^{r-1}  ~({\rm mod}~ N) \rangle 
 =
  \vert a^r ~({\rm mod}~ N) \rangle
  =
  \vert 1 \rangle 
\nonumber\\[5pt]
%  &\cdots&
%\nonumber\\
  U^{r+1} \vert 1 \rangle 
  &=&
  U   \vert a^{r}  ~({\rm mod}~ N) \rangle 
  =
  \vert a^{r+1} ~({\rm mod}~ N) \rangle
  =
  \vert a ~({\rm mod}~ N) \rangle
\nonumber\\
  U^{r+2} \vert 1 \rangle 
  &=&
  U   \vert a^{r+1}  ~({\rm mod}~ N) \rangle 
  =
 \vert a^{r+2} ~({\rm mod}~ N) \rangle
 =
  \vert a^2 ~({\rm mod}~ N) \rangle
\nonumber\\
  &\cdots&  ~~ \ ,
\nonumber
\end{eqnarray}
which can be summarized by
\begin{eqnarray}
  U^x  \vert 1 \rangle = \vert f(x) \rangle 
  ~~\text{and}~~
  \vert f(x + 1) \rangle = U\, \vert f(x) \rangle 
\end{eqnarray}
for any non-negative integer $x$.  We shall use these relations
later in the manuscript when constructing the ME operators
$U$.  Note that we can express the front-end output state 
from (\ref{eq_psi2_akmodN}) in the form
\begin{eqnarray}
  \vert \psi_2 \rangle
  &=&
   \frac{1}{\sqrt{M}}
   \sum_{k=0}^{M - 1} \vert k \rangle \otimes 
   U^k \, \vert 1 \rangle
%  =
%   \frac{1}{\sqrt{M}}
%   \sum_{k=0}^{M - 1} \vert k \rangle \otimes 
%   \vert f(k) \rangle
  \ ,
\label{eq_psi2_akmodN_x}
\end{eqnarray}
which shows that the ME operators between
point-1 and point-2 of Fig.~\ref{fig_shor_2} have the effect 
of augmenting the work register by $U^k \vert 1 \rangle$ for
every mode $k \in \{0, 1, \cdots, M-1\}$.
We have almost finished analyzing the circuit.  Now that 
we have state $\vert \psi_2 \rangle$ of (\ref{eq_psi2_us}) 
in hand,  we apply the inverse quantum Fourier transform 
$QFT^\dagger$ to the control register,  producing the 
final state 
\begin{eqnarray}
  \vert \psi_3 \rangle
  &=&
   \sum_{\ell=0}^{M-1}\sum_{s=0}^{r-1}  \,
   A_\ell(\phi_s) \, \vert \ell \rangle   \otimes \vert u_s \rangle
  \ ,
\end{eqnarray}
where the amplitudes $A_\ell(\phi_s)$ are given by 
(\ref{eq_Aktheta_def_a}). We now have a  linear superposition 
of the states $\vert \ell \rangle \otimes \vert u_s \rangle$,  
and as described above,  a measurement on the control 
register will produce any one of them, 
\begin{eqnarray}
   \sum_{\ell=0}^{M-1}\sum_{s=0}^{r-1}  \,
   A_\ell(\phi_s) \, \vert \ell \rangle   \otimes \vert u_s \rangle
\to
 \vert \tilde \ell \rangle \otimes \vert \tilde u_s \rangle 
\label{eq_es_ks_x1}
\end{eqnarray}
with probability $P_{\tilde\ell,s} = \vert A_{\tilde\ell}(\phi_s) 
\vert^2$. The measured phase $\tilde \phi_\ell$ is given in 
terms of the measured output index $\tilde \ell$ of the 
control register by \hbox{$\tilde\phi_\ell = \tilde \ell/2^m$}, 
and Shor's algorithm is designed to give dominant peaks 
close to the exact phases $\phi_s = s/r$. We will have more 
to say about the details of this procedure in later sections.  
For now, we summarize the action of the Shor circuit in 
Table~\ref{table_quantum_period}. Note that when the 
exact phases $\phi_s = s/r$ can be represented by
$m$-bit fractions, the phase histogram simplifies 
considerably: There are exactly $r$ equally likely 
peaks, each corresponding to one of the phases 
$\phi_s$. This situation is relatively rare, and does 
not occur in most cases, thereby allowing for more 
complex phase histograms.

\begin{table}[h!]
\caption{\footnoteskip 
   Quantum Period Finding
  }
\label{table_quantum_period}
\end{table}
\begin{enumerate}
  \baselineskip 10pt plus 1pt minus 1pt
  \setlength{\itemsep}{5pt} % single spacing
  \setlength{\parskip}{1pt} %
  \setlength{\parsep}{0pt}  %
\item[1.] Initialize the state:
\\[5pt]
Prepare the state $\vert 0 \rangle^{\otimes m} \otimes \vert 0 
\rangle^{\otimes n}$,   and apply  $X$ to the lowest order 
qubit of the work register to produce the initial state
\begin{eqnarray}
  \vert \psi_0 \rangle
  &=&
  \vert 0 \rangle^{\otimes m} \otimes \vert 1 \rangle
  \ ,
\end{eqnarray}
where the number of work qubits is given by $n = \lceil \log_2 N \rceil$.
\item[2.] Randomize the control register:
\\[5pt]
Apply $H^{\otimes m}$ to the control register to give 
\begin{eqnarray}
  \vert \psi_1 \rangle
  &=&
  \frac{1}{\sqrt{M}}
  \sum_{k=0}^{M-1} \vert k \rangle \otimes \vert 1 \rangle
\\[5pt]
  &=&
  \frac{1}{\sqrt{r M}} 
  \sum_{k=0}^{M-1} \sum_{s=0}^{r-1}  \, 
  \vert k \rangle \otimes \vert u_s \rangle
  \ ,
\end{eqnarray}
where the total number of quantum states is $M = 2^m$. 

\item[3. ] Modular exponentiation (ME):
\\[5pt]
Conditionally apply the ME operators $U^p$ for $p \in \{2^0,  
2^1,  \cdots,  2^{m-1} \}$ successively to the control qubits to 
produce the state
\begin{eqnarray}
  \vert \psi_2 \rangle
  &=&
   \frac{1}{\sqrt{ r\, M}}
   \sum_{k=0}^{M-1}  \sum_{s=0}^{r-1} 
  e^{2\pi i k \,  \phi_s} \, \vert k \rangle
  \otimes\,  \vert u_s \rangle 
  ~~~\text{where}~~ \phi_s = \frac{s}{r}
\\[5pt]
    &=&
  \frac{1}{\sqrt{M}}
  \sum_{k=0}^{M-1} \vert k \rangle \otimes 
  \vert a^k ~({\rm mod}~N) \rangle
  \ .
\end{eqnarray}
\item[4. ] Perform the inverse Fourier transform:
\begin{eqnarray}
  \vert \psi_3 \rangle
  &=&
  QFT^\dagger \vert \psi_2 \rangle
   =
  \sum_{\ell=0}^{M-1}\sum_{s=0}^{r-1}  \,
   A_\ell(\phi_s)  \, \vert \ell \rangle   \otimes \vert u_s \rangle
  \ ,
\end{eqnarray}
where the amplitudes $A_\ell(\phi_s)$ are given by 
(\ref{eq_Aktheta_def_a}). 

\item[5. ] Perform a measurement of the control register: 
\\[5pt]
The state then collapses into an Eigenstate, 
\begin{eqnarray}
  \sum_{\ell=0}^{M-1}\sum_{s=0}^{r-1}  \,
  A_\ell(\phi_s) \, \vert \ell \rangle   \otimes \vert u_s \rangle
  \to 
  \vert \tilde \ell \, \rangle   \otimes \vert \tilde u_s \rangle
  \ ,
\end{eqnarray}
with probability $P_{\tilde\ell,s}= \vert A_{\tilde\ell}(\phi_s) 
\vert^2$. The probability peaks at values of $\tilde\ell$ 
close to the exact phases $\phi_s = s/r$, and the outcomes 
of the measurement are equally distributed between the 
values of \hbox{$s \in \{0, 1, \cdots, r-1\}$}.

\item[6. ] Apply the method of continued fractions to
extract the exact phase $\phi_s = s/r$ from the 
approximately measured $m$-bit phase $\tilde\phi_\ell$. 
This provides the exact period $r$,  and therefore the
factors of $N$. 
\end{enumerate}
%%
%\label{table_quantum_period}
%\end{table}
%%

%\vfill
%\clearpage
%\pagebreak

\subsection{Extracting the Exact Period from 
the Measured Phase}
\label{sec_extract_phase}

We turn now to the (non-trivial) task of extracting 
the {\em exact} phase $\phi_s = s/r$ from the {\em 
approximately} measured phase $\tilde\phi_\ell$ of the 
control resister using the method of {\em continued 
fractions} outlined in Section~\ref{sec_cont_frac}. 
For ease of notation, we shall henceforth drop the 
superfluous $\ell$-subscript from $\tilde\phi_\ell$
and simply write $\tilde\phi$. 
Since the measured phase $\tilde\phi$ is a positive 
rational number less than one,  it can be expressed  
as a finite continued fraction of the form 
\begin{eqnarray}
  \tilde \phi
  =
  \displaystyle\frac{1}{a_1 +  \displaystyle\frac{1}
  {a_2 +   \displaystyle\frac{1}{a_3 
  + \ddots + 
  \frac{1}{a_\smR}}}}
\label{eq_a0dotsaN}
\end{eqnarray}
for some integer $R$,  where $a_1,  a_2,  \cdots,  a_\smR$ 
are all positive integer coefficients.  We shall drop the zero 
digit $a_0=0$,  and simply write $\tilde\phi = [a_1, a_2,  
\cdots,  a_\smR]$ to denote the form of the continued fraction.  
Suppose now that $s$ and $r$ are two relatively prime 
positive integers that satisfy the inequality
\begin{eqnarray} 
  \left\vert \frac{s}{r} - \tilde \phi \right\vert 
  \le \frac{1}{2 r^2}
  \ .
\label{eq_cf_1}
\end{eqnarray}
Then by Theorem~\ref{thm_c} of the Section 
\ref{sec_cont_frac},   the ratio $s/r$ is necessarily 
a {\em convergent} of the continued fraction 
(\ref{eq_a0dotsaN}) for~$\tilde\phi$,  so that $s/r 
= [a_1,  a_2,  \cdots,  a_q]$ for some $q \le R$.  We 
shall exploit this fact to extract the exact period~$r$ 
of the modular exponential function $f_{a,  \smN}
(x) = a^x ~({\rm mod} ~N)$.  

Continued fractions and their convergents can be 
calculated efficiently using a variation of the Euclidean 
algorithm for finding the greatest common divisor of 
two integers.  Let us call  the convergents of
$\tilde\phi$ by $s_0/r_0,  s_1/r_1,  \cdots,  s_\ell/
r_\ell,  \cdots,  s_q/r_q$.  We then cycle through 
these convergents,  from the smallest to the largest 
values of the $r$'s,  checking to ensure that  
\begin{eqnarray}
 r \text{ is even}
\label{eq_c1} 
\\
  a^{r/2} \ne \pm 1 ~({\rm mod}~N)
\label{eq_c2}
\\
  a^{r} = 1 ~({\rm mod}~N) 
\label{eq_c3}
\end{eqnarray}
for each $r = r_\ell$.  Condition (\ref{eq_c2}) simply 
means that $b = a^{r/2}$ is not a trivial root of unity.   
In passing,  we note that $r$ can in fact be odd, provided 
that $a$ is a {\em perfect square},  so that $b = a^{r/2}$ 
is still an integer\,\cite{twoone}.  Apart from this caveat, 
if any of  
the conditions (\ref{eq_c1})--(\ref{eq_c3}) are not met
for $r_\ell$,  then the trial fails,  and we move on to 
the next convergent.  The special cases $s_\ell = 0$ and 
$s_\ell =1$ correspond to the phases $\phi_0 =0/r_\ell 
= 0$ and $\phi_1 = 1/r_\ell$.  Technically, we cannot use 
the method of continued fractions for these cases since 
$0/r_\ell$ and $1/r_\ell$ are not the ratio of two primes 
(as 0 and 1 are not prime).  In the former case,  $\phi_0 =0$ 
yields no information,  and we must move on to the next 
iteration.  However,  $\phi_1 = 1/r_\ell$ yields an integer 
$r_\ell$,  which could in principle be the correct period 
that we are seeking.  Therefore we shall check the case
$s_\ell = 1/r_\ell$ just to make sure,  even though $1/r_\ell$ 
does  not satisfy the conditions of Theorem~\ref{thm_c}.  
In any event,  a detailed analysis shows that the probability 
of achieving a solution after just a few trials is quite high.   
The solution for the smallest value of $r_\ell$ gives the 
period $r = r_\ell$ that we are seeking,  and the factors 
of $N$ are then given by ${\rm gcd}(a^{r/2} \pm 1,  N)$.

In closing this section,  we should call attention to an 
important feature of the method.  We must somehow 
ensure that inequality (\ref{eq_cf_1}) always holds,  as 
it would be quite cumbersome if we had to check this 
by hand every time.  However,  this requirement can be 
hard-wired into the algorithm itself by choosing an 
appropriate number of qubits $m$ for the control 
register,  one that is based on the inequality (\ref{eq_cf_1}) 
itself.  Recall that the work register has $n = \lceil 
\log_2 N \rceil$ qubits,  and thus for any phase 
$\phi_s = s/r$,  we have $r < N \le 2^n$.  This leads 
to the inequality 
\begin{eqnarray}
  \frac{1}{2^{2n+1}} \le \frac{1}{2r^2 }
  \ . 
\end{eqnarray}
Therefore,  if we take the control register to have 
$m = 2n+ 1$ qubits,  then a measurement of
the phase $\tilde \phi$ will necessarily have
sufficient precision to ensure that 
\begin{eqnarray}
   \left\vert \frac{s}{r} - \tilde \phi \right\vert 
 =
  \left\vert \phi_s - \tilde\phi \right\vert
  \le 
  \frac{1}{2^{2n+1}}  \le  \frac{1}{2r^2}
  \ ,
\end{eqnarray}
and then inequality (\ref{eq_cf_1}) is automatically
satisfied.  This is the source of our previous requirement
that $m = 2n + 1$.  Of course we must use even
more control qubits to account for machine error.
For a probability of success at least as large as $1 
- \varepsilon$,  where $\varepsilon > 0$ is a small 
probability of failure,  we must use $m = 2n + 1 
+ n_\varepsilon$ qubits in the control register,  
where \hbox{$n_\varepsilon \equiv \lceil  {\rm log
}_2\big(2 + 1/(2 \varepsilon) \big) \rceil$}.  With this 
choice of $m$,  upon taking a measurement of the 
control register we find:  (i) inequality (\ref{eq_cf_1}) 
will always be satisfied,  (ii) the exact phase will be 
of the form $\phi_s  = s/r$,  where $s \in \{0, 1,  \cdots,  
r-1\}$ is randomly selected,  and (iii) the ratio $s/r$ 
will be a convergent of the continued fraction for 
$\tilde\phi$ (provided that $s$ and $r$ are relatively
prime).  However,  we should emphasize that conditions 
(\ref{eq_c1})--(\ref{eq_c3}) must also be satisfied.  If 
they are not,  then the method will fail,  and we must 
move on to another iteration of Shor's algorithm.   
However,  a complete error analysis shows that the 
probability of success is quite high,  and typically only 
a few iterations will be required before obtaining a factor.  
We summarize Shor's algorithm in Table~\ref{tab_shor}. 
\begin{table}[h!]
\caption{\footnoteskip 
Shor's Factorization Algorithm}
\label{tab_shor}
\end{table}
\begin{enumerate}
  \baselineskip 10pt plus 1pt minus 1pt
  \setlength{\itemsep}{3pt} % single spacing
  \setlength{\parskip}{1pt} %
  \setlength{\parsep}{0pt}  %
\item[1.] Chose a random number $a \in \{2,  3, \cdots, N-1\}$ 
  for the base.   If ${\rm gcd}(a, N) = 1$,  then proceed to the 
  next step (otherwise we have found a non-trivial factor of 
  $N$ as required). 

\item[2.] Use quantum phase estimation (QPE) to  
  measure the phase $\tilde\phi$ of the modular 
  exponentiation operator $U_{a \smN}$ defined by 
  $$
  U_{a \smN} \vert w \rangle = \vert a \cdot w
  ~({\rm  mod}~N)  \rangle 
  ~~\text{with}~~ U_{a \smN} \vert u_s \rangle 
  = e^{2\pi i\, s/r} \, \vert u_s \rangle
  \ ,
  $$ 
  where $s \in \{0, 1, \cdots, r-1\}$. 
  The QPE is the only  quantum component of Shor's 
  algorithm.  This is also the bottleneck of the algorithm,
  as (i) most of the quantum resources are deployed here,
  and (ii) a different operator $U_{a \smN}$ is required 
  for every choice of $a$ and~$N$. 

\item[3.] We then use the method of continued fractions
  to extract the {\em exact} period $r$ from the {\em 
  approximately} measured phase $\tilde\phi$.  To do this,  
  we examine all convergents $\phi_s= s/r$ of $\tilde\phi$ 
  such that
  $$
  \left\vert \frac{s}{r} - \tilde\phi \right\vert \le \frac{1}{2r^2} 
  \ ,
  $$
  which is achieved by requiring the number of control 
  qubits to be $m = 2 n +1$.  We then check the convergents 
  from the smallest to the largest values of $r$.  If $r$ is odd,  
  then return to step 1.   If $a^{r/2} = \pm 1~({\rm mod}~N)$,  
  then return to step 1.   If $a^r = 1~({\rm mod}~N)$, then we 
  have found a solution for $r$,  and we proceed to the next step;  
  otherwise return to step 1. 

\item[4.]  Factors of $N$ are now given by  ${\rm gcd}(a^{r/2} 
  \pm 1,  N)$.  Finding the greatest common divisor can be done 
  very efficiently on a classical computer using Euclid's algorithm. 
\end{enumerate}

\vfill
\pagebreak
\clearpage
\section{Verifying the Theory: Factoring N = 15} % $\bm{N=15}$}
\label{sec_examples}

\begin{figure}[h!]
\begin{centering}
\includegraphics[width=\textwidth]{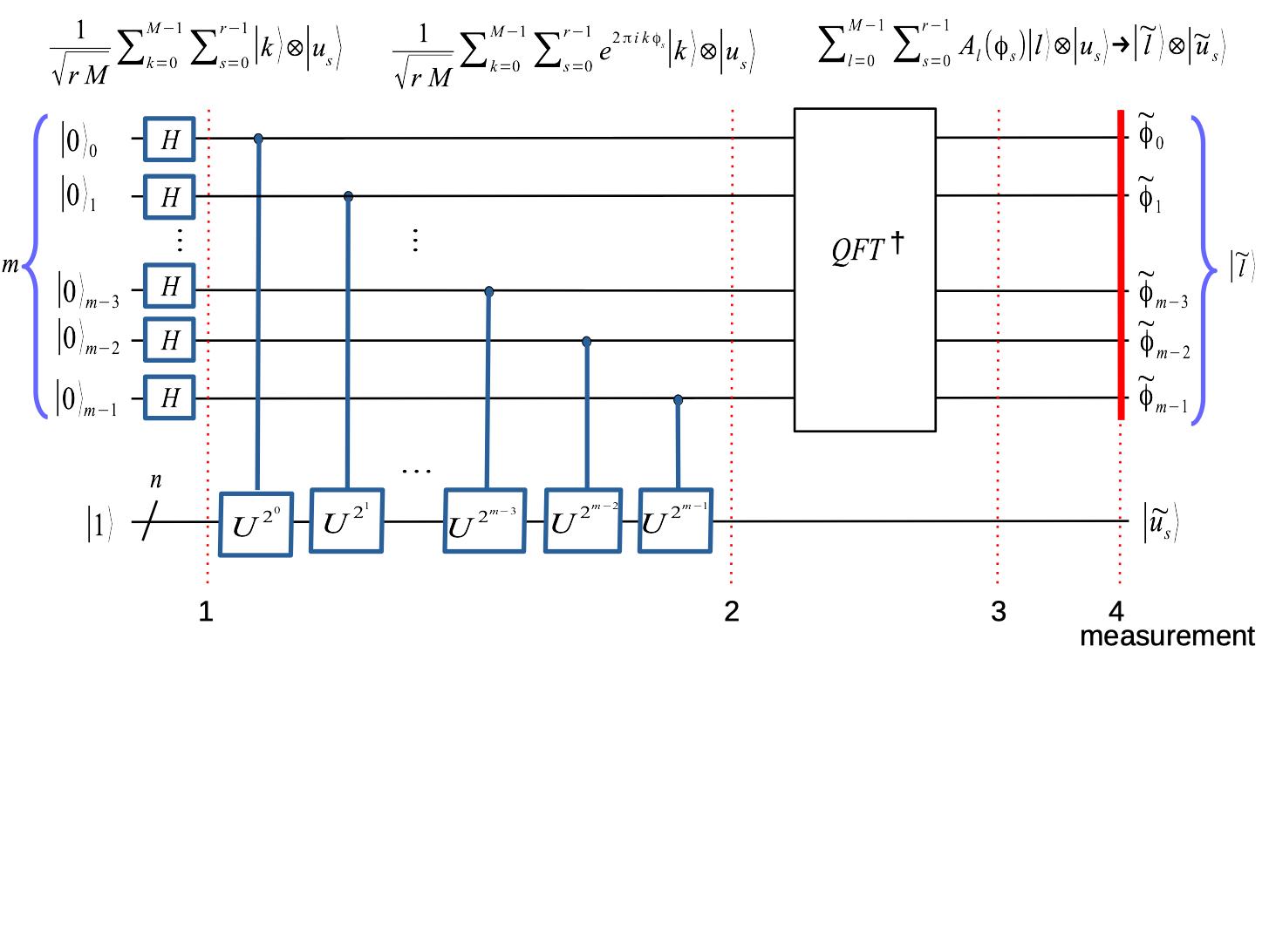}
\par\end{centering}
%\includegraphics[scale=0.40]{04_shor_4_three.jpg}
% 04_shor_N15_1.py
\vskip-3.8cm
\caption{\footnoteskip
Shor factorization algorithm for Qiskit 
convention~1.  In comparison,  Fig.~\ref{fig_shor_2}
uses the physics Convention 2. 
}
\label{fig_shor_4}
\end{figure}

To highlight the principal aspects of Shor's algorithm,  we now
build the computational machinery to factor the number 
$N = 15$.  In the next section we will employ the scripts 
developed here to factor larger and more complex numbers.
We shall employ IBM's circuit simulator Qiskit. This means
that we must use the Qiskit qubit ordering convention in 
which the upper 0-th qubit corresponds to the lowest order 
bit.  We developed the Shor factorization circuit in 
Section~\ref{sec_factor_circuit},  which is illustrated in 
Fig.~\ref{fig_shor_2}.   However, this analysis used the 
physics and mathematics ordering convention rather
than the Qiskit convention. Consequently, we must  
convert to the Qiskit ordering displayed in Fig.~\ref{fig_shor_4}.  
In this  convention,  the measurement of the control 
register (position 4 in the Figure,  and indicated by the 
bold red line across the register) gives the output state 
$\vert \tilde \ell\,\rangle =\vert \tilde\phi_{m-1} \cdots 
\tilde\phi_1 \, \tilde\phi_0 \rangle$.  
Just as in the previous section, we
will employ the convention in which the {\em measured} 
($m$-bit) phase in the control register is  written with 
a tilde,  and where the output state is indexed by the binary 
integer 
\begin{eqnarray}
  \tilde\ell
  &\equiv&
  \tilde\phi_{m-1} \, \tilde\phi_{m-2} \, \cdots \, \tilde\phi_1 \, 
  \tilde\phi_0
  ~~~\text{where}~~ \tilde\phi_k \in\{0, 1\}
\\[5pt]
  &=&
  2^{m-1}\,\tilde\phi_{m-1} +   2^{m-2}\,\tilde\phi_{m-2}  
  + \cdots + 
  2^1 \, \tilde\phi_{1} + 2^0 \, \tilde\phi_{0}
  \ .
\label{eq_ell_phi_m}
\end{eqnarray}
One must keep in mind that there is always an implicit 
$m$-bit resolution associated with any measured quantity in 
the control register, and in particular, the corresponding angular 
phase is given by the $m$-bit fraction
\begin{eqnarray}
 \tilde\phi_\ell
  \equiv
 \frac{\tilde\ell}{2^m}
  &=&
  \frac{\tilde\phi_{m-1}}{2^1} + \frac{\tilde\phi_{m-2}}{2^2} 
  +   \cdots + 
  \frac{\tilde\phi_1}{2^{m-1}} + \frac{\tilde\phi_0}{2^m} 
\\[5pt]
  &=&
  0.\tilde\phi_{m-1} \, \tilde\phi_{m-2} \, \cdots \,
  \tilde\phi_1 \, \tilde\phi_0
%  ~~~\text{where}~~ \tilde\phi_j \in\{0, 1\}
  \ .
\label{eq_phi_n}
\end{eqnarray}
For ease of notation,  we shall drop the $\ell$-subscript
and denote the measured phase by $\tilde\phi$. 

Since we will employ both binary and decimal numbers 
in this section,  we will often denote binary numbers using 
a bracket with a 2-subscript, writing $\tilde\phi= [0.\tilde
\phi_{m-1} \, \tilde\phi_{m-2} \, \cdots \,  \tilde\phi_1 \, \tilde
\phi_0]_2$.  We will always place a tilde over measured 
quantities,  so that $\tilde\phi$ denotes the phase as 
determined by a measurement of the $m$-bit control 
register, as opposed to the (as yet undetermined) exact 
phase $\phi_s = s/r$.  If the control register contains a 
sufficient number of qubits,  then the measurement 
should be quite accurate,  and the measured phase will 
be very close to the exact phase, $\tilde\phi \approx \phi_s
= s/r$.  Since $\tilde\phi$ and $\phi_s = s/r$ are approximately 
equal,  we can extract the {\em exact} value of the integers 
$s$ and $r$ using the method of continued fractions 
(provided that $s$ and $r$ are relatively prime). The 
integer $r$ is the period of the modular exponential 
function $f_{a \, \smN}(x) = a^x ~({\rm mod}~N)$ that we 
seek.

\begin{figure}[t!]
\begin{centering}
\includegraphics[width=\textwidth]{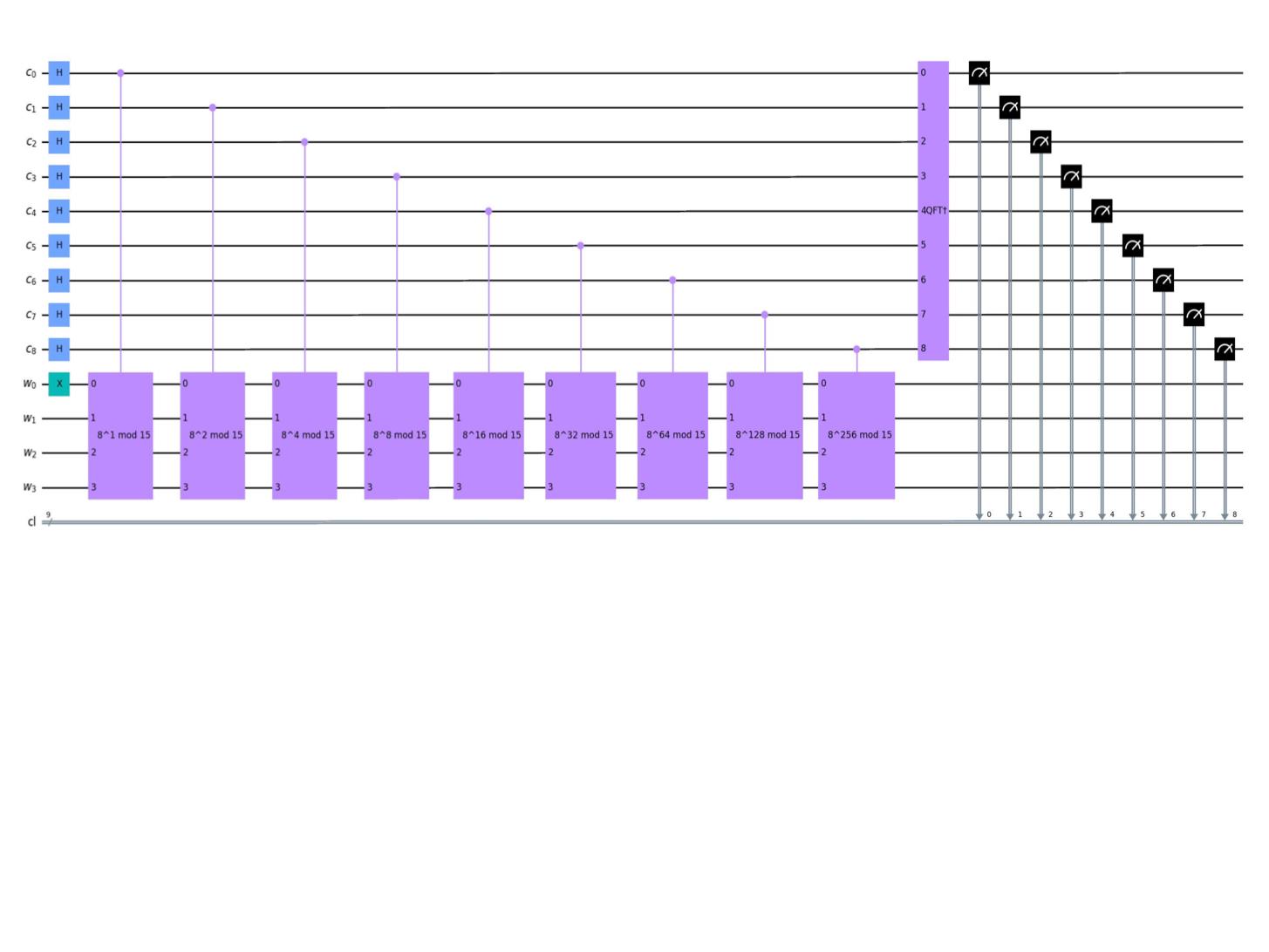}
\par\end{centering}
%\includegraphics[scale=0.70]{04_shor_N15_1_8_circuit_post.jpg}
% 04_shor_N15_1.py
\vskip-5.3cm
\caption{\footnoteskip
Qiskit version of Shor's factoring circuit for 
$N = 15$ and $a = 8$,  with $m=9$ qubits in the control 
register and $n =4$ qubits in the work register. 
}
\label{fig_15N_ex}
\end{figure}

Let us now concentrate on the specific example of
$N = 15$.  As we have established,  the work space 
requirement is $n = \lceil \log_2 15 \rceil  = 4$ qubits,  
and the control register must therefore contain 
\hbox{$m = 2 n + 1 = 9$} qubits (for simplicity we 
consider only perfect measurements in which 
\hbox{$n_\epsilon =0$}).  For a given integer~$N$,  
we choose the base $a$ to be a random integer such
that \hbox{$1 < a < N$} and ${\rm gcd}(a,N)=1$. 
Therefore,  for $N = 15$ we can only choose  $a \in 
\{ 2,4,7,8,11,  13, 14\}$.  It turns out that $a = 14$
gives a trivial root of unity, so we can neglect this
choice. In fact, in this example we shall take either 
$a = 4$ or $a = 8$,  where the former gives the period 
$r = 2$ and the latter gives $r = 4$. 
Figure~\ref{fig_15N_ex} illustrates the Qiskit circuit 
for Shor's algorithm with the base $a = 8$. The first 
9~qubits comprise the control register and are labeled 
by the index $c$,  and the last~4 qubits are the work 
register and are labeled by $w$.  We will therefore 
denote quantum states of the work register by  $\vert 
w_3 w_2 w_1 w_0 \rangle$.  In the Qiskit qubit convention,  
the work register is initially populated by the state 
\begin{eqnarray}
  \vert 1 \rangle  
  = 
  \vert 0 0  0 1 \rangle
  = \vert 1 \rangle_0 \otimes \vert 0 \rangle_1 \otimes
 \vert 0 \rangle_2 \otimes  \vert 0 \rangle_3 \ ,
\end{eqnarray}
with the least significant bit being $w_0=1$.  The controlled 
modular exponentiation operators $CU_{8, 15}^p$ for 
$p \in \{2^0,  2^1,  \cdots, 2^8\}$ are represented by 
purple boxes attached to their respective control qubits,  
and the operator $QFT^\dagger$ is represented by 
the large purple rectangle on the far right of the control 
register.  At the end of the circuit,  the control register 
undergoes a measurement on all $m$ qubits.  The 
work register might or might not undergo a measurement,  
and we shall return to this point later in the section. 

\subsection{Modular Exponentiation Operators}

\begin{figure}[b!]
\begin{minipage}[c]{0.4\linewidth}
\vskip-0.4cm
\includegraphics[scale=0.45]{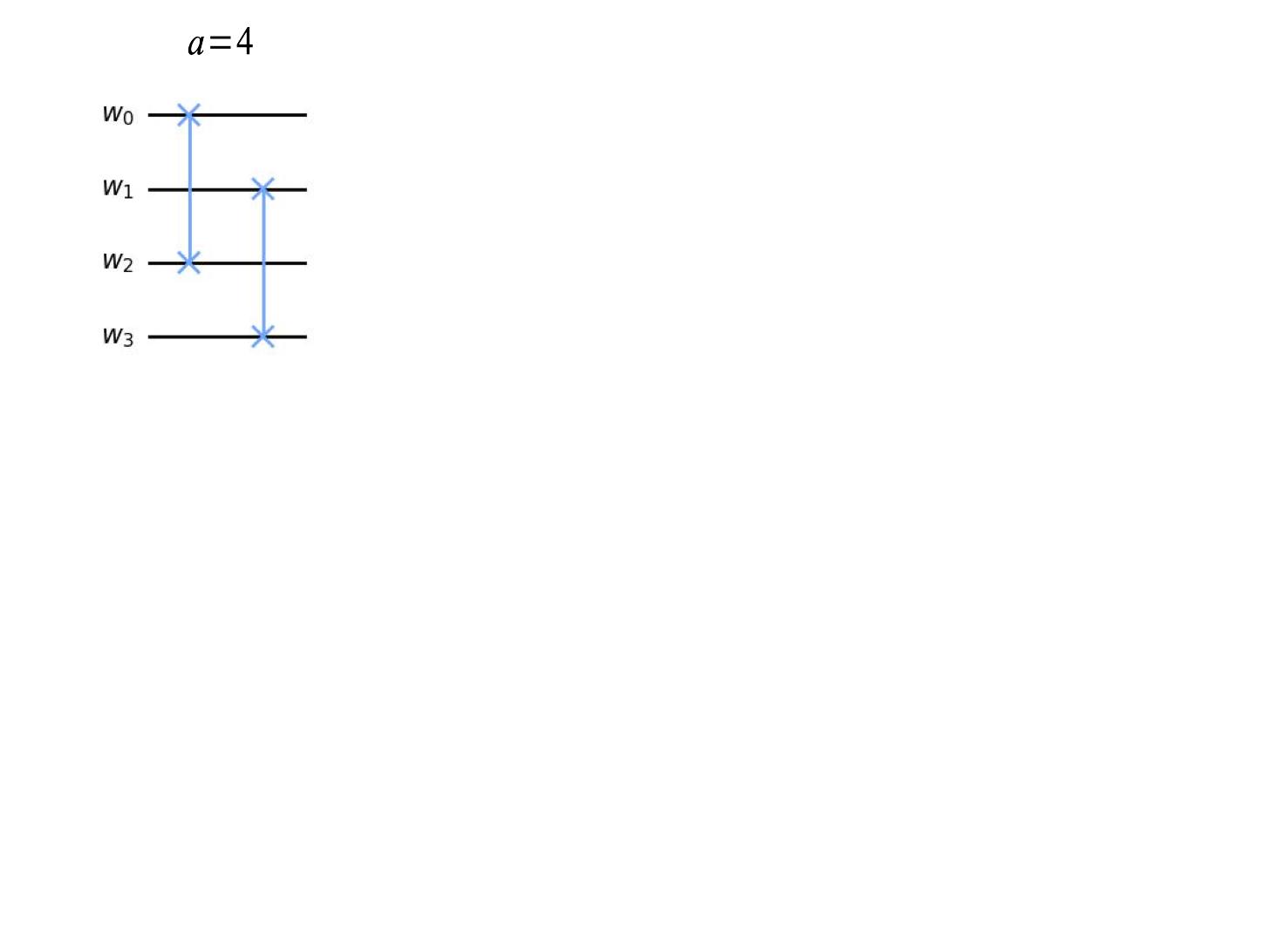} 
\end{minipage}
\begin{minipage}[c]{0.4\linewidth}
\vskip-0.4cm
\includegraphics[scale=0.45]{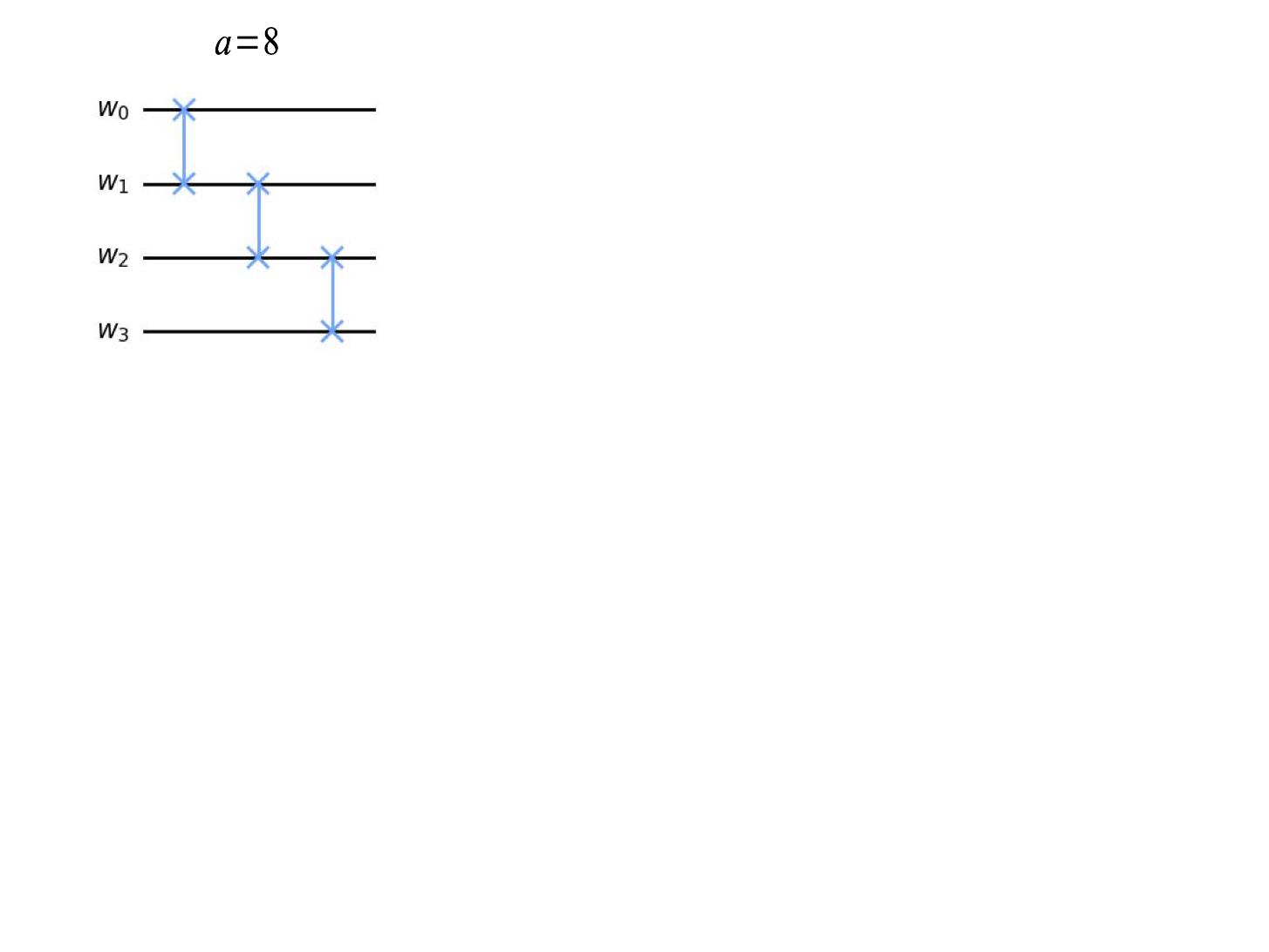} 
\end{minipage}
\vskip-5.8cm
\caption{\footnoteskip
Modular exponentiation operators $U_{a , 15}$ for 
$a=4$ in the left panel and $a = 8$ in the right panel.   
Their action on the work space is given by $U_{4,15} 
\vert w_3 w_2 w_1 w_0 \rangle = \vert w_1 w_0 w_3 
w_2 \rangle$ and $U_{8,15}\vert w_3 w_2 w_1 w_0 
\rangle = \vert w_0 w_3 w_2 w_1 \rangle$,  which can 
be reproduced by a sequence of SWAP gates. 
}
\label{eq_N15_circuit}
\end{figure}

\begin{figure}[h!]
\begin{minipage}[c]{0.4\linewidth}
\includegraphics[scale=0.40]{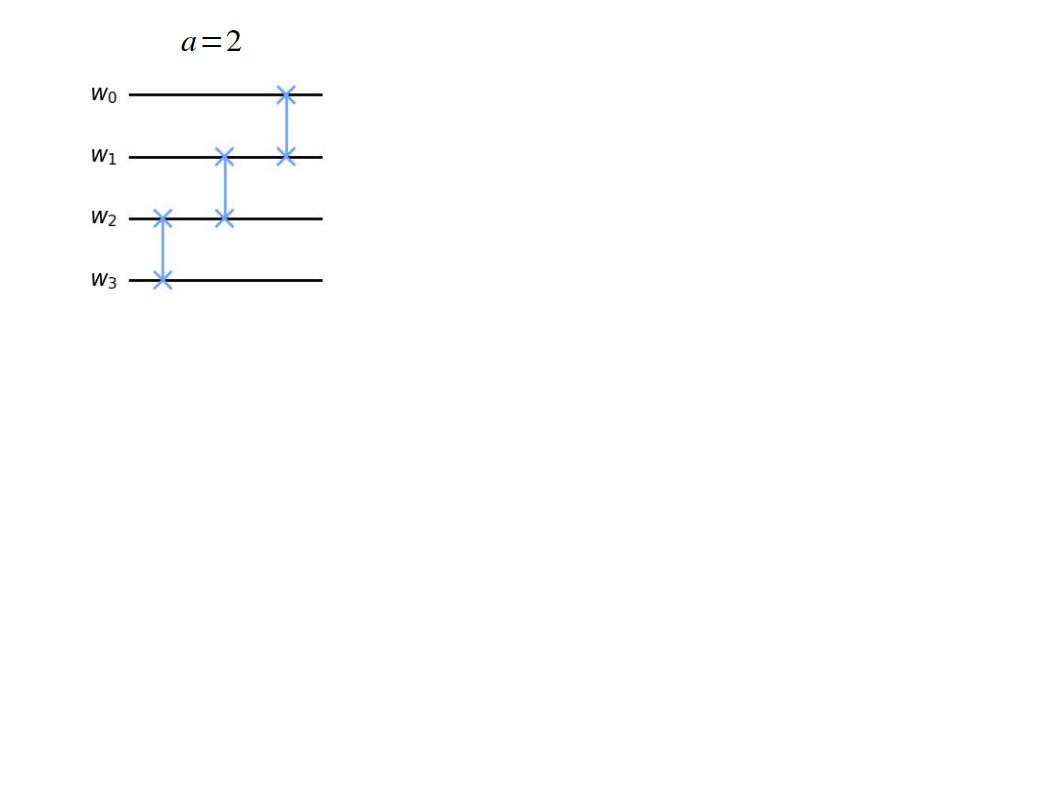} 
\end{minipage}
\begin{minipage}[c]{0.4\linewidth}
\includegraphics[scale=0.40]{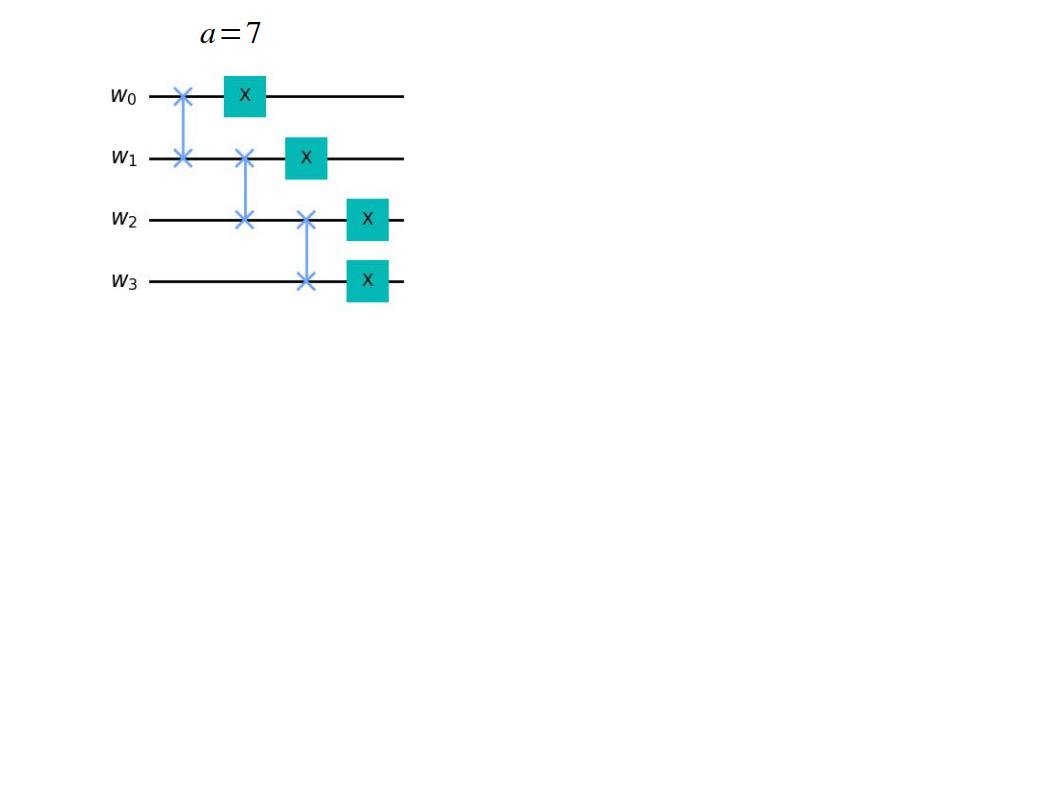} 
\end{minipage}
\vskip-5.0cm
\begin{minipage}[c]{0.4\linewidth}
\includegraphics[scale=0.40]{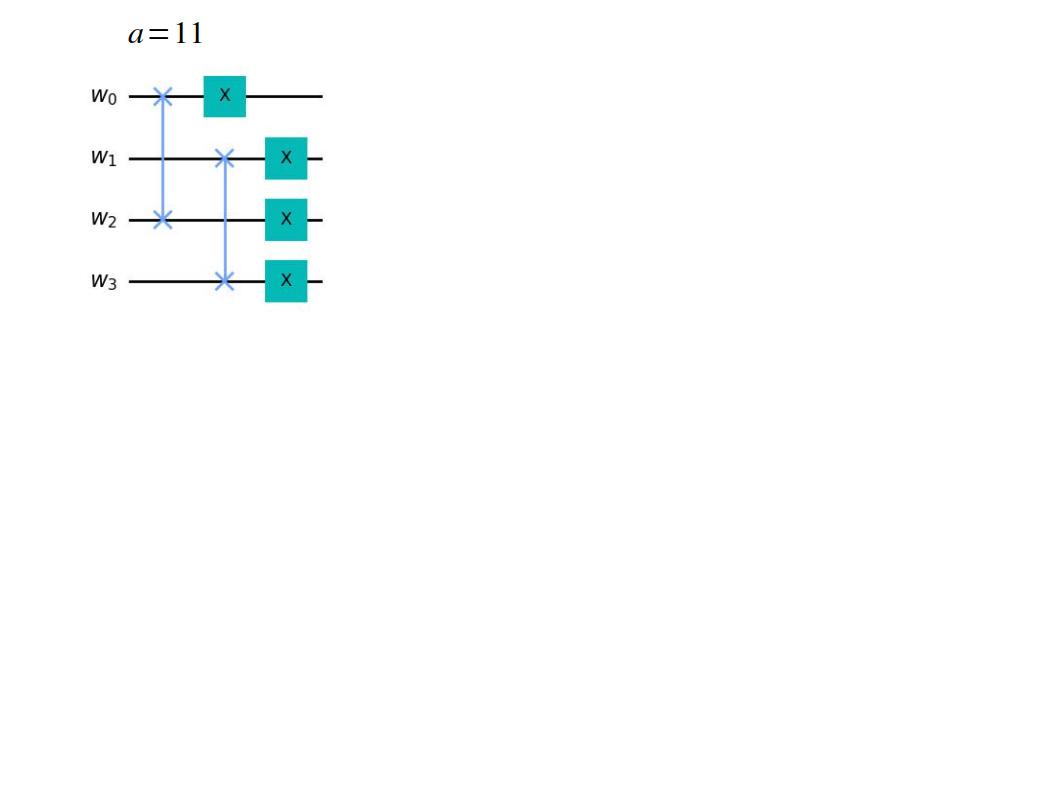} 
\end{minipage}
\begin{minipage}[c]{0.4\linewidth}
\includegraphics[scale=0.40]{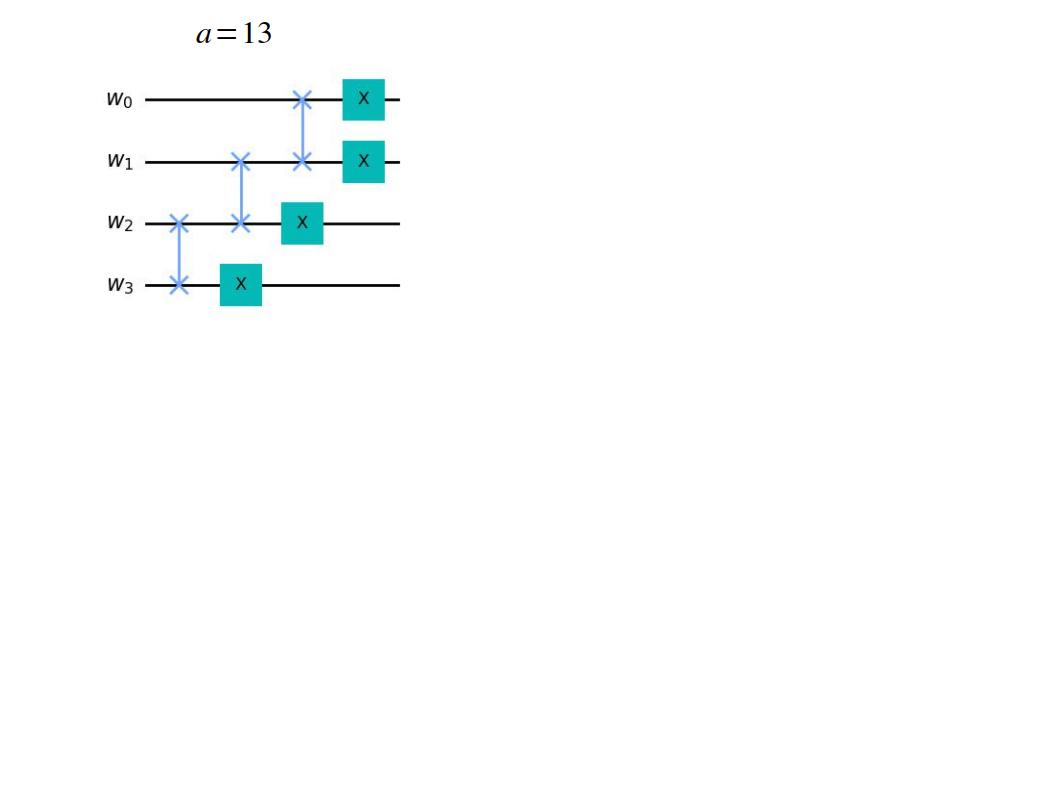} 
\end{minipage}
\vskip-5.0cm
\caption{\footnoteskip
Modular exponentiation operators $U_{a , 15}$
for $a = 2, 7, 11, 13$ as labeled.
}
\label{eq_N15_circuit_b}
\end{figure}

Let us now explore the modular exponentiation (ME) 
operators $U_{a \, \smN}$ in more detail.   For every 
choice of the number $N$ and the base $a$,  we 
must design a separate implementation of the operator 
$U_{a \, \smN}$,  and this is in fact the real bottleneck 
of Shor's algorithm.  Indeed,  this bottleneck occurs in 
two senses: (i) the ME operators consume the greatest 
majority of the quantum resources of the algorithm,  
and in the general case this will be of order $72\, n^3$ 
gates\,\cite{gen},   and (ii) even specialized cases of 
the ME operators are often highly non-trivial to construct.  
The ME operators for $N=15$ with $a=4$ and $a = 8$ 
are illustrated in Fig.~\ref{eq_N15_circuit},  
%% a = 4, 8
%% 04_U15.py
and Table~\ref{tab_U} gives the corresponding operations 
\hbox{$U_{a ,  15} \vert w \rangle = \vert a \cdot w ~({\rm 
mod}~15) \rangle$} for every basis element $\vert w 
\rangle$ in the work register.
Each ME operator $U_{a ,15}$ has two columns 
in the Table: One for the decimal representation of the 
basis elements $\vert w \rangle$,  and another for the 
corresponding binary representation $\vert w_3 w_2 
w_1 w_0 \rangle$, where in binary form $w = [w_3 w_2 
w_1 w_0]_2$. The operators $U_{a , 15}$ 
in Fig.~\ref{eq_N15_circuit} were determined through 
simple inspection of the results of the binary columns 
in Table~\ref{tab_U}.   For example,  the $a=8$ operator 
performs a permutation of the 4-bit binary work states,  
$U_{8,15}\vert w_3 w_2 w_1 w_0 \rangle = \vert w_0 w_3 
w_2 w_1 \rangle$,   which can be implemented by the 
three SWAP gates in the right 
panel of Fig.~\ref{eq_N15_circuit}.  Similarly,  the ME 
operator for $a = 4$ performs two SWAP operations 
on the work register,  so that $U_{4,15} \vert w_3 w_2 
w_1 w_0 \rangle = \vert w_1 w_0 w_3 w_2 \rangle$. The 
other ME operators for $a \in \{2, 7, 11, 13\}$ act similarly,  
and are illustrated in Fig.~\ref{eq_N15_circuit_b}.  
\vskip-0.4cm
\begin{table}[h!]
\caption{\footnoteskip 
\label{tab_U}
Modular Exponentiation Operators $U = U_{a , 15}$ for 
$a=4$ and $a=8$.}
\vskip0.5cm
\begin{tabular}{|c|c||c|c|} \hline
 \multicolumn{2}{|c|}{~$U_{4,15}\vert w \rangle = \vert 4 \cdot w ~({\rm mod}~15) \rangle$~}   &  
  \multicolumn{2}{|c|}{~$U_{8,15}\vert w \rangle = \vert 8 \cdot w ~({\rm mod}~15) \rangle$~}   \\ \hline
$U\vert 1 \rangle ~= \vert 4 \rangle$~&~~$U\vert 0001 \rangle = \vert 0100 \rangle$~
& $U\vert 1 \rangle = \vert 8 \rangle$  & ~~$U\vert 0001 \rangle = \vert 1000 \rangle$~ \\[-5pt]
$U\vert 2 \rangle ~= \vert 8 \rangle$~&~$U\vert 0010 \rangle = \vert 1000 \rangle$ 
& ~$U\vert 2 \rangle = \vert 1 \rangle$~ & ~$U\vert 0010 \rangle = \vert 0001 \rangle$  \\[-5pt]
~$U\vert 3 \rangle ~= \vert 12 \rangle$~&~$U\vert 0011 \rangle = \vert 1100 \rangle$ 
& $U\vert 3 \rangle = \vert 9 \rangle$~&~$U\vert 0011 \rangle = \vert  1001 \rangle$ \\[-5pt]
$U\vert 4 \rangle ~= \vert 1 \rangle$~&~$U\vert 0100 \rangle = \vert 0001 \rangle $ 
& $U\vert 4 \rangle = \vert 2 \rangle$~&~$U\vert 0100 \rangle = \vert 0010 \rangle$ \\[-5pt]
$U\vert 5 \rangle ~= \vert 5 \rangle$~&~$U\vert 0101 \rangle = \vert 0101 \rangle$ 
& ~$U\vert 5 \rangle = \vert 10 \rangle$~&~$U\vert 0101 \rangle = \vert 1010 \rangle$ \\[-5pt]
$U\vert 6 \rangle ~= \vert 9 \rangle$~&~$U\vert 0110 \rangle = \vert 1001 \rangle$ 
& $U\vert 6 \rangle = \vert 3 \rangle$~&~$U\vert 0110 \rangle = \vert 0011 \rangle$ \\[-5pt]
~$U\vert 7 \rangle ~= \vert 13 \rangle$~&~$U\vert 0111 \rangle = \vert 1101 \rangle$ 
& ~$U\vert 7 \rangle = \vert 11 \rangle$~&~$U\vert 0111 \rangle = \vert 1011 \rangle $\\[-5pt]
$U\vert 8 \rangle ~= \vert 2 \rangle$~&~$U\vert 1000 \rangle = \vert 0010 \rangle$
&$U\vert 8 \rangle = \vert 4 \rangle$~&~$U\vert 1000 \rangle = \vert 0100 \rangle$ \\[-5pt]
$U\vert 9 \rangle ~= \vert 6 \rangle$~&~$U\vert  1001 \rangle = \vert 0110 \rangle$
& ~$U\vert 9 \rangle ~= \vert 12 \rangle$~&~$U\vert 1001 \rangle = \vert 1100 \rangle$\\[-5pt]
~\,$U\vert 10 \rangle = \vert 10 \rangle$~&~$U\vert 1010 \rangle = \vert 1010 \rangle$
& $U\vert 10 \rangle = \vert 5 \rangle$~&~$U\vert 1010 \rangle = \vert 0101 \rangle$\\[-5pt]
~~$U\vert 11 \rangle = \vert 14 \rangle$~&~$U\vert  1011\rangle = \vert 1110 \rangle$
& ~$U\vert 11 \rangle = \vert 13 \rangle$~&~$U\vert 1011 \rangle = \vert 1101 \rangle$ \\[-5pt]
$U\vert 12 \rangle ~= \vert 3 \rangle$~&~$U\vert  1100\rangle = \vert 0011 \rangle$
& $U\vert 12 \rangle = \vert 6 \rangle$~&~$U\vert 1100 \rangle = \vert 0110 \rangle$\\[-5pt]
$U\vert 13 \rangle ~= \vert 7 \rangle$~&~$U\vert 1101\rangle = \vert 0111 \rangle$
& ~$U\vert 13 \rangle = \vert 14 \rangle$~&~$U\vert 1101 \rangle = \vert 1110 \rangle$\\[-5pt]
~$U\vert 14 \rangle ~= \vert 11 \rangle$~&~$U\vert  1110 \rangle = \vert 1011 \rangle$
& $U\vert 14 \rangle = \vert 7 \rangle$~&~$U\vert 1110 \rangle = \vert 0111 \rangle$\\\hline
\end{tabular} 
\end{table}
%% 04_U15.py
%%
\vskip-0.5cm
%% 04_U15.py

\clearpage
\subsection{The Factorization Circuit}
\label{sec_fact_circuit}

Our next goal  is to construct the Python script used 
in creating the Shor circuit in Fig.~\ref{fig_15N_ex},  
after which we shall run this code on the Qiskit simulator 
to factor $N =15$.  We will construct the requisite script 
slowly in sequences of Python code segments,  each one
illustrating an essential element of the algorithm.  First,  
we must import the necessary python packages:

\vskip0.4cm 
\noindent
\baselineskip 10pt
\begin{verbatim}
# import basics
import numpy as np
from random import randint
from math import gcd

# import Qiskit tools
from qiskit import Aer, transpile, assemble
from qiskit import QuantumCircuit, ClassicalRegister, QuantumRegister

# import plot tools
from qiskit.visualization import plot_histogram
import matplotlib.pyplot as plt
\end{verbatim}

\noindent
\bodyskip 
The first few imports are for basic mathematical functionality.  
For example,  we can create random integers with ${\tt
randint}()$,  and find the greatest common divisor with 
${\tt gcd}()$.  We have also imported packages to run 
Qiskit,  and for post-processing the Qiskit data.  We now 
select $N = 15$ and the corresponding base $a$:

\vskip0.4cm 
\noindent
\baselineskip 10pt
\begin{verbatim}
# number to factor
N = 15

# random number a in [2,N-1] wtih gcd(a,N)=1
n = 0
while n == 0:
    a = randint(2, N-1)
    if gcd(a, N) == 1: n = 1
    print("**:", a, N, gcd(a, N))
\end{verbatim}
\noindent
\bodyskip 
This code segment chooses a random integer between 
$2$ and $N-1$ inclusive,  and makes sure that the  
choice does not contain a non-trivial factor in common 
with $N$ (otherwise we have found a sought after 
factor of $N$).   This  piece of code can be omitted 
if we wish to work only with a specific value of $a$
(provided  of course that we set the values of $N$
and $a$ here).  

The next code segment defines the ME operators 
$CU_{a, \smN}^p$ for $N=15$ for all permissible 
choices of~$a$.   This is really the heart of Shor's 
algorithm.  For a general value of $N$,  we would 
not be able to implement {\em all} values of $a$ 
(we do this here only for purposes of illustration),  
as there are an exponentially large number of them.  
Finally,  we define a subroutine for the inverse Fourier 
transform $QFT^\dagger$.  The code is given below:

\vskip1.0cm 
\noindent
\baselineskip 10pt
\begin{verbatim}
# modular exponentiation gates: p = 2^0, 2^1, .... , 2^(m-1)
def c_Uamod15(a, p):
    U = QuantumCircuit(4)
    # concatenate U-factors to form U^p
    for iteration in range(p):
        if a in [2,13]:
            U.swap(0,1)
            U.swap(1,2)
            U.swap(2,3)
        if a in [7,8]:
            U.swap(2,3)
            U.swap(1,2)
            U.swap(0,1)
        if a in [4, 11]:
            U.swap(1,3)
            U.swap(0,2)
        if a in [7,11,13]:
            for q in range(4):
                U.x(q)
    U = U.to_gate()
    U.name = "{0}^{1} mod {2}".format(a, p, N)
    c_U = U.control()
    return c_U

# inverse QFT
def qft_dagger(n):
    qc = QuantumCircuit(n)
    for q in range(n//2):
        qc.swap(q, n-q-1)
    for j in range(n):
        for m in range(j):
            qc.cp(-np.pi/float(2**(j-m)), m, j)
        qc.h(j)
    qc.name = "QFT†"
    return qc
\end{verbatim}
\bodyskip

%# QFT
%def qft(n):
%    qc = QuantumCircuit(n)
%    for j in range(n)[::-1]:
%        qc.h(j)        
%        for m in range(j)[::-1]:
%            qc.cp(pi/float(2**(j-m)), m, j)
%    for qubit in range(n//2):
%        qc.swap(qubit, n-qubit-1)
%    circuit.name = "QFT"        
%    return qc

\noindent
Next we construct the quantum circuit itself.  We  must
set the work register to $4$ qubits and the control register
to $9$ qubits.  We must also apply a Hadamard gate to 
every qubit in the control register,  and we must populate 
the work register with the state $\vert 1 \rangle$ (using 
the Qiskit conventions).  We then construct the ME gates 
to form the operators $CU^p$ for the powers \hbox{$p
\in \{2^0,  2^1,  \cdots,  2^8 \}$}.  Finally,  we perform the 
inverse $QFT$ operation,   after which we make the final 
measurements on the control register.  We also draw the 
circuit and save it as a JPG file.  This leads to the following 
code segment:

%\vskip0.2cm
\noindent
\baselineskip 10pt
\begin{verbatim}
# Initialize registers and the quantum circuit
n_work = 4 # L
n_control = 2 * n_work + 1 # 2*L+1
c = QuantumRegister(n_control, name='c')
w = QuantumRegister(n_work, name='w')
cl  = ClassicalRegister(n_control, name='cl')
qc = QuantumCircuit(c, w, cl)



# Initialize control qubits
for q in range(n_control):
    qc.h(q)
    
# Populate work register with state |1>
qc.x(n_control)

# Controlled-U^p operations formed by concatenation
for k in range(n_control):
    qc.append(c_Uamod15(a, 2**k), 
             [k] + [i+n_control for i in range(n_work)])

# Inverse-QFT
qc.append(qft_dagger(n_control), range(n_control))

# Measure control register
qc.measure(c, cl)
qc.draw(fold=-1)
plt.savefig('circuit_{0}.jpg'.format(a))
plt.show()
\end{verbatim}
\bodyskip

\noindent
In constructing the modular exponentiation operators 
$U^p$ for $p \in \{2^1, 2^2,  \cdots,  2^8 \}$,  we have 
simply concatenated the operator $U$.  
This procedure will not do for general values of
$m$, as it leads to an exponentially large number 
of gates.  For a general $m$,  we must produce 
$m$ {\em distinct} operators $U^p$ for each power 
\hbox{$p \in \{2^0,   2^1,  \cdots,  2^{m-1} \}$}. This 
reduces the gate count to a polynomial order,  and
we will have more to say about this in the next section.
In any event,  this is the Qiskit code that produced 
Fig.~\ref{fig_15N_ex}, and it is adequate for any 
(small-ish)~$N$,  assuming of course that we modify 
the ME operators \verb+c_Uamod15+ accordingly.  
Finally, we must run the circuit on the Aer simulator 
(the QASM simulator has been deprecated): 

%\vskip0.2cm 
\noindent
\baselineskip 10pt
\begin{verbatim}
# simulate
aer_sim = Aer.get_backend('aer_simulator')
t_qc = transpile(qc, aer_sim)
obj = assemble(t_qc)
results = aer_sim.run(obj, shots=1024).result()
counts = results.get_counts()
plot_histogram(counts, title='N = {0} a = {1}'.format(N, a), figsize=(6,8))
plt.savefig('hist_{0}.jpg'.format(a))
plt.show()
\end{verbatim}
\bodyskip

%\pagebreak
%%
\begin{figure}[t!]
\includegraphics[scale=0.45]{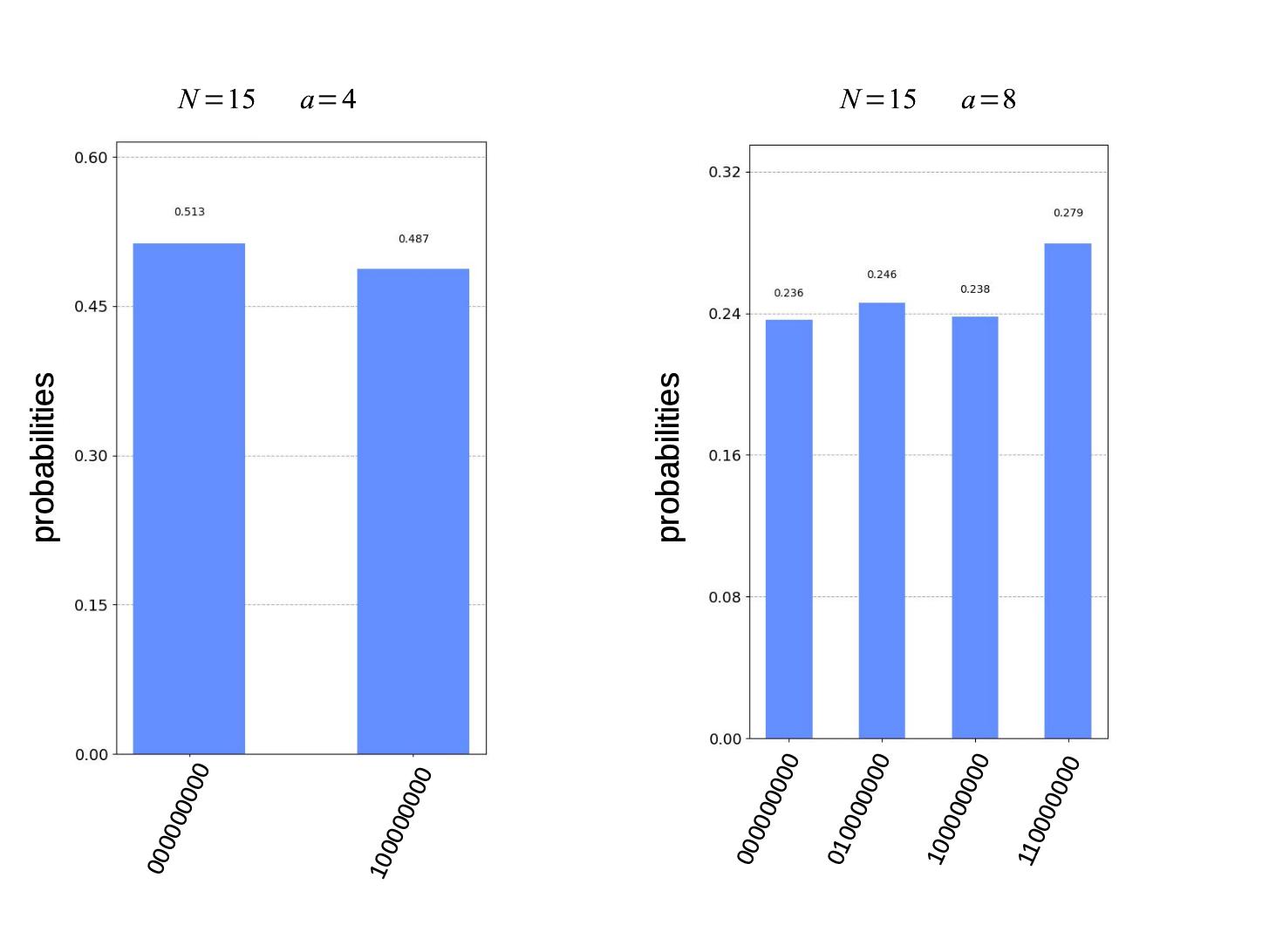}
\vskip-1.0cm
\caption{\footnoteskip
The phase histograms for $N = 15$.  The left and right 
panels show the output of Shor's algorithm for two Qiskit
 simulations in which the control register has $m=9$ 
qubits and the work register has $n = 4$ qubits. The 
results for the base $a = 4$ are illustrated in the left panel,  
while the $a=8$ simulation is shown in the right panel.  
The histograms peak at specific evenly spaced values
$\tilde\ell = [\tilde\phi_{m-1} \cdots \tilde\phi_0]_2$,  
and the corresponding phases are given by  $\tilde\phi 
=\tilde\ell/2^m =  [0.\tilde\phi_{m-1} \cdots \tilde
\phi_0]_2$, where $m = 9$. The period $r$ of the modular 
exponential function $f_{a \, \smN}(x) = a^ x ~({\rm mod}~N)$ 
is encoded in the phase $\tilde\phi \approx\phi_s = s/r$ 
for $s \in \{0, 1,  \cdots,  r-1\}$. 
}
\label{eq_N15_hist}
\end{figure}
Figure~\ref{eq_N15_hist} illustrates the output phase
histogram of an ensemble of 1024 Qiskit runs for both 
$a = 4$ and \hbox{$a = 8$}.  The histograms count the 
output measurements of the control register, and they 
consist of a series of well defined peaks at specific (binary 
integer) values $\tilde\ell = [\tilde\phi_{m-1}
\tilde\phi_{m-2} \cdots \tilde\phi_1 \tilde\phi_0]_2$.  
For example,  $a = 4$ gives two peaks in the histogram,  
while $a = 8$ gives four peaks.  From the peak values 
$\tilde\ell$,  we then construct the measured 
phases $\tilde\phi = \tilde\ell/2^m = [0.\tilde
\phi_{m-1}\tilde\phi_{m-2} \cdots \tilde\phi_1 \tilde
\phi_0]_2$, from which the exact period $r$ can be 
extracted by the method of continued fractions.

\subsection{The Spectrum for a = 4} % $\bm{a=4}$}

Let us examine the  $a=4$ histogram in the left panel 
of Fig.~\ref{eq_N15_hist}.  Recall that the control register 
has $m = 9$ qubits,  and therefore the binary integers 
$\tilde\ell$ are 9 bits long.  We see that there
are two peaks at locations 
\begin{eqnarray}
\nonumber
 \tilde \ell_0 &=& [000000000]_2 = 0
\\
  \tilde\ell_1 &=& [100000000]_2  = 2^8 = 256
  \ ,
\label{eq_hist1}
\end{eqnarray}
where we are now using subscripts to distinguish
the different measurements.
It is of course no accident that there are two peaks,
as the modular exponential function $f_{4,15}(x)$ 
has a period $r = 2$,  as verified by the upper-left 
panel of Fig.~\ref{fig_ax_modN_a}.  The peaks 
$\tilde\ell_n$  (for $n = 0, 1$)
correspond to positive (and rational) 
phase angles $\tilde\phi_n =\tilde\ell_n/2^m$,  
which take the values
\begin{eqnarray}
\nonumber
  \tilde\phi_0 &=& [0.000000000]_2 = 0
\\[3pt]
  \tilde\phi_1 &=& [0.100000000]_2  = 1/2 
  \ .
\label{eq_hist2}
\end{eqnarray}
These are the measured phases of the ME operator 
$U_{4, 15}$.   Since the control register consists of 9 
qubits, and all bits except the most significant bit are 
zero,  the measurements can be regarded as exact.  
Consequently,  there is no need to employ continued 
fractions for this example.  The first peak at $\tilde
\phi_0 =  0$ is guaranteed {\em not} to provide a factor,  
so we move on to the second peak at $\tilde\phi_1 
= 1/2$.  This gives an even period of $r = 2$,  so that 
condition (\ref{eq_c1}) is met.   Furthermore,  
conditions (\ref{eq_c2}) and (\ref{eq_c3}) are also
satisfied,  since
\begin{eqnarray}
  &&
  a^{r/2} ~({\rm mod}~N) =  4^{1} ~({\rm mod}~15) 
  = 4  \ne  \pm 1 ~({\rm mod}~15)
  ~~~~
\\
  &&
  a^r ~({\rm mod}~N) =
  4^{2} ~({\rm mod}~15) = 16 ~({\rm mod}~15) 
  = 1 
  \ .
\end{eqnarray}
The factors of $N=15$ are therefore given by ${\rm 
gcd}(a^{r/2}-1,N) = {\rm gcd}(3, 15) = 3$ and \hbox{
${\rm gcd}(a^{r/2} + 1,N)$}$ = {\rm gcd}(5, 15) = 5$.  

\subsection{The Spectrum for a = 8} % $\bm{a=8}$}

We now turn to the $a = 8$ phase histogram in the right 
panel of Fig.~\ref{eq_N15_hist}.  Again,  upon changing 
subscript notation slightly,  there are four peaks at locations
\begin{eqnarray}
\nonumber
  \tilde\ell_0 &=& [000000000]_2 = 0
\\
\nonumber
  \tilde\ell_1 &=& [010000000]_2 = 128
\\
\label{eq_ell_M9}
 \tilde \ell_2 &=& [100000000]_2 = 256
\\
  \tilde\ell_3 &=& [110000000]_2 = 384
  \ ,
\nonumber
\end{eqnarray}
which correspond to the phase angles
\begin{eqnarray}
\nonumber
  \tilde\phi_0 &=& [0.000000000]_2 = 0
\\[3pt]
\nonumber
  \tilde\phi_1 &=& [0.010000000]_2 = 1/4 
\\[3pt]
\label{eq_phi_M9}
  \tilde\phi_2 &=& [0.100000000]_2 = 1/2 
\\[3pt]
  \tilde\phi_3 &=& [0.110000000]_2 = 3/4 
  \ .
\nonumber
\end{eqnarray}
Again,  these angles can be regarded as exact,  and 
we can immediately extract the period $r$.  We 
must,  however,   check every potential $r$ to make 
sure that conditions (\ref{eq_c1})--(\ref{eq_c3}) 
hold.  We can skip the first peak at $\tilde\phi_0 
= 0$,  so let us now consider the third peak at 
$\tilde\phi_2 = 1/2$.  The period $r = 2$ is even,  
but it does not satisfy requirement (\ref{eq_c3}): 
\begin{eqnarray}
  a^r ~({\rm mod}~N) =  8^{2} ~({\rm mod}~15) 
  =  4  \ne 1 
  \ .
\end{eqnarray}
This illustrates that Shor's 
algorithm can fail for a given phase measurement 
$\tilde\phi$.  However,  the probability of success is 
quite high,  and the algorithm usually requires at most 
a few tries before finding a factor.  Let us move on 
to the second and  fourth peaks,   whose  phases
$\tilde\phi_1=1/4$ and $\tilde\phi_3=3/4$ give 
the period $r = 4$.   Note that conditions 
(\ref{eq_c1})--(\ref{eq_c3}) are indeed satisfied,  
since $r = 4$ is even,  and
\begin{eqnarray}
  &&
  a^{r/2} ~({\rm mod}~N) 
  =
  8^2 ~({\rm mod}~15) 
  = 
  4  \ne  \pm 1 ~({\rm mod}~15)
  ~~~~
\\
  &&
  a^r ~({\rm mod}~N) 
  =
  8^{4} ~({\rm mod}~15) 
  =  1 
  \ .
\end{eqnarray}
Thus,  $r = 4$ is the exact period that we seek,  which
is confirmed by the upper-right panel of Fig.~\ref{fig_ax_modN_a}. 
The factors of $N=15$ are therefore  determined by 
$a^{r/2} = 8^2 = 64$,  so that ${\rm gcd}(a^{r/2}-1,N) 
= {\rm gcd}(63, 15) = 3$ and ${\rm gcd}(a^{r/2} +1,N) 
= {\rm gcd}(65, 15) = 5$.   

The results of this simulation are free from machine
error,  which would not be the case on a real quantum 
computer.  One could build noise models for the various 
gates in Shor's algorithm,  and then place acceptable 
error bounds on the circuit.  This would require taking 
$n_\epsilon > 0$,  which would increase the number
of control qubits.   In this document,  we shall instead 
perform a simplified error analysis by just adding 
a 1 to the least significant bit of the phases $\tilde 
\phi$.  That is to say,  let us suppose the measurements 
are given by 
\begin{eqnarray}
\nonumber
  \tilde\ell_0 &=& [000000001]_2  = 1
\\[5pt]
\nonumber
  \tilde\ell_1   &=&   [010000001]_2  = 129
\\[5pt]
\nonumber
 \tilde\ell_2   &=&   [100000001]_2   = 257
\\[5pt]
  \tilde\ell_3   &=&   [110000001]_2   = 385
  \ ,
\end{eqnarray}
which produces the phases
\begin{eqnarray}
\nonumber
  \tilde\phi_0 
  &=& 
  [0.000000001]_2 
  =
  1/2^9
  =
  0.001953125
  =
  1/512
\\[5pt]
\nonumber
  \tilde\phi_1 
  &=& 
  [0.010000001]_2 
  =
  1/2^2 + 1/2^9
  = 
  0.251953125
  =
  129/512
\\[5pt]
\nonumber
  \tilde\phi_2 
  &=& 
  [0.100000001]_2 
  =
  1/2^1 +  1/2^9
  =
  0.501953125
  =
  257/512 
\\[5pt]
  \tilde\phi_3 
  &=& 
  [0.110000001]_2 
  = 
  1/2^1 + 1/2^2 + 1/2^9
  = 
  0.751953125
  =
  385/512  
  \ .
\end{eqnarray}
The method of continued fractions will now be 
required.  We recommend a useful Python package 
called \verb+contfrac+, which can be installed as 
follows\,\cite{cf}: 

\vskip0.1cm
\noindent
\baselineskip 10pt
\begin{verbatim}
$ pip install contfrac
\end{verbatim}
\bodyskip

\vskip0.1cm
\noindent
We can pick a phase $\tilde\phi$ at random,  or we can
examine every phase sequentially.  Let us concentrate 
on $\tilde\phi_3 = 385/512$ as an example.  With the
above Python package,  one can effortlessly find the 
continued fraction representation of the measured 
phase and its various convergents using the following 
code segment (with output included):

%\vskip0.4cm 
\noindent
\baselineskip 10pt
\begin{verbatim}

# import packages
import contfrac

#
phi = (385, 512) # phi3=[0.110000001]_2=0.751953125=385/512
coefficients = list(contfrac.continued_fraction(phi))
convergents = list(contfrac.convergents(phi))
#
print("cont frac of phi:",coefficients)
print("convergents of phi:", convergents)
\end{verbatim}
\bodyskip
\noindent
output:
\vskip-0.25cm
\noindent
\baselineskip 10pt
\begin{verbatim}
cont frac of phi: [0, 1, 3, 31, 1, 3]
convergents of phi: [(0,1),(1,1),(3,4),(94,125),(97,129),(385,512)]
\end{verbatim}
\bodyskip

\noindent
Therefore,   we can express the phase by the following 
continued fraction, 
\begin{eqnarray}
  \tilde\phi_3 
  &=& [0.110000001]_2 
%  = 
%  0.751953125
  =
  \frac{385}{512}
  =
  [0;1, 3, 31, 1, 3]   
  \ .
\end{eqnarray}
The convergents of $\tilde\phi_3$  have also been calculated:
\begin{eqnarray}
\nonumber
  s_0/r_0 
  &=&
  [0]
  = 
  0/1 
\\
\nonumber
  s_1/r_1 
  &=& 
  [0, 1]
   =
  1/1 
\\
  s_2/r_2 
  &=& 
 [0, 1, 3]
  =
  3/4
  ~~~~~\Leftarrow~~\text{solution:}~ r = 4
\\
\nonumber
  s_3/r_3 
  &=& 
  [0, 1, 3, 31]
  =
  94/125
\\
\nonumber
  s_4/r_4 
  &=& 
[0, 1, 3, 31, 1]
  =
  97/192
\\
\label{eq_conv_a8}
  s_5/r_5 
  &=& 
 [0, 1, 3, 31, 1, 3]
  =
  385/512 
  ~~~~~\Leftarrow~~\text{trivial solution:}~r = 512 = 
  4 \times 128
  \ .
\nonumber
\end{eqnarray}
%% 04_test_cont_fract_3.py
%% 
We must examine every convergent on the list,  but 
fortunately there are only a handful of them.  The 
convergents take the form $s_\ell/r_\ell$,  where $s_\ell$ 
and $r_\ell$ are relatively prime.  Since 
we are interested in the {\em smallest} value of $r_\ell$ such 
that (\ref{eq_c1})--(\ref{eq_c3}) are satisfied,  we must 
work our way {\em up} the list of convergents,  from the
{\em smallest} to the {\em largest} values of~$r_\ell$, 
testing every~$r_\ell$.  This determines the {\em exact} 
period $r =r_\ell$ from the {\em approximately} 
measured phases.  The first two convergents $s_0/r_0
= 0$ and $s_1/r_1=1$ are unacceptable,  so we continue 
on to  the convergent $s_2/r_2=3/4$,  which gives the 
period $r_2 = 4$.  This value indeed satisfies 
(\ref{eq_c1})--(\ref{eq_c3}),  giving the factors 3 and 5
as we have seen.  We can stop here,  but it is pedagogically 
useful to consider the other convergents.  Note that $r_3
=125$ and $r_4 = 192$ do not satisfy (\ref{eq_c3}),
and are therefore ruled out as possible periods.  In
contrast,  note that $r_5 = 512$ is even and it does 
satisfies equation (\ref{eq_c3}),  
\begin{eqnarray}
  a^{r_5} ~({\rm mod}~N)= 8^{512} ~({\rm mod}~15) 
  = 4096 ~({\rm mod}~15)  = 1 \ .
\end{eqnarray}
This is because $512 = 128 \times 4$ is a multiple
of 4,  and $r = 4$ satisfies (\ref{eq_c3}).  Note,  however,  
that 
\begin{eqnarray}
  a^{r_5/2} ~({\rm mod}~N)= 8^{256} ~({\rm mod}~15) 
  = 1 \ ,
\end{eqnarray}
and therefore $b = a^{r_5/2} = 8^{256}$ is a trivial root of unity,  
contrary to condition (\ref{eq_c2}).  This analysis can be 
automated using the following Python script. The first part 
of the script takes the binary input of the peak $\tilde\ell$, 
denoted by \verb+l_phi+, and then converts it to a fraction 
$\tilde\phi_\ell = s_\ell/r_\ell$ with no common factors other than 
unity:

%\vskip0.4cm 
\noindent
\baselineskip 10pt
\begin{verbatim}
# import basics
import contfrac
from numpy import gcd

# construct decimal value of l_phi
n = 0
l_tilde = 0
for l in l_phi[::-1]:
    n += 1
    l_tilde = l_tilde + 2**(n-1) * int(l)
print("l_measured   :", l_phi, l_tilde)

# construct decimal value of phi
n = 0
phi_tilde = 0
for l in l_phi:
    n -= 1
    phi_tilde = phi_tilde + 2**n * int(l)
print("phi_phase_bin :", "0."+l_phi)
print("phi_phase_dec:", phi_tilde)

# express phi_tilde as a fraction
res = len(str(phi_tilde)) - 2 # subtract 2 for "0."
print("res:", res)
scale = 10**res # automated scale set by res
num = int(phi_tilde*scale) 
den = int(scale)
# in lowest terms
c = gcd(num, den) 
num = int(num / c)
den = int(den / c)
phi = (num, den)
print("phi:", phi)
\end{verbatim}
%% 04_test_cont_fract_4.py
\bodyskip

\noindent
We now pass the measured phase  {\tt phi} into the 
continued fraction package to find the convergents,
and then we check each convergent to confirm
that conditions (\ref{eq_c1})--(\ref{eq_c3}) are
satisfied.  If they are not,  we move on to the next
peak in the histogram:

\noindent
\baselineskip 10pt
\begin{verbatim}
# construct convergents for phi
coefficients = list(contfrac.continued_fraction(phi))
convergents = list(contfrac.convergents(phi))
print("cont frac of phi:",coefficients)
print("convergents of phi:", convergents)



# check convergents for solution
for conv in convergents:
    r = conv[1]
    test1 = r % 2 # 0 if r is even
    test2 = (a**int(r/2)-1) % N # 0 if a^r/2 is a trivial root
    test3 = (a**int(r/2)+1) % N # 0 if a^r/2 is a trivial root
    test4 = a**r % N # 1 if r is a solution
    if (test1==0 and test2!=0 and test3!=0 and test4==1):
        print("conv:", conv, "r =", r, ": factors")
        print("factor1:", gcd(a**int(r/2)-1, N))
        print("factor2:", gcd(a**int(r/2)+1, N))
    else:
        print("conv:", conv, "r =", r, ": no factors found")        

\end{verbatim}
%% 04_test_cont_fract_4.py
% 
\bodyskip

\noindent
As an example of the script,  we use the
peak $\tilde\ell_3 = [110000001]_2$.  As we see,
this reproduces the previous analysis.

\vskip0.5cm
\noindent
$\tilde\ell_3 = [110000001]_2 = 385$:

%\vskip-0.5cm
\noindent
\baselineskip 10pt
\begin{verbatim}
l_measured   : 110000001 385
phi_phase_bin: 0.110000001
phi_phase_dec: 0.751953125
res: 9
phi: (385, 512)
cont frac of phi  : [0, 1, 3, 31, 1, 3]
convergents of phi: [(0, 1), (1, 1), (3, 4), (94, 125), (97, 129), (385, 512)]
conv: (0, 1) r = 1 : no factors found
conv: (1, 1) r = 1 : no factors found
conv: (3, 4) r = 4 : factors
factor1: 3
factor2: 5
conv: (94, 125) r = 125 : no factors found
conv: (97, 129) r = 129 : no factors found
conv: (385, 512) r = 512 : no factors found
\end{verbatim}
\bodyskip

%\vskip0.4cm 
\noindent
For completeness,  we use the script to analyze 
the other three peaks of the histogram,  starting
with the 0-th peak.

\vskip0.5cm
\noindent
$\tilde\ell_0 = [000000001]_2 = 1$:

\vskip-0.5cm
\baselineskip 10pt
\begin{verbatim}
l_measured   : 000000001 1
phi_phase_bin: 0.000000001
phi_phase_dec: 0.001953125
res: 9
phi: (1, 512)
cont frac of phi  : [0, 512]
convergents of phi: [(0, 1), (1, 512)]
conv: (0, 1) r = 1 : no factors found
conv: (1, 512) r = 512 : no factors found
\end{verbatim}
\bodyskip

\pagebreak
\noindent
$\tilde\ell_2 = [010000001]_2 = 129$:

\baselineskip 10pt
\begin{verbatim}
l_measured   : 010000001 129
phi_phase_bin: 0.010000001
phi_phase_dec: 0.251953125
res: 9
phi: (129, 512)
cont frac of phi  : [0, 3, 1, 31, 4]
convergents of phi: [(0, 1), (1, 3), (1, 4), (32, 127), (129, 512)]
conv: (0, 1) r = 1 : no factors found
conv: (1, 3) r = 3 : no factors found
conv: (1, 4) r = 4 : factors
factor1: 3
factor2: 5
conv: (32, 127) r = 127 : no factors found
conv: (129, 512) r = 512 : no factors found
\end{verbatim}
\bodyskip

%\vskip0.4cm 
\noindent
$\tilde\ell_3 = [100000001]_2 = 257$:

\baselineskip 10pt
\begin{verbatim}
l_measured   : 100000001 257
phi_phase_bin: 0.100000001
phi_phase_dec: 0.501953125
res: 9
phi: (257, 512)
cont frac of phi  : [0, 1, 1, 127, 2]
convergents of phi: [(0, 1), (1, 1), (1, 2), (128, 255), (257, 512)]
conv: (0, 1) r = 1 : no factors found
conv: (1, 1) r = 1 : no factors found
conv: (1, 2) r = 2 : no factors found
conv: (128, 255) r = 255 : no factors found
conv: (257, 512) r = 512 : no factors found

\end{verbatim}
\bodyskip

\noindent
We see that $\tilde\ell_1$ and $\tilde\ell_3$ give the period 
$r = 4$,  which results in the factors 3 and 5,  while 
the other two peaks do not pass the requisite tests.

%\vfill
%\pagebreak
\subsection{Analysis of the Phase Histogram}

\begin{figure}[h!]
\includegraphics[scale=0.40]{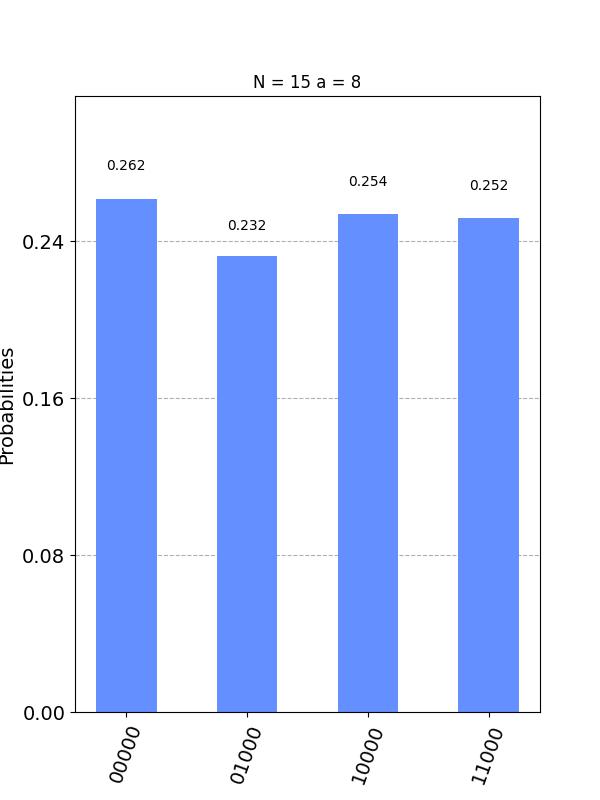}
\vskip0.2cm
\caption{\footnoteskip
  The phase histogram from Qiskit for $N=15$ and $a = 8$, 
  as in the right panel of Fig.~\ref{eq_N15_hist},  except 
  that the control register has $m=5$ qubits.  There are 
  a total of $M = 2^5 = 32$ states,  and the peaks occur 
  at $\tilde\ell_0=0$,  $\tilde\ell_1 = 8$, $\tilde\ell_2 = 16$,  
  and $\tilde\ell_3 = 24$.  The corresponding phases are 
  $\tilde\phi_0 = 0$,  
  $\tilde\phi_1 = 1/4$,  $\tilde\phi_2 = 1/2$,  and $\tilde\phi_3 
  = 3/4$.  From these phases,  we can infer that $r = 4$.  
  Note that Fig.~\ref{eq_N15_hist} uses $m = 9$ qubits,  
  and the peaks therefore occur at different values of 
  $\tilde\ell_n$,  although the phases turn out to be the same.   
}
\label{eq_N15_hist_5}
\end{figure}

We close this section with a theoretical analysis 
of the phase histogram  of the control register for a 
general number of qubits $m$.  Our main focus will 
be calculating the locations of the peaks $\tilde\ell_n$.  
We have already examined the situation for $m=9$,  
in which Fig.~\ref{eq_N15_hist} illustrates the Qiskit 
output histograms for $N=15$ 
with the bases $a = 4$ and $a = 8$.  In this section 
we will primarily concentrate on the $a = 8$ histogram 
with four peaks.  Before looking at a general value of $m$,  
however,  let us first examine the simpler case of $m 
= 5$ qubits (with $a = 8$),  where the Qiskit phase
histogram is illustrated in Fig.~\ref{eq_N15_hist_5}.   
%% 04_shor_N15_1.py
We do this because the $m=5$ case can be calculated
quite easily using a minimum of algebra.  The histogram 
peaks now lie at different values of $\tilde\ell_n$ from those
of the $m=9$ simulation since the value of $m$ differs,  
but the phases $\tilde\phi_n = \tilde\ell_n/2^m$ are identical.  
For $m = 5$ there are $M = 2^5 = 32$ quantum states,   
and the corresponding output phases are given by 

\begin{minipage}[c]{0.45\linewidth}
\begin{eqnarray}
\nonumber
  \tilde\ell_0 &=& [00000]_2 = 0
\\
\nonumber
  \tilde\ell_1 &=& [01000]_2 = 8
\\
\nonumber
  \tilde\ell_2 &=& [10000]_2 = 16
\\
  \tilde\ell_3 &=& [11000]_2 = 24
\nonumber
\end{eqnarray}
\end{minipage}
\begin{minipage}[c]{0.5\linewidth}
\begin{eqnarray}
\nonumber
  \tilde\phi_0 &=& [0.00000]_2 = 0
\\[3pt]
\nonumber
  \tilde\phi_1 &=& [0.01000]_2 = 1/4
\\[3pt]
\label{eq_phi_M5}
  \tilde\phi_2 &=& [0.10000]_2 = 1/2
\\[3pt]
  \tilde\phi_3 &=& [0.11000]_2 = 3/4
  \ .
\nonumber
\end{eqnarray}
%% 
%\caption{Image A}
\end{minipage}
\vskip0.5cm

\begin{figure}[h!]
\begin{centering}
\includegraphics[width=\textwidth]{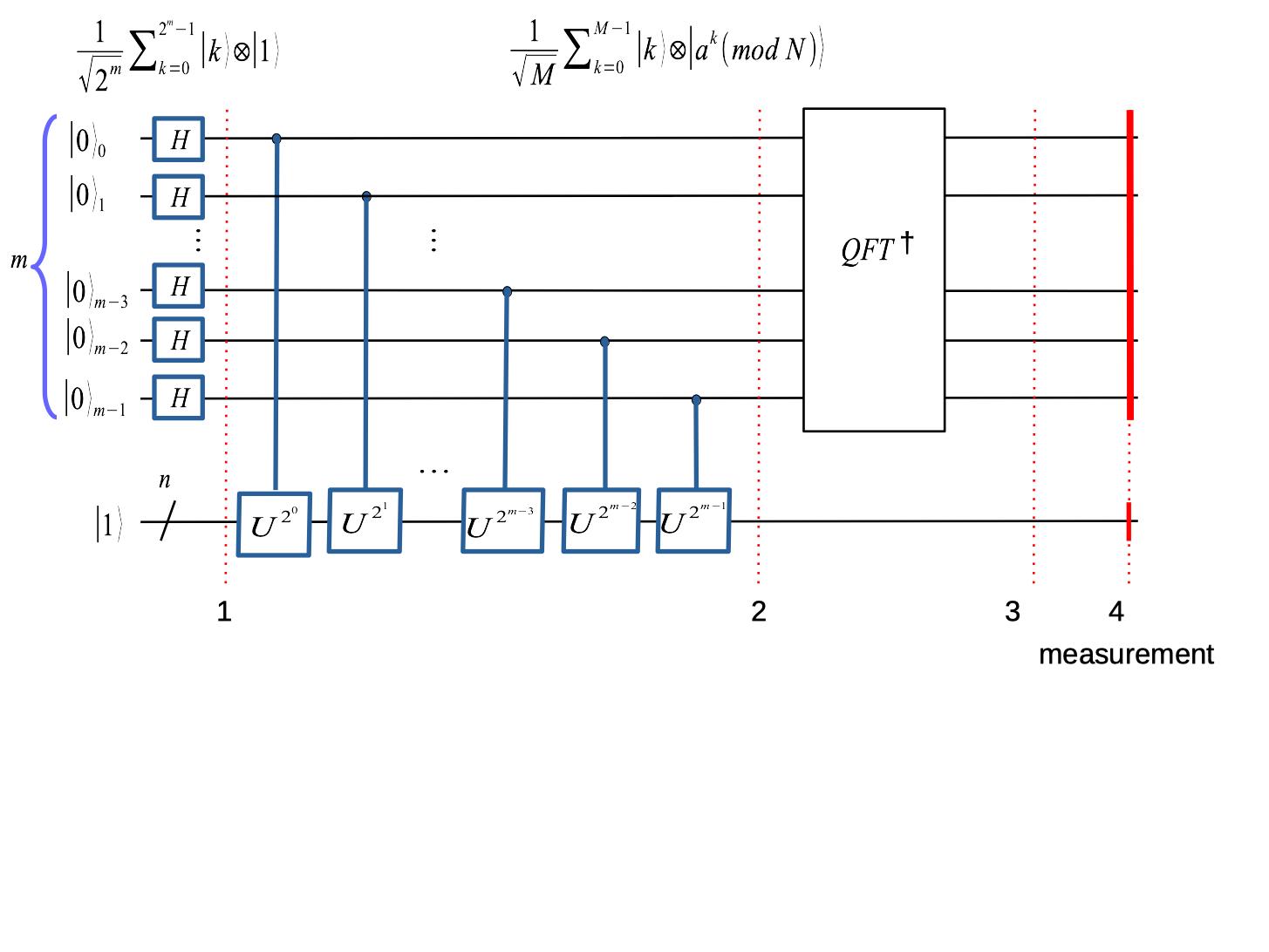} % n
\par\end{centering}
\vskip-2.5cm
\begin{centering}
\includegraphics[width=\textwidth]{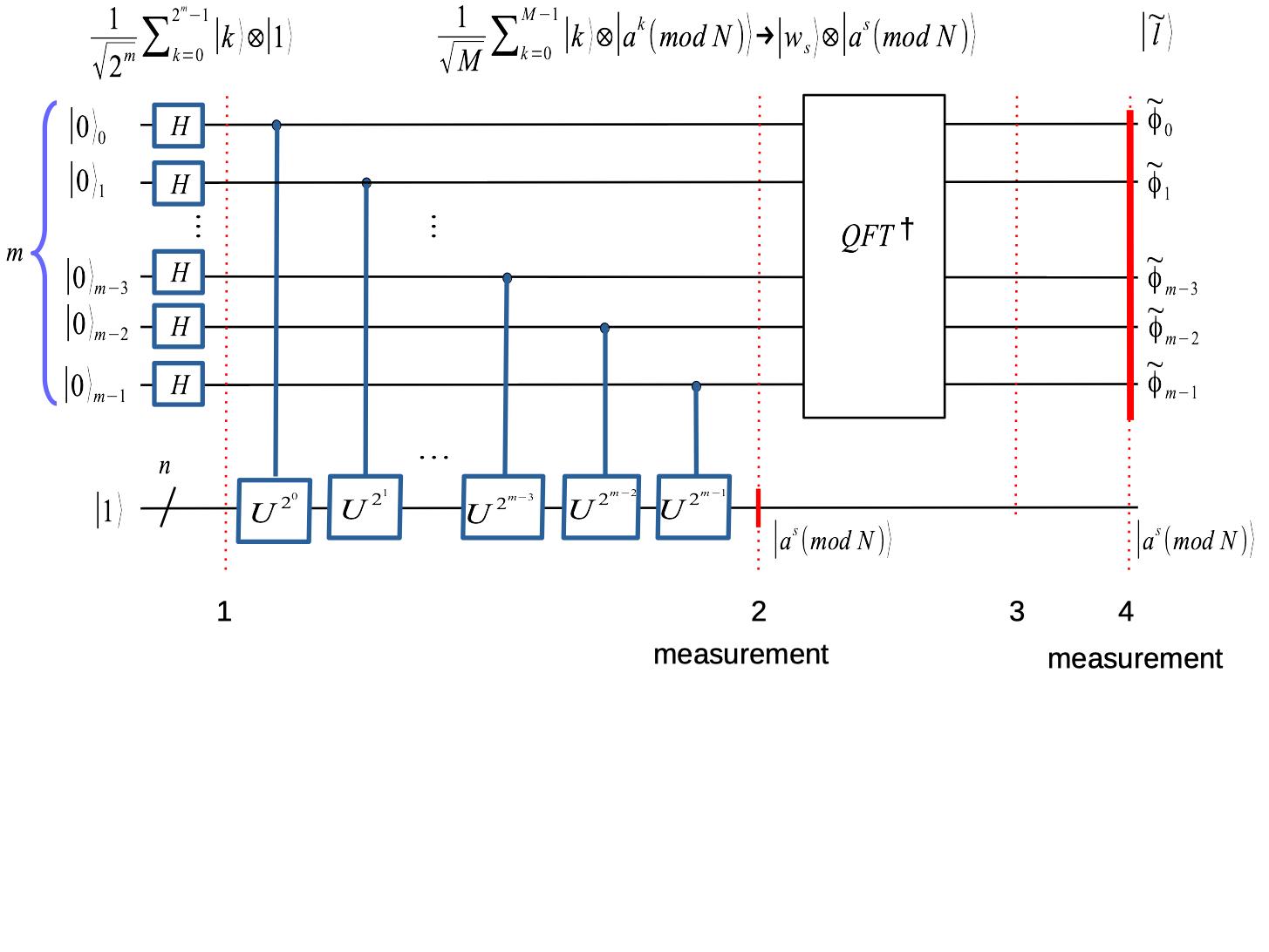} % n
\par\end{centering}
% 04_shor_N15_1.py
\vskip-4.0cm
\caption{\footnoteskip
Changing the order of measurements.
}
\label{fig_shor_5_6}
\end{figure}

\noindent
Just as in Fig.~\ref{fig_shor_4},  we measured the 
$m = 5$ control registers after the inverse Fourier 
transform $QFT^\dagger$ was applied (at position 
4 in the Figure).  Note that the control and work 
registers are entangled at position 2 (just before 
the $QFT^\dagger$ operator) because of the 
action of the ME operators $CU^p$.  The measurement 
of the control register at position~4 therefore collapses 
the quantum state of the work register.   While we 
have not yet talked about work register measurements,  
there is no reason why we should not be able to 
simultaneously measure the work register and 
the control register,  as illustrated in position~4 
in the top panel of Fig.~\ref{fig_shor_5_6}.   The work
register measurement is indicated by the short red 
bar across the register at position~4,  right after 
the $QFT^\dagger$ operation acts on the control 
register.  In fact,  we could measure the work register 
{\em before} the $QFT^\dagger$ operation,  at 
position~3 in the bottom panel of~Fig.~\ref{fig_shor_5_6}.  
When the work register is measured in this way,  
the state of the control register collapses at position~2,  
but we must obtain the same result as in previous
measurement at position~4.

% 
%\vskip-15.0cm
%%

\pagebreak
Let us examine this situation in more detail.  We have 
in fact essentially  performed this calculation near the 
end of Section~\ref{sec_gen_input}, but the following 
method brings out additional physics. Note that the 
wavefunction of the control- and work-register system 
at position~2 takes the form (before the measurement), 
%\vfill
%\pagebreak
%\clearpage

%%
\begin{eqnarray}
  \vert \psi_2 \rangle
  &=&
    \frac{1}{\sqrt{32}}
  \sum_{k=0}^{31} \vert k \rangle \otimes 
  \big\vert 8^k ~({\rm mod}~15) \big\rangle
  \ .
\label{eq_psitwo_before}
\end{eqnarray}
As we have seen,  the modular exponential function
$f(x) = 8^x ~({\rm mod}~15)$ has a period $r = 4$,  
and it takes the values $f(0)=1$,  $f(1)=8$,   $f(2)=4$,   
and $f(3)=2$.   We can therefore express the state 
$\vert \psi_2 \rangle$ in (\ref{eq_psitwo_before}) 
by the following:

%\vfill
%\pagebreak

%%
\begin{eqnarray}
  \vert \psi_2 \rangle
  &=&
    \frac{1}{\sqrt{32}}
  \sum_{k=0}^{31} \vert k \rangle \otimes 
  \underbrace{\vert 8^k ~({\rm mod}~15) 
  \rangle}_{1, 8, 4, 2}
\\[5pt]
  &=&
\nonumber
  \frac{1}{\sqrt{32}}
  \Big[
  \,\vert 0 \rangle \otimes \,\vert 1 \rangle
  +
  \,\vert 1 \rangle \otimes \,\vert 8 \rangle
  +
  \,\vert 2 \rangle \otimes \,\vert 4 \rangle
  +
  \,\vert 3 \rangle \otimes \,\vert 2 \rangle
  + 
\\ && \hskip1.0cm
  \,\vert 4 \rangle \otimes \,\vert 1 \rangle
  +
  \,\vert 5 \rangle \otimes \,\vert 8 \rangle
  +
  \,\vert 6 \rangle \otimes\, \vert 4 \rangle
  +
  \,\vert 7 \rangle \otimes \,\vert 2 \rangle
  +
\nonumber\\ && \hskip1.0cm
  \,\vert 8 \rangle \otimes \,\vert 1 \rangle
  +
  \,\vert 9 \rangle \otimes \,\vert 8 \rangle
  +
  \vert 10 \rangle \otimes \,\vert 4 \rangle
  +
  \vert 11 \rangle \otimes \,\vert 2 \rangle
  +
\\ && \hskip1.0cm
  \vert 12 \rangle \otimes \vert 1 \rangle
  +
  \vert 13 \rangle \otimes \vert 8 \rangle
  +
  \vert 14 \rangle \otimes \vert 4 \rangle
  +
  \vert 15 \rangle \otimes \vert 2 \rangle
  +
\nonumber\\ && \hskip1.0cm
  \vert 16 \rangle \otimes \vert 1 \rangle
  +
  \vert 17 \rangle \otimes \vert 8 \rangle
  +
  \vert 18 \rangle \otimes \vert 4 \rangle
  +
  \vert 19 \rangle \otimes \vert 2 \rangle
  +
  \nonumber
%\end{eqnarray}
%\begin{eqnarray}
\nonumber\\ && \hskip1.0cm
  \vert 20 \rangle \otimes \vert 1 \rangle
  +
  \vert 21 \rangle \otimes \vert 8 \rangle
  +
  \vert 22 \rangle \otimes \vert 4 \rangle
  +
  \vert 23 \rangle \otimes \vert 2 \rangle
  +
\nonumber\\ && \hskip1.0cm
  \vert 24 \rangle \otimes \vert 1 \rangle
  +
  \vert 25 \rangle \otimes \vert 8 \rangle
  +
  \vert 26 \rangle \otimes \vert 4 \rangle
  +
  \vert 27 \rangle \otimes \vert 2 \rangle
  +
\nonumber\\ && \hskip1.0cm
  \vert 28 \rangle \otimes \vert 1 \rangle
  +
  \vert 29 \rangle \otimes \vert 8 \rangle
  +
  \vert 30 \rangle \otimes \vert 4 \rangle
  +
  \vert 31 \rangle \otimes \vert 2 \rangle
  ~
  \Big]
  \ .
 \nonumber
\end{eqnarray}
Since $M = 2^5 = 32$ is so small,  we have been 
able to write down every term in the wavefunction.
Upon collecting like states in the work register,  we 
find
\begin{eqnarray}
  \vert \psi_2 \rangle 
  &=&
  \frac{1}{\sqrt{32}}
  \Big[
  \Big(  \vert 0 \rangle + \vert 4 \rangle + \vert 8 \rangle +
  \vert 12 \rangle + \vert 16 \rangle +  \vert 20 \rangle + 
  \vert 24 \rangle + \vert 28 \rangle 
  \Big) \otimes \vert 1 \rangle +
\nonumber \\ && \hskip1.0cm
  \Big(  \vert 1 \rangle + \vert 5 \rangle + \vert 9 \rangle +
  \vert 13 \rangle + \vert 17 \rangle +  \vert 21 \rangle + 
  \vert 25 \rangle + \vert 29 \rangle 
  \Big) \otimes \vert 8 \rangle +
\\ && \hskip1.0cm
  \Big(  \vert 2 \rangle + \vert 6 \rangle + \vert 10 \rangle +
  \vert 14 \rangle + \vert 18 \rangle +  \vert 22 \rangle + 
  \vert 26 \rangle + \vert 30 \rangle 
  \Big) \otimes \vert 4 \rangle +
\nonumber \\ && \hskip1.0cm
  \Big(  \vert 3 \rangle + \vert 7 \rangle + \vert 11 \rangle +
  \vert 15 \rangle + \vert 19 \rangle +  \vert 23 \rangle + 
  \vert 27 \rangle + \vert 31 \rangle 
  \Big) \otimes \vert 2 \rangle ~
  \Big]
\nonumber\\[5pt]
  &=&
  \frac{1}{\sqrt{4}}
  \Big[
  \vert w_0 \rangle \otimes \vert f(0) \rangle +
  \vert w_1 \rangle \otimes \vert f(1) \rangle +
  \vert w_2 \rangle \otimes \vert f(2) \rangle +
  \vert w_3 \rangle \otimes \vert f(3) \rangle
  \Big]
  \ ,
\label{eq_w_fs}
\end{eqnarray}
where the four control register states are defined by 
\begin{eqnarray}
  \vert w_0 \rangle
  &=&
  \sqrt{\frac{4}{32}} \, 
  \Big(
  \vert 0 \rangle + \vert 4 \rangle + \vert 8 \rangle +
  \vert 12 \rangle + \vert 16 \rangle +  \vert 20 \rangle + 
  \vert 24 \rangle + \vert 28 \rangle
  \Big) 
\label{eq_w0}
\\[5pt]
    \vert w_1 \rangle
  &=&
  \sqrt{\frac{4}{32}} \, 
  \Big(
   \vert 1 \rangle + \vert 5 \rangle + \vert 9 \rangle +
  \vert 13 \rangle + \vert 17 \rangle +  \vert 21 \rangle + 
  \vert 25 \rangle + \vert 29 \rangle 
  \Big) 
\label{eq_w1}
\\[5pt]
  \vert w_2 \rangle
  &=&
  \sqrt{\frac{4}{32}} \, 
  \Big(
  \vert 2 \rangle + \vert 6 \rangle + \vert 10 \rangle +
  \vert 14 \rangle + \vert 18 \rangle +  \vert 22 \rangle + 
  \vert 26 \rangle + \vert 30 \rangle 
  \Big) 
\label{eq_w2}
\\[5pt]
    \vert w_3 \rangle
  &=&
  \sqrt{\frac{4}{32}} \, 
  \Big(
  \vert 3 \rangle + \vert 7 \rangle + \vert 11 \rangle +
  \vert 15 \rangle + \vert 19 \rangle +  \vert 23 \rangle + 
  \vert 27 \rangle + \vert 31 \rangle 
  \Big) 
  \ .
\label{eq_w3}
\end{eqnarray}
We can re-express (\ref{eq_w_fs}) in the more general form 
\begin{eqnarray}
  \vert \psi_2 \rangle
    &=&
  \frac{1}{\sqrt{r}}\sum_{s=0}^{r-1} 
  \vert w_s \rangle \otimes \vert f(s) \rangle  
  \ ,
\end{eqnarray}
which also holds for period $r = 2$,  although 
the states $\vert w_s \rangle$ and $\vert f(s)
\rangle$ will be different.   
Returning to $r = 4$,  we can now generalize 
(\ref{eq_w0})--(\ref{eq_w3}) to arbitrary $m$, 
where $M = 2^m$: 
\begin{eqnarray}
  \vert w_s \rangle
  =
  \sqrt{\frac{4}{M}} \, 
  \sum_{k=0}^{M/4-1}  \vert s+4 k \rangle 
  ~~~\text{for}~~ s \in \{0, 1, 2, 3\}
  \ .
\label{eq_ws}
\end{eqnarray}
This form relies on a special feature of $N = 15$,
namely that $M/r = 2^m/4 = 2^{m-2}$ is an integer
for $r=4$.  Also note that $M/r= M/2= 2^{m-1}$ is 
also an integer for $r = 2$. This leads to the further 
generalization
\begin{eqnarray}
  \vert w_s \rangle
  =
  \sqrt{\frac{r}{M}} \, 
  \sum_{k=0}^{M/r-1}  \vert s+r k \rangle 
  ~~~\text{for}~~ s \in \{0,  \cdots,  r-1\}
  \ .
\label{eq_ws_rgen}
\end{eqnarray}
When $r=2$ we have  $s \in \{0, 1\}$,  and 
when $r = 4$ we have $s \in \{0, 1, 2, 3\}$.

Returning again to $r=4$,   let us now measure the 
state $\vert \psi_2 \rangle$ at position 2,   as illustrated 
by the lower panel of Fig.~\ref{fig_shor_5_6}. This 
leads to wavefunction collapse,  so that 
\begin{eqnarray}
 \vert \psi_2 \rangle
  =
  \frac{1}{\sqrt{4}} \,
  \sum_{s=0}^3
  \vert w_s \rangle \otimes \vert f(s) \rangle  
  \rightarrow
  \vert w_s \rangle \otimes \vert f(s) \rangle  
  \ ,
\label{eq_meas}
\end{eqnarray}
where $s \in \{0, 1, 2, 3\}$ is randomly selected with
a uniform probability of $1/4$.  We must now apply 
the inverse Fourier transform to the control register,  
thereby giving the state
\begin{eqnarray}
  QFT^\dagger \,  \vert w_s \rangle
  \equiv
  \sum_{\ell=0}^{M-1} A_{\ell, s} \, \vert \ell \rangle
  \ .
\end{eqnarray}
The next step is to find the amplitudes $A_{\ell, s}$ 
by performing the inverse Fourier transform on the
state $\vert w_s \rangle$,  which can be calculated 
exactly:
\begin{eqnarray}
  QFT^\dagger\,  \vert w_s \rangle 
  &=&
  \sqrt{\frac{4}{M}} \,  \sum_{k=0}^{M/4-1}\, 
  QFT^\dagger \,  \vert s+4 k \rangle 
  = 
  \sqrt{\frac{4}{M}} \,  \sum_{k=0}^{M/4-1}\, 
  \frac{1}{\sqrt{M}}\sum_{\ell = 0}^{M-1} 
  e^{-2\pi i \,\ell (s + 4 k)/M}\,
  \vert \ell \rangle
  ~~~~~~~~~
\\[5pt]
  &=& 
  \frac{2}{M} \,    \sum_{\ell = 0}^{M-1}
  e^{-2\pi i \, \ell s/M}\,   \sum_{k=0}^{M/4-1}\,  e^{-8\pi i \, k \ell /M}\,
  \vert \ell \rangle
\\[5pt]
  &=&  
  \frac{2}{M} \,    \sum_{\ell = 0}^{M-1} 
  e^{-2\pi i \, \ell s /M}\,   
  \frac{1 - \big[ e^{-8\pi i \, \ell /M}\big]^{M/4}}{1 - e^{-8\pi i \, \ell/M}}\,
  \vert \ell \rangle
\\[5pt]
  &=&
  \frac{2}{M} \,    \sum_{\ell = 0}^{M-1} 
  e^{-2\pi i \, \ell s/M}\,   
  \frac{1 - e^{-2\pi i \, \ell}}{1 - e^{-8\pi i \, \ell/M}}\,
  \vert \ell \rangle
  \ .
\label{eq_qftdagger_ws}
\end{eqnarray}
We therefore find 
\begin{eqnarray}
  A_{\ell, s}
   =
   e^{-2\pi i \, \ell s/M}\,    \frac{2}{M} \,  
  \frac{1 - e^{-2\pi i \, \ell}}{1 - e^{-8\pi i \, \ell/M}}
  ~~~\text{for}~~  \ell \in \{0, 1,  \cdots,  M-1\}
  \ .
\end{eqnarray}
Note that the $s$-dependence lies only in the complex
phases,  and therefore the probabilities $P_\ell = \vert 
A_{\ell, s} \vert^2$ can be expressed by 

\begin{eqnarray}
  P_\ell
  &=&
  \frac{4}{M^2}\, 
  \left\vert 
  \frac{1 - e^{-2\pi i \, \ell}}{1 - e^{-8\pi i \, \ell/M}}\,
  \right\vert^2
  =
  \frac{4}{M^2}\, 
  \frac{1 - \cos(2\pi \ell)}{1 - \cos(8\pi\ell/M)}
 \\[5pt]
  &=&
  \frac{4}{M^2}\, 
  \frac{\sin^2(\pi \ell)}{\sin^2(4\pi\ell/M)}
  ~~~~\text{for}~~ \ell \in \{0,  1,  \cdots,  M-1\}
  \ .
\label{eq_Pell}
\end{eqnarray}
Also note that the numerator vanishes for every integer 
$\ell$, so the only way we can obtain a non-zero probability 
is when the denominator also vanishes (so that we have the 
indeterminate form $0/0$).  However, the denominator 
vanishes only for $\ell$ such that 
\begin{eqnarray}
  \frac{4\pi \ell}{M} = n \pi 
  ~~\text{for}~~ n \in \mathbb{Z}
  \ ,
\label{eq_elln_four1}
\end{eqnarray}
or equivalently for 
\begin{eqnarray}
  \ell_n = \frac{n M}{4}
  ~~\text{for}~~ n \in \{0, 1, 2, 3\}
  \ .
\label{eq_elln_four}
\end{eqnarray}
We have dropped the tilde over $\ell_n$ since this is a 
theoretical prediction and not a measurement. The value 
of $n$ is restricted to $\{0, 1, 2, 3\}$ because $\ell = \ell_n$
must be a non-negative integer that cannot exceed $M-1$.  
Note that expression (\ref{eq_elln_four1}) means that 
\begin{eqnarray}
  \sin(4\pi\ell_n/M) &=& 0
\label{eq_sin_M}
\\[5pt]
  \cos(4\pi\ell_n/M) &=& \pm 1
  \ ,
\label{eq_cos_M}
\end{eqnarray}
relations that we shall use momentarily.  From the probability 
(\ref{eq_Pell}),  we see that $P_\ell$ vanishes for all values
of $\ell$ except for $\ell = \ell_n$. Thus,  $P_\ell$ vanishes 
at all but four of its $2^m$ possibilities!  We must next calculate 
the corresponding probabilities at the four poles $\ell = \ell_n$, 
and it should come as no surprise that they are all equally likely 
with probability $P = 1/4$.

Let us now turn to calculating these probabilities. In the language 
of quantum field theory,  we must perform a {\em regularization} 
procedure on the function (\ref{eq_Pell}),  thereby eliminating 
the poles at $\ell_n = n M/4$.  To do this,  we shall (i) displace 
each pole $\ell_n$ by a small distance $\varepsilon$,   (ii)~evaluate 
this {\em regularized} probability exactly for a non-zero $\varepsilon$, 
and (iii) only afterward take the limit of zero displacement 
$\varepsilon \to 0$.  Therefore, let us make the shift
\begin{eqnarray}
  \ell_n \to \ell_n + \varepsilon
\end{eqnarray}
in expression (\ref{eq_Pell}). Upon using the more suggestive 
notation $P(\ell) \equiv P_\ell$, we thereby {\em define} the 
regularized probabilities by 
\begin{eqnarray}
  P_n 
  \equiv 
  \lim_{\varepsilon \to 0 } P({\ell_n + \varepsilon})
  \ .
\end{eqnarray}
We can now calculate the probabilities at the poles $\ell=\ell_n$:
\begin{eqnarray}
  P_n
  &=&
  \frac{4}{M^2}\, 
  \lim_{\varepsilon \to 0}
  \left[\frac{\sin (\pi \ell_n + \pi \varepsilon)}
  {\sin (4 \pi \ell_n / M + 4 \pi \varepsilon/M)}
  \right]^2
\\[5pt]
  &=&
  \frac{4}{M^2}\, 
  \lim_{\varepsilon \to 0}
  \left[\frac{\sin (\pi \ell_n)  \cos(\pi\varepsilon)
  + 
  \cos (\pi \ell_n)  \sin(\pi\varepsilon)}
  {\sin (4 \pi \ell_n / M) \cos(4\pi\varepsilon/M)
  + 
  \cos (4 \pi \ell_m / M) \sin(4\pi \varepsilon/M)}
  \right]^2
\\[5pt]
  &=&
  \frac{4}{M^2}\, 
  \lim_{\varepsilon \to 0}
  \left[\frac{\sin(\pi\varepsilon)}{ \sin(4\pi \varepsilon/M)}
  \right]^2
  = 
    \frac{4}{M^2}\, 
  \lim_{\varepsilon \to 0}
  \left[\frac{
    \pi \varepsilon}{4\pi \varepsilon/M}
  \right]^2
  =
  \frac{1}{4}
  \ ,
\end{eqnarray}
\begin{figure}[h!]
\includegraphics[scale=0.40]{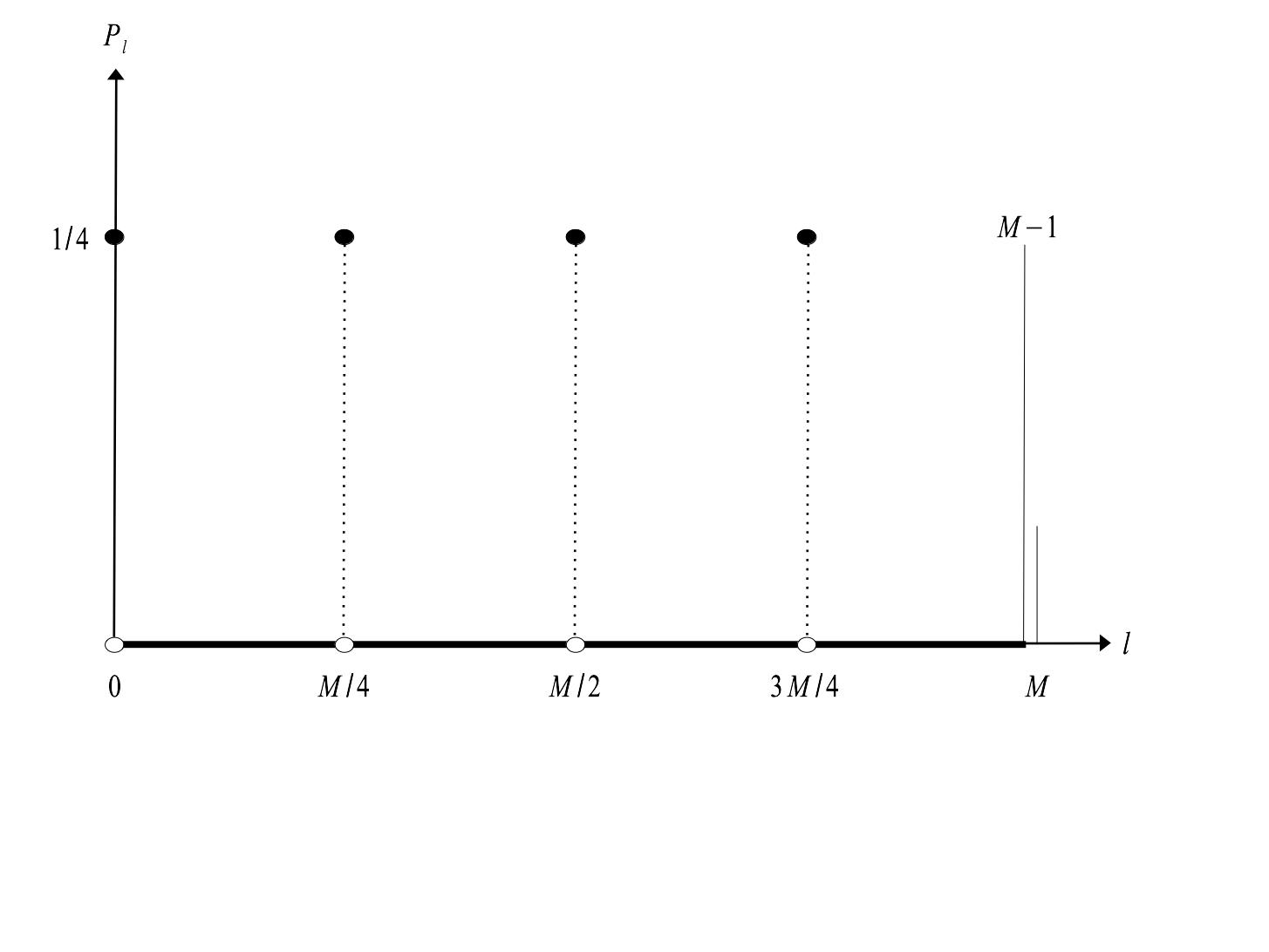}
\vskip-1.5cm
\includegraphics[scale=0.40]{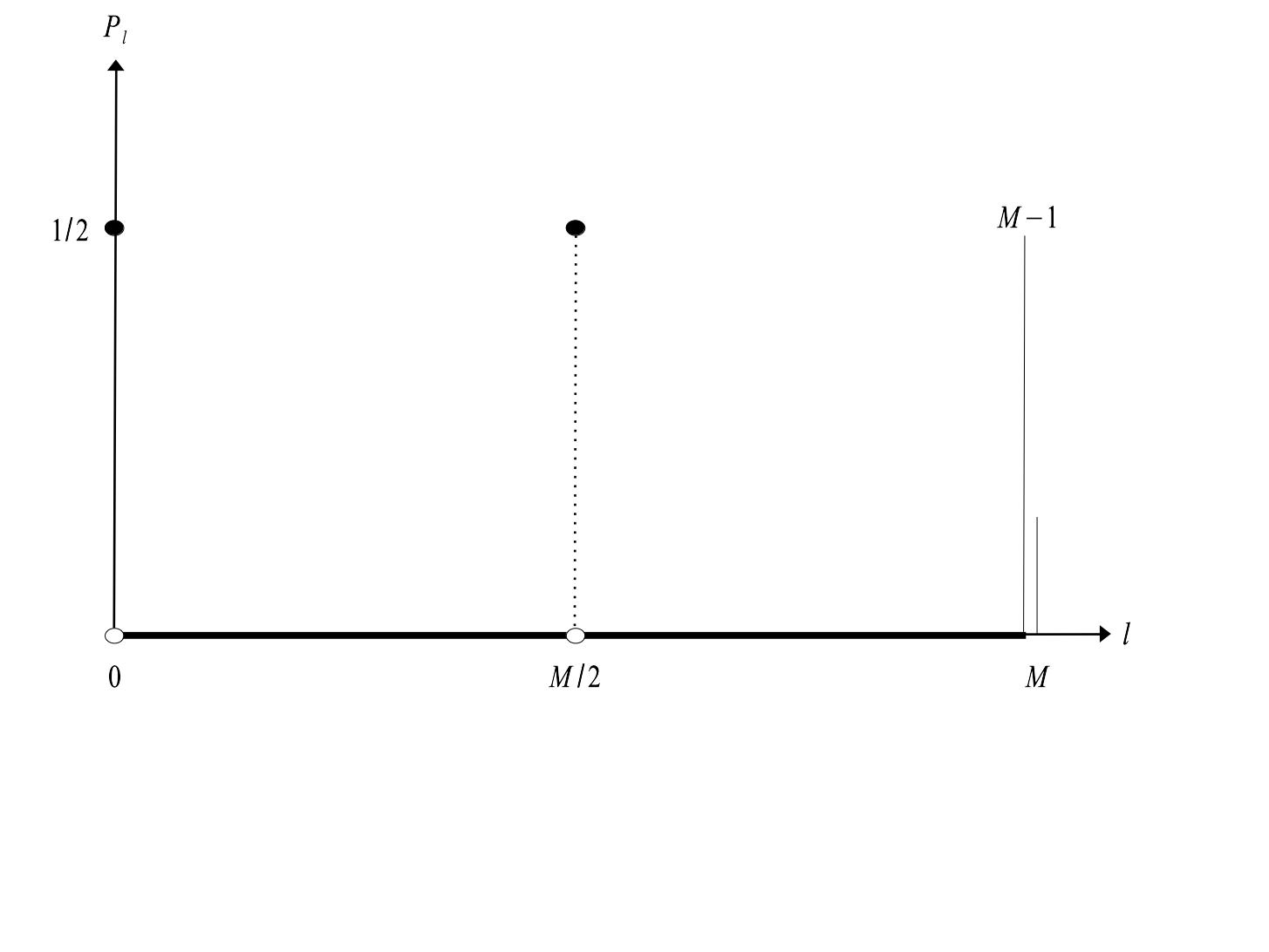}
\vskip-1.5cm
\caption{\footnoteskip
  Probabilities $P_\ell$ for $\ell \in \{0, 1, \cdots, M-1\}$.
  Top panel is for $a =8$ (period $r = 4$) ,  and the
  bottom panel is for $a =4$ (period $r = 2$). 
}
\label{eq_N15_gen}
\end{figure}

\noindent
where we have used relations (\ref{eq_sin_M}) and 
(\ref{eq_cos_M}). These are the only non-zero values of 
$P_\ell$,  and the graph  of the probabilities for $a =8$ 
is shown in the top panel of Fig.~\ref{eq_N15_gen}.  For 
the period $r = 2$,  we would  find 
\begin{eqnarray}
  \ell_n = \frac{n M}{2}
  ~~\text{for}~~ n \in \{0, 1\}
  \ ,
\end{eqnarray}
with $P_n = 1/2$,  which is illustrated in the bottom panel 
of Fig.~\ref{eq_N15_gen}.  

 Let us now check this calculation
against the previous Qiskit output for $m = 5$ and $m =9$.
\vskip0.4cm 
\begin{enumerate}
  \baselineskip 10pt plus 1pt minus 1pt
  \setlength{\itemsep}{3pt} % single spacing
  \setlength{\parskip}{1pt} %
  \setlength{\parsep}{0pt}  %
\item[(i)] $a = 8$ or $r =4$:  For $M = 2^5 = 32$,  we find 
  $\ell_n = 8 n$,  or $\ell_0 = 0,   \ell_1 = 8,  \ell_2 = 16,  \ell_3 =24$,  
  in agreement with (\ref{eq_phi_M5}).  For $M = 2^9$,   we have 
  $\ell_n = 128 n$,  or $\ell_0 = 0,   \ell_1 = 128,  \ell_2 = 256,  
  \ell_3 = 384$,  in agreement with (\ref{eq_ell_M9}). 

\item[(ii)] $a = 4$ or $r =2$: For $m=9$ we have $M = 2^9$,
and therefore $\ell_n = 256 n$.  Thus, $\ell_0 = 0$ and 
$\ell_1 =256$, in agreement with~(\ref{eq_hist1}). 
\end{enumerate}
This analysis is actually a special case of a result that we 
have already derived. In Section~\ref{sec_gen_input} we 
showed that when the phase angle $\theta $ is such that 
$\ell_\theta \equiv 2^m \theta$ is an integer for $m$ control 
qubits, then the amplitude simplifies to $A_\ell(\theta) 
= \delta_{\ell,\ell_\theta}$. This is exactly what we have 
shown here for $N=15$ and $a = 4, 8$, since the phases 
are $\phi_s = s/r$ with the periods $r = 2,  4$,  which are 
just powers of 2.

\vfill

\pagebreak
\clearpage
\section{Further Examples}
\label{sec_further_ex}

\subsection{Factoring Larger Numbers: N = 21 = 3 $\times$ 7, a = 2,
r = 6}
%$\bm{N=21 = 3 \times 7}$, $\bm{a=2}$, $\bm{r=6}$}

We now turn to factoring numbers larger than $N = 15$, where we 
will need to construct more complex {\em modular exponentiation}
(ME) operators $U_{a \, \smN}$ for an appropriate base $a$. While 
the $N=15$ operators were rather easy to construct, this is not the 
case for larger values of~$N$.  The ME operators $U_{a, 15}$ were 
completely general,  valid for any permissible base $a$, and acting
on any computational basis element in the work-state Hilbert space. 
In contrast, creating such general operators for larger values of $N$ 
appears to be extremely difficult.  However,  we do not require the 
general structure of the ME operators.  This is because the first 
operation of $U_{a \, \smN}$ acts on the work-state $\vert 1 \rangle 
= \vert 0 \cdots 0 1 \rangle$,  and the next operation acts on the 
output of the first,  and so on.  Since
\begin{eqnarray}
   U_{a \, \smN}^x \vert 1 \rangle 
  = 
  \big\vert f_{a \, \smN}(x)  \big\rangle 
   ~~\text{for any}~ x \in \{0, 1, 2,  \cdots \} \ ,
\end{eqnarray}
we therefore only need to find the operation of $U_{a \,\smN}$
on the states $\vert f_{a\,\smN}(x) \rangle$ for $x = 0, 1, \cdots,$
$r-1$,  where $r$ is the period of $f_{a \, \smN}(x)$.
Let us return momentarily to a general number of work qubits 
$n$.  We see that the work space,  which we shall denote by 
${\cal W}_n$, has dimension~$2^n$, and a general ME operator 
$U$ can act on this entire Hilbert space. Consider now the 
$r$-dimensional subspace defined by 
\begin{eqnarray}
  {\cal U}_r 
  \equiv
  {\rm Span}\Big\{ \,\big\vert f(x) \big\rangle\, \,\Big\vert\, 
  x \in \{0, 1,  \cdots,  r-1\}  \Big\} \subseteq {\cal W}_n
  \ .
\end{eqnarray}
As discussed above, the  $U$ operator transforms one basis 
element of ${\cal U}_r$ into another basis element,  that is to 
say, the ME operator $U$ leaves the $r$-dimensional space 
${\cal U}_r$ invariant, so that $U[{\cal U}_r] = {\cal U}_r$. 
Thus, as the $U$ operator acts successively, the states in the 
work register only vary over the $r$-dimensional subspace 
${\cal U}_r$. The operators $CU^p$ therefore entangle the 
control register and the subspace ${\cal U}_r$ (and not 
the entire work space), and this plays a crucial role in the 
exponential speedup of Shor's algorithm.  We have reduced 
the problem to the action of ME operator $U$ on the lower
dimensional subspace ${\cal U}_r$ of the exponentially large 
work space ${\cal W}_n$, and we can henceforth restrict 
our attention to this subspace ${\cal U}_r$.  This is quite 
similar to Grover's search algorithm that reduces to movement 
within a \hbox{2-dimensional} subspace.

\begin{figure}[t!]
\hskip-1.0cm
\begin{minipage}[c]{0.4\linewidth}
\includegraphics[scale=0.48]{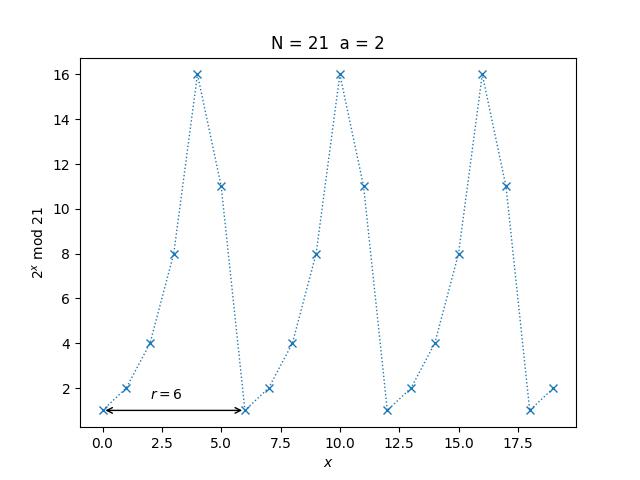}
\end{minipage}
\hskip1.6cm
\begin{minipage}[c]{0.4\linewidth}
\begin{tabular}{|c|c|} \hline
 \multicolumn{2}{|c|}{~$U\vert w \rangle = 
\big\vert 2 \cdot w ~({\rm mod}~21) \big\rangle$~}  
\\\hline
$~~~U\vert 1 \rangle ~= \vert 2 \rangle$~~~~~&
~~$U\vert 00001 \rangle = \vert 00010 \rangle$~~~~~\\[-5pt]
$~~~U\vert 2 \rangle ~= \vert 4 \rangle$~~~~~&
~~$U\vert 00010 \rangle = \vert 00100 \rangle$~~~~~\\[-5pt]
$~~~U\vert 4 \rangle ~= \vert 8 \rangle$~~~~~&
~~$U\vert 00100 \rangle = \vert 01000 \rangle$~~~~~\\[-5pt]
$~~~~U\vert 8 \rangle ~= \vert 16 \rangle$~~~~~&
~~$U\vert 01000 \rangle = \vert 10000 \rangle$~~~~~\\[-5pt]
$~~~U\vert 16 \rangle ~= \vert 11 \rangle$~~~~~&
~~$U\vert 10000 \rangle = \vert 01011 \rangle$~~~~~\\[-5pt]
$~~~U\vert 11 \rangle~ = \vert 1  \rangle$~~~~~~&
~~$U\vert 01011 \rangle = \vert 00001 \rangle$~~~~~\\\hline
\end{tabular} 
\end{minipage}
\caption{\footnoteskip
$N = 21$, $a = 2$, $r = 6$:
The left panel illustrates the modular exponential function 
$f_{2, 21}(x) = 2^x ~({\rm mod}~21)$, while the right panel 
shows the action of the ME operator $U_{2, 21}$ on the
 closed sequence $[1, 2, 4, 8, 16, 11, 1 ]$.  
}
\label{fig_fN21a2x}
\end{figure}

In the case of $N = 21$ with base $a = 2$, the left panel of 
Fig.~\ref{fig_fN21a2x} illustrates the modular exponential 
function $f_{2, 21}(x)$.  The period is observed to be $r = 6$,  
with the closed cycle $[1,  2,  4,  8,  16,  11,  1]$, as illustrated 
in the right panel of the Figure. Thus, we only need to consider 
$U_{2 , 21}$ on the states $\vert 1 \rangle$, $\vert 2 \rangle$, 
$\vert 4 \rangle$, $\vert 8 \rangle$, $\vert 16 \rangle$,  and 
$\vert 11 \rangle$. We must employ $n = \lceil \log_2 21 
\rceil = 5$ qubits to enumerate these states in the work register.
The work states can therefore be indexed by a binary 
string of the form $w_4 w_3 w_2 w_1 w_0$ with $w_k 
\in \{0, 1\}$, and we can use a collection of multi-control
$C^n X$ and $X$ gates to transform this string into the 
next string in the sequence. Furthermore,  we must 
measure the phase in the \hbox{$m$-bit} control register 
with sufficient accuracy to extract the correct period. For 
$r=6$,  the permissible Eigen-phases of the ME operator 
are $\phi_s = s/6$ for $s  \in \{ 0,  1, \cdots,  5\}$,  and 
we must therefor be able to resolve a phase difference 
of $\Delta \phi = 1/6 \approx 0.16666$.  We showed 
earlier in the text that the continued fractions method 
requires $m = 2 n + 1 =  11$ control qubits (for $n
=5$); however,  since $r = 6$ is so small,  it turns out 
that we can get by with only $m = 5$. Therefore, the 
Shor circuit for $N=21$ and $a=2$ will have the same 
structure as the one illustrated in Fig.~\ref{fig_15N_ex}, 
except  that the number of control qubits will be 
reduced to $m=5$. 

We will have to construct the appropriate ME operators 
$U_{2, 21}^p$ for $p \in \{ 2^0, 2^1, 2^2, 2^3, 2^4\}$, namely
$U,  U^2,  U^4,  U^8, U^{16}$,  since there are five control 
qubits. Later in this section we shall consider 
$m = 6$ control qubits, in which case we will also require 
$U^{32}$. For the time being, we will only construct the 
$U$ operator, and then concatenate this operator to form  
the composite operators $U^p$ for $p>1$.  We will refer
to this procedure by version number $\tt{u\_ver}=0$.  It 
is clear that this method will not work for general $N$, 
as it requires an exponentially large number of 
concatenations.  We will shortly illustrate how to directly 
construct the set of composite operators $U^p$ for 
$p > 1$, but for the time being we shall continue with 
our current line of development using simple concatenation. 
\begin{figure}[h!]
\hskip-1.3cm
\begin{centering}
\includegraphics[width=\textwidth]{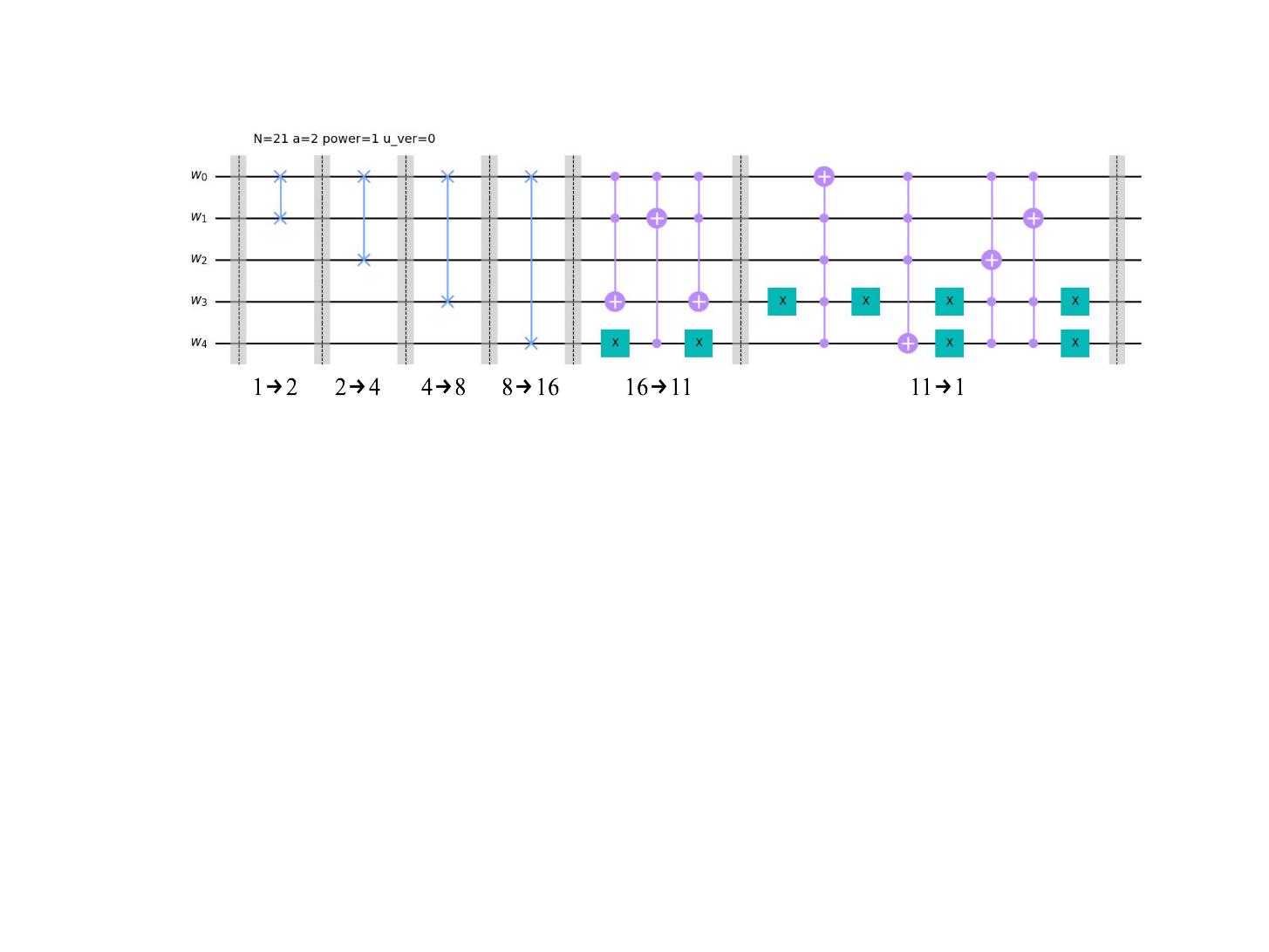} 
\par\end{centering}
\vskip-7.0cm
\caption{\footnoteskip  
$N = 21$, $a = 2$, $r = 6$:
The quantum circuit for the modular exponentiation (ME) operator  
$U_{2 , 21}$.   The quantum gates between the barriers transform 
the state from one value of $f_{2 , 21}(x)$ to the next in the closed
sequence $[1, 2, 4, 8, 16, 11, 1]$.  We will call this version of the 
ME operator by $\tt{u\_ver}=0$. 
}
\label{fig_U21a2_sub}
\end{figure}
\begin{figure}[b!]
\begin{centering}
\includegraphics[width=\textwidth]{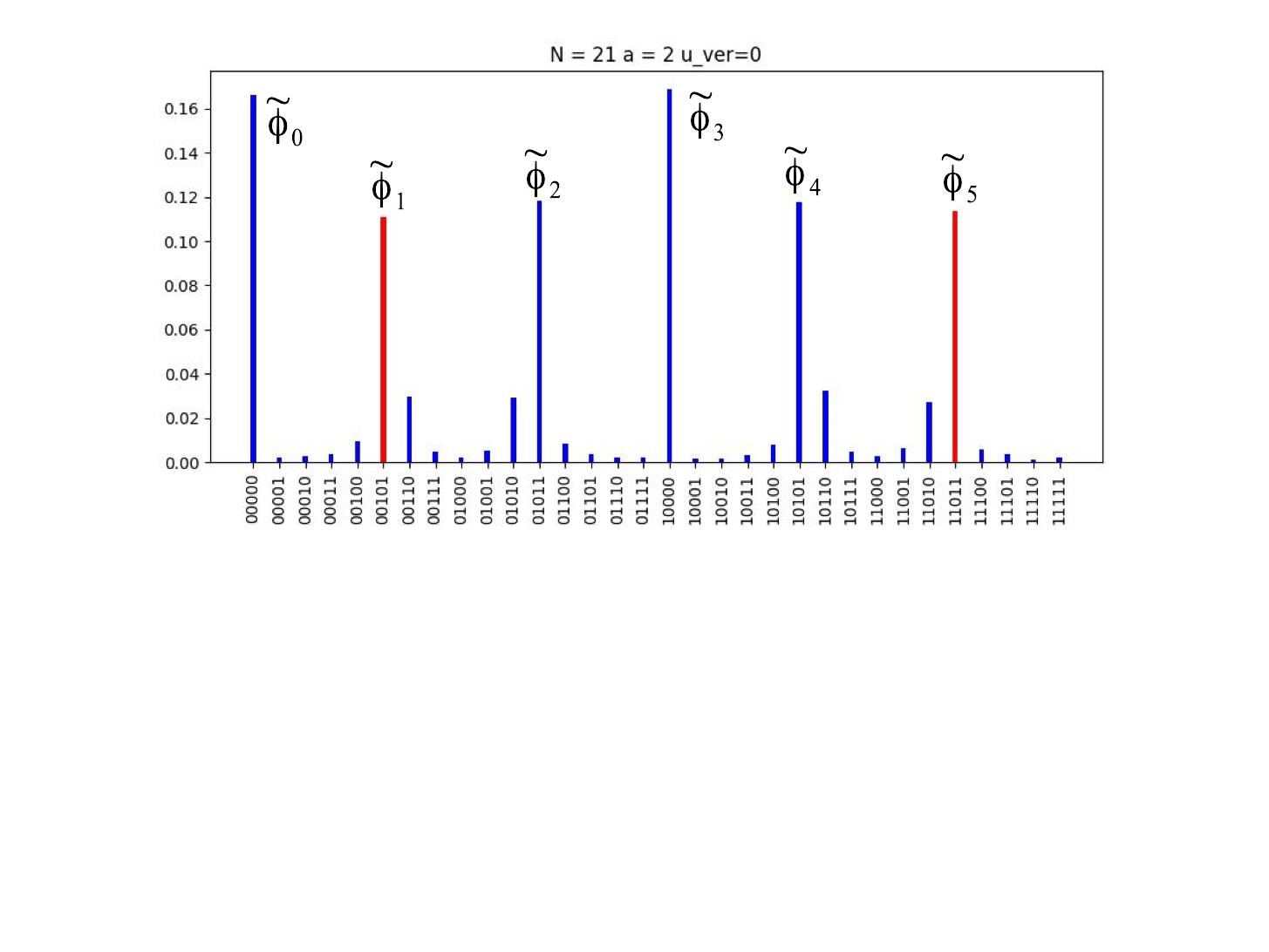} 
\par\end{centering}
\vskip-5.0cm 
\caption{\footnoteskip  
The phase histogram for $N=21$, $a=2$ and $m=5$ for $\tt{u\_ver}=0$
from a Qiskit simulation with 4096 runs. The six dominant peaks 
of the histogram occur very close to the six Eigen-phases $\phi_s 
= s/6$ of the ME operator $U_{2,21}$, where $s \in \{0, 1, \cdots, 5\}$.
The phases that produce factors are shown in red, occurring at 
the (binary)  values $\tilde\phi_1 = [0.00101]_2 \approx \phi_1 
= 1/6$ and $\tilde\phi_5 = [0.11011]_2 \approx \phi_5  = 5/6$. 
Note that these phase peaks lie well above the noise. 
}
\label{fig_N21a2_shor_hist}
\end{figure}
Figure~\ref{fig_U21a2_sub} illustrates the ME  operator
$U = U_{2, 21}$.  This operator is constructed only from  
multi-control-NOT gates $C^n X$ and single-qubit 
NOT gates $X$. The operator $U_{2, 21}$ is partitioned 
into six sections, where each section is indexed by an integer
$x = 0, 1, \cdots, 5$, and the gates in section $x$ transform
the work-state $\vert f(x) \rangle \equiv \vert w_4 \cdots
w_0 \rangle $ into  $\vert f(x+1) \rangle \equiv \vert 
w_4^\prime \cdots w_0^\prime \rangle$. For example,  
since $U \vert 1 \rangle = \vert 2 \rangle$, the SWAP gate 
between the first barrier changes the initial state $\vert 
1 \rangle = \vert 00001 \rangle$ into the next state $\vert 
2 \rangle = \vert 00010 \rangle$.  The second SWAP operation 
transforms $\vert 2 \rangle = \vert 00010 \rangle$ into 
$\vert 4 \rangle = \vert 00100 \rangle$,  and so on.  This is 
illustrated in the right panel of Fig.~\ref{fig_fN21a2x}, and 
by the annotations under the barriers in Fig.~\ref{fig_U21a2_sub}.  
We shall use this restricted version of the ME operator $U_{2, 21}$  
in Shor's algorithm.  For $m=5$ control qubits and $n = 5$ 
work qubits, Fig.~\ref{fig_N21a2_shor_hist} illustrates the 
output phase histogram from a Qiskit simulation using 4096 
shots. The abscissa indexes the possible phases, and the ordinate 
provides their corresponding probabilities. In agreement with
the phase histogram of Fig.~\ref{fig_N21a2_shor_hist}, Shor's 
algorithm is designed so that the most dominant phases 
correspond to the Eigen-phases of $U_{2, 21}$,  which take 
the simple form  $\phi_s = s/6$ for \!\!\hbox{$s \in \{0, 1, 
\cdots, 5\}$}. The phases for which ${\rm gcd}(s, 6) =1$ 
(namely $s=1$ and $s=5$) lead  to the factors of $N=21$, 
and these peaks are plotted in red. The code output for 
these phases is detailed in Table~\ref{table_N21a2_shor_table}.  
Note that the phase histogram for $N=21$ is considerably 
more complex than the $N = 15$ histogram in 
Fig.~\ref{eq_N15_hist_5}.

\begin{table}[h!]
\caption{\baselineskip 13pt
  The output of Shor's algorithm for $N=21$, $a=2$ and $m=5$
  for version $\tt{u\_ver}=0$. Only the two  phase values that 
  produced factors are listed. The variable $\tt{l\_measured}$ 
  corresponds to the control register state indexed by the integer 
  $\tilde\ell = \tilde\phi_4\cdots \tilde\phi_0$,  while 
  $\tt{phi\_phase\_bin}$ corresponds to the $5$-bit binary 
  (measured) phase $\tilde\phi = \tilde\ell/2^5 = 0.\tilde\phi_4
  \cdots \tilde\phi_0$. The decimal representation of the phase 
  is also provided for convenience. The continued fraction 
  representation, and the associated convergents (from the 
  Python package $\tt{contfrac}$)  are also given, where each 
  convergent $c = s/r$ is represented by an ordered pair $(s, r)$. 
  The code checks to see if the denominator $r$ is a solution 
  to (\ref{eq_c1})--(\ref{eq_c3}). If $r$ is a solution, then the 
  factors are given by ${\rm gcd}(a^{r/2} \pm 1, N)$. 
}
\baselineskip 10pt
\begin{verbatim}
l_measured   : 00101 5 frequency: 466
phi_phase_bin: 0.00101
phi_phase_dec: 0.15625
phi: (5, 32)
cont frac of phi  : [0, 6, 2, 2]
convergents of phi: [(0, 1), (1, 6), (2, 13), (5, 32)]
conv: (0, 1) r = 1 : no factors found
conv: (1, 6) r = 6 : factors
factor1: 7
factor2: 3
conv: (2, 13) r = 13 : no factors found
conv: (5, 32) r = 32 : no factors found

l_measured   : 11011 27 frequency: 458
phi_phase_bin: 0.11011
phi_phase_dec: 0.84375
phi: (27, 32)
cont frac of phi  : [0, 1, 5, 2, 2]
convergents of phi: [(0, 1), (1, 1), (5, 6), (11, 13), (27, 32)]
conv: (0, 1) r = 1 : no factors found
conv: (1, 1) r = 1 : no factors found
conv: (5, 6) r = 6 : factors
factor1: 7
factor2: 3
conv: (11, 13) r = 13 : no factors found
conv: (27, 32) r = 32 : no factors found
\end{verbatim}
\label{table_N21a2_shor_table}
\end{table}
%%

%\pagebreak
Let us examine the phase histogram in Fig.~\ref{fig_N21a2_shor_hist} 
in a little more detail. As noted above, the abscissa gives the phases
of the ME operator $U_{2, 21}$. They are represented by $5$-bit 
integers $\tilde \ell = \tilde\phi_4\cdots \tilde\phi_0$, and correspond
to the binary phases $\tilde\phi =\tilde\ell/2^5 = 0.\tilde\phi_4\cdots 
\tilde\phi_0$,  where $\tilde\phi_k \in \{0, 1\}$ is the measured value 
of qubit $k$ in the control register. The ordinate gives the probability 
that the given phase will be observed during a measurement of the 
control register. As noted in the previous paragraph, we expect the 
phase histogram to have large peaks close to the six phases of the  
ME operator,

%\vfill
%\pagebreak

%%
\begin{eqnarray}
\nonumber
  \phi_0 &=&  ~0 ~~= 0.00000\cdots
  \\[-3pt]
  \nonumber
  \phi_1 &=& 1/6 = 0.16666\cdots
  \\[-3pt]
    \phi_2 &=&2/6 = 0.33333\cdots
  \\[-3pt]
   \nonumber
    \phi_3 &=&3/6 = 0.50000\cdots
  \\[-3pt]
   \nonumber
    \phi_4 &=&4/6 = 0.66666\cdots
  \\[-3pt]
   \nonumber
    \phi_5 &=&5/6 = 0.83333\cdots
    \ .
\end{eqnarray}
And indeed it does, as  the six dominant peaks in 
Fig.~\ref{fig_N21a2_shor_hist} occur at
\begin{eqnarray}
  \nonumber
  \tilde\ell_0 &=& [00000]_2 = 0
  \hskip2.0cm
  \tilde\phi_0 = [0.00000]_2 = 0.00000 = \phi_0
  \\[-3pt]
    \nonumber
  \tilde\ell_1 &=& [00101]_2 = 5
  \hskip2.0cm
  \tilde\phi_1 = [0.00101]_2 = 0.15625 \approx \phi_1
  ~\Leftarrow~\text{factors}: 3, 7
  \\[-3pt]
      \nonumber
  \tilde\ell_2 &=&[ 01011]_2 = 11
  \hskip1.8cm
  \tilde\phi_2 =[ 0.01011]_2 = 0.34375 \approx \phi_2
  \\[-3pt]
  \tilde\ell_3 &=& [10000]_2 = 16
  \hskip1.8cm
  \tilde\phi_3 = [0.10000]_2 = 0.50000 = \phi_3
  \\[-3pt]
    \nonumber
  \tilde\ell_4 &=& [10101]_2 = 21
  \hskip1.8cm
  \tilde\phi_4 = [0.10101]_2 = 0.65625 \approx \phi_4
  \\[-3pt]
    \nonumber
  \tilde\ell_5 &=& [11011]_2 = 27
  \hskip1.8cm
  \tilde\phi_5 = [0.11011]_2 = 0.84375 \approx \phi_5
  ~\Leftarrow~\text{factors}: 3, 7
 ~ \ . ~~~
\end{eqnarray}
A phase measurement $\tilde\phi_s$ will typically not 
exactly equal the associated Eigen-phase $\phi_s$ 
because of the finite resolution of the  control register. 
However, each peak $\tilde\phi_s$ lies very close to an 
actual phase $\phi_s = s/6$,  although only the phases 
$s=1$ and $s = 5$ produce the factors of 3 and 7. The 
probability of finding a factor during each shot is 
about $2/6 \approx 30\%$,  and we are therefore almost 
guaranteed to find a factor after only a few iterations.

Let us next examine the first entry in Table~\ref{table_N21a2_shor_table}
in some detail.  This entry corresponds to the first red peak
in the phase histogram,  $\tilde\ell_1 = [00101]_2 = 5$, which
gives a measured phase of  $\tilde\phi_1 = [0.00101]_2 = 
0.15625$. This lies very close to $\phi_1 = 1/6 = 0.16666
\cdots$. In general, the difference between each measured 
phase and the exact ME phase is of order 0.01, which is less 
than the resolution $2^{-5} =  0.03125$.  We also list the 
frequency of occurrence out of the total number of shots 
of 4096. The Python continued fraction module \verb+contfrac+  
prefers fractional inputs, so we convert the decimal value 
of the phase to a fraction, $\tilde\phi_1 = 5/32$, and the
corresponding continued fraction is found to be
\begin{eqnarray}
  \tilde\phi_1 = 0.15625 = 5/32 = [0; 0, 6, 2, 2]
  \ .
\end{eqnarray}
The module \verb+contfrac+ then calculates all possible 
convergents of the rational number $\tilde\phi_1 = 5/32$, 
thereby giving $c_0 = 0/1$, $c_1 = 1/6$, $c_2 = 2/13$, $c_3 
= 5/32$. Note that these fractions are represented by the 
ordered pairs $(0, 1)$, $(1, 6)$, $(2, 13)$ and $(5, 32)$ in the 
continued fraction package. We then sequence through 
this small list of convergents $c_\ell = s_\ell/r_\ell$, testing 
each value of $r = r_\ell$ for a solution to 
(\ref{eq_c1})--(\ref{eq_c3}).  As we expect, only $r_1 = 6$ 
provides such a solution.  As a matter of completeness, 
we provide below the code output for phase $\tilde\phi_2$, 
which does not produce factors. 

%\pagebreak
\noindent
Code output for $N=21$, $a=2$, $r =6$, $m=5$, 
peak $\tilde\phi_2 = [0.01011]_2=0.34375$ (no factors):
\vskip0.5cm
\baselineskip 10pt
\begin{verbatim}
l_measured   : 01011 11 frequency: 480
phi_phase_bin: 0.01011
phi_phase_dec: 0.34375
phi: (11, 32)
cont frac of phi  : [0, 2, 1, 10]
convergents of phi: [(0, 1), (1, 2), (1, 3), (11, 32)]
conv: (0, 1) r = 1 : no factors found
conv: (1, 2) r = 2 : no factors found
conv: (1, 3) r = 3 : no factors found
conv: (11, 32) r = 32 : no factors found
\end{verbatim}
\bodyskip

\begin{figure}[t!]
\begin{centering}
\includegraphics[width=\textwidth]{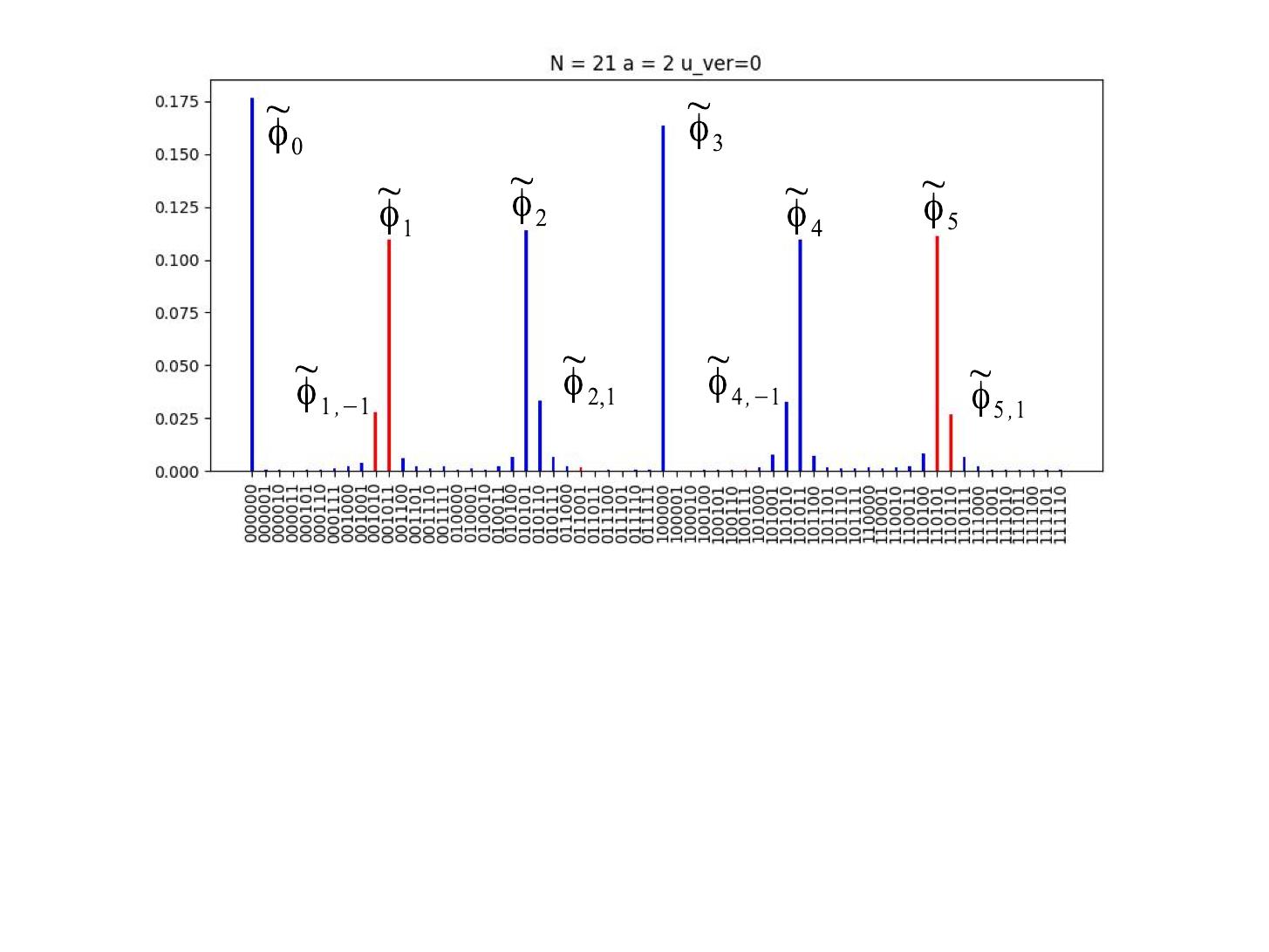} 
\par\end{centering}
\vskip-5.0cm
\begin{centering}
\includegraphics[width=\textwidth]{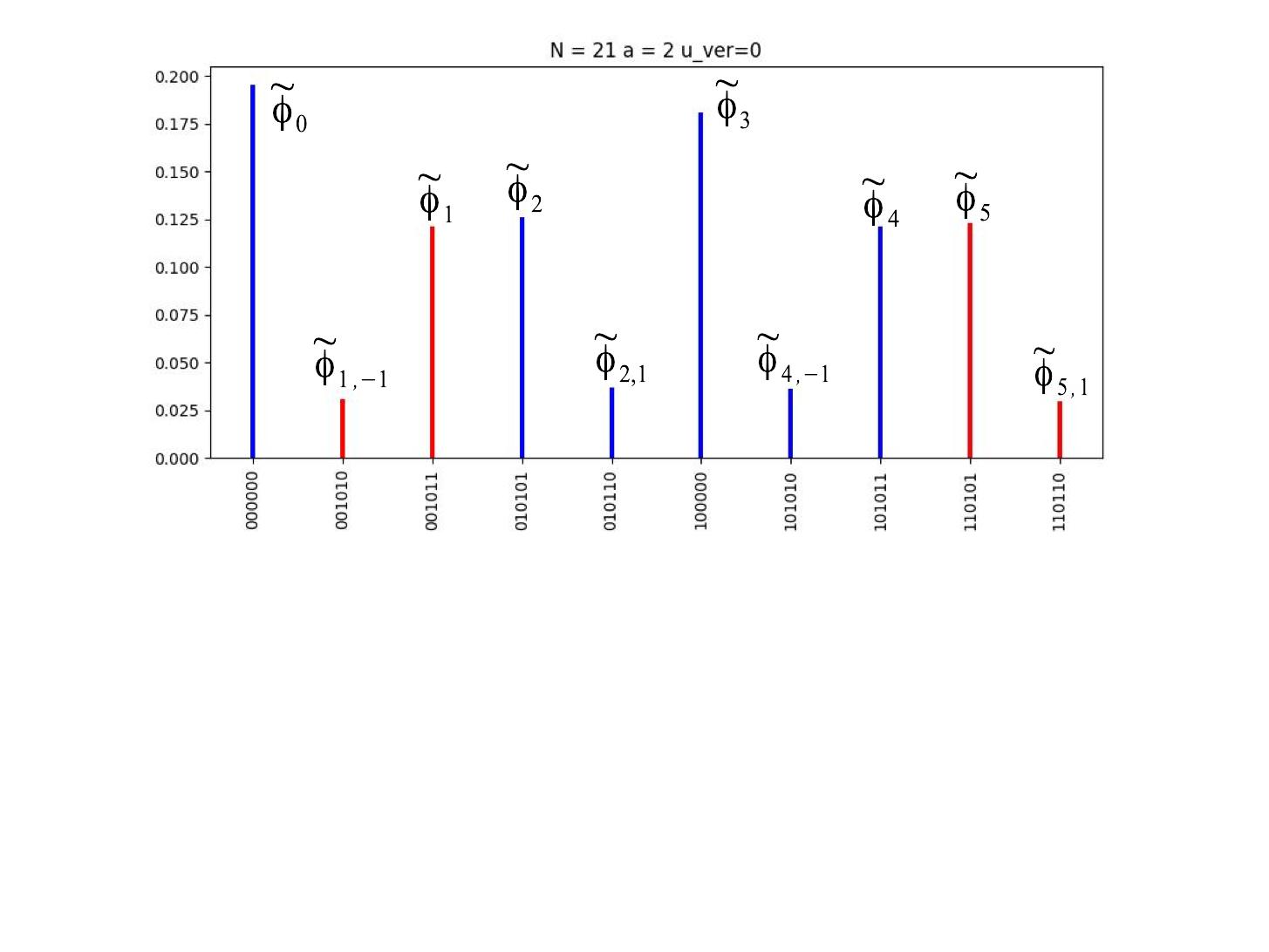} 
\par\end{centering}
\vskip-5.3cm 
\caption{\footnoteskip  
The corresponding phase histogram for $m = 6$. There is a dominant
and sub-dominate peak for each ME phase $\phi_s$. Note that sub-dominant
peaks can also produce factors.
}
\label{fig_N21a2_shor_m6_hist}
\end{figure}

It is interesting to increase the number of control qubits to 
$m = 6$. This involves the operator $U_{2, 21}^{32}$, which 
can be constructed from $U_{2, 21}$ by simple concatenation, 
albeit with the price of increasing the number of gates substantially. 
Figure~\ref{fig_N21a2_shor_m6_hist} illustrates the output
phase histogram from another Qiskit run with 4096 shots.
The top panel of the Figure shows all output phases, while
the bottom panel gives only the most likely phases. 
The measured phases are peaked near each of the exact 
phases~$\phi_s$, and they lie within the resolution $2^{-6} 
= 0.015625$ of these phases. Note, however, that most phases 
have a dominant and a sub-dominant peak, labeled by $\tilde\phi_\ell$ 
and $\tilde\phi_{\ell, k}$ respectively, where $k = \pm 1$ depending on 
whether the sub-dominant peak lies to the left or right of the 
dominant peak. Note that the sub-dominant peaks for $s=1$ 
and $s=5$ also produce factors. The measured values of each 
of these peaks are given below:
\begin{eqnarray}
    \nonumber
  \tilde\ell_0 &=& [000000]_2 = 0
  \hskip1.8cm
  \tilde\phi_0 ~~~= [0.000000]_2 = 0.000000 = \phi_0
  \\[-3pt]
  \nonumber
  \tilde\ell_{1, -1} &=& [001010]_2 = 10
  \hskip1.60cm
  \tilde\phi_{1,-1} = [0.001010]_2 = 0.156250 \approx \phi_1
    ~\Leftarrow~\text{factors}: 3, 7
  \\[-3pt]
  \nonumber
  \tilde\ell_1 &=& [001011]_2 = 11
  \hskip1.65cm
  \tilde\phi_1 ~~~= [0.001011]_2 = 0.171875 \approx \phi_1
    ~\Leftarrow~\text{factors}: 3, 7
    \\[-3pt]
  \nonumber
  \tilde\ell_2 &=& [010101]_2 = 21
  \hskip1.65cm
  \tilde\phi_2 ~~~= [0.010101]_2 = 0.328125 \approx \phi_2
  \\[-3pt]
    \nonumber
    \tilde\ell_{2,1} &=& [010110]_2 = 22
  \hskip1.65cm
  \tilde\phi_{2,1} ~\,= [0.010110]_2 = 0.343750 \approx \phi_2
 \label{eq_N21a3m6_phases}
   \\[-3pt]
    \tilde\ell_3 &=& [100000]_2 = 32
  \hskip1.7cm
  \tilde\phi_3 ~~~= [0.100000]_2 = 0.500000 = \phi_3
  \\[-3pt]
    \nonumber
    \tilde\ell_{4, -1} &=& [101010]_2 = 42
  \hskip1.65cm
  \tilde\phi_{4, -1} = [0.101010]_2 = 0.656250 \approx \phi_4
  \\[-3pt]
    \nonumber
    \tilde\ell_4 &=& [101011]_2 = 43
  \hskip1.7cm
  \tilde\phi_4 ~~~= [0.101011]_2 = 0.671875 \approx \phi_4
  \\[-3pt]
    \nonumber
    \tilde\ell_5 &=& [110101]_2 = 53
  \hskip1.7cm
  \tilde\phi_5 ~~~= [0.110101]_2 = 0.828125 \approx \phi_5
    ~\Leftarrow~\text{factors}: 3, 7
  \\[-3pt]
    \nonumber
    \tilde\ell_{5, 1} &=& [110110]_2 = 54
  \hskip1.65cm
  \tilde\phi_{5, 1}   ~\,= [0.110110]_2 = 0.843750 \approx \phi_5
    ~\Leftarrow~\text{factors}: 3, 7
    \ .
\end{eqnarray}
\begin{figure}[t!]
\includegraphics[scale=0.45]{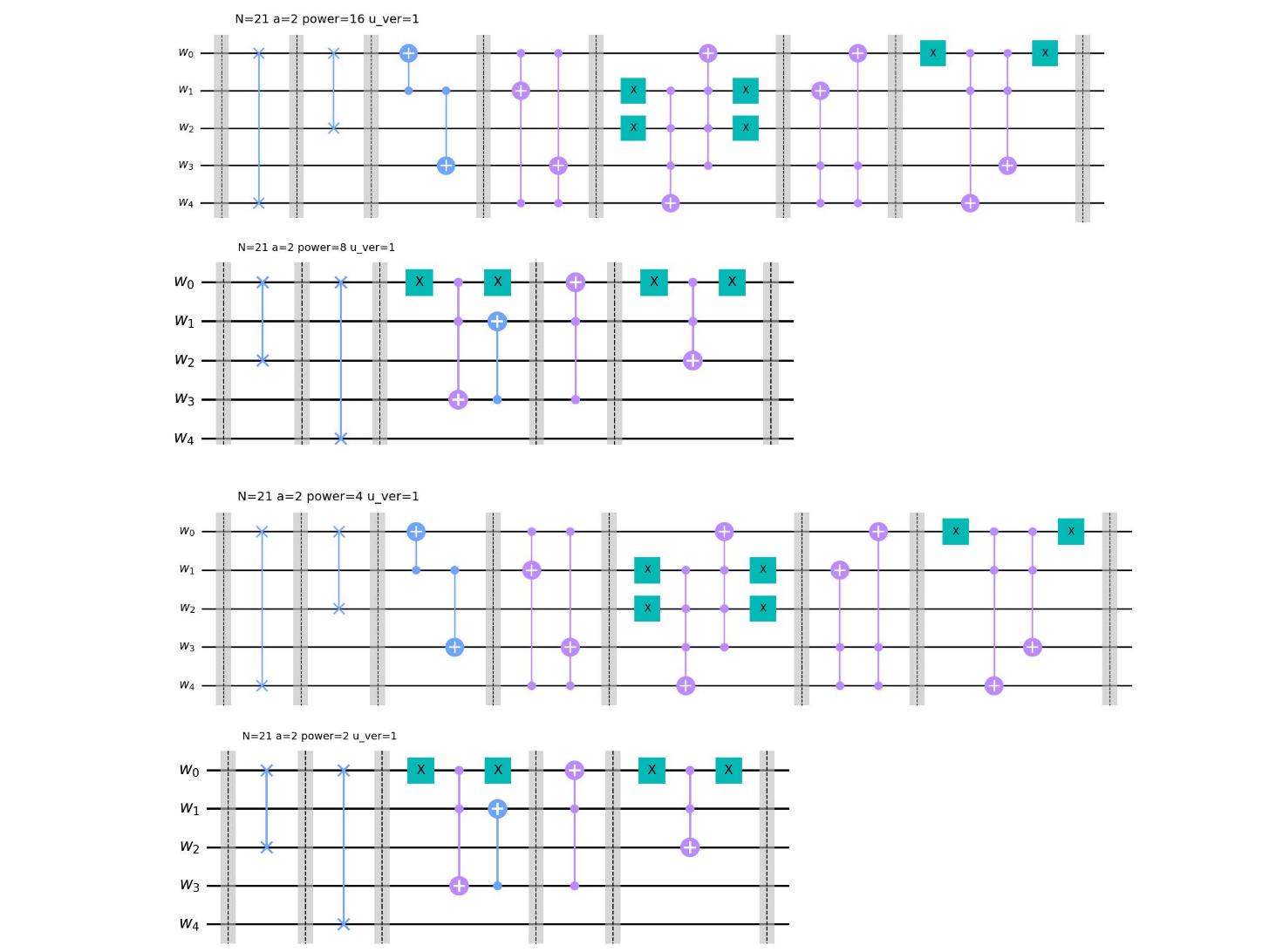}
\caption{\footnoteskip
$N = 21$, $a = 2$, $r = 6$:
The ME operators $U^2,  U^4,  U^8$ and 
$U^{16}$ for version $\tt{u\_ver}=1$ given in expression 
(\ref{eq_Up_N21a2_seq}). 
}
\label{fig_UpN21a2}
\end{figure}
\begin{figure}[h!]
\begin{centering}
\includegraphics[width=\textwidth]{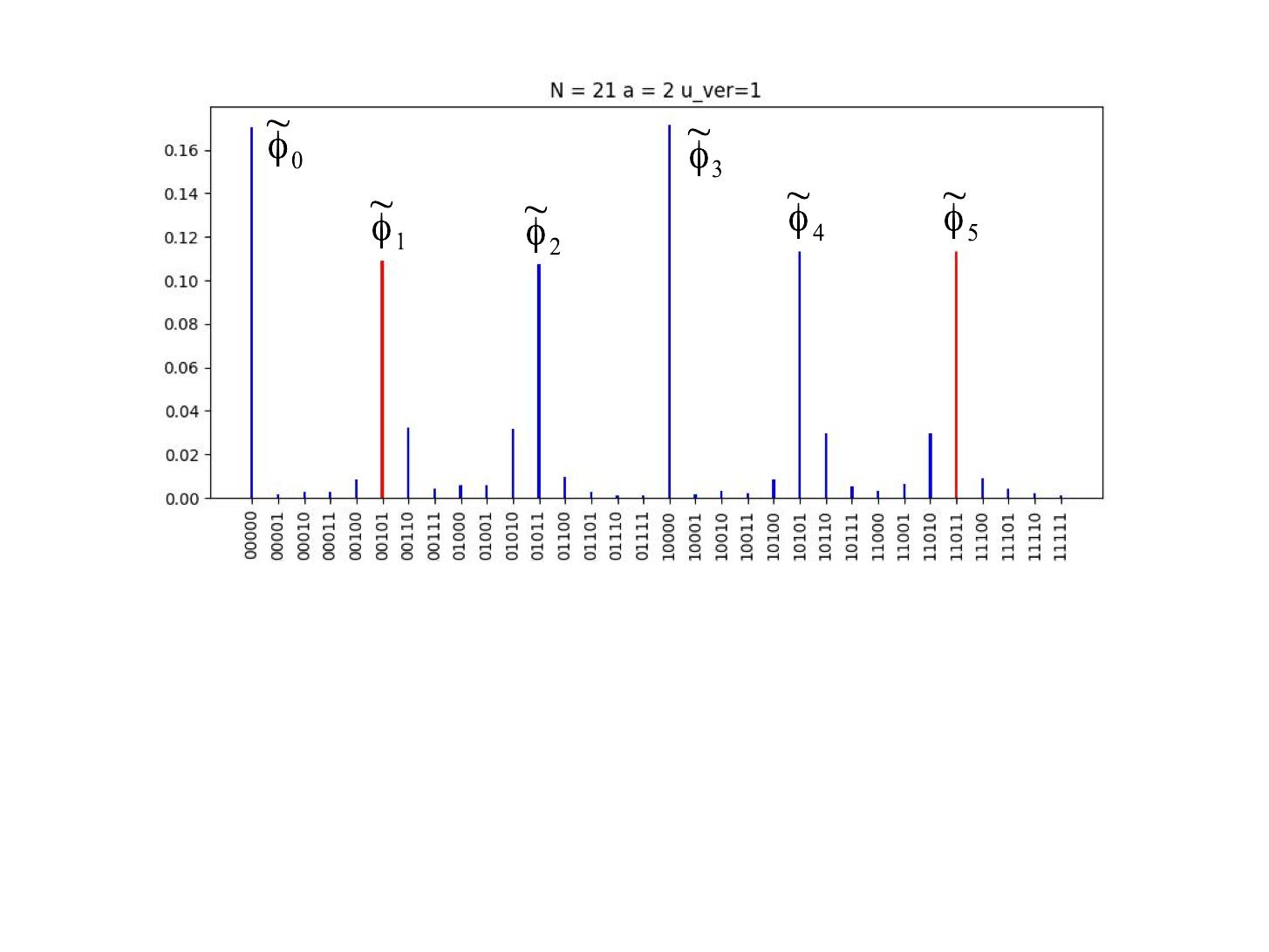}
\par\end{centering}
\vskip-5.0cm
\caption{\footnoteskip
$N = 21$, $a = 2$, $r = 6$, $m=5$:
Phase histogram for ME operator version $\tt{u\_ver}=1$ 
agrees with the original phase histogram of 
Fig.~\ref{fig_N21a2_shor_hist}. As before, the phases 
that produce factors are shown in red. 
}
\label{fig_phase_uver1}
\end{figure}
We are now ready to address the concatenation issue.  
Let us return to $m = 5$ control qubits.  Up to now we have 
produced $U^2,  U^4,  U^8$ and $U^{16}$ by simply 
concatenating the operator $U$ from Fig.~\ref{fig_U21a2_sub}.
However,  we only require the five operators $U^p$ on the 
6-dimensional subspace ${\cal U}_{r=6} \subseteq {\cal W}_{
n=5}$.  Note that the operator $U^2$ acts on every other 
element of the sequence $[1, 2, 4, 8, 16, 11, 1]$,  producing 
two closed sub-sequences $[1, 4, 16, 1]$ and $[2, 8, 11, 2]$.  
Similarly,  $U^4$ chooses every 4-th element of the sequence
and so on,  so that the ME operators $U^2$,  $U^4$, $U^8$, and 
$U^{16}$ all act on pairs of closed sub-sequences:
\vskip-1.0cm
\begin{eqnarray}
  U_{2, 21} && ~:~~~ [1, 2, 4, 8, 16, 11, 1]
\nonumber\\[-3pt]
  U^2_{2, 21} && ~:~~~ [1, 4, 16, 1] ~~\text{and}~~ [2, 8, 11, 2]
\nonumber\\[-3pt]
  U^4_{2, 21} && ~:~~~  [1, 16, 4, 1] ~~\text{and}~~ [2, 11, 8, 2]
\label{eq_Up_N21a2_seq}
\\[-3pt]
  U^8_{2, 21} && ~:~~~ [1, 4, 16, 1]  ~~\text{and}~~ [2, 8, 11, 2]
\nonumber\\[-3pt]
  U^{16}_{2, 21} && ~:~~~ [1, 16, 4, 1] ~~\text{and}~~ [2, 11, 8, 2]
\nonumber
  \ ,
\end{eqnarray}
where we have restored the $N=21$ and $a=2$ subscripts
on the ME operator $U = U_{2, 21}$ for clarity. The corresponding 
circuits that produce these sequences are given in 
Fig.~\ref{fig_UpN21a2},  and we will refer to this collection 
as version number $\tt{u\_ver}=1$.  Figure~\ref{fig_phase_uver1} 
illustrates the phase histogram from Shor's algorithm with these
ME operators.  The graph is identical to that of 
Fig.~\ref{fig_N21a2_shor_hist} (as it should be). We see that 
the phases $\tilde\phi_1= [0.00101]_2 \approx 1/6 $ and 
$\tilde\phi_5 = [0.11011]_2 \approx 5/6$ still lie well above 
the noise,  producing the correct factorization.

\begin{figure}[t!]
\hfill
\begin{centering}
\includegraphics[width=\textwidth]{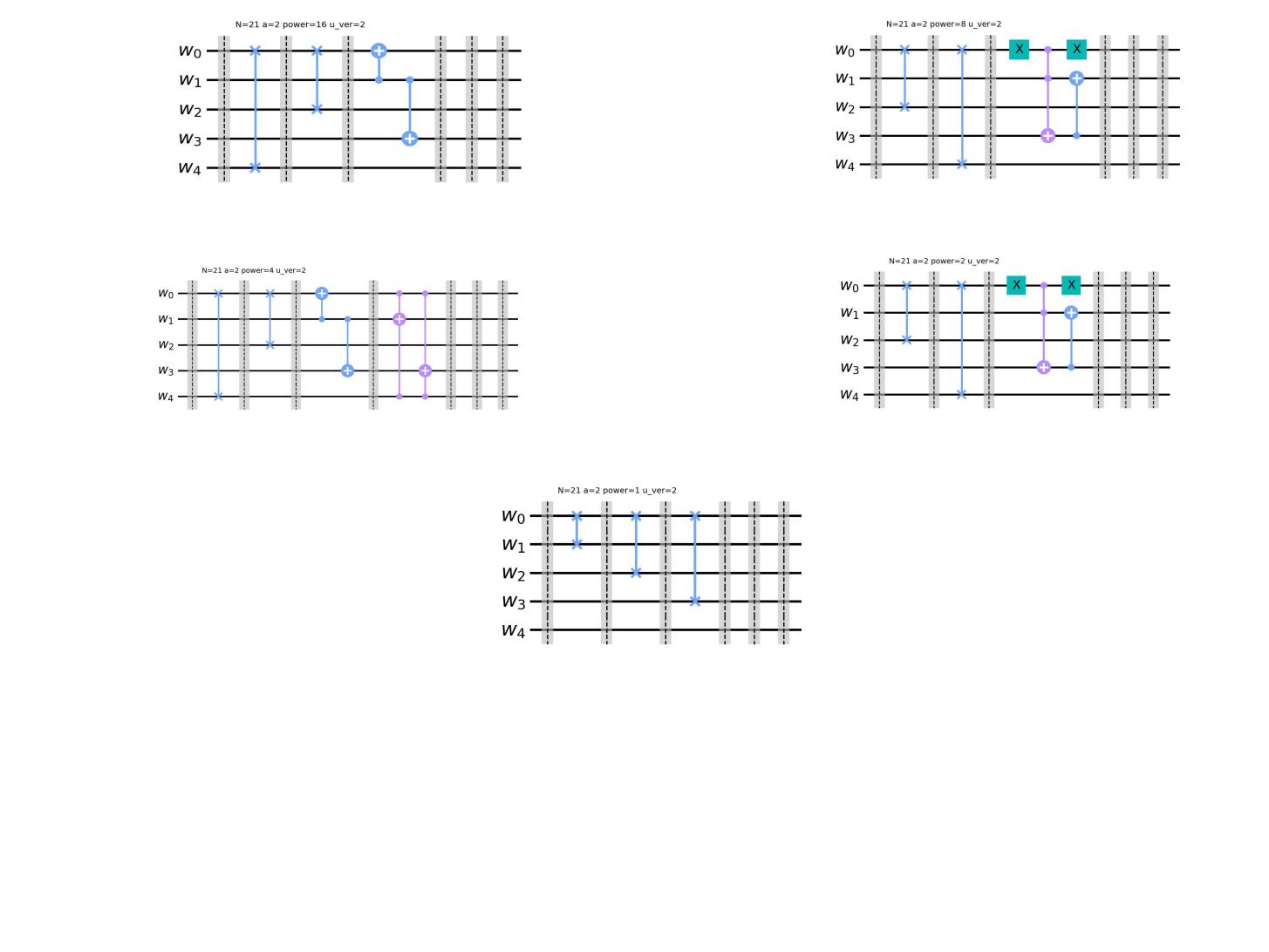}
\par\end{centering}
\vskip-3.8cm
\caption{\footnoteskip
$N =21$ and $a=2$: 
Truncated ME operators $U,  U^2, U^4,  U^8$ and $U^{16}$
for ME operator version $\tt{u\_ver}=2$.  
}
\label{fig_UpN21a2_truncated}
\end{figure}

\begin{figure}[t!]
\begin{centering}
\includegraphics[width=\textwidth]{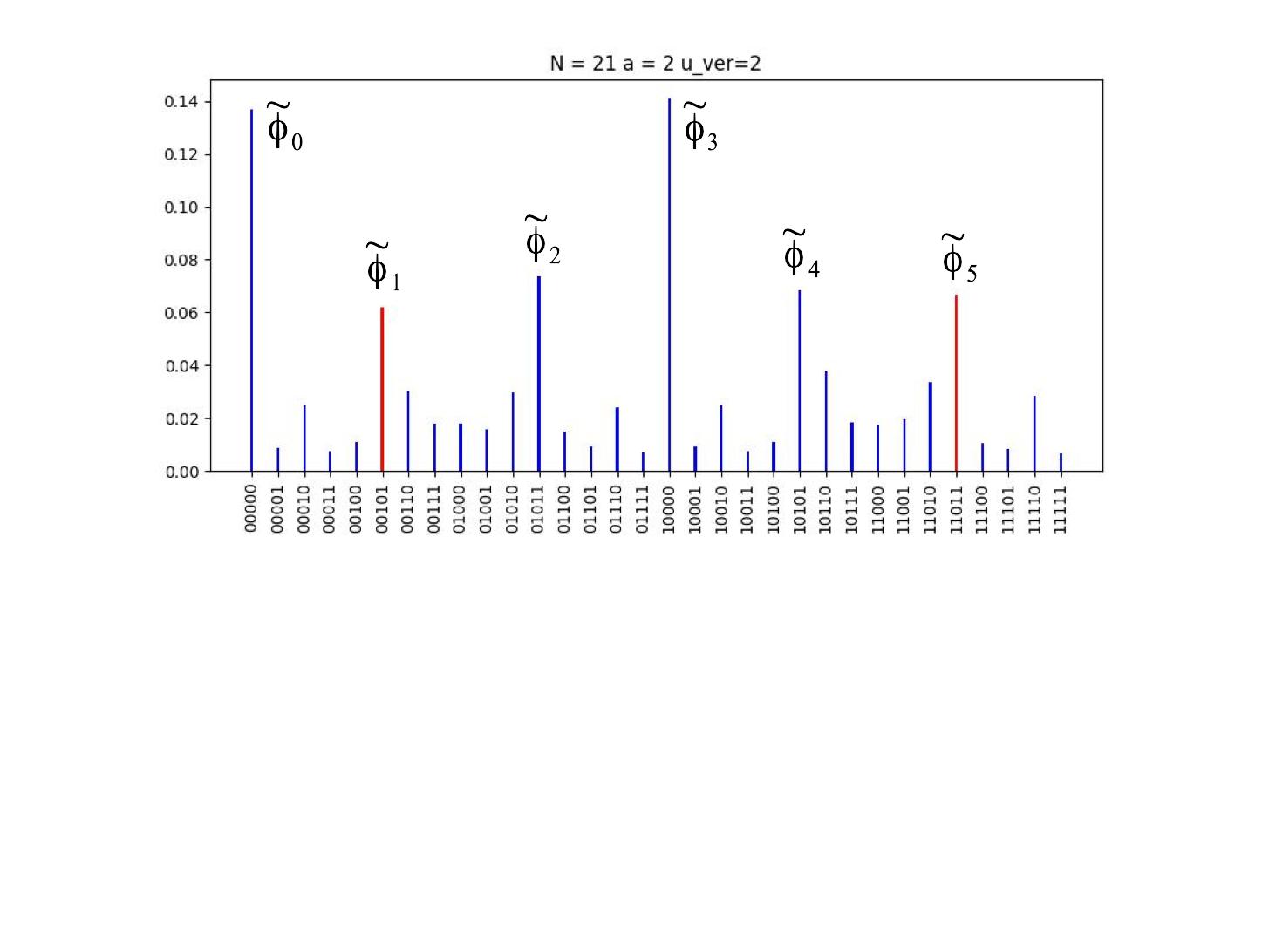}
\par\end{centering}
\vskip-5.5cm
\caption{\footnoteskip
$N = 21$, $a = 2$, $r = 6$, $m=5$:
Phase histogram for version $\tt{u\_ver}=2$,  which employs 
the truncated ME operators in Fig.~\ref{fig_UpN21a2_truncated}. 
The signal agrees with the previous two versions, with only slightly 
more noise, and the peaks in red correspond to phases that produce
the factors of 3 and 7. 
}
\label{fig_uver2}
\end{figure}

%\pagebreak
At this point,  one can (and should) levy another charge against 
this procedure:
we have used the {\em entire} cycle $[1, 2, 4, 8, 16, 11, 1]$ for the
ME operator $U_{2, 21}$,  which means that we know {\em a priori} 
that the period of the modular exponential function $f_{2, 21}(x)$
is $r=6$.  In other words, if we knew the complete closed-sequence
for a general $N$,  then this is equivalent to knowing the period 
$r$,  so there is no need for Shor's algorithm. However,  we do
{\em  not} require the complete sequence!  This is because when
employing the method of continued fractions,  it is not necessary
to know the {\em exact} phase, but only an {\em approximate} 
phase. Therefore,  we require only as much resolution in the phases
$\tilde\phi_s$ as to  extract the corresponding convergents $s/r$ 
using continued fractions. Figure~\ref{fig_UpN21a2_truncated}
illustrates a {\em truncated} version of the operators $U,  U^2,  
U^4, U^8$, and $U^{16}$,  in which we have omitted several 
stages from each ME operator $U^p$.   We shall refer to this 
as version $\tt{u\_ver}=2$.  We see from the phase histogram
in  Fig.~\ref{fig_uver2} that employing these operators in 
Shor's algorithm still permits one to extract the appropriate 
phases,  and therefore the correct factors.  Not surprisingly,  
the phase histogram has more noise,  but this does not 
overwhelm the signal. We have explored a number of 
truncation procedures, and they all produce similar results. 
We will see in the next section that these methods continue
to work for even larger values of~$N$. 

\vfill

\pagebreak%%
\begin{figure}[b!]
\hskip-1.0cm
\begin{minipage}[c]{0.4\linewidth}
\includegraphics[scale=0.45]{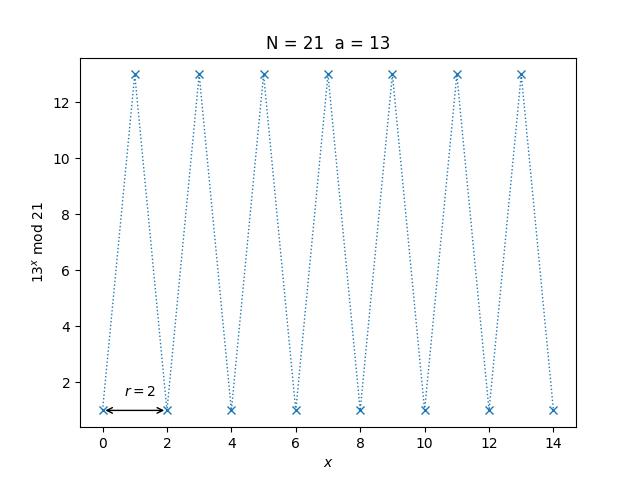}
\end{minipage}
\hskip1.6cm
\begin{minipage}[c]{0.4\linewidth}
\begin{tabular}{|c|c|} \hline
 \multicolumn{2}{|c|}{~$U\vert w \rangle = 
\big\vert 14 \cdot w ~({\rm mod}~21) \big\rangle$~}  
\\\hline
$~~~~~U\vert 1 \rangle~~  =~ \vert 13 \rangle$~~~~~&
~~$U\vert 00001 \rangle = \vert 01101 \rangle$~~~~~~\\[-5pt]
$~~~U\vert 13 \rangle ~=~ \vert 1 \rangle$~~~~~&
~~$U\vert 01101 \rangle = \vert 00001 \rangle$~~~~~\\\hline
\end{tabular} 
\end{minipage}
\vskip-0.3cm
\includegraphics[scale=0.35]{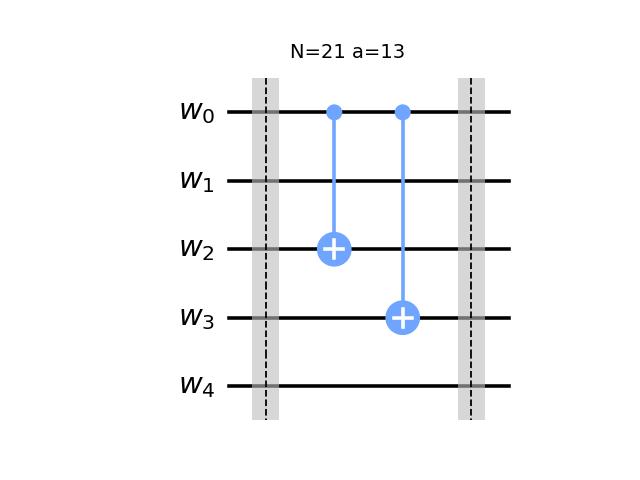}
\vskip-0.5cm
\caption{\footnoteskip
$N = 21$, $a = 13$, $r = 2$:
The left panel illustrates the modular exponential function 
$f_{2, 13}(x) = 2^x ~({\rm mod}~21)$, while the right panel 
shows the action of the ME operator $U_{2, 13}$ on the
 closed sequence $[1, 13, 1 ]$ .
}
\label{fig_fN21a13x}
\end{figure}
\begin{figure}[h!]
\includegraphics[scale=0.40]{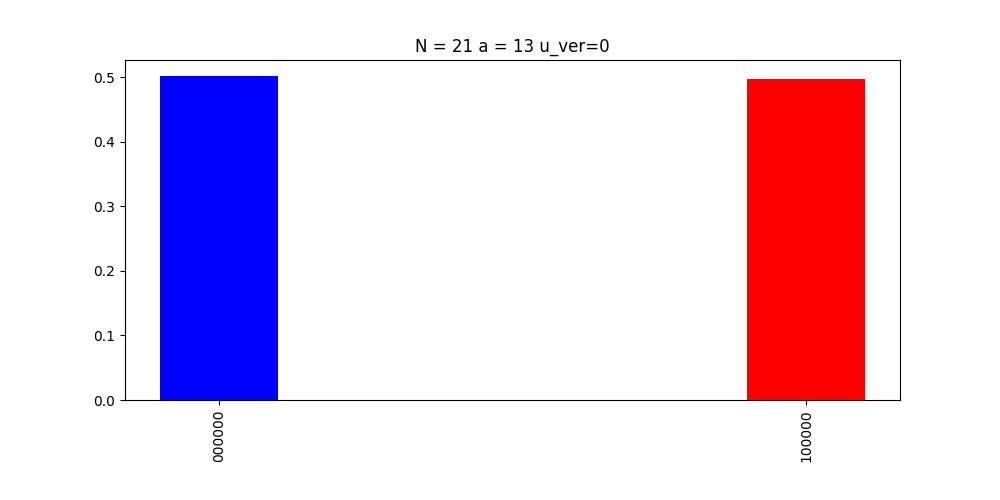}
\vskip-0.2cm
\caption{\footnoteskip
$N = 21$, $a = 13$, $r = 2$: The  corresponding phase
histogram for $m=6$ and $\tt{u\_ver}=0$.  The red peak
produces the factors of 3 and 7.
}
\label{fig_fN21a13_hist}
\end{figure}

Before continuing on to larger numbers,  however,  let us confirm
a result derived in the previous sections. We have shown that
when $\ell_s = 2^m \phi_s$ is an integer for all \hbox{$s \in \{0, 1, \cdots, 
r-1\}$},  then the final state amplitudes are non-zero only for $\ell 
= \ell_s$.  In the
upper panel of Fig.~\ref{fig_fN21a13x}, we see that the modular
exponential function for $N = 21$ and $a =13$ has a period of
$r = 2$. The lower panel gives the corresponding ME operator,
while Fig.~\ref{fig_fN21a13_hist} illustrates the phase histogram
for $m = 6$. There are exactly two peaks at the values $\tilde\ell_0 
= 000000$ and $\tilde\ell_1 =  100000$, as expected. These peaks
correspond to the phase angles $\tilde\phi_0 = 0.000000$ and 
$\tilde\phi_1 =  0.100000 = 1/2$, and the latter produces the 
correct factors of 3 and 7.

%%%
%\begin{figure}[h!]
%\hskip-1.0cm
%\begin{minipage}[c]{0.4\linewidth}
%\includegraphics[scale=0.45]{02_period_v1_N21_a13.jpg}
%\end{minipage}
%\hskip1.6cm
%\begin{minipage}[c]{0.4\linewidth}
%\begin{tabular}{|c|c|} \hline
% \multicolumn{2}{|c|}{~$U\vert w \rangle = 
%\big\vert 14 \cdot w ~({\rm mod}~21) \big\rangle$~}  
%\\\hline
%$~~~U\vert 1 \rangle~~  ~~=~ \vert 13 \rangle$~~~~~&
%~~$U\vert 00001 \rangle = \vert 01101 \rangle$~~~~~\\[-5pt]
%$~~~U\vert 13 \rangle ~=~ \vert 1 \rangle$~~~~~&
%~~$U\vert 01101 \rangle = \vert 00001 \rangle$~~~~~\\\hline
%\end{tabular} 
%\end{minipage}
%\vskip-0.3cm
%\includegraphics[scale=0.35]{12_shor_v2_N21_a13_m6_uver0_U1.jpg}
%\vskip-0.2cm
%\caption{\footnoteskip
%$N = 21$, $a = 13$, $r = 2$:
%The left panel illustrates the modular exponential function 
%$f_{2, 13}(x) = 2^x ~({\rm mod}~21)$, while the right panel 
%shows the action of the ME operator $U_{2, 13}$ on the
% closed sequence $[1, 13, 1 ]$ .
%}
%\label{fig_fN21a13x}
%\end{figure}
%%%

%\vskip-1.0cm
%%
%%

%\vfill
%\clearpage
%\pagebreak
\subsection{More Factoring}

\begin{figure}[h!]
\includegraphics[scale=0.50]{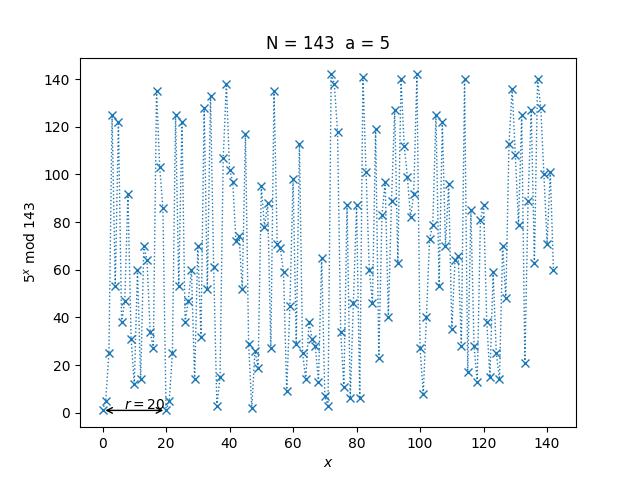}
%\vskip-0.3cm
\caption{\footnoteskip
The range of $f_{5, 143}(x)$ has been extended.  The 
function appears random,  and the period $r=20$ is
not apparent at first glance. 
}
\label{fig_fxaN_all}
\end{figure}

We now turn to factoring even large numbers $N$.  As pointed out 
in Ref.~\cite{pretend},  the difficulty for Shor's algorithm lies 
not in the size of the number $N$,  but in the length of the 
period~$r$.  As illustrated in Table~\ref{table_factors},  we 
have therefore chosen a collection of composite numbers,
$N = 21, 35, 33, 143, 247$,  and corresponding bases $a$, 
that cover a wide range of periods from $r=6$ to $r=36$.
\begin{table}[b!]
\caption{\footnoteskip 
Composite numbers $N = p \times q$ for primes $p$ and $q$, 
together with the corresponding  Shor parameters: the work-space
length $n = \lceil \log_2 N \rceil$,  the base $a$,  the period
$r$ of $f_{a \, \smN}(x)$,  and the length $m$ of the control 
register.
}
\begin{tabular}{|l||c|c|c|l|} \hline
   ~$N = p \times q$~ &  ~$n$~    &  ~$a$~    &  ~$r$~  & ~$m$~  \\ \hline
   ~$21=3\times 7$~                &   ~5~  &  ~2~ &   ~6~      &    ~~5, 6~ \\[-5pt]
   ~$35 = 5 \times 7$~           &   6       &  4      &   ~6~      &    ~~5 \\[-5pt] 
   ~$33 = 3 \times 11$~        &   6       &  7      &   ~10~    &   ~~6 \\[-5pt]
   ~$143 = 13 \times 11$~   &   8       &  5      &   ~20~   &    ~~8, 9, 10~ \\[-5pt]
   ~$247 = 13 \times 19$~   &   8       & 2       &   ~36~   &    ~~9, 10 \\ \hline
\end{tabular} 
\label{table_factors}
\end{table}
\begin{figure}[h!]
\begin{minipage}[c]{0.5\linewidth}
\includegraphics[scale=0.50]{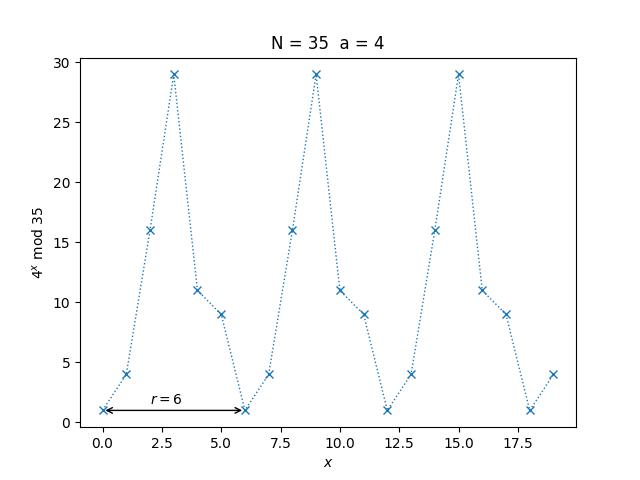}
\end{minipage}
\begin{minipage}[c]{0.4\linewidth}
\includegraphics[scale=0.50]{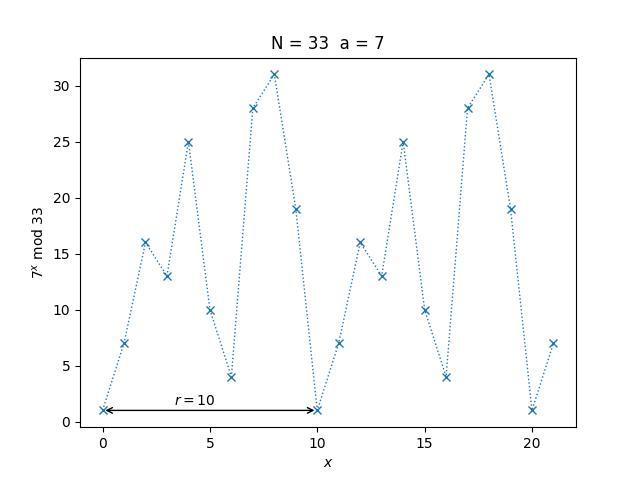}
\end{minipage}
\begin{minipage}[c]{0.5\linewidth}
\includegraphics[scale=0.50]{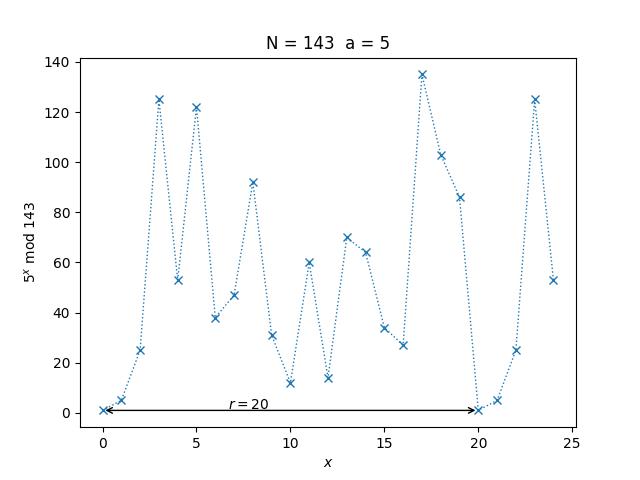}
\end{minipage}
\begin{minipage}[c]{0.4\linewidth}
\includegraphics[scale=0.50]{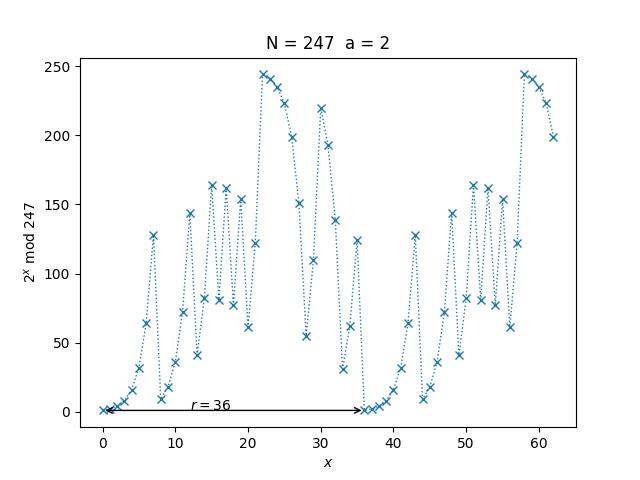}
\end{minipage}
\caption{\footnoteskip
The modular exponential functions $f_{a, \smN}(x) = a^x ~ 
({\rm mod}~N)$ for the numbers $N$ and bases $a$ in 
Table~\ref{table_factors}.  We shall use these in Shor's 
algorithm: the top left corresponds to $N=35$ and $a=4$, 
the top right is for $N = 33$ and $a=7$, the bottom left is
for $N=143$ and $a=5$, and the bottom right is for $N=247$ 
and $a=2$.
}
\label{fig_fxaNmodN_four}
\end{figure}
\hskip-0.2cm
Note that we require larger values of $m$ for the control 
register with increasing period $r$.  We plot the corresponding 
modular exponential functions $f_{a \, \smN}(x)$ in 
Fig.~\ref{fig_fxaNmodN_four} for each value of $N$ and its 
respective base $a$. For readability,  the plots in 
Fig.~\ref{fig_fxaNmodN_four} are restricted to small values 
of the domain variable $x$. In general,  however,  the function 
$f_{a \, \smN}(x)$ looks highly random over the entire domain 
of $x$ (although it is not),  as illustrated in Fig.~\ref{fig_fxaN_all},  
where we plot the exponential function $f_{5, 143}(x)$ over 
an extended domain of $x$-values.  
%%

%\vfill
%\clearpage
\subsubsection{$N=35 = 5 \times 7$,  $a=4$,  $r=6$ }

\begin{figure}[b!]
\begin{minipage}[c]{0.4\linewidth}
\includegraphics[scale=0.50]{02_period_v1_N35_a4.jpg}
\end{minipage}
\hskip1.6cm
\begin{minipage}[c]{0.4\linewidth}
\begin{tabular}{|c|c|} \hline
 \multicolumn{2}{|c|}{~$U\vert w \rangle = 
\big\vert 4 \cdot w ~({\rm mod}~35) \big\rangle$~}  
\\\hline
$~~~U\vert 1 \rangle = \vert 4 \rangle$~~~~~&
~~$U\vert 000001 \rangle = \vert 000100 \rangle$~~~~~\\[-5pt]
$~~~~U\vert 4 \rangle = \vert 16 \rangle$~~~~~&
~~$U\vert 000100 \rangle = \vert 010000 \rangle$~~~~~\\[-5pt]
$~~~U\vert 16 \rangle = \vert 29 \rangle$~~~~~&
~~$U\vert 010000 \rangle = \vert 011101 \rangle$~~~~~\\[-5pt]
$~~~~U\vert 29 \rangle = \vert 11 \rangle$~~~~~&
~~$U\vert 011101 \rangle = \vert 001011 \rangle$~~~~~\\[-5pt]
$~~~U\vert 11 \rangle = \vert 9 \rangle$~~~~~&
~~$U\vert 001011 \rangle = \vert 001001 \rangle$~~~~~\\[-5pt]
$~~~~U\vert 9 \rangle ~= \vert 1  \rangle$~~~~~~&
~~$U\vert 001001 \rangle = \vert 000001 \rangle$~~~~~\\\hline
\end{tabular} 
\end{minipage}
\begin{minipage}[c]{1.0\linewidth}
\includegraphics[scale=0.35]{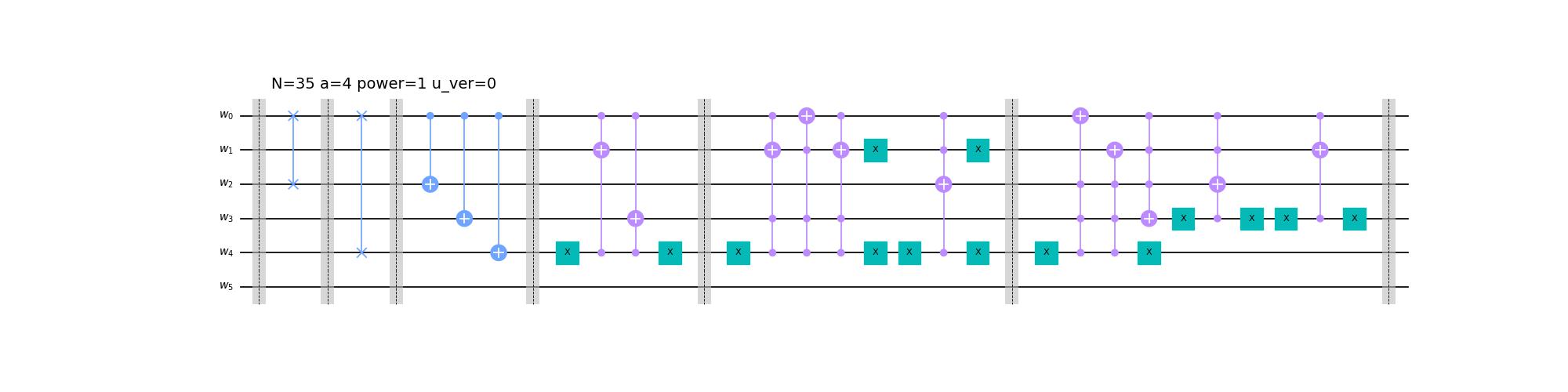} 
\end{minipage}
\vskip-0.8cm
\caption{\footnoteskip
$N=35$,  $a=4$,  $r=6$:
The top-left panel shows the modular exponential function 
$f_{4,  35}(x) = 4^x ~ ({\rm mod}~35)$, which is seen  to 
have period $r = 6$.  The top-right panel gives the action 
of the ME operator $U_{4, 35}$ on the closed sequence 
$[1, 4, 16, 11, 9, 1 ]$.  The bottom panel  illustrates the 
circuit formulation of $U_{4, 35}$. Note that qubit $w_5$
is not used. 
}
\label{fig_fxN35a4}
\end{figure}

Having examined $N=21$, we now address 
$N=35 = 5 \times 7$ with the base $a=4$.  As illustrated 
in the top panel of Fig.~\ref{fig_fxN35a4} (and the top-left 
of Fig~\ref{fig_fxaNmodN_four}),  these parameters give 
a modular exponential function $f_{4, 35}(x)$ with period 
$r = 6$,  just as with $N=21$ and $a=2$.   We can therefore 
get by with $m = 5$ control register qubits.  
%%
%\clearpage
%%
\begin{figure}[t!]
\begin{centering}
\includegraphics[width=\textwidth]{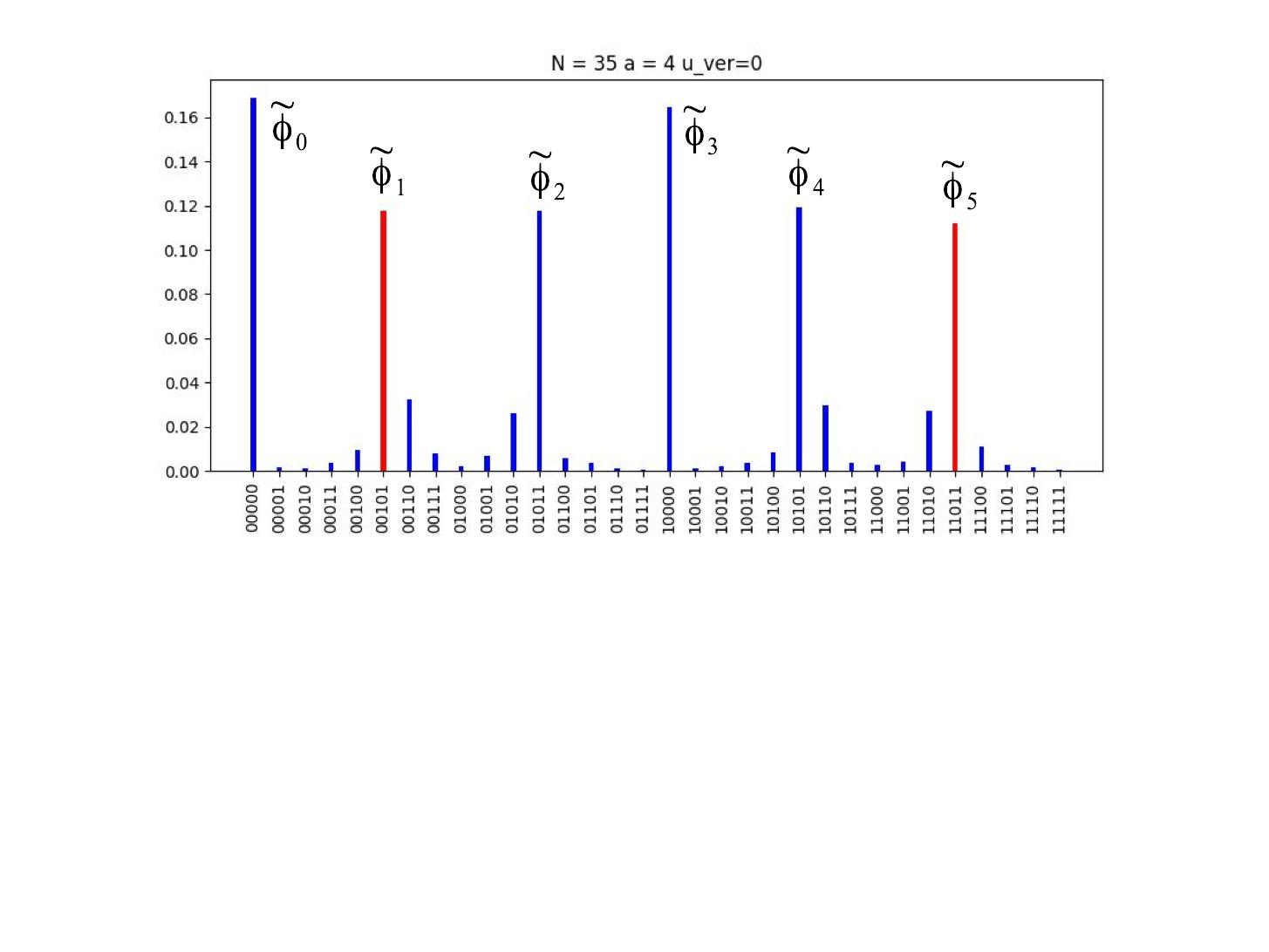} 
\par\end{centering}
\vskip-5.3cm 
\caption{\footnoteskip  
  $N=35$, $a=4$, $r=6$, $m=5$: The six dominant peaks 
  of the phase histogram occur very close to the six phases
  $\phi_s = s/6$ of the ME operator $U_{4,35}$, where $s \in 
  \{0, 1, \cdots, 5\}$. The phases that produce factors are 
  shown in red, and and occur at $\tilde\phi_1 = [0.00101]_2 
  \approx 1/6$ and $\tilde\phi_5 = [0.11011]_2 \approx 5/6$, 
  each providing the factors of 3 and 7.  These peaks are 
  amplified above the noise by Shor's algorithm.  
}
\label{fig_hist_N35_a4}
\end{figure}
\noindent
However,  we must increase the work register to $n = \lceil 
\log_2 35 \rceil = 6$ qubits.  We see that the ME operator 
$U_{4, 35}$ only acts on the closed cycle $[1, 4, 16,  29,  11,  
9,  1]$, and the  corresponding circuit representation of 
$U_{4, 35}$  is given in the lower panel of Fig~\ref{fig_fxN35a4}.
Note that qubit $w_5$ is not used. 
Finally, the phase histogram of Shor's algorithm from a Qiskit 
simulation of 4096 runs  is presented in Fig.~\ref{fig_hist_N35_a4}. 
Again,  we call this version $\tt{u\_ver=0}$,  as the operators 
$U_{4, 35}^p$ are formed by simple concatenation of $U_{4,35}$.  
The two peaks in red occur at $\tilde\ell_1 =  [00101]_2$ and 
$\tilde\ell_5 = [11011]_2$,  which correspond to the phases  
$\tilde\phi_1 = [0.00101]_2 \approx 1/6$ and $\tilde\phi_5 = 
[0.11011]_2\approx 5/6$.  Each such phase provides the 
factors of 5 and 7. Note that the two red phase peaks occur 
at the same values as for $N=21$ and $a=2$, although these 
peaks give different factors because $N$ and $a$ differ.

%\clearpage
Recall that by concatenating the ME operator $U_{4, 35}$ to 
form the composite operators $U^p_{4, 35}$,  we will eventually 
employ an exponential number of terms,  and this procedure 
will consequently break down for large values of $N$. As before, 
we can address this problem by noting that the ME operators 
$U^p_{4, 35}$ for $p=1,  2,  4,  8, 16$ possess the following 
closed cycles:
\vskip-0.7cm
\begin{eqnarray}
  U_{4, 35} && ~:~~ [1, 4, 16, 29, 11, 9, 1]
\nonumber\\[-3pt]
  U^2_{4, 35} && ~:~~ [1, 16, 11, 1]   ~~\text{and}~~ [4, 29, 9, 4]
\nonumber\\[-3pt]
  U^4_{4, 35} && ~:~~  [1, 11, 16, 1] ~~\text{and}~~ [4, 9, 29, 4]
\label{eq_Up_N35a4_seq}
\\[-3pt]
  U^8_{4, 35} && ~:~~ [1, 16, 11, 1] ~~\text{and}~~ [4, 29, 9, 4]
\nonumber\\[-3pt]
  U^{16}_{4, 35} && ~:~~ [1, 11, 16, 1] ~~\text{and}~~ [4, 9, 29, 4]
\nonumber
  \ .
\end{eqnarray}
%%
%\vskip-2.0cm
\begin{figure}[t!]
\includegraphics[scale=0.55]{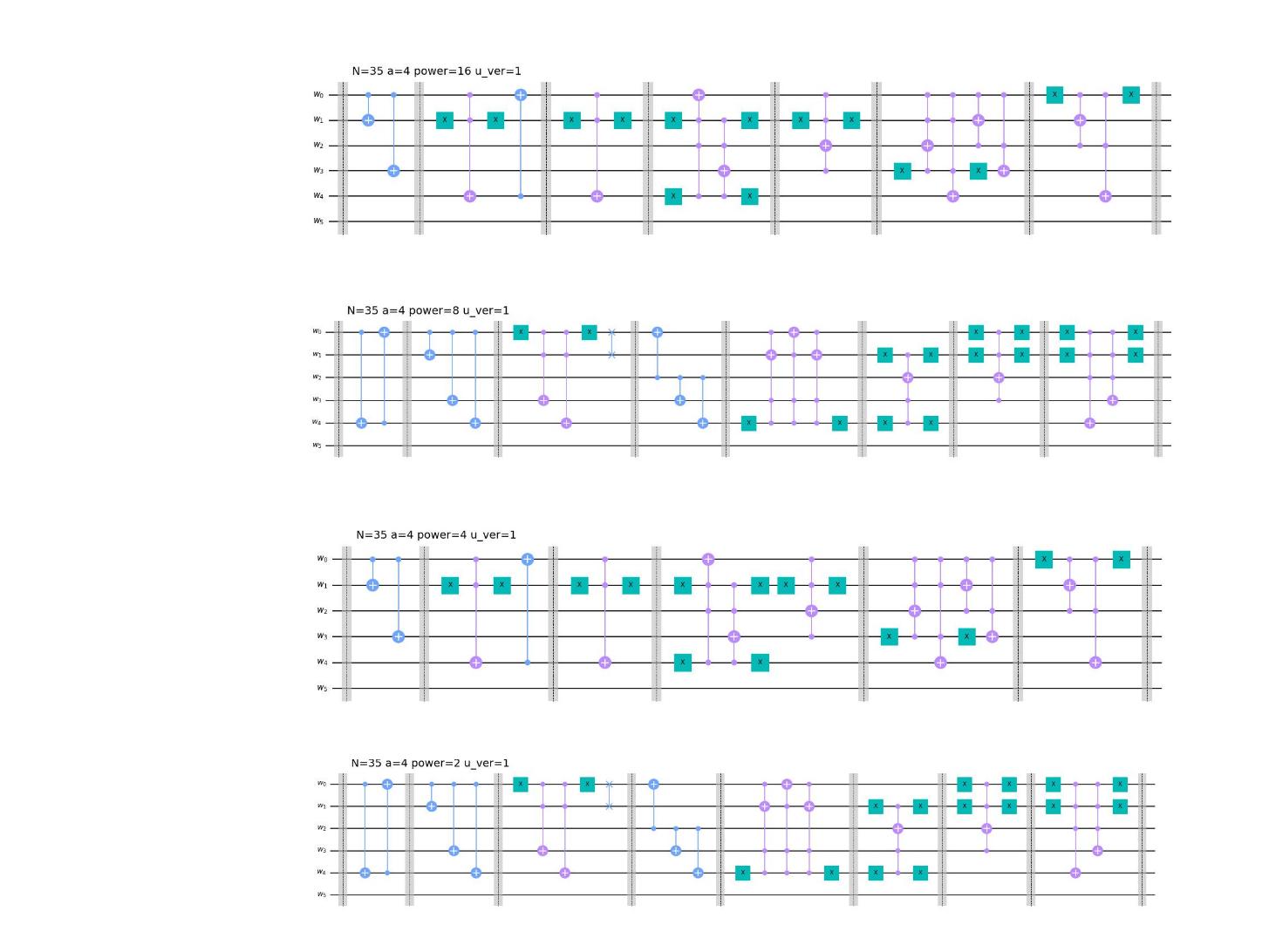}
\vskip-0.7cm
\caption{\footnoteskip
$N=35$,  $a=4$,  $r=6$: 
The ME operators $U^2,  U^4,  U^8$ and $U^{16}$ for 
version $\tt{u\_ver}=1$.   
}
\label{fig_UpN35a4}
\end{figure}
\begin{figure}[h!]
\vskip-3.0cm
\begin{centering}
\includegraphics[width=\textwidth]{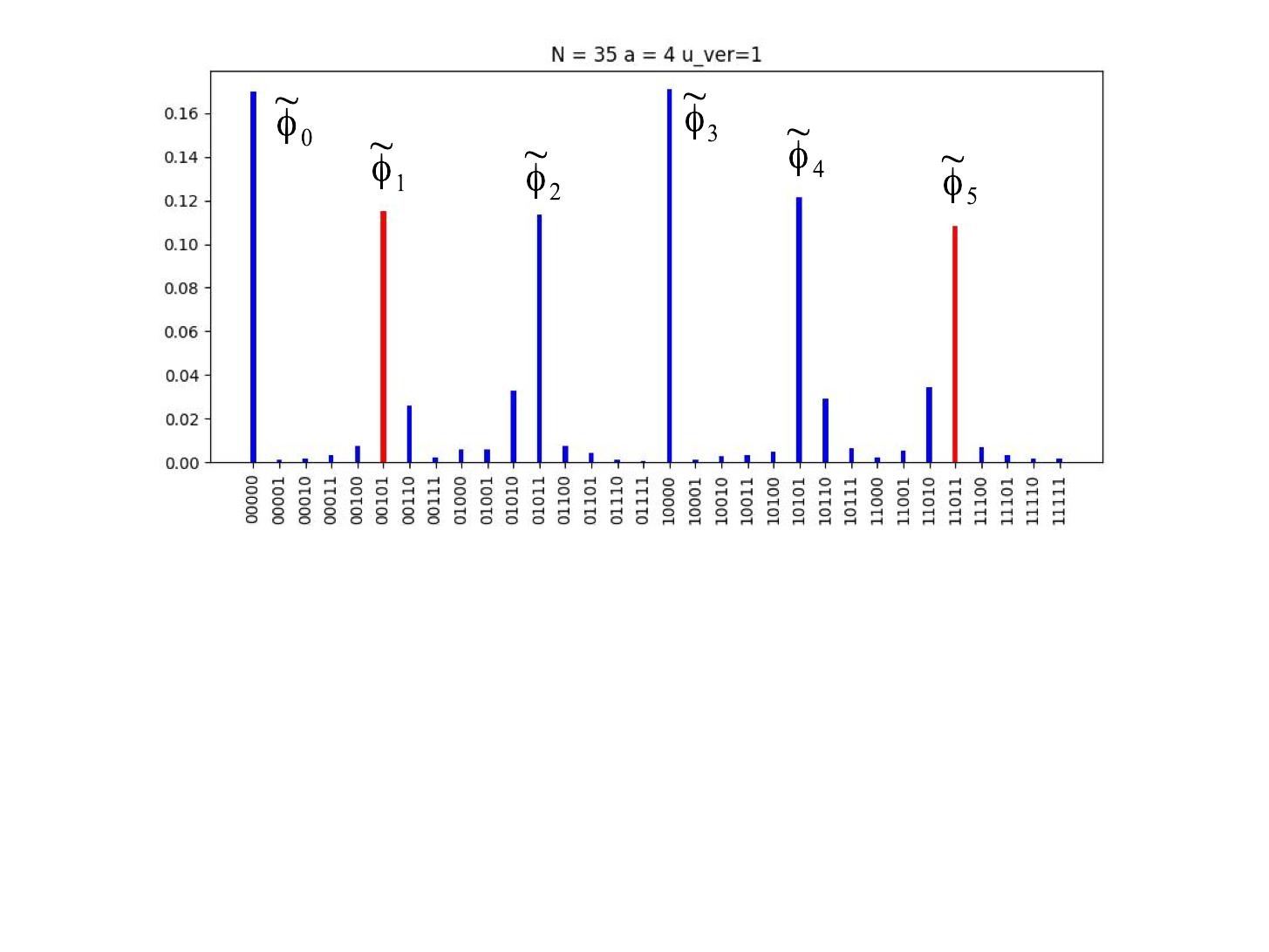}
\par\end{centering}
\vskip-5.5cm 
\caption{\footnoteskip  
$N=35$,  $a=4$,  $r=6$, $m=5$: 
Phase histogram for ME operator version $\tt{u\_ver}=1$
from Fig.~\ref{fig_UpN35a4}. 
}
\label{fig_hist_N35a4_uver1}
\end{figure}

\vskip-0.6cm
\noindent
We can now construct operators $U^p_{4, 35}$ that reproduce
these cycles by concatenating commensurate cycle-pairs together; 
for example, we must ensure that $U^2_{4, 35}$ reproduces the 
double cycle $ [1, 16, 11, 1, 4, 29, 9, 4]$, and we refer to this 
procedure as version number $\tt{u\_ver}=1$. The composite 
operators $U^p_{4, 35}$ for $p > 1$ are illustrated in  
Fig.~\ref{fig_UpN35a4}, while the corresponding phase 
histogram from a Qiskit simulation with 4096 runs is given 
in Fig.~\ref{fig_hist_N35a4_uver1}.  We see that the results 
still agree with the previous version from Fig.~\ref{fig_hist_N35_a4},
in which we concatenated the $U_{4, 35}$ operator to form 
the $U^p_{4, 35}$ operators.  

\begin{figure}[h!]
\includegraphics[scale=0.28]{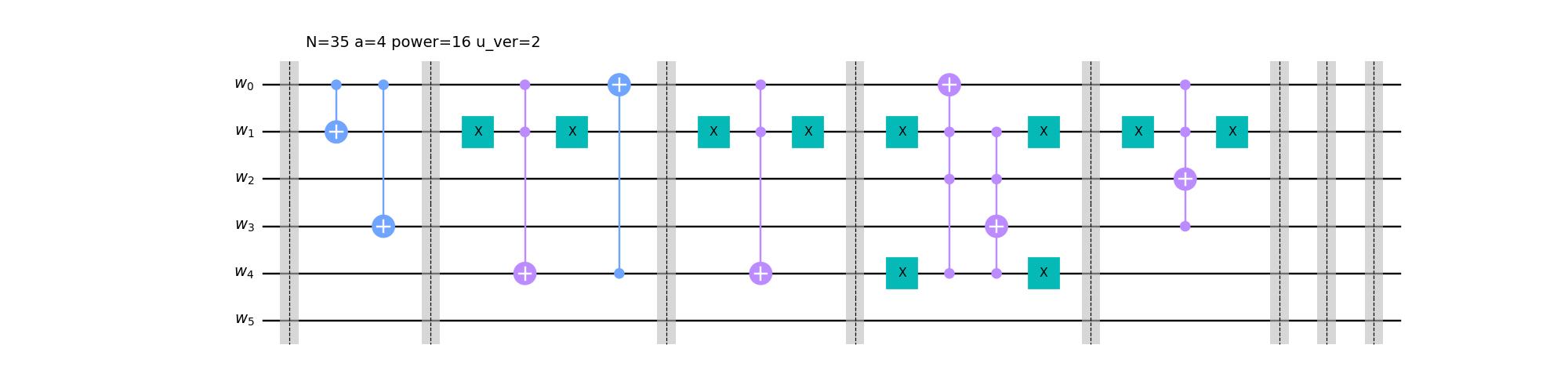}
\vskip-0.5cm
\includegraphics[scale=0.28]{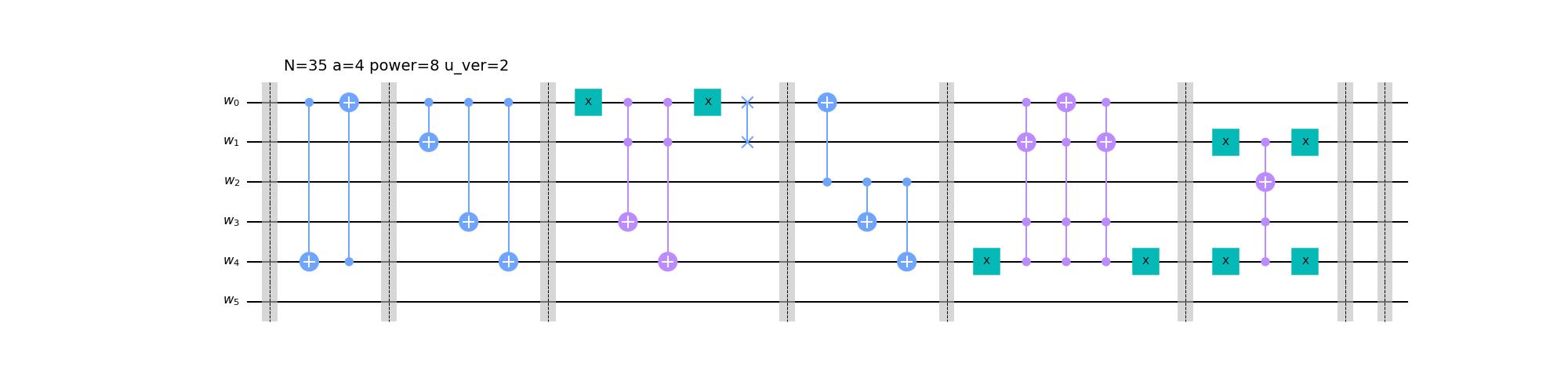}
\caption{\footnoteskip
$N=35$, $a=4$, $r=6$:  The ME operators $U^8, U^{16}$ for 
version $\tt{u\_ver}=2$. 
}
\label{fig_UpN35a4_truncate1}
\end{figure}

\begin{figure}[h!]
\includegraphics[scale=0.28]{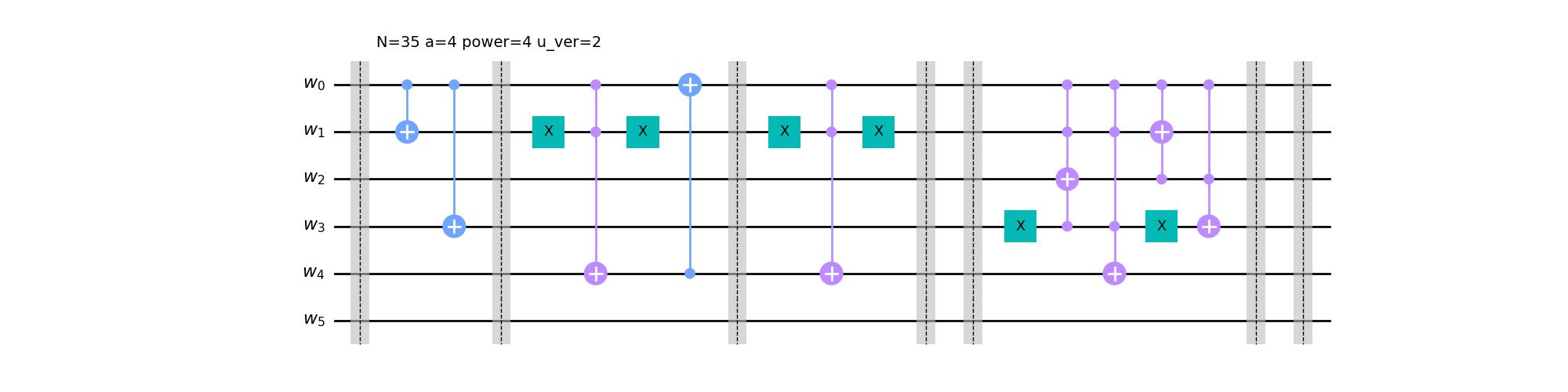}
\vskip-0.5cm
\includegraphics[scale=0.28]{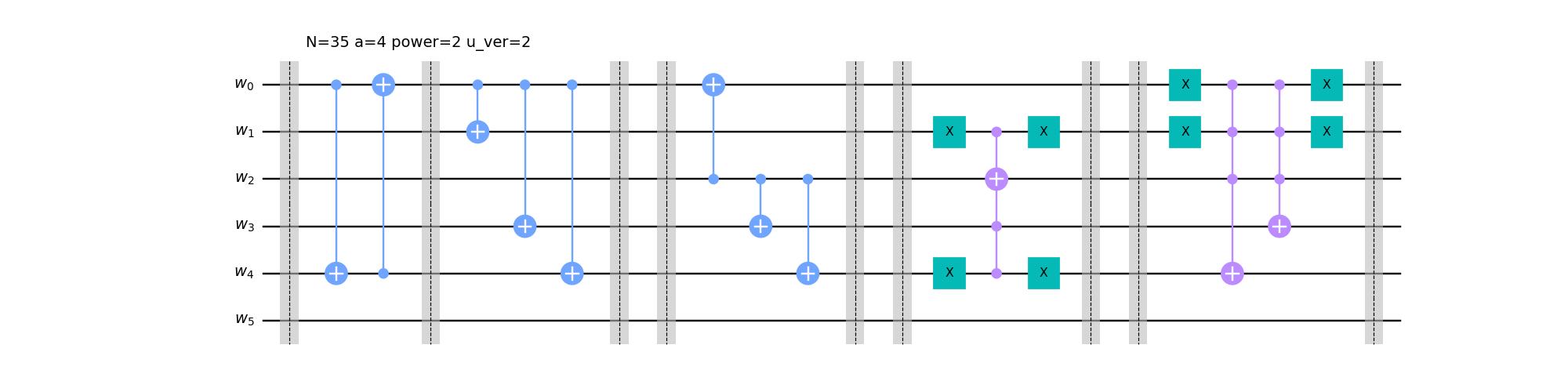}
\vskip-0.5cm
\includegraphics[scale=0.28]{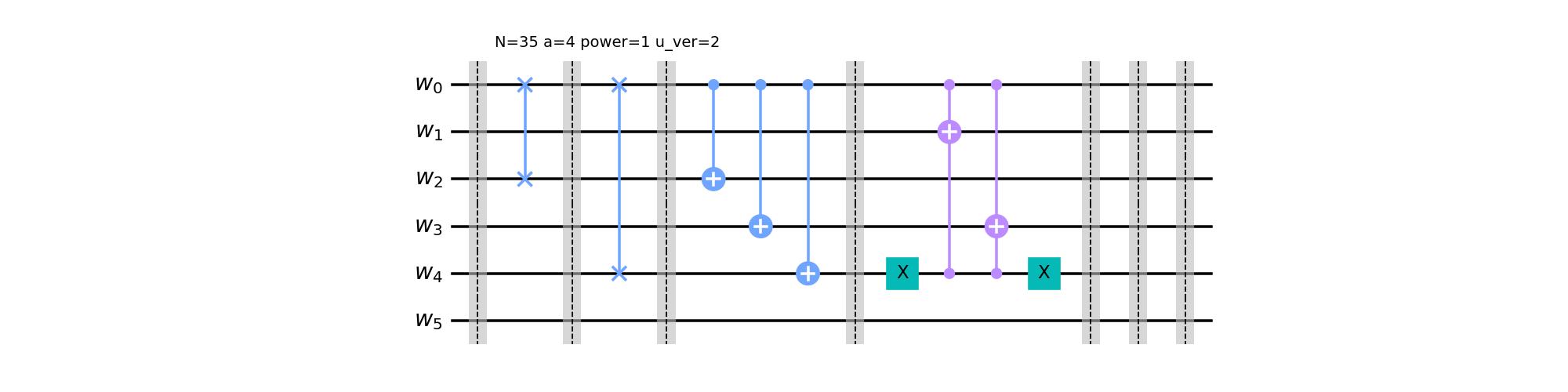}
%\vskip-1.7cm
\caption{\footnoteskip
$N=35$, $a=4$, $r=6$:  The ME operators $U,  U^2,  U^4$ for 
version $\tt{u\_ver}=2$. 
}
\label{fig_UpN35a4_truncate2}
\end{figure}

\begin{figure}[h!]
\begin{centering}
\includegraphics[width=\textwidth]{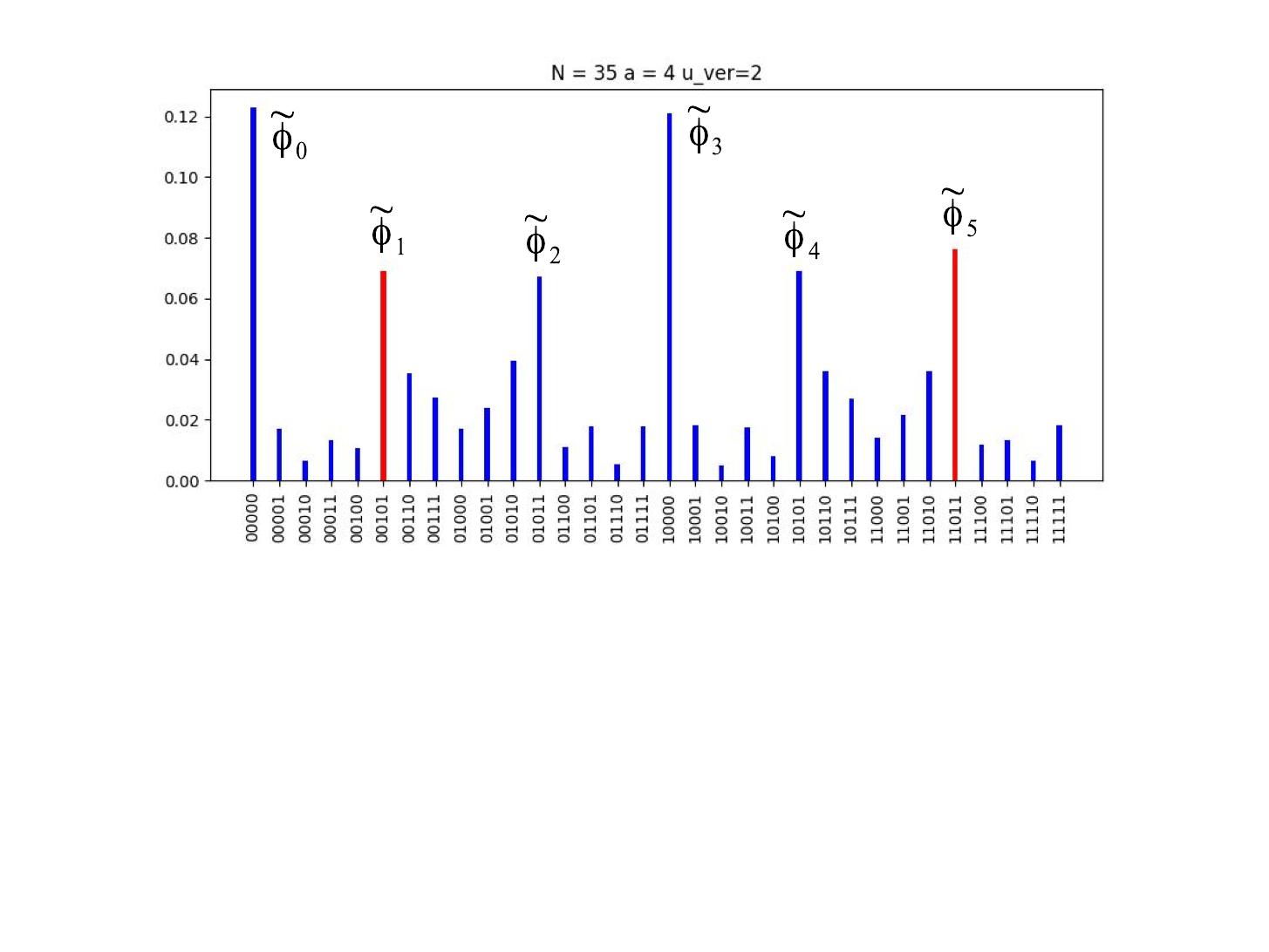} 
\par\end{centering}
\vskip-5.5cm 
\caption{\footnoteskip  
$N=35$, $a=4$, $r=6$, $m=5$:  Phase histogram for ME operator 
version $\tt{u\_ver}=2$. 
}
\label{fig_hist_N35a4_uver2_truncate}
\end{figure}

\vfill
\clearpage
Finally, Figs.~\ref{fig_UpN35a4_truncate1} and \ref{fig_UpN35a4_truncate2} 
 illustrates a truncated 
version of the ME operators $U_{4, 35}^p$, with the corresponding 
phase histogram given in Fig.~\ref{fig_hist_N35a4_uver2_truncate}.
We see that the truncated operators $U_{4, 35}^p$ still provide the correct
peaks in the phase histogram, albeit with more noise. 
%%
%\vskip-0.5cm
%%
%%
%\vskip-0.5cm
%%

%\vfill
%\clearpage
%\pagebreak
\subsubsection{$N=33 = 3 \times 11$, $a=7$,  $r=10$}

\begin{figure}[h!]
\includegraphics[scale=0.65]{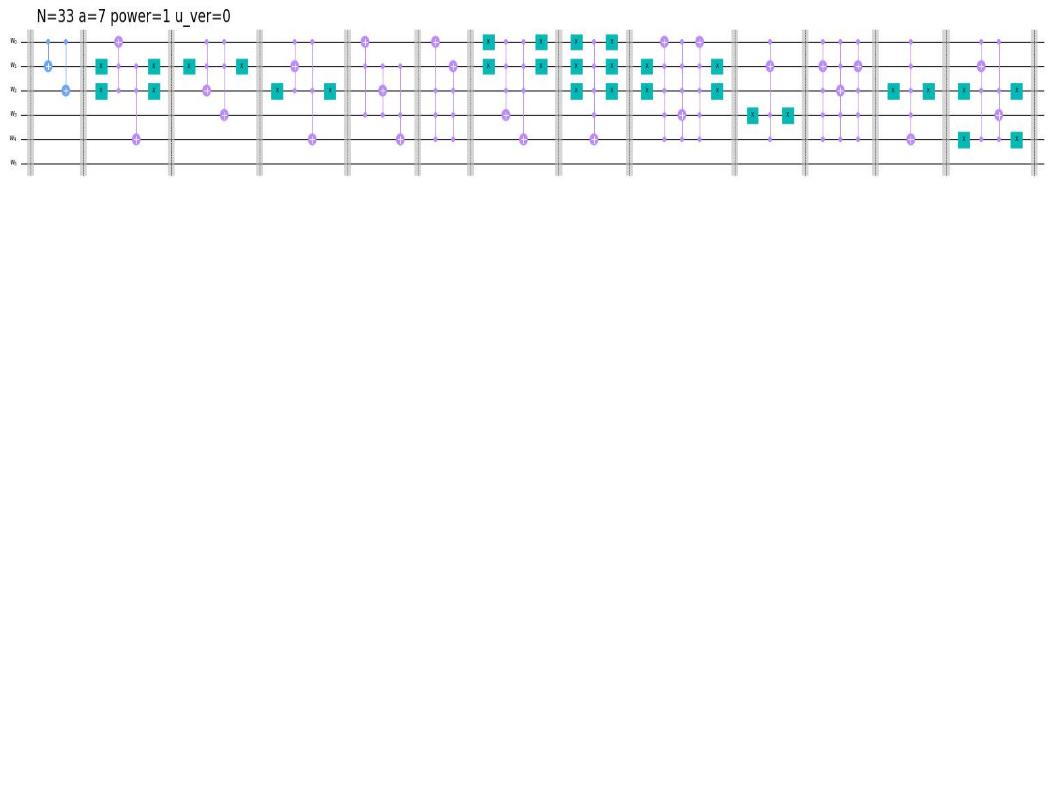} 
\vskip-11.0cm 
\caption{\footnoteskip  
The circuit formulation of $U_{7, 33}$. 
}
\label{fig_fxN33a7_a}
\end{figure}

\begin{figure}[h!]
\begin{minipage}[c]{0.3\linewidth}
\includegraphics[scale=0.50]{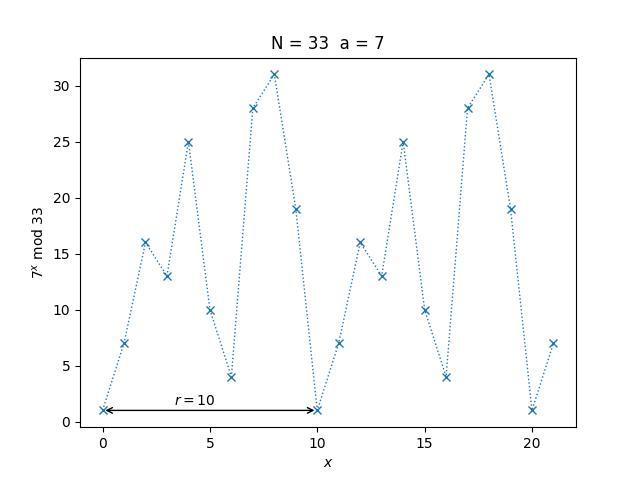}
\end{minipage}
\hskip3.0cm
\begin{minipage}[c]{0.5\linewidth}
\begin{tabular}{|c|c|} \hline
 \multicolumn{2}{|c|}{~$U\vert w \rangle = 
\big\vert 7 \cdot w ~({\rm mod}~33) \big\rangle$~}  
\\\hline
$~~~U\vert 1 \rangle = \vert 7 \rangle$~~~~~&
~~$U\vert 00001 \rangle = \vert 000111 \rangle$~~~~~\\[-5pt]
$~~~~U\vert 7 \rangle = \vert 16 \rangle$~~~~~&
~~$U\vert 000111 \rangle = \vert 010000 \rangle$~~~~~\\[-5pt]
$~~~U\vert 16 \rangle = \vert 13 \rangle$~~~~~&
~~$U\vert 010000 \rangle = \vert 001101 \rangle$~~~~~\\[-5pt]
$~~~~U\vert 13 \rangle = \vert 25 \rangle$~~~~~&
~~$U\vert 001101 \rangle = \vert 011001 \rangle$~~~~~\\[-5pt]
$~~~U\vert 25 \rangle = \vert 10 \rangle$~~~~~&
~~$U\vert 011001 \rangle = \vert 001010 \rangle$~~~~~\\[-5pt]
$~~~~U\vert 10 \rangle = \vert 4  \rangle$~~~~~~&
~~$U\vert 001010 \rangle = \vert 000100 \rangle$~~~~~\\[-5pt]
$~~~~U\vert 4 \rangle ~= \vert 28  \rangle$~~~~~~&
~~$U\vert 000100 \rangle = \vert 011100 \rangle$~~~~~\\[-5pt]
$~~~~U\vert 28 \rangle ~= \vert 31  \rangle$~~~~~~&
~~$U\vert 011100 \rangle = \vert 011111 \rangle$~~~~~\\[-5pt]
$~~~~U\vert 31 \rangle ~= \vert 19  \rangle$~~~~~~&
~~$U\vert 011111 \rangle = \vert 010011 \rangle$~~~~~\\[-5pt]
$~~~~U\vert 19 \rangle ~= \vert 1  \rangle$~~~~~~&
~~$U\vert 010011 \rangle = \vert 000001 \rangle$~~~~~\\\hline
\end{tabular} 
\end{minipage}
%\includegraphics[scale=0.65]{12_shor_v2_N33_a7_uver0_U1_odp.jpg} 
%\vskip-10.0cm
\caption{\footnoteskip
$N=33$, $a=7$, $r=10$: 
The left panel gives the modular exponential function $f_{7,  33}(x) 
= 7^x ~ ({\rm mod}~33)$,  and the right panel shows the action of 
the ME operator $U_{7, 33}$ on the closed sequence $[1,  7, 16, 13, 
25, 10, 4, 28, 31, 19, 1]$.  
}
\label{fig_fxN33a7_b}
\end{figure}

\begin{figure}[h!]
\vskip-0.8cm
\begin{centering}
\includegraphics[width=5.7in]{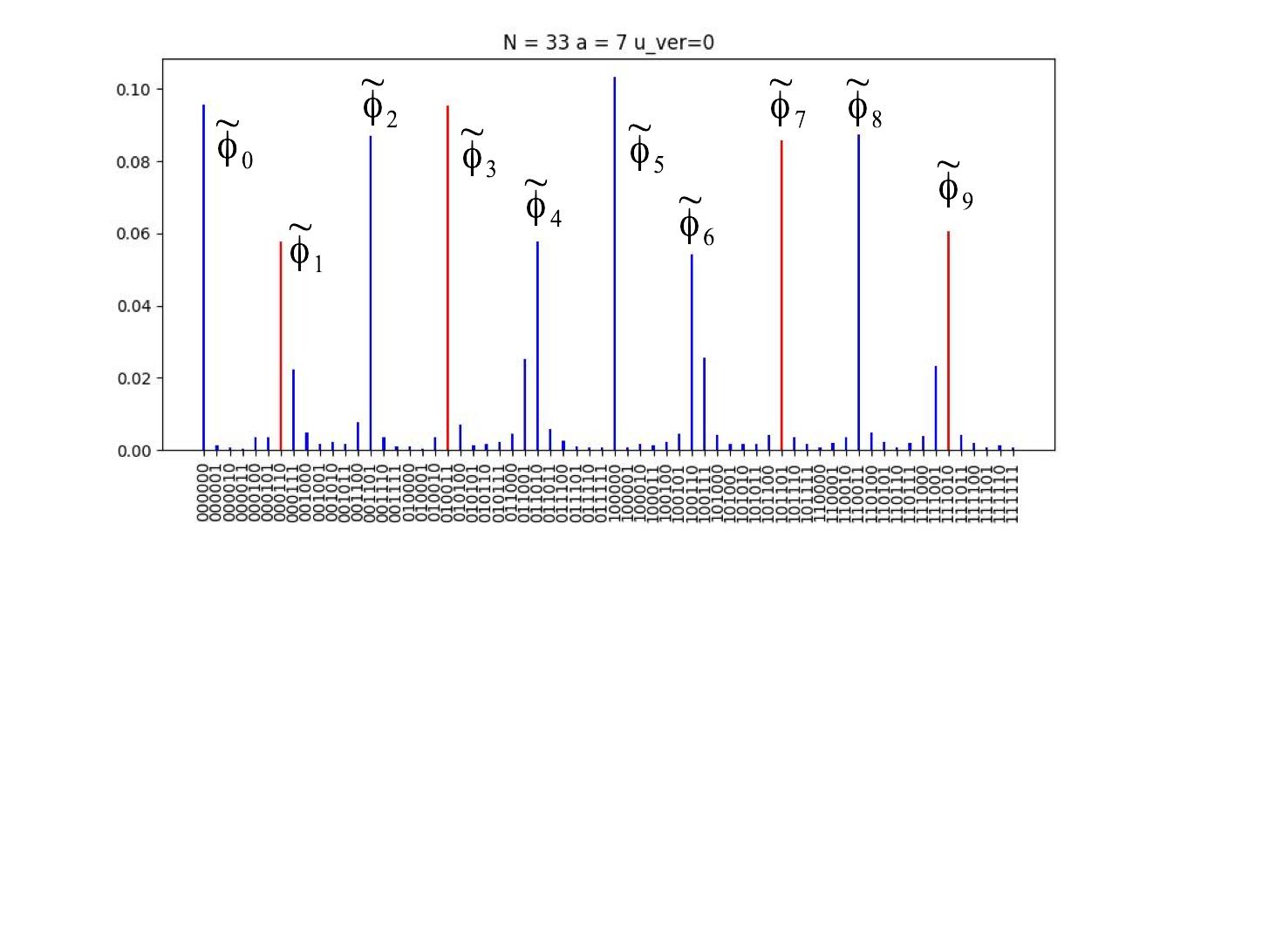} 
\par\end{centering}
\vskip-4.8cm
\begin{centering}
\includegraphics[width=5.7in]{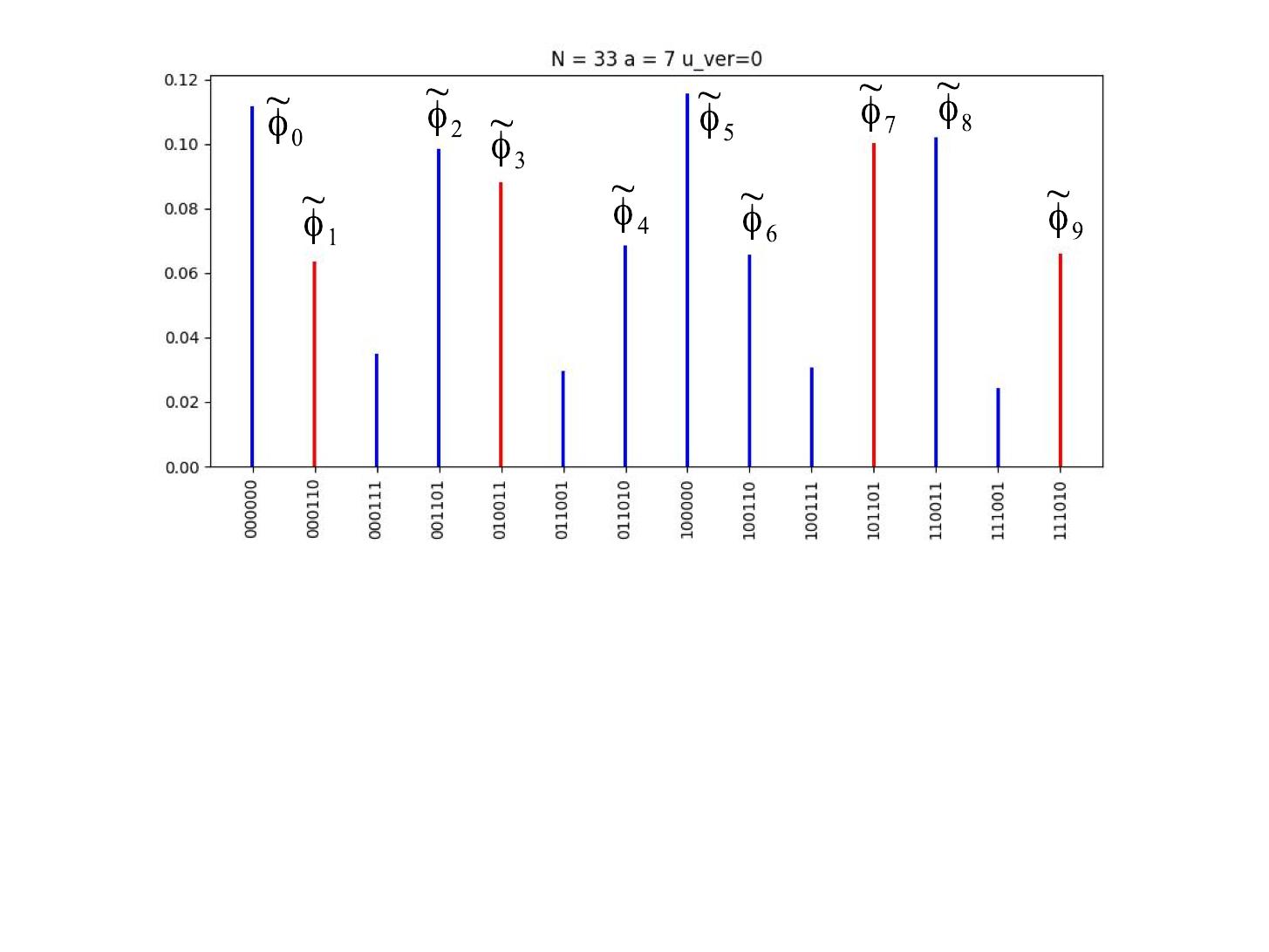} 
\par\end{centering}
\vskip-4.5cm 
\caption{\footnoteskip  
  $N=33$, $a=7$, $r=10$, $m=6$: 
  Phase histogram for the ME operator of version $\tt{u\_ver}=0$. 
  The top panel illustrates the histogram over the full range of
  phases,  while the bottom panel shows only the most frequent 
  peaks. 
%The four peaks in red correspond to the phases that produce
% factors: $\tilde\phi_1 = [0.000110]_2 \approx 1/10$, $\tilde\phi_3 =
% [0.010011]_3\approx 3/10$, $\tilde\phi_7 =[0.101101]_2 \approx 
% 7/10$, and $\tilde\phi_9 =[0.111010]_2 \approx 9/10$. 
%%
}
\label{fig_hist_N33a7_uver0}
\end{figure}

We now move on to factoring a number with a larger period:
$N = 33 = 3 \times 11$ with base $a = 7$ has period $r = 10$.
It turns out that $m = 6$ control qubits give sufficient resolution 
for $r = 10$, while the number of work qubits must be set 
to $n = \lceil \log_2 33 \rceil = 6$,  as with the previous example. 
The circuit representation of the ME 
operator is shown in Fig.~\ref{fig_fxN33a7_a}, while
the corresponding modular exponential function $f_{7, 33}(x)$ 
is plotted in the  left panel of Fig.~\ref{fig_fxN33a7_b}, with 
the action of the ME operator $U_{7, 33}$ on the closed sequence 
$[1, 7, 16, 13, 25, 10, 4, 28, 31, 19, 1]$ given in the right panel.  As 
usual,  we form the 
composite operators $U^p$ for $p >1$ by concatenation,  
and call this version $\tt{u\_ver} = 0$. The phase histogram 
for 4096 runs is illustrated in Fig.~\ref{fig_hist_N33a7_uver0}, 
where the top panel gives the histogram over the full range 
of phases from the Qiskit simulation, while the bottom panel 
shows only the most frequent peaks. 
%%
%\vskip-2.5cm 
%%

%\vfil
%\pagebreak

%%
%%

% 
%\clearpage

\clearpage
%\pagebreak
Note that the ten dominant peaks in Fig.~\ref{fig_hist_N33a7_uver0} 
lie close to the ME Eigen-phases $\phi_s =s/10$ for $s \in \{0, 1,
\cdots, 9\}$,  as they should.  Furthermore, the peaks corresponding 
to the factors of 3 and 11 occur only when  ${\rm gcd}(s, 10) = 1$, 
or for $s = 1, 3, 7, 9$, and they are plotted in red.  The phase values
of all ten dominant peaks are listed below: 
%\vfill
%\pagebreak
%%
\begin{eqnarray}
   \tilde\phi_0 &=&   [0.000000]_2 = 0.000000  = 0
   \nonumber\\[-1pt]
   \tilde\phi_1 &=&   [0.000110]_2 = 0.093750  \approx 1/10
   ~~~\Leftarrow~\text{factors}: 3, 11
   \nonumber\\[-1pt]
  \tilde\phi_2 &=&   [0.001101]_2 = 0.203125  \approx 2/10
%\end{eqnarray}
%\begin{eqnarray}
   \nonumber\\[-1pt]
   \tilde\phi_3 &=&[0.010011]_2 = 0.296875 \approx  3/10  
    ~~~\Leftarrow~\text{factors}: 3, 11
 \nonumber\\[-1pt]
 \tilde\phi_4 &=&   [0.011010]_2 = 0.406250  \approx 4/10
 \label{eq_peaks1_N33a7} 
 \\[-1pt]
 \tilde\phi_5 &=&   [0.100000]_2 = 0.500000  = 5/10
  \nonumber\\[-1pt]
 \tilde\phi_6 &=&   [0.100110]_2 = 0.593750  \approx 6/10
  \nonumber\\[-1pt]
  \tilde \phi_7 &=&  [0.101101]_2 = 0.703125 \approx 7/10 
  ~~~\Leftarrow~\text{factors}: 3, 11
  \nonumber\\[-1pt]
  \tilde\phi_8 &=&   [0.110011]_2 = 0.796875  \approx 8/10
  \nonumber  \\[-1pt]
  \tilde\phi_9  &=&[0.111010]_2 = 0.906250  \approx  9/10  
   ~~~\Leftarrow~\text{factors}: 3, 11
    \ .
\nonumber
\end{eqnarray}

\begin{figure}[h!]
\includegraphics[scale=0.35, center]{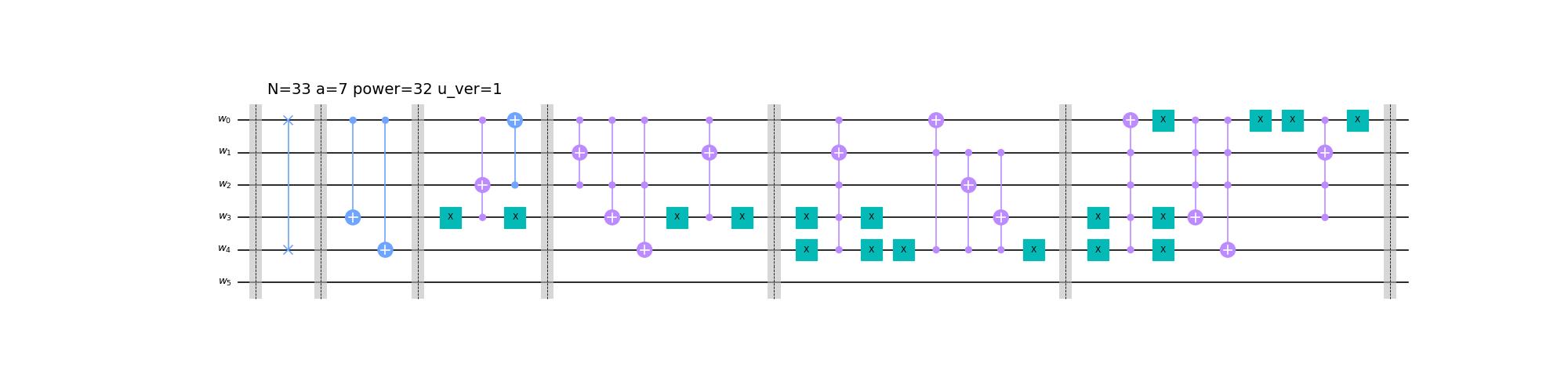}
%\vskip-1.0cm
%\includegraphics[scale=0.40, center]{12_shor_v2_N33_a7_uver1_U16.jpg}
%\vskip-1.0cm
%\includegraphics[scale=0.49, center]{12_shor_v2_N33_a7_uver1_U8.jpg}
%\vskip-2.0cm
%\includegraphics[scale=0.50, center]{12_shor_v2_N33_a7_uver1_U4.jpg}
%\vskip-1.5cm
%\includegraphics[scale=0.36, center]{12_shor_v2_N33_a7_uver1_U2.jpg}
\vskip-1.5cm
\caption{\footnoteskip
$N=33$, $a=7$, $r=10$: 
The ME operators $U^{32}$ for version $\tt{u\_ver}=1$. 
}
\label{fig_UpN33a7_truncate1_a}
\end{figure}

\begin{figure}[h!]
%\includegraphics[scale=0.38, center]{12_shor_v2_N33_a7_uver1_U32.jpg}
%\vskip-1.0cm
\includegraphics[scale=0.43, center]{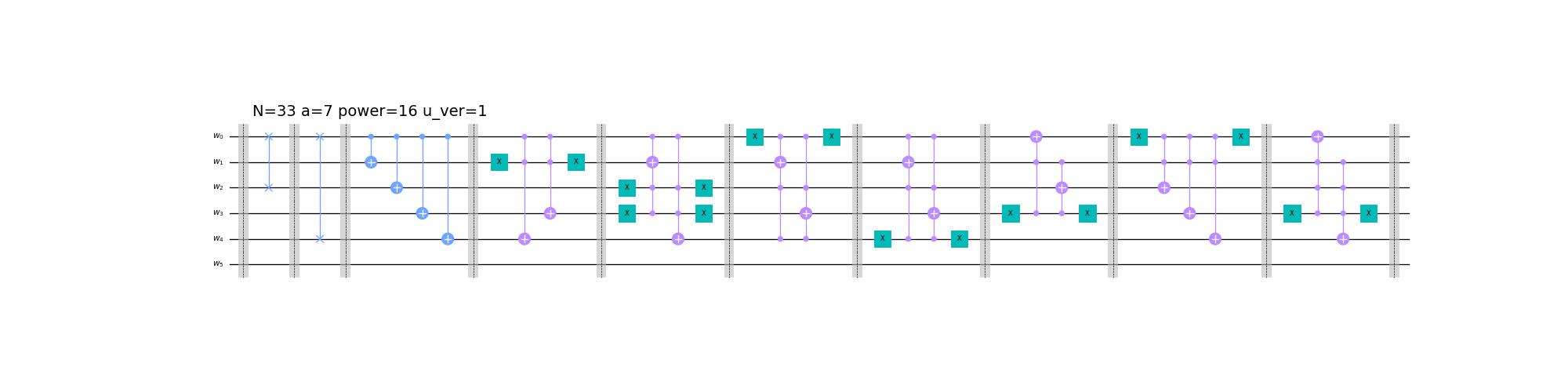}
\vskip-1.0cm
\includegraphics[scale=0.45, center]{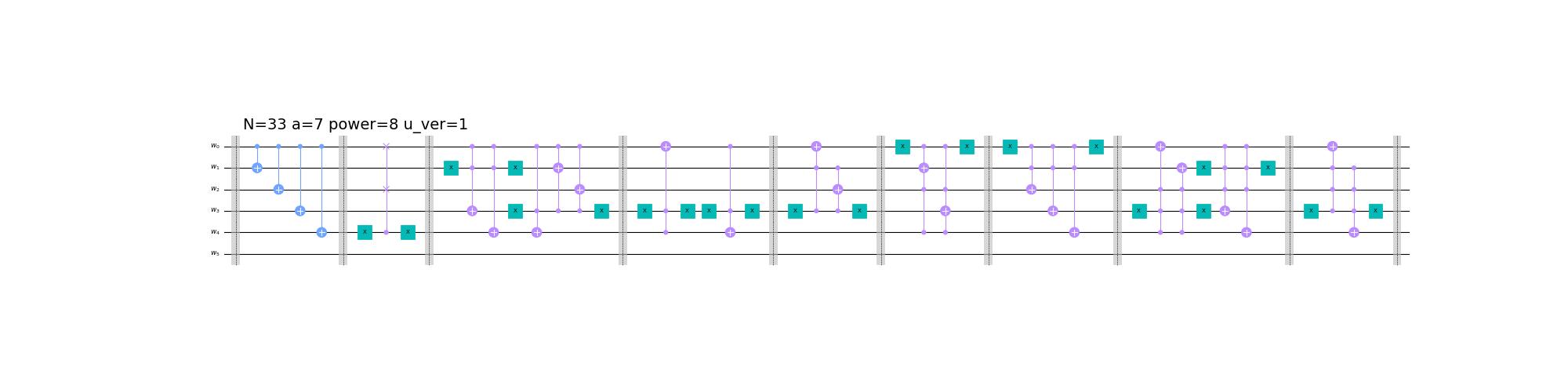}
\vskip-2.0cm
\includegraphics[scale=0.48, center]{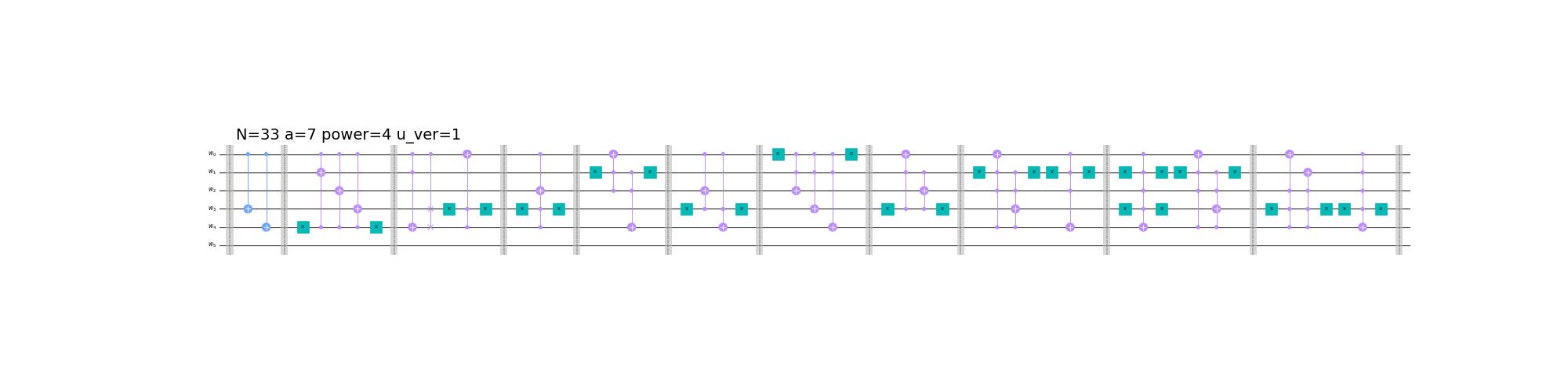}
\vskip-1.8cm
\includegraphics[scale=0.35, center]{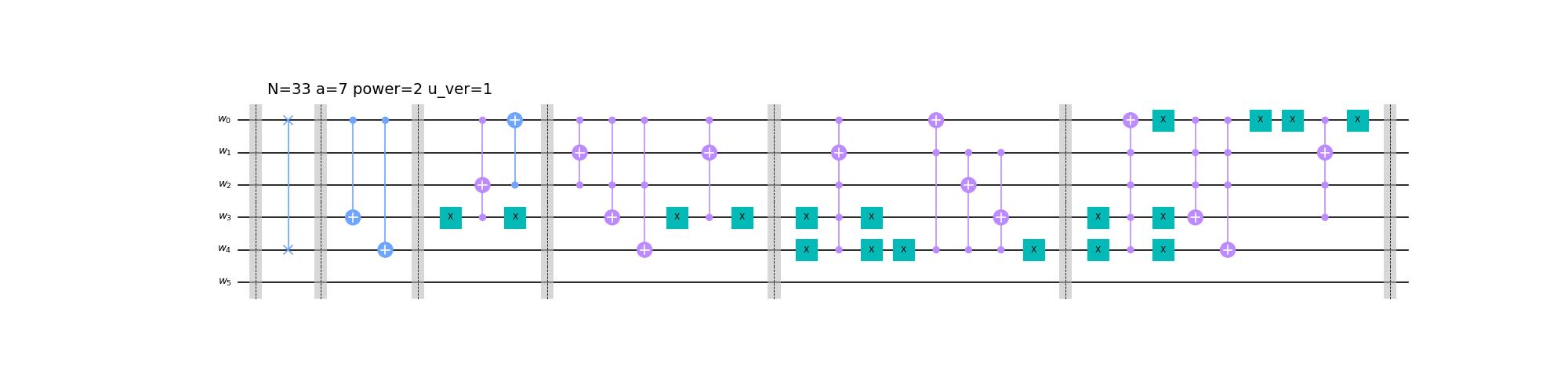}
\vskip-1.0cm
\caption{\footnoteskip
$N=33$, $a=7$, $r=10$: 
The ME operators $U^2,  U^4, U^8$ and $U^{16}$ for version $\tt{u\_ver}=1$. 
}
\label{fig_UpN33a7_truncate1_b}
\end{figure}

\begin{figure}[h!]
\begin{centering}
\includegraphics[width=\textwidth]{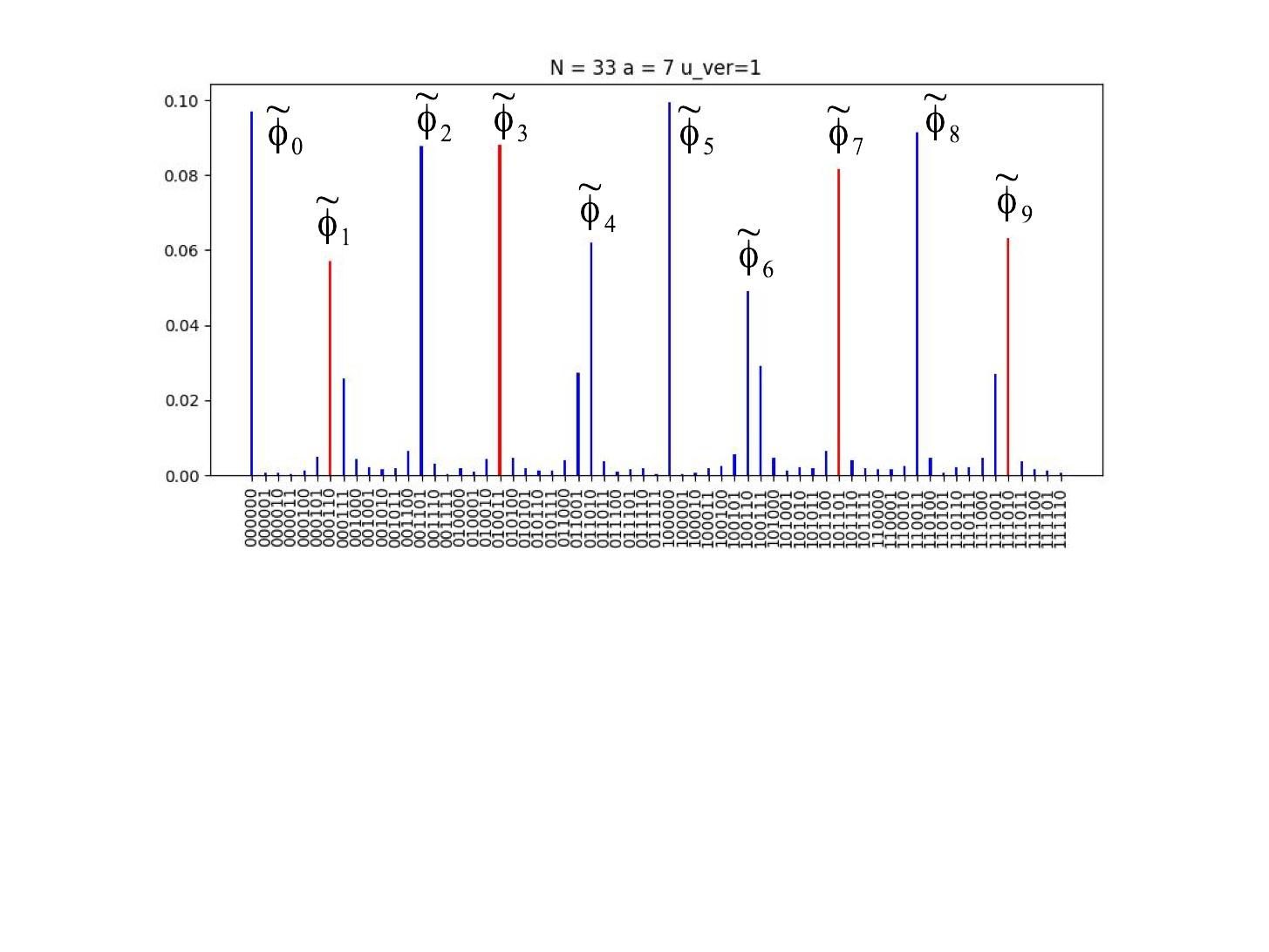} 
\par\end{centering}
\vskip-5.0cm 
\caption{\footnoteskip  
  $N=33$,  $a=7$,  $r=10$, $m=6$: 
  Phase histogram for  ME operator version $\tt{u\_ver}=1$. 
}
\label{fig_hist_N33a7_uver1_truncate}
\end{figure}

Let us now construct the composite operators $U^p$
for $p=2^0, 2^1,  \cdots, 2^5$. Note that these operators 
possess the following cycles:
\begin{eqnarray}
  U_{7, 33} && ~:~ [1, 7, 16, 13, 25, 10, 4, 28, 31, 19, 1]
\nonumber\\[-3pt]
  U^2_{7, 33}  && ~:~ [1, 16, 25, 4, 31, 1]
  ~~\text{and}~~  [7, 13, 10, 28, 19, 7]
\nonumber\\[-3pt]
  U^4_{7, 33} && ~:~  [1, 25, 31, 16, 4, 1]
  ~~\text{and}~~ [7, 10, 19, 13, 28, 7]
\label{eq_Up_N33a7_seq}
\\[-3pt]
  U^8_{7, 33} && ~:~ [1, 31, 4, 25, 16, 1]
  ~~\text{and}~~  [7, 19, 28, 10, 13, 7]
\nonumber\\[-3pt]
  U^{16}_{7, 33}  && ~:~ [1, 4, 16, 31, 25, 1]
  ~~\text{and}~~ [7, 28, 13, 19, 10, 7]
\nonumber\\[-3pt]
  U^{32}_{7, 33}  && ~:~ [1, 16, 25, 4, 31, 1]
  ~~\text{and}~~  [7, 13, 10, 28, 19, 7]
    \ .
\nonumber
\end{eqnarray}
The operators $U^p$ for $p > 1$ are given 
in Figs.~\ref{fig_UpN33a7_truncate1_a} and 
\ref{fig_UpN33a7_truncate1_b},  and the corresponding 
phase histogram for 4096 runs is shown in
Fig.~\ref{fig_hist_N33a7_uver1_truncate}. We will 
call this version $\tt{u\_ver}=1$, and it agrees with 
the previous result. 
%%
%\vskip-0.5cm
%%
%% 
%\vskip-0.5cm
\clearpage
%%

\begin{figure}[h!]
\includegraphics[scale=0.30, center]{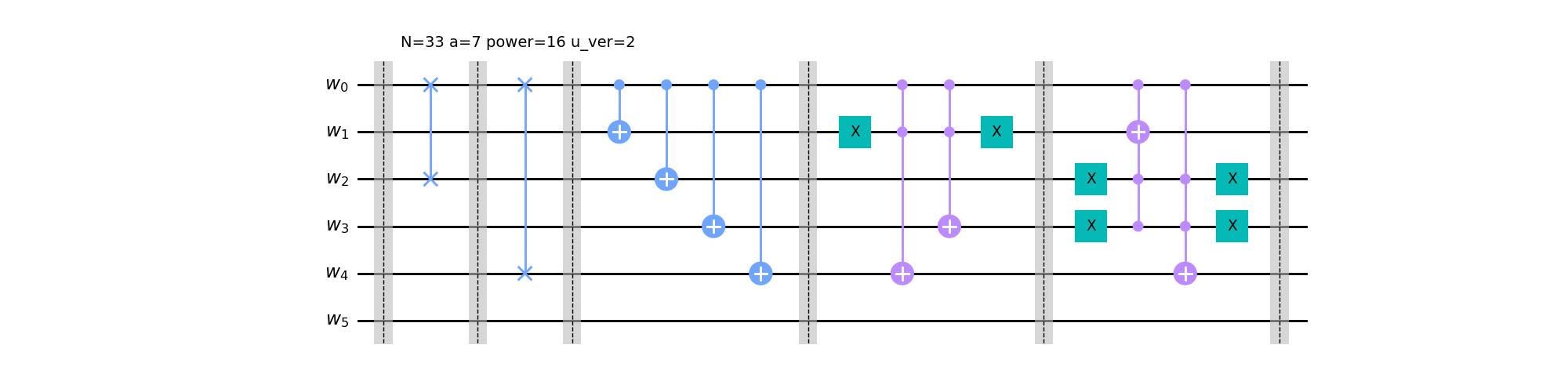}
\includegraphics[scale=0.30, center]{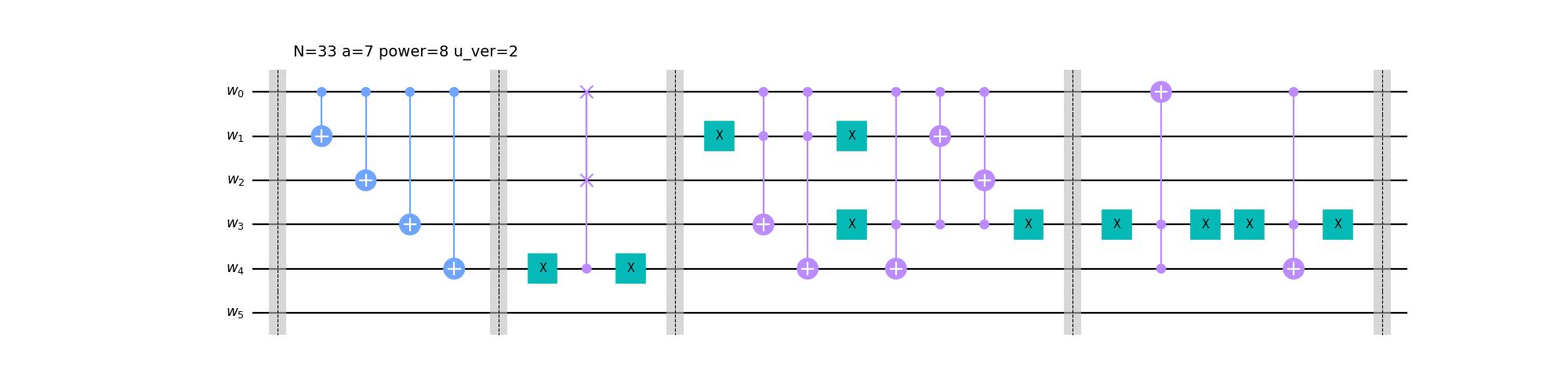}
\vskip-0.5cm
\includegraphics[scale=0.30, center]{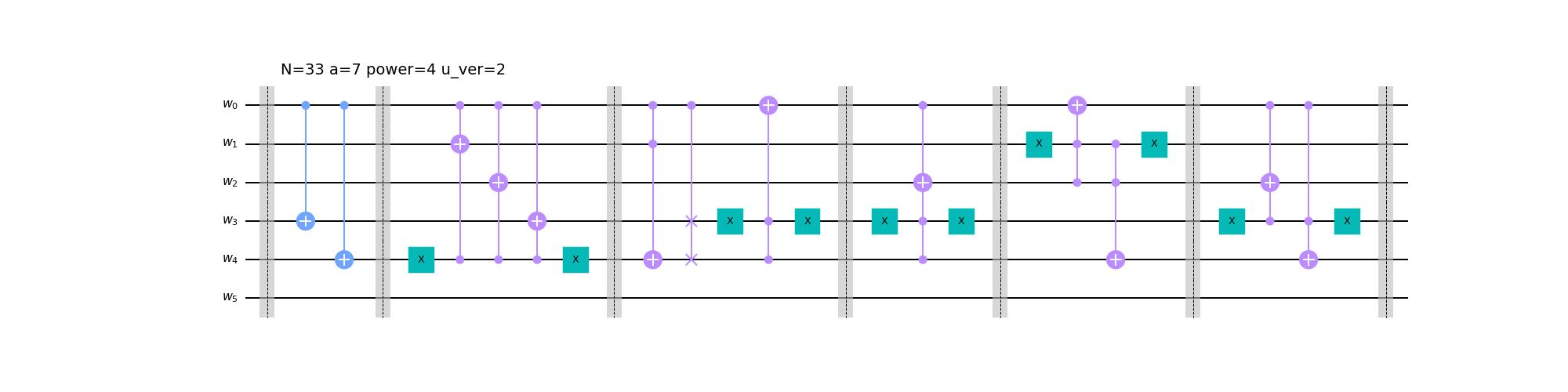}
\vskip-0.3cm
\includegraphics[scale=0.30, center]{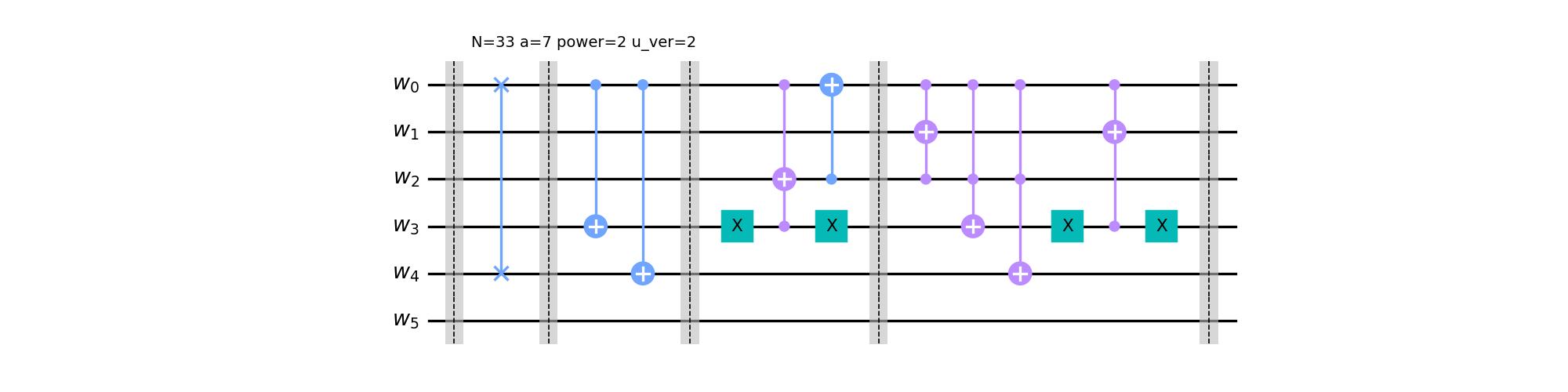}
\vskip-0.3cm
\includegraphics[scale=0.35, center]{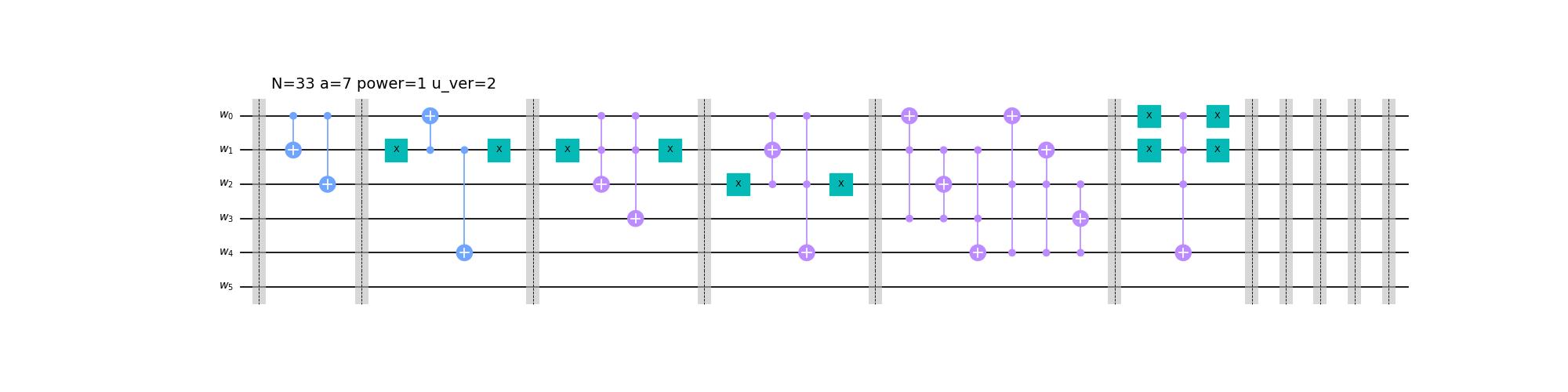}
\vskip-0.8cm
\caption{\footnoteskip
$N=33$, $a=7$, $r=10$: 
The ME operators $U, U^2,  U^4,  U^8, U^{16}$ and $U^{32}$ for version 
$\tt{u\_ver}=2$.  
}
\label{fig_UpN33a7_truncate2_b}
\end{figure}

\begin{figure}[h!]
\begin{centering}
\includegraphics[width=\textwidth]{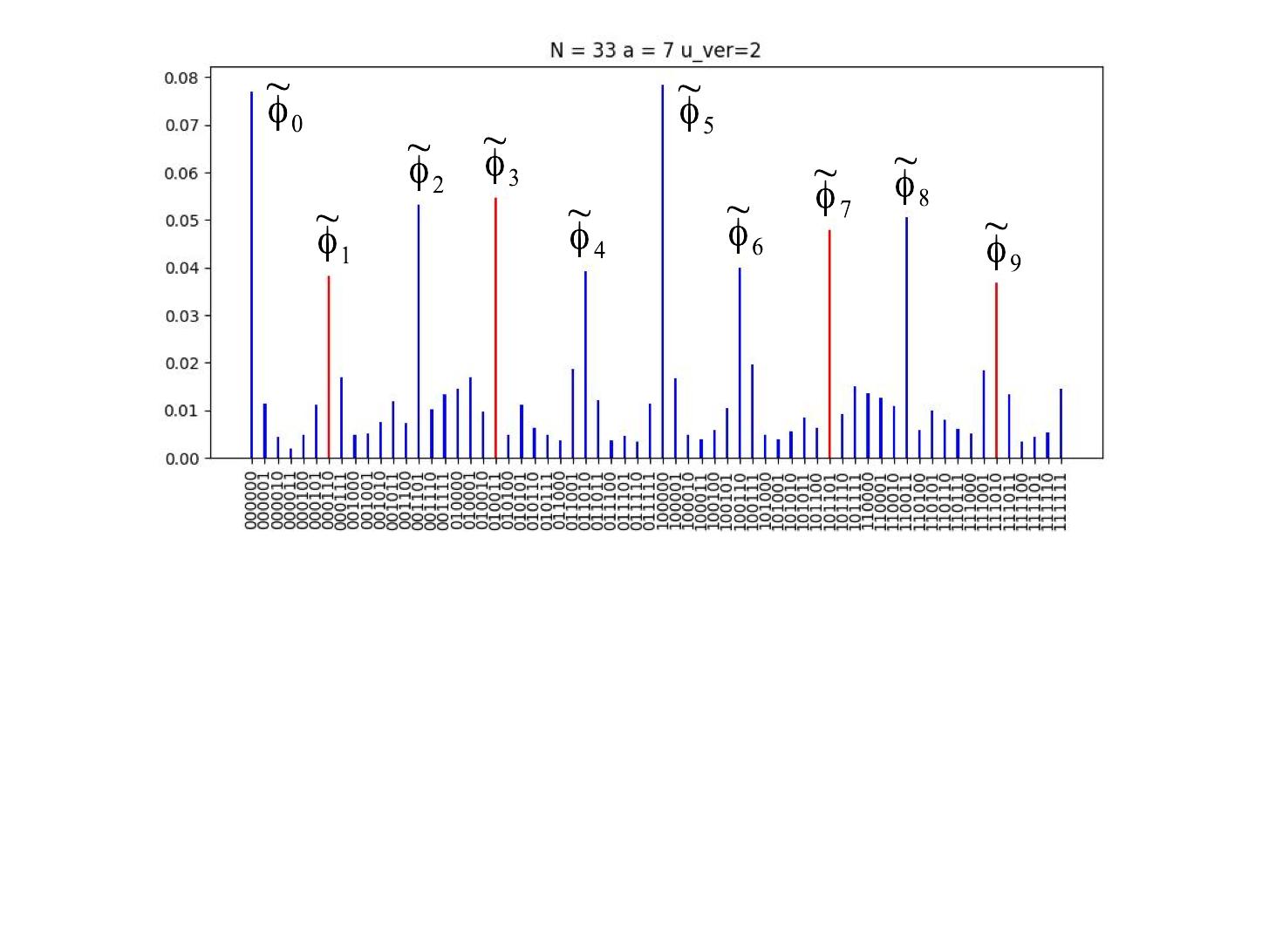} 
\par\end{centering}
\vskip-5.0cm
\begin{centering}
\includegraphics[width=\textwidth]{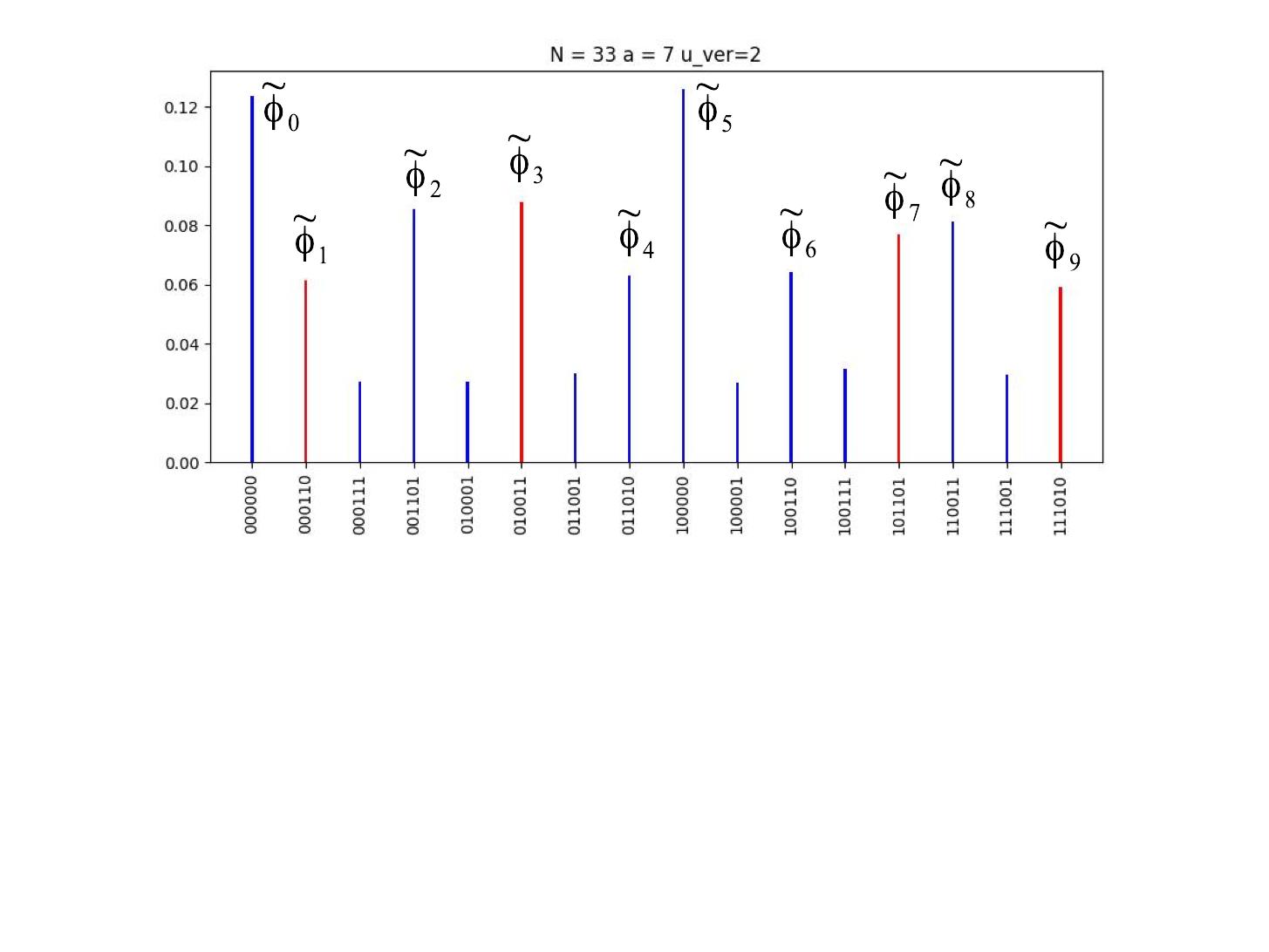}
\par\end{centering}
\vskip-5.0cm 
\caption{\footnoteskip  
  $N=33$, $a=7$, $r=10$, $m=6$: 
  Phase histogram for  truncated ME operator version $\tt{u\_ver}=2$.  
}
\label{fig_hist_N33a7_uver2_truncate_c}
\end{figure}

%\clearpage
%\pagebreak
%\vskip-0.25cm
As we have seen, 
the choice of ME operators seems to be rather forgiving,  as 
long as they encode sufficient correlations to yield an approximate 
phase for which the continued fractions algorithm can be
employed.  For example, consider the case in which we ignore
half of the cycles in the previous version:
%\vfill
%\pagebreak
%%
\begin{eqnarray}
  U_{7, 33} && ~:~ [1, 7, 16, 13, 25, 10, 4] % 28, 31, 19, 1]
\nonumber\\[-3pt]
  U^2_{7, 33}  && ~:~ [1, 16, 25, 4, 31, 1]
\nonumber\\[-3pt]
  U^4_{7, 33} && ~:~  [1, 25, 31, 16, 4, 1]
%\label{eq_Up_N33a7_seq}
\\[-3pt]
  U^8_{7, 33} && ~:~ [1, 31, 4, 25, 16, 1]
\nonumber\\[-3pt]
  U^{16}_{7, 33}  && ~:~ [1, 4, 16, 31, 25, 1]
\nonumber\\[-3pt]
  U^{32}_{7, 33}  && ~:~ [1, 16, 25, 4, 31, 1]
    \ .
\nonumber
\label{eq_Up1_N33a7}
\end{eqnarray}
%%
%%
%\vskip-0.5cm
\begin{figure}[b!]
\vskip-0.7cm
\includegraphics[scale=0.30, center]{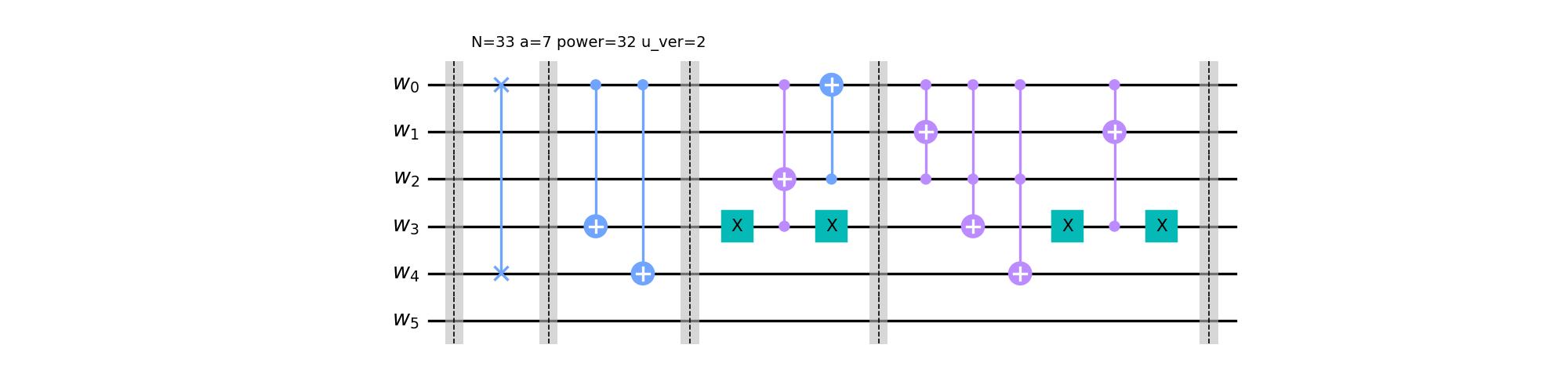}
\vskip-0.7cm
\caption{\footnoteskip
$N=33$, $a=7$, $r=10$: 
The ME operators $U^{32}$ for version $\tt{u\_ver}=2$.  
}
\label{fig_UpN33a7_truncate2_a}
\end{figure}
\noindent
We shall call this version $\tt{u\_ver}=2$. The operators are given 
in  Figs.~\ref{fig_UpN33a7_truncate2_a} and \ref{fig_UpN33a7_truncate2_b}, 
and the phase histogram given by Fig.~\ref{fig_hist_N33a7_uver2_truncate_c}
agrees with the previous results (although there is a bit more noise).  
The lesson here is that much freedom is permitted when constructing 
the ME operators.

%\vskip-0.5cm

\vfill
%\clearpage
\pagebreak

\phantom{set-mark}
\vskip2.0cm
\subsubsection{$N=143 = 13 \times 11$, $a=5$, $r=20$}

%%%%
\begin{figure}[h!]
\begin{minipage}[c]{0.4\linewidth}
\includegraphics[scale=0.50]{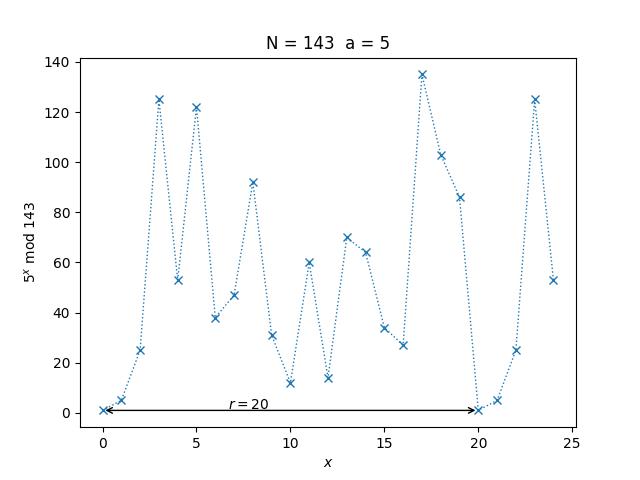}
\end{minipage}
\hskip1.5cm
\begin{minipage}[c]{0.5\linewidth}
\begin{tabular}{|c|c|} \hline
 \multicolumn{2}{|c|}{~$U\vert w \rangle = 
\big\vert 5 \cdot w ~({\rm mod}~143) \big\rangle$~}  \\\hline
$~~~U\vert 1 \rangle = \vert 5 \rangle$~~~~~&
~~$U\vert 00000001 \rangle = \vert 00000101 \rangle$~~~~~\\[-5pt]
$~~~~U\vert 5 \rangle = \vert 25 \rangle$~~~~~&
~~$U\vert 00000101 \rangle = \vert 00011001 \rangle$~~~~~\\[-5pt]
$~~~U\vert 25 \rangle = \vert 125 \rangle$~~~~~&
~~$U\vert 00011001 \rangle = \vert 01111101 \rangle$~~~~~\\[-5pt]
$~~~U\vert 125 \rangle = \vert 53 \rangle$~~~~~&
~~$U\vert 01111101 \rangle = \vert 00110101 \rangle$~~~~~\\[-5pt]
$~~~~U\vert 53 \rangle = \vert 122 \rangle$~~~~~&
~~$U\vert 00110101 \rangle = \vert 01111010 \rangle$~~~~~\\[-5pt]
$~~~U\vert 122 \rangle = \vert 38 \rangle$~~~~~&
~~$U\vert 01111010 \rangle = \vert 00100110 \rangle$~~~~~\\[-5pt]
$~~~~U\vert 38 \rangle = \vert 47  \rangle$~~~~~~&
~~$U\vert 00100110 \rangle = \vert 00101111 \rangle$~~~~~\\[-5pt]
$~~~~U\vert 47 \rangle ~= \vert 92  \rangle$~~~~~~&
~~$U\vert 00101111 \rangle = \vert 01011100 \rangle$~~~~~\\[-5pt]
$~~~~U\vert 92 \rangle ~= \vert 31  \rangle$~~~~~~&
~~$U\vert 01011100 \rangle = \vert 00011111 \rangle$~~~~~\\[-5pt]
$~~~~U\vert 31 \rangle ~= \vert 12  \rangle$~~~~~~&
~~$U\vert 00011111 \rangle = \vert 00001100 \rangle$~~~~~\\[-5pt]
$~~~~U\vert 12 \rangle ~= \vert 60  \rangle$~~~~~~&
~~$U\vert 00001100 \rangle = \vert 00111100 \rangle$~~~~~\\[-5pt]
$~~~U\vert 60 \rangle = \vert 14 \rangle$~~~~~&
~~$U\vert 00111100 \rangle = \vert 00001110 \rangle$~~~~~\\[-5pt]
$~~~~U\vert 14 \rangle = \vert 70 \rangle$~~~~~&
~~$U\vert 00001110 \rangle = \vert 01000110 \rangle$~~~~~\\[-5pt]
$~~~U\vert 70 \rangle = \vert 64 \rangle$~~~~~&
~~$U\vert 01000110 \rangle = \vert 01000000 \rangle$~~~~~\\[-5pt]
$~~~~U\vert 64 \rangle = \vert 34 \rangle$~~~~~&
~~$U\vert 01000000 \rangle = \vert 00100010 \rangle$~~~~~\\[-5pt]
$~~~U\vert 34 \rangle = \vert 27 \rangle$~~~~~&
~~$U\vert 00100010 \rangle = \vert 00011011 \rangle$~~~~~\\[-5pt]
$~~~~U\vert 27 \rangle = \vert 135  \rangle$~~~~~~&
~~$U\vert 00011011 \rangle = \vert 10000111 \rangle$~~~~~\\[-5pt]
$~~~~U\vert 135 \rangle ~= \vert 103  \rangle$~~~~~~&
~~$U\vert 10000111 \rangle = \vert 01100111 \rangle$~~~~~\\[-5pt]
$~~~~U\vert 103 \rangle ~= \vert 86  \rangle$~~~~~~&
~~$U\vert 01100111 \rangle = \vert 01010110 \rangle$~~~~~\\[-5pt]
$~~~~U\vert 86 \rangle ~= \vert 1  \rangle$~~~~~~&
~~$U\vert 01010110 \rangle = \vert 00000001 \rangle$~~~~~\\\hline
\end{tabular} 
\end{minipage}
%\vskip0.5cm
\caption{\footnoteskip
$N=143$, $a=5$, $r=20$: 
The left panel gives the modular exponential function $f_{5,  143}(x) = 
5^x ~ ({\rm mod}~143)$, and the right gives the action of the ME operator
$U_{5, 143}$ on the closed sequence $[1,  5, 25, 125, 53, 122,  38, 47, 92, 31, 
12, 60, 14, 70, 64, 34, 27, 135, 103, 86, 1]$.  
}
\label{fig_fxN143a5}
\end{figure}
\begin{figure}[h!]
\includegraphics[scale=0.55]{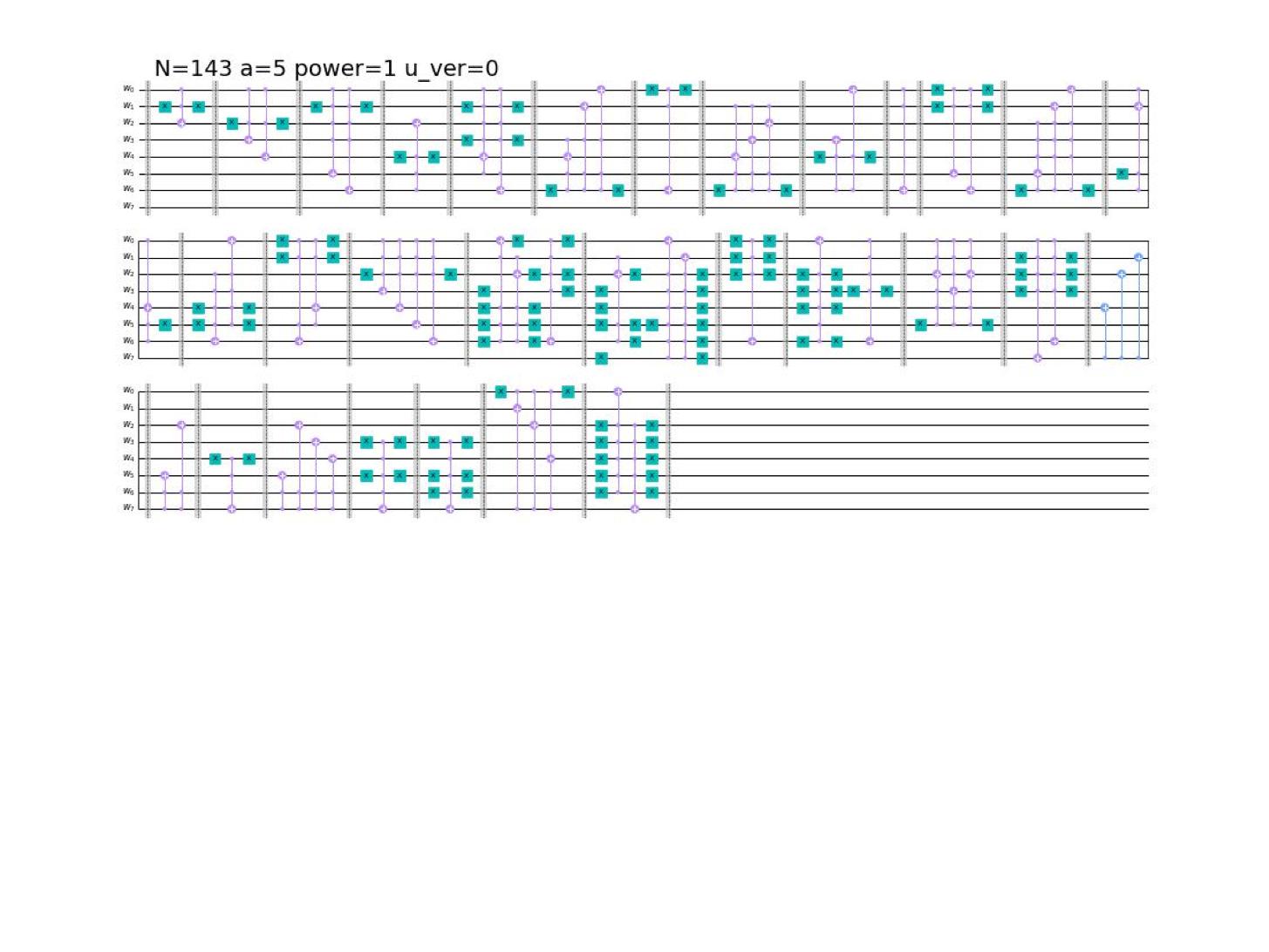} 
\vskip-5.5cm
\caption{\footnoteskip
$N=143$, $a=5$, $r=20$: 
The modular exponentiation operator $U_{5, 143}$.  
}
\label{fig_fxN143a5_U}
\end{figure}
\begin{figure}[h!]
%\vskip-2.0cm
\begin{centering} % 
\includegraphics[width=5.5in]{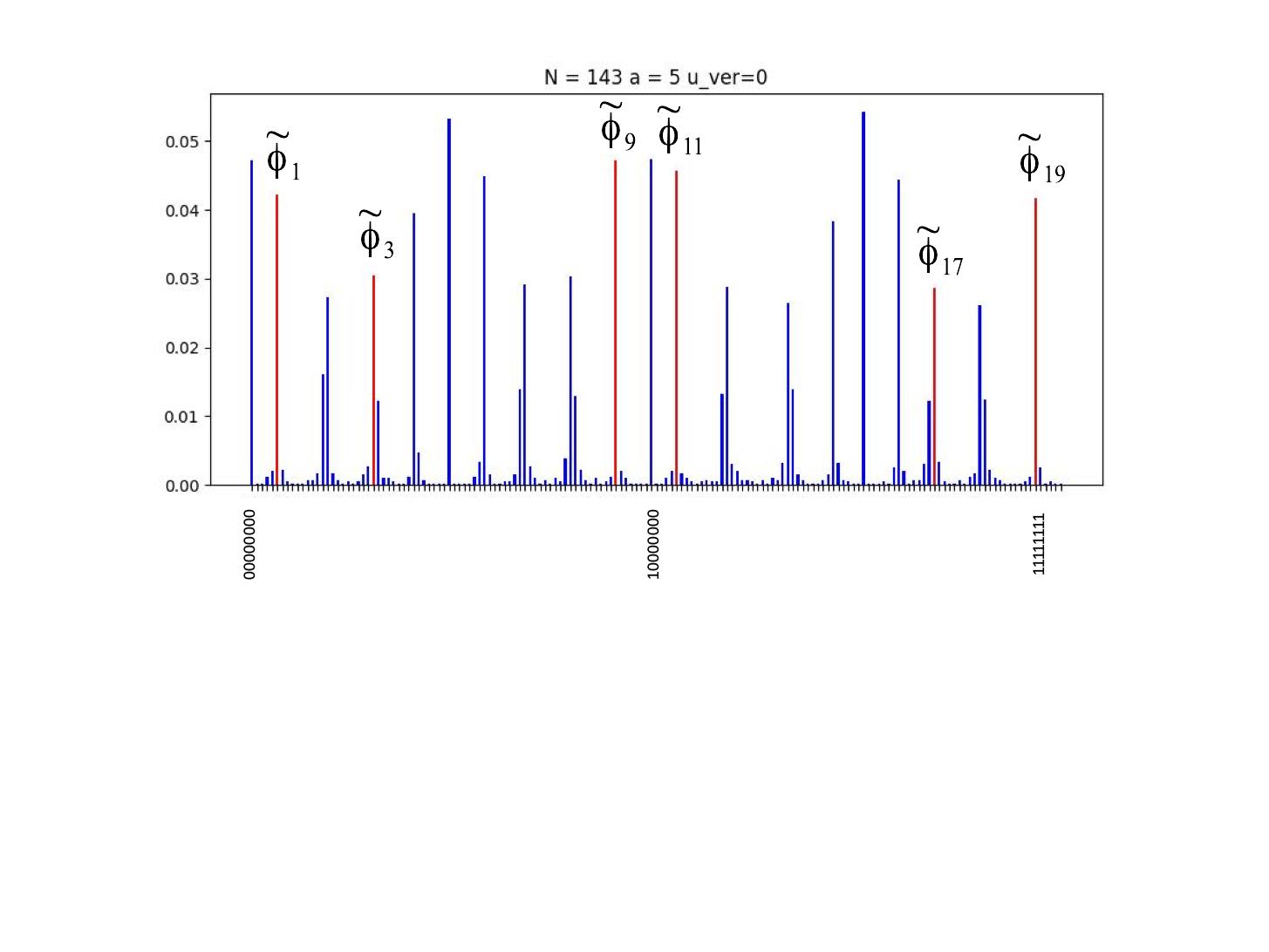} 
\par\end{centering}
\vskip-5.0cm
\begin{centering}
\includegraphics[width=5.5in]{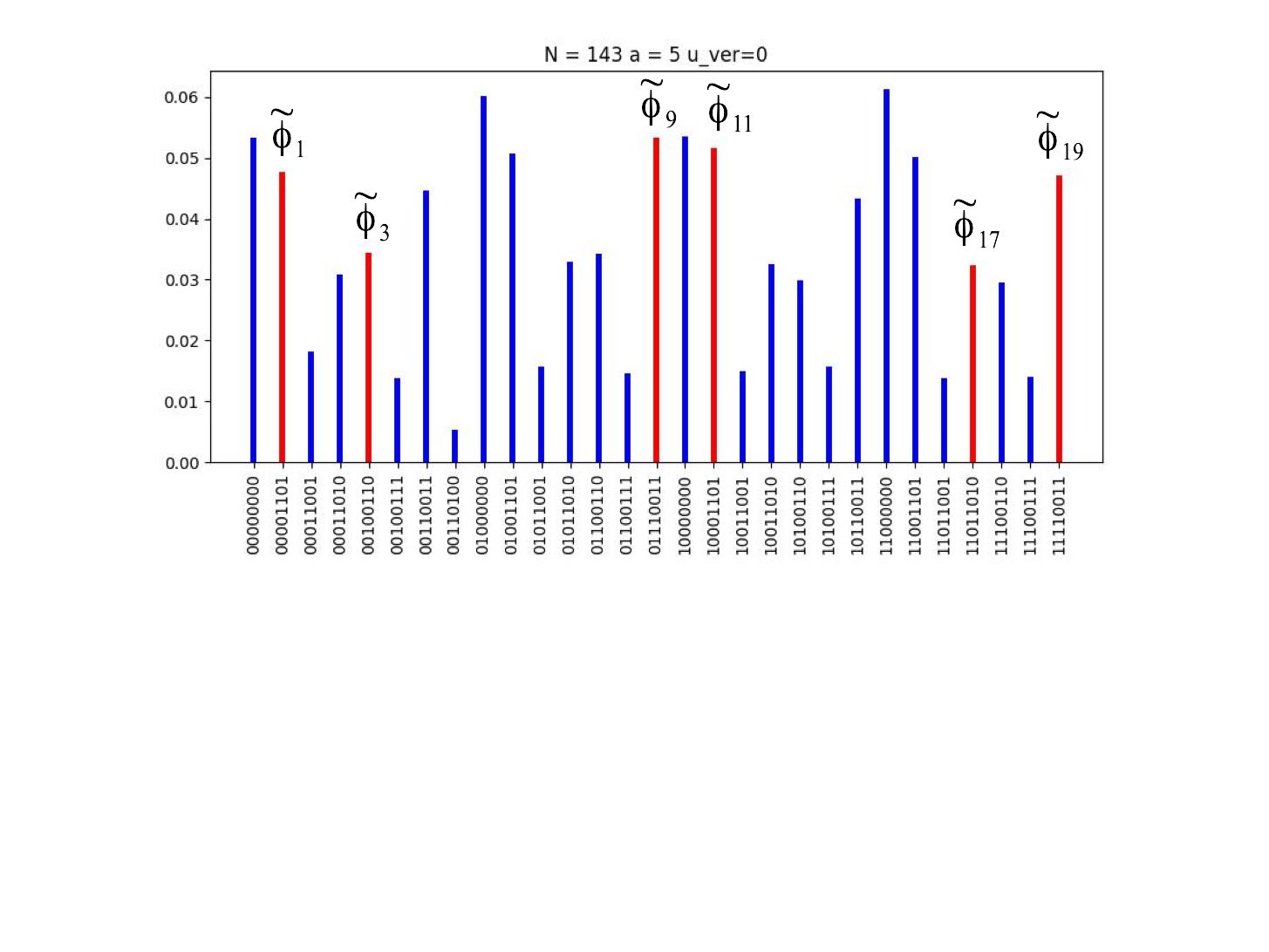} 
\par\end{centering}
\vskip-4.5cm 
\caption{\footnoteskip  
 $N=143$, $a=5$, $r=20$, $m=8$:
  Phase histogram for ME operator version $\tt{u\_ver} = 0$
  The peaks in red correspond to the ME phases $\phi_s = s/20$ 
  with $s \in \{0, 1, \cdots, 19\}$ and ${\rm gcd}(s, 20) = 1$. Thus
  the eight phases $s = 1, 3, 7, 9, 11, 13, 17, 19$ provide the factors
  of 11 and 13. Note, however, that the peaks for $s=7, 13$ are 
  missing.  This is because $m=8$ does not provide sufficient 
  resolution. 
}
\label{fig_fxN143a5m8_hist}
\end{figure}

The next number we shall factor is $N=143 = 11 \times 13$.  As
illustrated in the  left panel of Fig.~\ref{fig_fxN143a5}, the base 
$a= 5$ gives a modular exponential function $f_{5, 143}(x)$ with 
a  period of $r = 20$.  The work register must have $n = \lceil
\log_2 143 \rceil = 8$ qubits, and the corresponding ME operator
$U_{5, 143}$ is given in Fig.~\ref{fig_fxN143a5_U}. We will 
perform a resolution study on the control register by taking 
$m = 8, 9,10$. We will start our analysis with $m = 8$, so we 
must  implement the ME operators $U_{5, 143}^p$ for $p = 2^0, 
2^1, \cdots, 2^7$, {\em i.e.} we require the operators 
$U_{5, 143}, U_{5, 143}^2, U_{5, 143}^4, U_{5, 143}^8, 
U_{5, 143}^{16}, 
U_{5, 143}^{32}, U_{5, 143}^{64}$ and $U_{5, 143}^{128}$.  
For $m = 9$ control qubits, we will also require the operator 
$U_{5, 143}^{256}$, and for $m =10$ we must implement 
$U_{5, 143}^{512}$. As always, we refer to the concatenated 
operators by $\tt{u\_ver}=0$. The phase histogram for $m=8$
is given in Fig.~\ref{fig_fxN143a5m8_hist}.  The top panel of 
the Figure gives the histogram over the full range of phases, 
while the bottom panel only plots the most frequent phases, 
with red phases providing factors.

%%
%\vskip-2.5cm 

\clearpage
The 20 phases of the ME operator $U_{5, 143}$ are supposed to occur 
at $\phi_s = s/20$ for \hbox{$s \in \{0, 1, \cdots, 19\}$}, with the factors 
coming from the phases for which ${\rm gcd}(s,20) = 1$,  that is to 
say, at the eight phases $\phi_s = 1/20$, 3/20, 7/20, 9/20, 11/20, 
13/20, 17/20, 19/20. Note, however, that the red peaks of the 
phase histogram only contain six of the eight phases:
\begin{eqnarray}
   \tilde\phi_1 &=&  [0.00001101]_2 =  0.05078125 \approx \phi_1 ~= 1/20 
   \\
   \tilde\phi_3 &=& [ 0.00100110]_2 =  0.14843750 \approx \phi_3 ~= 3/20
   \\
   \tilde\phi_9 &=&    [0.01110011]_2 = 0.44921875 \approx \phi_9 ~= 9/20
   \\
  \tilde\phi_{11}  &=&   [ 0.10001101]_2 =  0.55078125 \approx \phi_{11} = 11/20
     \\
  \tilde\phi_{17}  &=& [0.11011010]_2 =  0.85156250 \approx  \phi_{17} =17/20
     \\
  \tilde\phi_{19}  &=&  [0.11110011]_2 = 0.94921875 \approx \phi_{19} = 19/20
    \ .
\end{eqnarray}
\begin{figure}[h!]
\vskip-1.0cm
\begin{centering}
\includegraphics[width=6.0in]{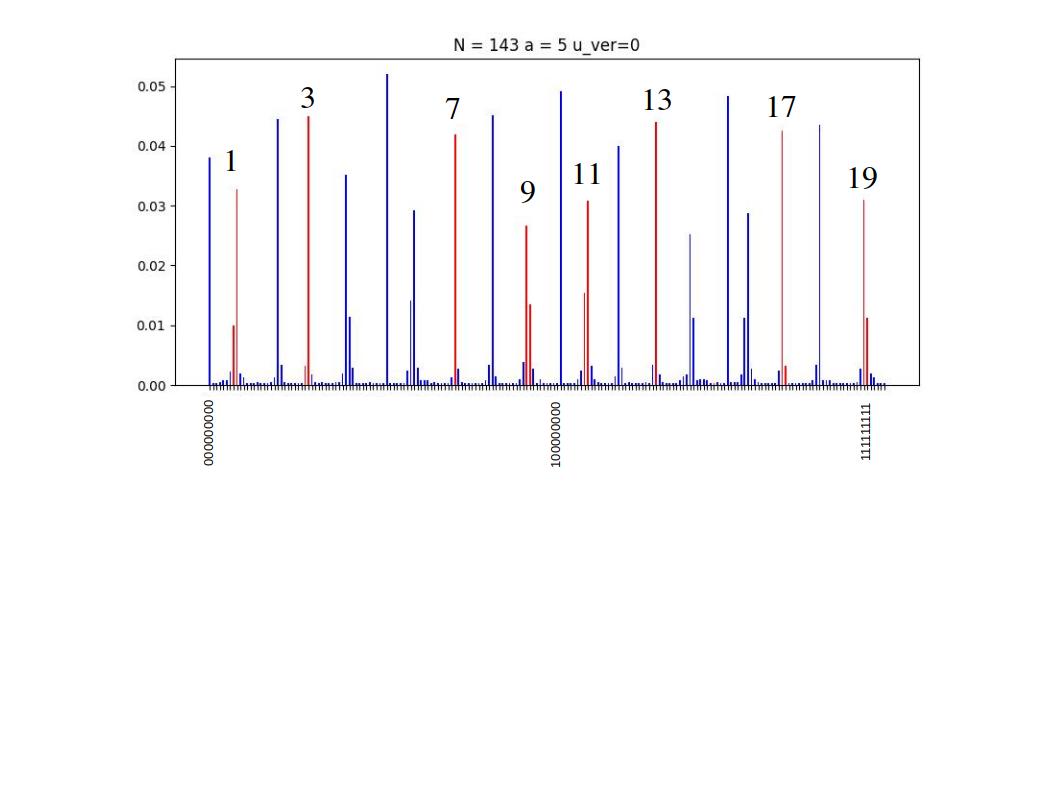}  
\par\end{centering}
\vskip-5.7cm
\begin{centering}
\includegraphics[width=6.0in]{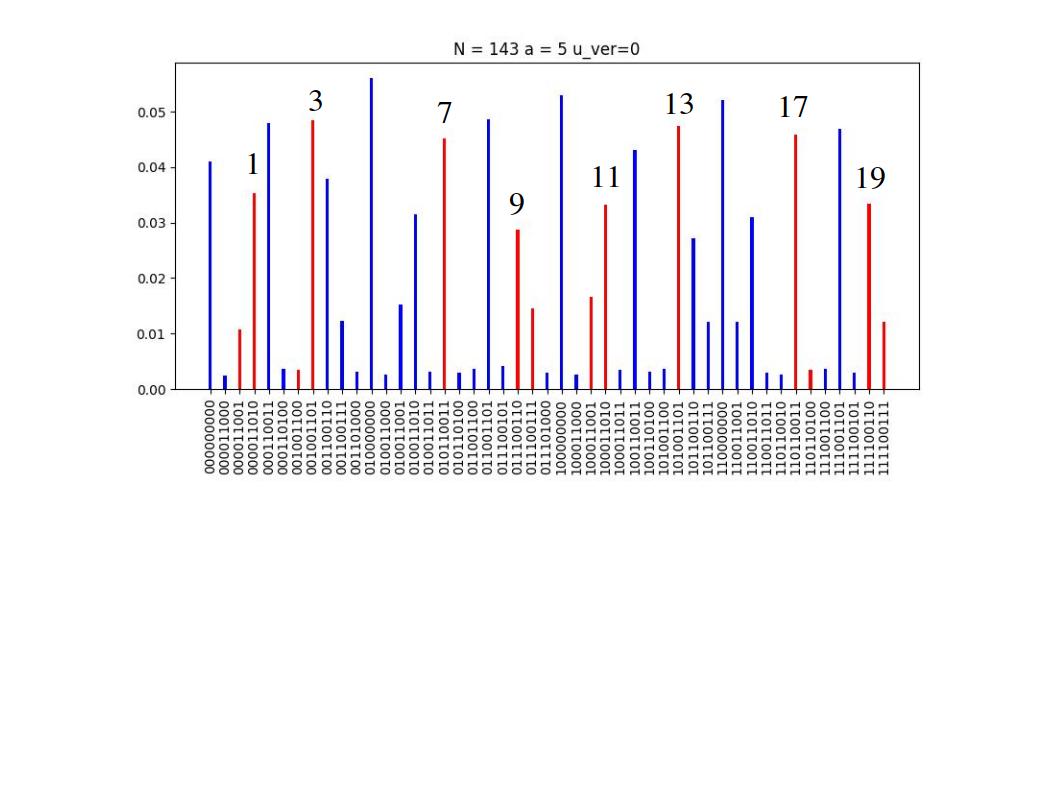} 
\par\end{centering}
\vskip-4.5cm 
\caption{\footnoteskip  
  $N=143$, $a=5$, $r=20$, $m=9$: 
  Phase histogram for ME operator version $\tt{u\_ver} = 0$.
  Increasing the phase resolution to $m = 9$ provides all ten
  phases associated with factors. 
}
\label{fig_fxN143a5m9_hist}
\end{figure}

\begin{figure}[h!]
\begin{centering}
\includegraphics[width=6.0in]{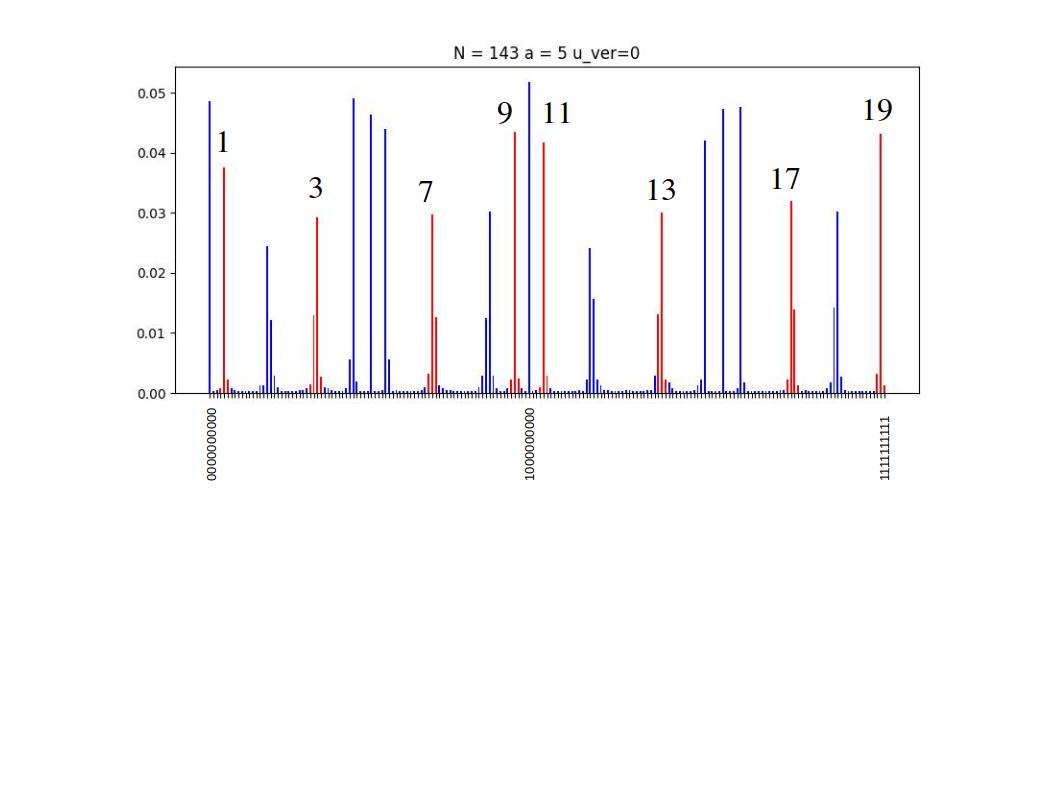} 
\par\end{centering}
\vskip-5.6cm
\begin{centering}
\includegraphics[width=6.0in]{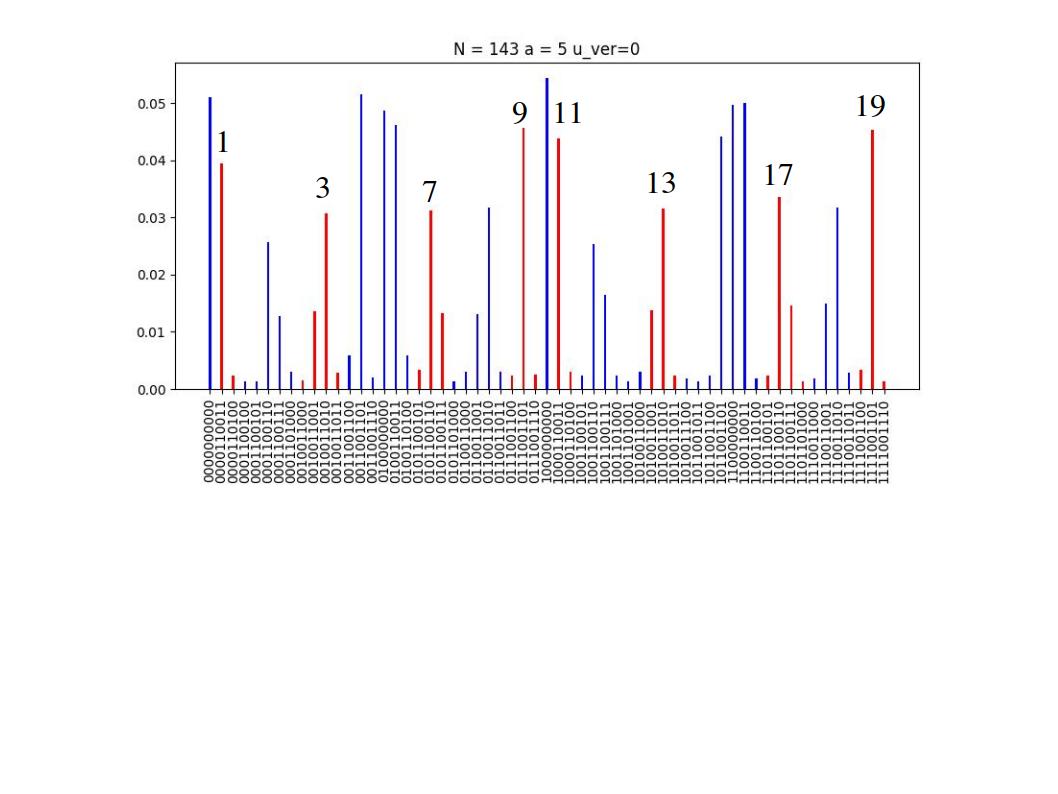} 
\par\end{centering}
\vskip-4.8cm 
\caption{\footnoteskip  
  $N=143$, $a=5$, $r=20$, $m=10$:
  Phase histogram for ME operator version $\tt{u\_ver} = 0$.
}
\label{fig_fxN143a5m10_hist}
\end{figure}

\noindent
The two phases corresponding to  $\phi_7=7/20$ and $\phi_{13} = 
13/20$ are absent. This is actually a resolution problem: when we
increase the control register to $m =9$ qubits, we obtain all eight 
phases, as the phase histogram of Fig.~\ref{fig_fxN143a5m9_hist} 
reveals.   Note that sub-dominant peaks have appeared, and they
too can provide factors. The situation is even more dramatic for 
$m = 10$, where further sub-dominant peaks emerge, as 
illustrated in the phase histogram in Fig.~\ref{fig_fxN143a5m10_hist}.  

%\vskip-2.5cm 
%%

%\clearpage
Finally, let us turn to constructing the composite operators
$U^p$ without using concatenation.  These operators have
the following pairs of cycles:
\begin{eqnarray}
  U_{5, 143} && ~:~  
  [1, 5, 25, 125, 53, 122, 38, 47, 92, 31, 12, 60, 14, 70, 64,
   34, 27, 135, 103, 86, 1]
\nonumber\\[-3pt]
  U^2_{5, 143}   && ~:~ [1, 25, 53, 38, 92, 12, 14, 64, 27, 103, 1] 
 ~~\text{and}~~ [5, 125, 122, 47, 31, 60, 70, 34, 135, 86, 5]
\nonumber\\[-3pt]
  U^4_{5, 143}  && ~:~ [1, 53, 92, 14, 27, 1] 
  ~~\text{and}~~ [5, 122, 31, 70, 135, 5]
\nonumber\\[-3pt]
  U^8_{5, 143}  && ~:~ [1, 92, 27, 53, 14, 1]
  ~~\text{and}~~ [5, 31, 135, 122, 70, 5]
\nonumber\\[-3pt]
  U^{16}_{5, 143}   && ~:~[1, 27, 14, 92, 53, 1]
  ~~\text{and}~~  [5, 135, 70, 31, 122, 5]
\label{eq_Up_N143a5}
\\[-3pt]
  U^{32}_{5, 143}   && ~:~ [1, 14, 53, 27, 92, 1]
  ~~\text{and}~~  [5, 70, 122, 135, 31, 5]
\nonumber\\[-3pt]
  U^{64}_{5, 143}   && ~:~  [1, 53, 92, 14, 27, 1] 
  ~~\text{and}~~ [5, 122, 31, 70, 135, 5]
\nonumber\\[-3pt]
  U^{128}_{5, 143}   && ~:~ [1, 92, 27, 53, 14, 1] 
  ~~\text{and}~~ [5, 31, 135, 122, 70, 5]
\nonumber
  \ .
\end{eqnarray}

\begin{figure}[h!]
%vskip-1.0cm
\begin{centering}
\includegraphics[width=6.0in]{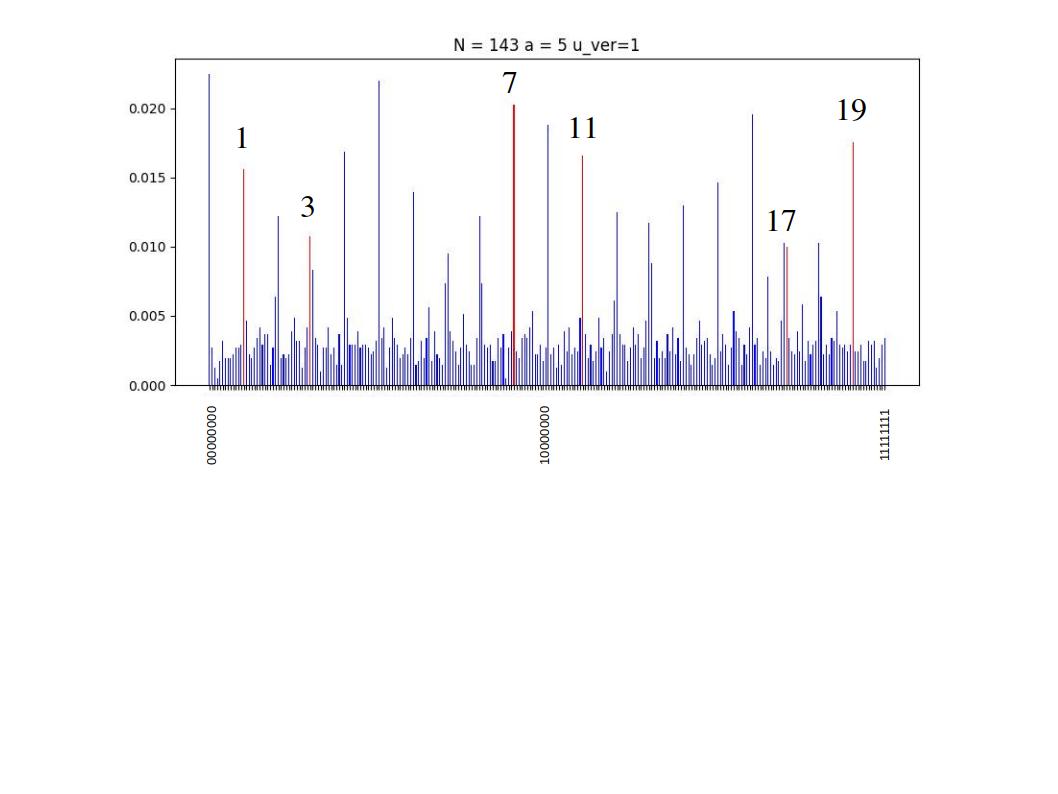} 
\par\end{centering}
\vskip-5.8cm
\begin{centering}
\includegraphics[width=6.0in]{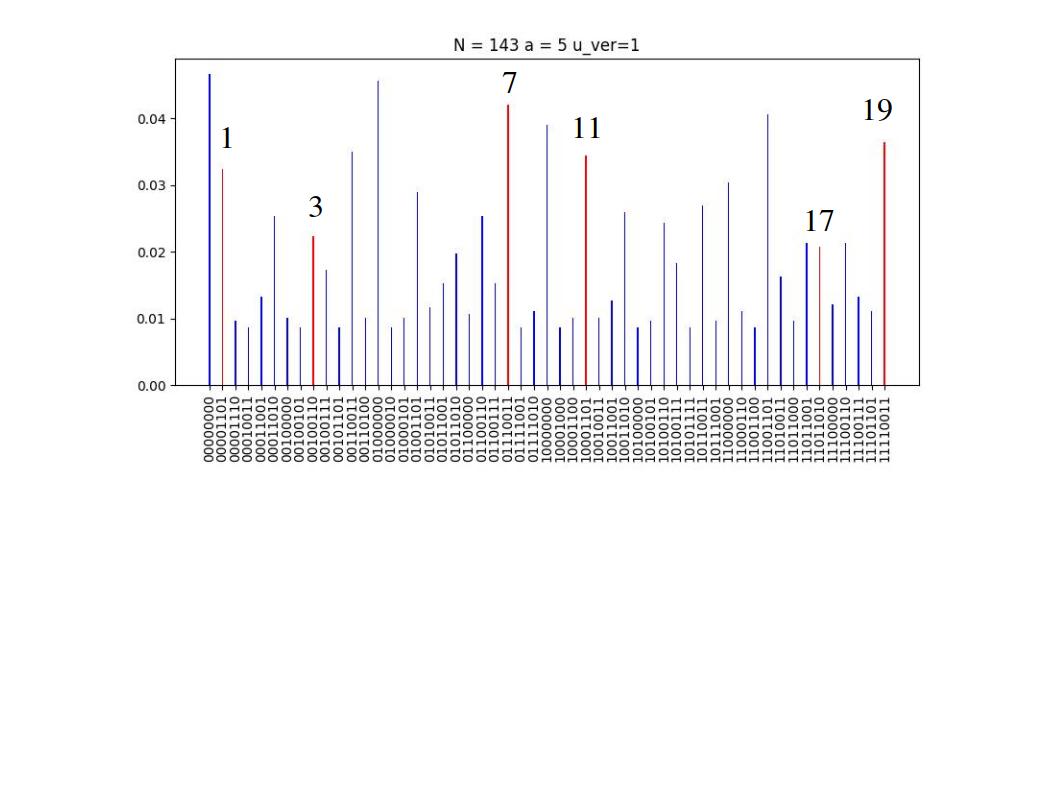} 
\par\end{centering}
\vskip-4.5cm 
\caption{\footnoteskip  
  $N=143$,  $a=5$,  $r=20$, $m=8$:
  Phase histogram for ME operator version $\tt{u\_ver} = 1$. 
  Despite the noise relative to the previous version  $\tt{u\_ver} 
  = 0$, the signal is quite discernible. 
}
\label{fig_fxN143a5m8_hist1}
\end{figure}

\noindent
For simplicity we have not included all possible cycles, and 
we will refer to the procedure given by (\ref{eq_Up_N143a5}) 
as version number $\tt{u\_ver}=1$ (so this can also be regarded
as a truncated version of the ME operators). The composite 
operators $U^p$  are given in Figs.~\ref{fig_fxN143a5_Up_a} 
and \ref{fig_fxN143a5_Up_b} in Appendix~\ref{sec_N143_uver1}, 
and the corresponding phase histogram from a Qiskit simulation 
with 4096 runs is given in Fig.~\ref{fig_fxN143a5m8_hist1}.
As usual, the top panel plots all output phases, and the bottom 
panel plots only the most frequent ones.  Note that the noise 
in the top Figure has increased significantly, but the signal still
dominates.

\clearpage
\subsubsection{$N=247 = 13 \times 19$, $a=2$, $r=36$}

\begin{figure}[h!]
\begin{minipage}[c]{0.4\linewidth}
\includegraphics[scale=0.50]{02_period_v1_N247_a2.jpg}
\end{minipage}
\vskip-0.4cm
\includegraphics[scale=0.65]{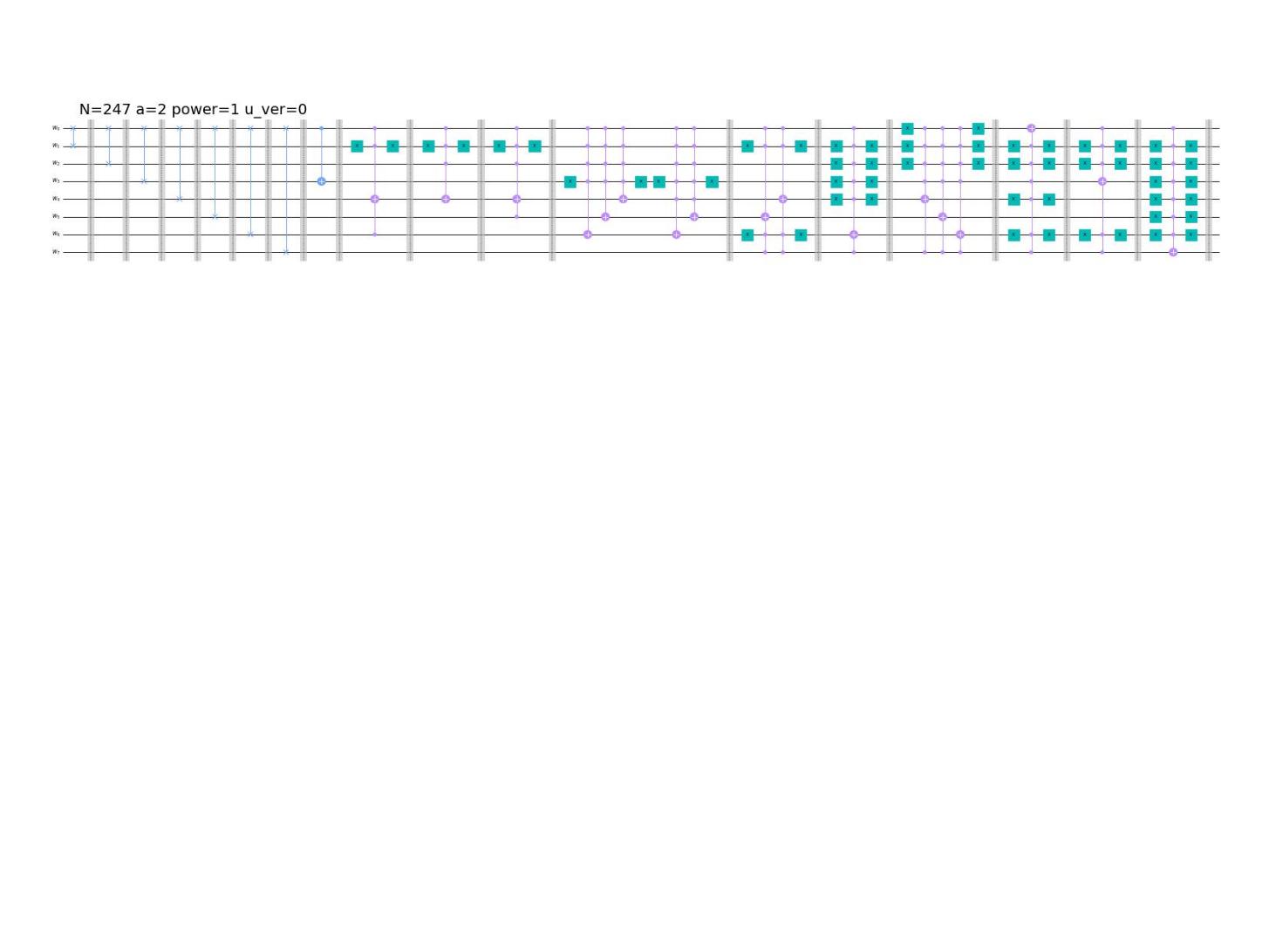} 
\vskip-10.0cm
\caption{\footnoteskip
$N=247$,  $a=2$,  $r=36$: 
The top panel is the modular exponential function $f_{2,  247}(x) = 
2^x ~ ({\rm mod}~247)$,  while the bottom panel shows the modular 
exponentiation operator $U_{2, 247}$. 
}
\label{fig_fxN247a2}
\end{figure}

\begin{figure}[h!]
\vskip-0.5cm
\begin{centering}\includegraphics[width=5.4in]{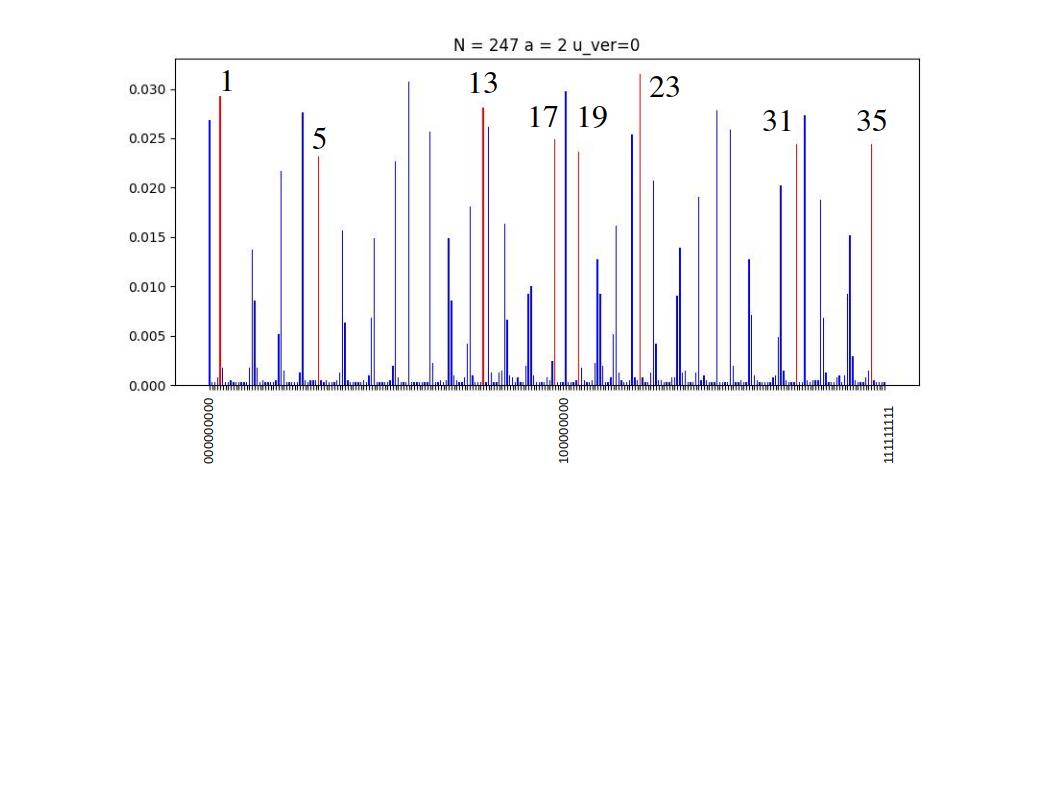} 
\par\end{centering}
\vskip-5.2cm
\begin{centering}
\includegraphics[width=5.4in]{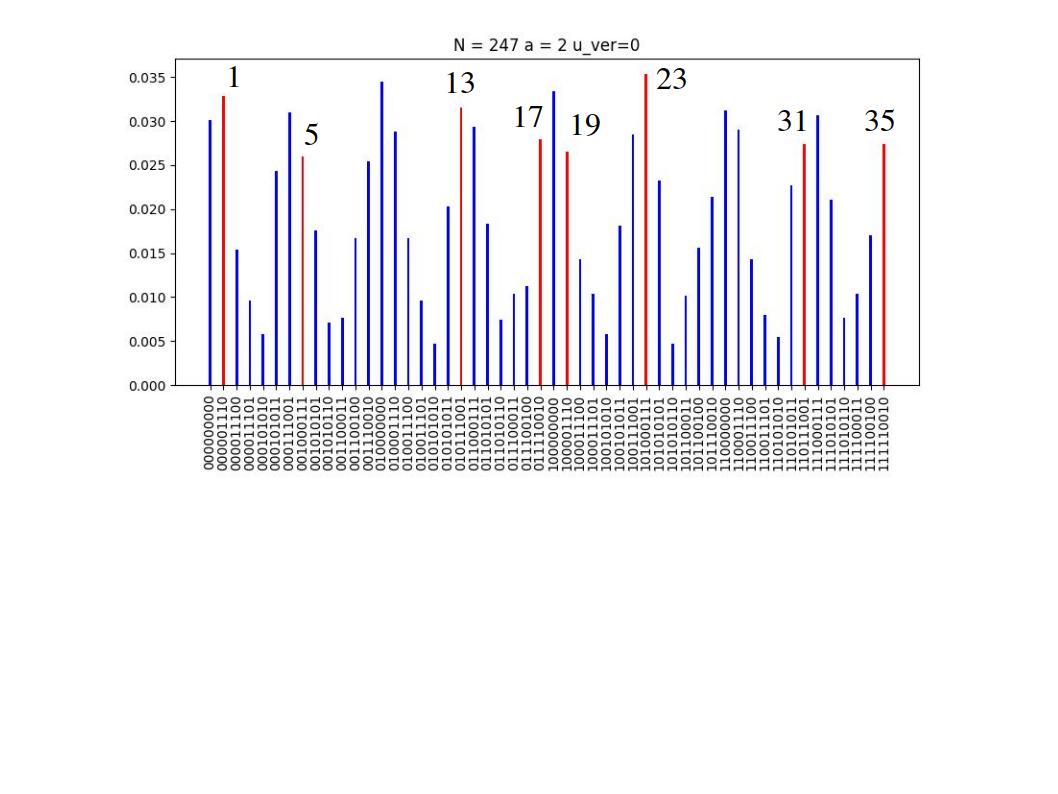} 
\par\end{centering}
\vskip-4.5cm 
\caption{\footnoteskip  
  $N=247$,  $a=2$,  $r=36$, $m=9$: 
  Phase histogram for ME operator version $\tt{u\_ver}=0$.
}
\label{fig_N247a2_phase_hist}
\end{figure}

\begin{figure}[b!]
%\vskip-1.0cm
\begin{centering}
\includegraphics[width=5.2in]{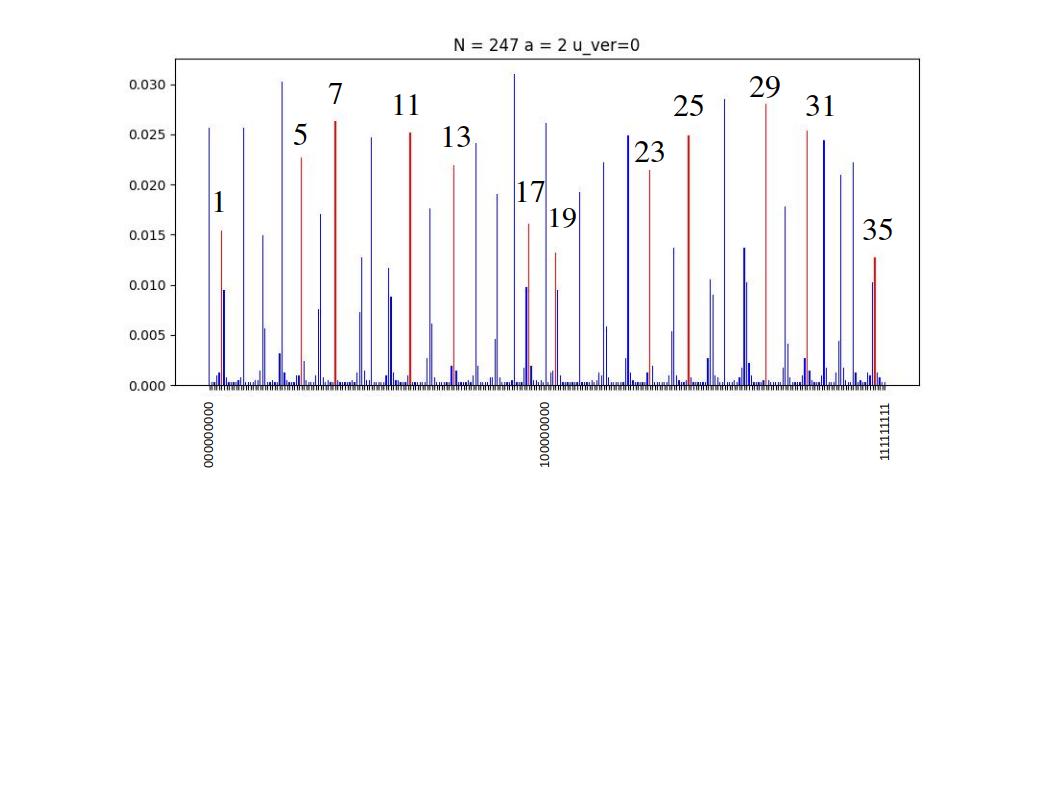}
\par\end{centering}
\vskip-5.0cm
\begin{centering}
\includegraphics[width=5.2in]{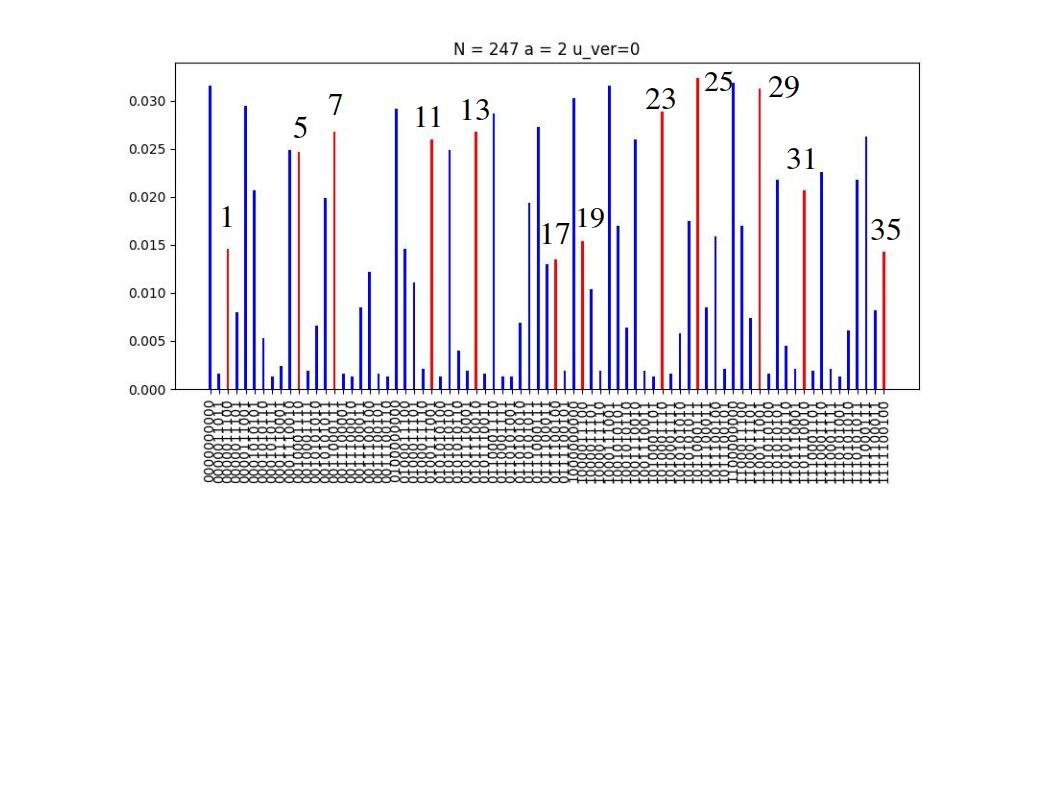} 
\par\end{centering}
\vskip-4.0cm 
\caption{\footnoteskip  
  $N=247$,  $a=2$,  $r=36$ $m=10$: 
  Phase histogram for ME operator version $\tt{u\_ver}=0$.
}
\label{fig_N247a2m10_phase_hist}
\end{figure}

\begin{minipage}[c]{0.50\linewidth}

\baselineskip 19pt
As our last example,  let us factor $N = 247$ into 13 and~19. 
For the base $a =2$, the top panel of  Fig.~\ref{fig_fxN247a2} 
shows that the period of the modular exponential function 
$f_{2,247}(x)$ is $r = 36$.  For this period, $m = 9$ control 
qubits is  sufficient to resolve the phase difference 
$\Delta\phi =1/36$.  Also note that we require $n = \lceil 
\log_2 246 \rceil = 8$ work qubits.  The action of the ME 
operator $U_{2, 247}$ on the work state $\vert 1 \rangle =
\vert 00000001 \rangle$ is illustrated in the table to the right,  
and its circuit representation is given in the bottom panel 
of Fig.~\ref{fig_fxN247a2}. Since $m = 8$, we shall also
require the ME operators $U_{2, 247}^p$ for $p =1, 2, 4,
\cdots, 256$.  As usual, we can construct these operators 
by concatenating $U_{2, 247}$,  and we will refer to this 
as version number $\tt{u\_ver}=0$. The phase histogram 
from Shor's algorithm is illustrated in 
Fig.~\ref{fig_N247a2_phase_hist}. The phases of the 
$U_{2, 247}$ operator that provide factors are supposed 
to occur at $\phi_s = s/36$ for $s \in \{ 0,  1,  \cdots, 35\}$, 
where $r=36$ and $s$ have no non-trivial common factors. 
This gives 12 possible phases: $\phi_s =$ 1/36, 5/36, 7/36,  
11/36, 13/36, 17/36, 19/36, 23/36, 25/36, 29/36, 31/36, 35/36. 
The phase histogram in Fig.~\ref{fig_N247a2_phase_hist}
exhibits eight of these phases. As 
before, the top panel shows every phase from the simulation,
while the bottom panel gives only the most frequent phases. 
However, if we increase the phase resolution to $m = 10$
control qubits, Fig~\ref{fig_N247a2m10_phase_hist} shows 
that we capture all 12 possible phases.   

\bodyskip
\end{minipage}
\begin{minipage}[c]{0.1\linewidth}
\end{minipage}
\begin{minipage}[c]{0.0\linewidth}
\begin{tabular}{|c|c|} \hline
 \multicolumn{2}{|c|}{~$U\vert w \rangle = 
\big\vert 2 \cdot w ~({\rm mod}~247) \big\rangle$~}  \\\hline
$~U\vert 1 \rangle = \vert 2 \rangle$~&
~$U\vert 00000001 \rangle = \vert 00000010 \rangle$~\\[-5pt]
$~U\vert 2 \rangle = \vert 4 \rangle$~&
~$U\vert 00000010 \rangle = \vert 00000100 \rangle$~\\[-5pt]
$~U\vert 4 \rangle = \vert 8 \rangle$~&
~$U\vert 00000100 \rangle = \vert 00001000 \rangle$~\\[-5pt]
$~U\vert 8 \rangle = \vert 16 \rangle$~&
~$U\vert 00001000 \rangle = \vert 00010000 \rangle$~\\[-5pt]
$~U\vert 16 \rangle = \vert 32 \rangle$~&
~$U\vert 00010000 \rangle = \vert 00100000 \rangle$~\\[-5pt]
$~U\vert 32 \rangle = \vert 64 \rangle$~&
~$U\vert 00100000 \rangle = \vert 01000000 \rangle$~\\[-5pt]
$~U\vert 64 \rangle = \vert 128  \rangle$~&
~$U\vert 01000000 \rangle = \vert 10000000 \rangle$~\\[-5pt]
$~U\vert 128 \rangle = \vert 9  \rangle$~&
~$U\vert 10000000 \rangle = \vert 00001001 \rangle$~\\[-5pt]
$~U\vert 9 \rangle ~= \vert 18  \rangle$~&
~$U\vert 00001001 \rangle = \vert 00010010 \rangle$~\\[-5pt]
$~U\vert 18 \rangle ~= \vert 36  \rangle$~&
~$U\vert 00010010 \rangle = \vert 00100100 \rangle$~\\[-5pt]
$~U\vert 36 \rangle ~= \vert 72  \rangle$~&
~$U\vert 00100100 \rangle = \vert 01001000 \rangle$~\\[-5pt]
$~U\vert 72 \rangle = \vert 144 \rangle$~&
~$U\vert 01001000 \rangle = \vert 10010000 \rangle$~\\[-5pt]
$~U\vert 144 \rangle = \vert 41 \rangle$~&
~$U\vert 10010000 \rangle = \vert 00101001 \rangle$~\\[-5pt]
$~U\vert 41 \rangle = \vert 82 \rangle$~&
~$U\vert 00101001 \rangle = \vert 01010010 \rangle$~\\[-5pt]
$~U\vert 82 \rangle = \vert 164 \rangle$~&
~$U\vert 01010010 \rangle = \vert 10100100 \rangle$~\\[-5pt]
$~U\vert 164 \rangle = \vert 81 \rangle$~&
~$U\vert 10100100 \rangle = \vert 01010001 \rangle$~\\[-5pt]
$~U\vert 81 \rangle = \vert 162  \rangle$~&
~$U\vert 01010001 \rangle = \vert 10100010 \rangle$~\\[-5pt]
$~U\vert 162 \rangle ~= \vert 77  \rangle$~&
~$U\vert 10100010 \rangle = \vert 01001101 \rangle$~\\[-5pt]
$~U\vert 77 \rangle ~= \vert 154  \rangle$~&
~$U\vert 01001101 \rangle = \vert 10011010 \rangle$~\\[-5pt]
$~U\vert 154 \rangle ~= \vert 61  \rangle$~&
~$U\vert 10011010 \rangle = \vert 10011010 \rangle$~\\[-5pt]
$~U\vert 61 \rangle ~= \vert 122  \rangle$~&
~$U\vert 10011010 \rangle = \vert 01111010 \rangle$~\\[-5pt]
$~U\vert 122 \rangle = \vert 244 \rangle$~&
~$U\vert 01111010 \rangle = \vert 11110100 \rangle$~\\[-5pt]
$~U\vert 244 \rangle = \vert 241 \rangle$~&
~$U\vert 11110100 \rangle = \vert 11110001 \rangle$~\\[-5pt]
$~U\vert 241 \rangle = \vert 235\rangle$~&
~$U\vert 11110001 \rangle = \vert 11101011 \rangle$~\\[-5pt]
$~U\vert 235 \rangle = \vert 223 \rangle$~&
~$U\vert 11101011 \rangle = \vert 11011111 \rangle$~\\[-5pt]
$~U\vert 223 \rangle = \vert 199 \rangle$~&
~$U\vert 11011111 \rangle = \vert 11000111 \rangle$~\\[-5pt]
$~U\vert 199 \rangle = \vert 151 \rangle$~&
~$U\vert 11000111 \rangle = \vert 10010111 \rangle$~\\[-5pt]
$~U\vert 151 \rangle = \vert 55  \rangle$~&
~$U\vert 10010111 \rangle = \vert 00110111 \rangle$~\\[-5pt]
$~U\vert 55 \rangle ~= \vert 110  \rangle$~&
~$U\vert 00110111 \rangle = \vert 01101110 \rangle$~\\[-5pt]
$~U\vert 110 \rangle ~= \vert 220  \rangle$~&
~$U\vert 01101110 \rangle = \vert 11011100 \rangle$~\\[-5pt]
$~U\vert 220 \rangle ~= \vert 193  \rangle$~&
~$U\vert 11011100 \rangle = \vert 11000001 \rangle$~\\[-5pt]
$~U\vert 193 \rangle ~= \vert 139  \rangle$~&
~$U\vert 11000001 \rangle = \vert 10001011 \rangle$~\\[-5pt]
$~U\vert 139 \rangle = \vert 31 \rangle$~&
~$U\vert 10001011 \rangle = \vert 00011111 \rangle$~\\[-5pt]
$~U\vert 31 \rangle = \vert 62 \rangle$~&
~$U\vert 00011111 \rangle = \vert 00111110 \rangle$~\\[-5pt]
$~U\vert 62 \rangle = \vert 124 \rangle$~&
~$U\vert 00111110 \rangle = \vert 01111100 \rangle$~\\[-5pt]
$~U\vert 124 \rangle = \vert 1 \rangle$~&
~$U\vert 01111100 \rangle = \vert 00000001 \rangle$~\\\hline
\end{tabular} 
\end{minipage}

%%
%\vskip-2.5cm 

\clearpage

%\vskip-2.5cm 

Let us examine the phase histogram of Fig.~\ref{fig_N247a2_phase_hist}
in a bit more detail. Note that the phases that produce factors of $N=247 
= 13 \times 19$ are given in red, and take the following values:
\begin{eqnarray}
   \tilde\phi_1 &=&    [0.000001110]_2 = 0.027343750 \approx 1/36
\nonumber\\[3pt]
   \tilde\phi_5 &=&    [ 0.001000111]_2 = 0.138671875 \approx 5/36
\nonumber\\[3pt]
   \tilde\phi_{13} &=&  [0.010111001]_2 = 0.361328125 \approx 13/36
\nonumber\\[3pt] 
  \tilde\phi_{17}  &=&  [ 0.011110010]_2 = 0.472656250 \approx 17/36
 \label{eq_phi_N247a2}\\[3pt]
  \tilde\phi_{19}  &=&   [0.100001110]_2 = 0.527343750 \approx 19/36
\nonumber\\[3pt]
  \tilde\phi_{23}  &=&  [0.101000111]_2 = 0.638671875 \approx 23/36
\nonumber\\[3pt]
  \tilde\phi_{31}  &=&  [0.110111001]_2 = 0.861328125 \approx 31/36
\nonumber\\[3pt]
  \tilde\phi_{35}  &=&   [0.111110010]_2 =  0.972656250 \approx 35/36
\nonumber
    \ .
\end{eqnarray}
This is in agreement with the theoretical predictions of $\phi_s 
= s/36$ for ${\rm gcd}(s, 36) = 1$, except that the phases for 
$s=7,11, 25, 29$ are missing.  As we have seen before, we 
can recover these phases by increasing the resolution of 
the control register. As noted above,  for $m = 10$ we find that all expected 
phases are observed, as shown in Fig.~\ref{fig_N247a2m10_phase_hist}.

\begin{figure}[h!]
\begin{centering}
\includegraphics[width=\textwidth]{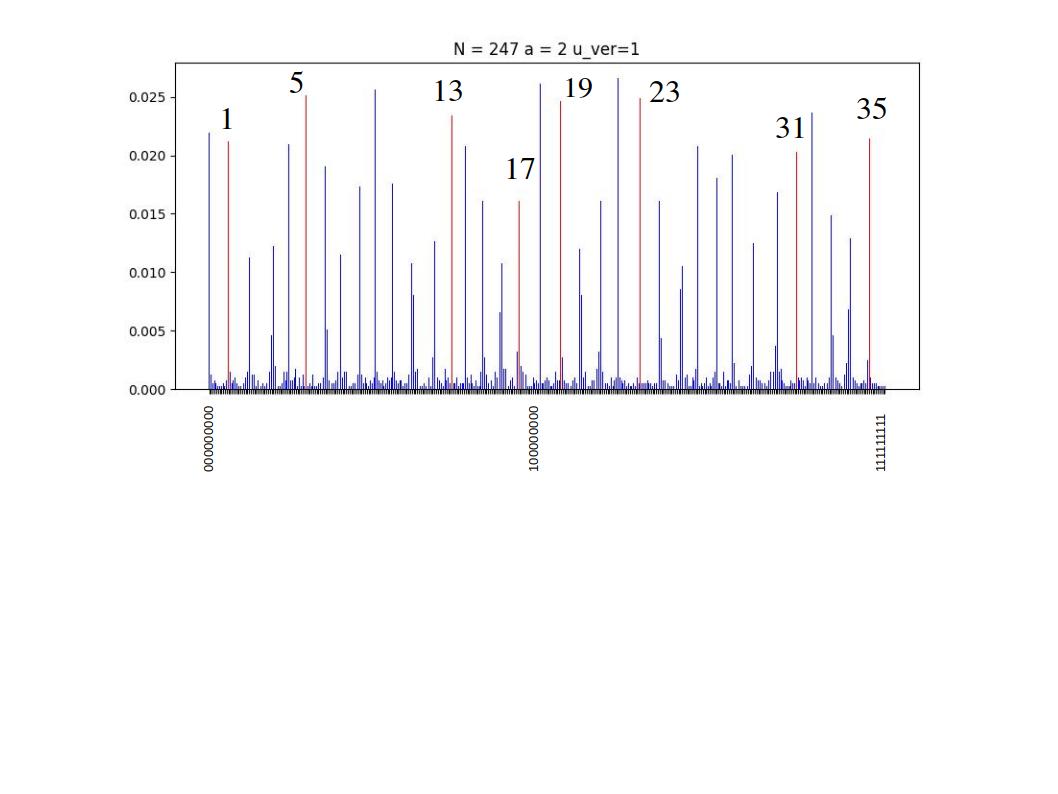}
\par\end{centering}
\vskip-5.0cm
\begin{centering}
\includegraphics[width=\textwidth]{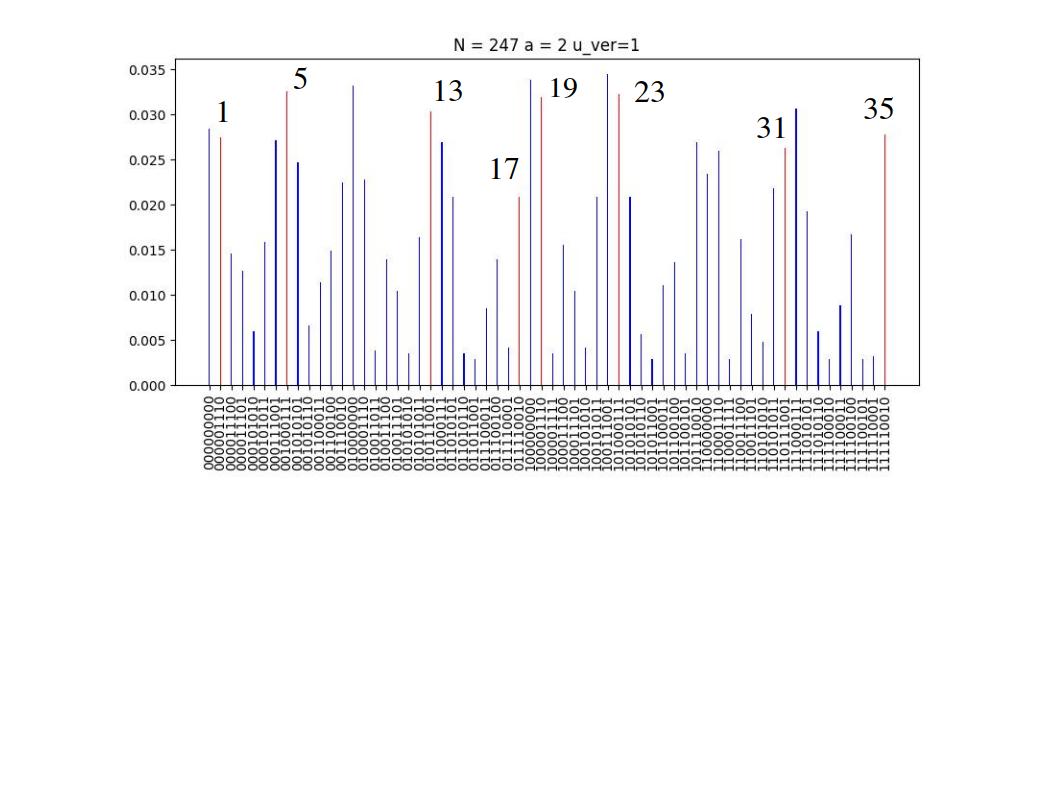} 
\par\end{centering}
\vskip-5.0cm 
\caption{\footnoteskip  
  $N=247$,  $a=2$,  $r=36$, $m=9$: 
  Phase histogram for ME operators version $\tt{u\_ver}=1$.
}
\label{fig_N247a2_phase1_hist}
\end{figure}

\clearpage
Let us now address the composite operator issue for $U^p_{
2, 247}$ with $p > 1$. Returning to $m = 9$, so that $p =  
1, 2, 4, \cdots,256 $, some of the closed cycles are given by
\begin{eqnarray}
  U_{2, 247} && ~:~
  [1, 2, 4, 8, 16, 32, 64, 128, 9, 18, 36, 72,
 144, 41, 82, 164, 81, 162, 77, 154, 
\label{eq_Up1_N247a2} 
\\[-5pt] && ~~~~\,
  61, 122, 244, 241, 235, 223, 199, 151, 
  55, 110, 220,193, 139, 31,   62, 124, 1]
\nonumber \\
  U^{128}_{2, 247}   && ~:~
     [1, 4, 16, 64, 9, 36, 144, 82, 81, 77, 61, 244,
     235, 199, 55, 220, 139, 62, 1]
      ~~\text{and}
\nonumber\\[-5pt] && ~~~~\, 
  [2, 8, 32, 128, 18, 72, 41, 164, 162, 154, 122,
   241, 223, 151, 110, 193, 31, 124, 2]
\nonumber \\
   U^2_{2, 247}  && ~:~
  [1, 16, 9, 144, 81, 61, 235, 55, 139, 1] 
    ~~\text{and}~~
  [2, 32, 18, 41, 162, 122, 223, 110, 31, 2]
  \nonumber \\ [-5pt] &&  ~~~~\, 
   [4, 64, 36, 82, 77, 244, 199, 220, 62, 4]
  ~~\text{and}~~
  [8, 128, 72, 164, 154, 241, 151, 193, 124, 8]
\nonumber\\
  U^4_{2, 247}  && ~:~
  [1, 16, 9, 144, 81, 61, 235, 55, 139, 1]
   ~~\text{and}~~
  [2, 32, 18, 41, 162, 122, 223, 110, 31, 2]
  \nonumber \\ [-5pt] &&  ~~~~\, 
  [4, 64, 36, 82, 77, 244, 199, 220, 62, 4]
   ~~\text{and}~~
  [8, 128, 72, 164, 154, 241, 151, 193, 124, 8]
\nonumber\\
  U^8_{2, 247}  && ~:~ 
  [1, 9, 81, 235, 139, 16, 144, 61, 55, 1]
   ~~\text{and}~~
  [2, 18, 162, 223, 31, 32, 41, 122, 110, 2]
  \nonumber \\ [-5pt] &&  ~~~~\, 
   [4, 36, 77, 199, 62, 64, 82, 244, 220, 4]
   ~~\text{and}~~
  [8, 72, 154, 151, 124, 128, 164, 241, 193, 8]
\nonumber\\
  U^{16}_{2, 247}   && ~:~
   [1, 81, 139, 144, 55, 9, 235, 16, 61, 1]
   ~~\text{and}~~
   [2, 162, 31, 41, 110, 18, 223, 32, 122, 2]
  \nonumber \\ [-5pt] &&  ~~~~\, 
   [8, 154, 124, 164, 193, 72, 151, 128, 241, 8]
   ~~\text{and}~~
   [4, 77, 62, 82, 220, 36, 199, 64, 244, 4]
% \end{eqnarray}
% \begin{eqnarray}
\nonumber\\
  U^{32}_{2, 247}   && ~:~
   [1, 139, 55, 235, 61, 81, 144, 9, 16, 1]
   ~~\text{and}~~
  [2, 31, 110, 223, 122, 162, 41, 18, 32, 2]
  \nonumber \\ [-5pt] &&  ~~~~\, 
  [8, 124, 193, 151, 241, 154, 164, 72, 128, 8]
   ~~\text{and}~~
  [4, 62, 220, 199, 244, 77, 82, 36, 64, 4]
\nonumber\\
  U^{64}_{2, 247}  && ~:~  
          [1, 55, 61, 144, 16, 139, 235, 81, 9, 1]
   ~~\text{and}~~
       [2, 110, 122, 41, 32, 31, 223, 162, 18, 2]
  \nonumber \\ [-5pt] &&  ~~~~\, 
      [8, 193, 241, 164, 128, 124, 151, 154, 72, 8]
   ~~\text{and}~~
        [4, 220, 224, 217, 79, 146, 126, 14,
 \nonumber \\ [-5pt] && ~~~~\,
                   29, 113, 40, 224, 4]
\nonumber\\
  U^{128}_{2, 247}   && ~:~
       [1, 61, 16, 235, 9, 55, 144, 139, 81, 1]
   ~~\text{and}~~
         [2, 122, 32, 223, 18, 110, 41, 31, 162, 2]
\nonumber \\ [-5pt] &&  ~~~~\, 
        [8, 241, 128, 151, 72, 193, 164, 124, 154, 8]
   ~~\text{and}~~
         [4, 244, 64, 199, 36, 220, 82, 62, 77, 4]
  \nonumber\\
  U^{256}_{2, 247}   && ~:~
       [1, 16, 9, 144, 81, 61, 235, 55, 139, 1]
   ~~\text{and}~~
        [2, 32, 18, 41, 162, 122, 223, 110, 31, 2]
\nonumber \\ [-5pt] &&  ~~~~\, 
      [4, 64, 36, 82, 77, 244, 199, 220, 62, 4]
   ~~\text{and}~~
        [8, 128, 72, 164, 154, 241, 151, 193, 124, 8]
\nonumber
  \ .
\end{eqnarray}
For simplicity, we have not included all closed sub-cycles, and 
we will refer to this by  version number $\tt{u\_ver}=1$ (this 
can be regarded as a truncated version of the ME operators). 
The composite operators $U^p$  are given in 
Figs.~\ref{fig_fxN247a2_Up1_a}, \ref{fig_fxN247a2_Up1_b}, 
and \ref{fig_fxN247a2_Up1_c} of Appendix~\ref{sec_N247_uver1},  
and the  corresponding phase histogram is presented in 
Fig.~\ref{fig_N247a2_phase1_hist}. We see that the results 
agree with the previous version, although there is more noise.

%%
%\vskip-2.5cm 

\vfill

\pagebreak
\clearpage

\section{Conclusions and Outlook}
\label{sec_conclusions}

It is an interesting mathematical fact that {\em factoring} 
is a notoriously difficult problem.  That is to say,   given an 
exponentially large integer,  it is exceedingly hard to find 
the corresponding prime factors.  In fact,  all {\em known} 
factorization algorithms that run on a  traditional 
classical (or digital) computer require an exponential time 
to factor such large numbers: A~typical digital 
computer would take the age of the universe to factor a 
several thousand bit number used for encryption.  Exponential 
computational costs are incurred because a classical computer 
must {\em sequentially} check almost every number less than 
the one being factored.  Indeed,  this is the basis upon which 
the security of many public key cryptographic protocols rests.  
More specifically, the  security of public key cyrto-systems 
relies on the principle that certain mathematical problems
are intrinsically difficult to solve.   For example,  the 
RSA\,\cite{rsa} system relies on the difficulty in factoring 
large numbers into their prime constituents,  while 
Diffie-Hellman\,\cite{df1,df2} public key distribution 
relies on the difficulty in solving the discrete logarithm 
problem (a problem that is closely related to factoring).   
However,  Shor's algorithm utilizes a quantum circuit that 
can factor exponentially large numbers in polynomial time 
(and a variant of the algorithm can also quickly solve the 
discrete logarithm problem)\cite{shor}. This is achieved by 
exploiting the massive parallelism inherent in quantum 
mechanics,  so that all possibilities can be tested 
simultaneously rather than sequentially,  thereby 
providing for a polynomial factorization process.   
Since Shor's algorithm  can solve these very hard 
problems so quickly, the implications are quite sobering 
for the security of public key cryptography in particular,  
and digital security in general.  

In this manuscript we have presented a rigorous and 
pedagogical presentation of Shor's factoring algorithm.
We have assumed no prior knowledge of the algorithm,
except a familiarity with the circuit model of quantum
computing,  and we have walked the reader through the
framework required to understand the algorithm,  
which is at the same time both elegant and complex.  
There are a number of moving parts to Shor's algorithm, 
and we have worked through each of them in turn, 
culminating in the requisite quantum factoring circuit. 

The mathematical basis for Shor's algorithm has no
connection with quantum mechanics,  but rather rests 
upon a deep but quite simple result from number 
theory.  Suppose we wish to factor an integer $N$,  
and we have a ``guess'' $a \in \{2,  3,  \cdots,  N-1\}$.  
We will usually refer to the guess $a$ as the {\em base}.  
Let us further assume that the base $a$ and the number 
$N$ that we wish to factor contain no {\em common} factors,  so that 
${\rm gcd}(a,N)=1$,  otherwise ${\rm gcd}(a,N)$ 
is one of the factors of $N$ that we seek (and the 
problem is solved).  Suppose now that we can find 
a non-trivial {\em modular square root of unity},
so that 
%that is to say,  an integer $b$ such that (i) $b \ne 
%\pm 1 ~({\rm mod}~N)$ and (ii) 
$b^2 = 1 ~({\rm mod}~N)$. The latter condition 
ensures that $b^2 - 1 = m N$ 
for some integer $m$.  We can write this expression 
in the form  $(b + 1)(b - 1) = m N$,  and we immediately 
see that factors of $N$ are given by ${\rm gcd}(b + 1,  N)$ 
and ${\rm gcd}(b - 1,  N)$. The greatest common divisor 
can be computed quickly on a classical computer in polynomial time. 
We can find $b$ by looking at the modular exponential 
function $f_{a, \smN}(x) = a^x ~({\rm mod}~N)$.  This 
function is periodic with some period $r$,  which means 
that $a^r = 1 ~({\rm mod}~N)$.  Therefore,  $b = a^{r/2}$ 
is a square root of unity,  and the factors of $N$ are 
thus ${\rm gcd}(a^{r/2} \pm 1,  N)$.  We have now
reduced the factoring problem to finding the period
of the function $f_{a, \smN}(x)$! However,  there are
several caveats: the conditions (\ref{eq_c1})--(\ref{eq_c3})
must all be met.  First,  equation (\ref{eq_c1}) requires 
that the period $r$ be even,  so that $b = a^{r/2}$ is
an integer.  Second, (\ref{eq_c3}) requires that $r$ 
be a solution to $a^r = 1 ~({\rm mod}~N)$, so that 
$b = a^{r/2}$ is indeed a square root of unity.  Third,
while $b = a^{r/2}$ is a square root of unity,  equation 
(\ref{eq_c2}) prohibits it from being a {\em trivial} square 
root,  in that $b \ne \pm 1 ~({\rm mod}~N)$.  In passing, 
we note that $r$ can in fact be odd,  provided that 
$a$ is a {\em perfect square},  in which case $b = 
a^{r/2}$ is still an integer\,\cite{twoone}.  

In contrast, the computational foundation for Shor's 
algorithm is a bit involved,  and rests upon two fundamental 
quantum algorithms:  the quantum Fourier transform (QFT) 
and quantum phase estimation (QPE).  The QFT implements 
the discrete Fourier transform on a quantum computer,  
and the QPE algorithm finds the complex phases or the 
Eigenvalues of an arbitrary unitary linear operator.   We spent 
Section~\ref{sec_QFT} developing the theory of the QFT,  
and Section~\ref{sec_QPE} covered the QPE,  deriving these
algorithms from scratch.  Shor's algorithm is just a special 
case of the QPE algorithm,  with a well-chosen unitary 
{\em modular exponentiation} (ME) operator,  denoted by 
$U_{a \, \smN}$. The ME operator is defined by its action 
on the computational basis by $U_{a \,\smN}\,\vert w
\rangle = \vert a \cdot w ~({\rm mod}~N) \rangle$,  and 
it is related to the modular exponential function by 
$U_{a \, \smN}^x \vert 1 \rangle =\vert f_{a \, \smN}(x) 
\rangle$. The Eigenvalue problem for the ME operator 
takes the form $U_{a \, \smN} \vert u_s \rangle = 
e^{2\pi i\,\phi_s}\, \vert u_s \rangle$,  where the phases 
are given by $\phi_s = s/r$ with \hbox{$s \in \{0, 1,  \cdots, 
r-1\}$}.  With the ME operators in hand, we combined the 
QFT and the QPE to construct Shor's factoring circuit in 
Section~\ref{sec_factoring}.  The result is a hybrid approach 
requiring both classical processing and quantum computation 
for the QPE analysis.  In the classical post-processing stage,  
the method of {\em continued fractions} allows one to 
extract the {\em exact} period $r$ from the {\em 
approximately} measured  phase $\tilde\phi$, thereby 
obtaining the period of the modular exponential function 
$f_{a \, \smN}(x)$. As we have seen,  this period is directly 
related to the factors of the number $N$,  and the QPE 
cleverly extracts $r$ to provide these sought after factors.  
Since continued fractions might not be familiar to the 
average reader, we gave a brief introduction to the 
subject in Section~\ref{sec_cont_frac}, proving a 
number of fundamental theorems.  More specifically,  Theorem~\ref{thm_c} 
ensures that the phases $\phi_s = s/r$ that we are 
seeking will necessarily be one of the convergents 
of the continued fraction representation of $\tilde\phi$.  
We therefore simply construct all such convergents 
$s_\ell/r_\ell$,  verifying that each value $r = r_\ell$ 
satisfies the necessary conditions (\ref{eq_c1})--(\ref{eq_c3}).  
If so,  then the smallest such value of $r_\ell$ is the exact
period that we are seeking, thus permitting us to 
calculate the factors of $N$ in polynomial time. 

In Section~\ref{sec_examples}, we presented a detailed 
example by  factoring the number $N = 15$ using the 
Qiskit simulator, providing the necessary Python source 
code to reproduce the results.  In particular,  we looked 
more closely at how continued fractions are utilized to 
extract the exact phase $\phi_s = s/r$ from the 
approximately measured $m$-bit phase $\tilde\phi$.
We also performed a theoretical analysis of the output
phase histograms for $N=15$ with bases $a=4, 8$,
calculating the expected histograms exactly 
for a general number of control qubits $m$. The Qiskit 
simulations agree precisely with the exact calculations. 
%Finally,  in Section~\ref{sec_factoring},  we put all the
%pieces together to construct the Shor factorization
%circuit. 

After verifying the formalism by factoring $N=15$, 
the smallest number accessible to Shor's algorithm, in 
Section~\ref{sec_further_ex} we moved on to factoring 
the more interesting composite numbers \hbox{$N=21, 
33, 35, 143, 247$}. The difficulty in factoring a number 
with Shor's algorithm does not lie in the size of the 
number itself, but in the magnitude of the period $r$ 
of the modular exponential function $f(x)$\,\cite{pretend}.  
The numbers $N$ have therefore been chosen,  along 
with their respective 
bases~$a$, to provide a wide range of periods,  running 
from $r = 2$ to $r = 36$. We go on to develop a general 
procedure that will find the appropriate modular 
exponentiation operator $U$ for any semi-prime 
$N = p \times q$, where $p$ and $q$ are prime. 
The principle behind this technique rests upon the 
fact that the modular exponential function $f(x)$ 
creates a sequence of states $\vert f(x) \rangle$ as 
we increment the argument $x$ successively over 
its range of permissible values $x = 0, 1, 2, \cdots, r-1$. 
These states are the basis elements of an invariant
$r$-dimensional subspace ${\cal U}_r$ of the 
exponentially large work space ${\cal W}_n$. 
To be more precise, note that the ME operator 
$U$ first acts on the work state $\vert 1 \rangle$, and 
the next operation acts on the output of the first, and 
so on.  Since $U^x \vert 1 \rangle = \vert f(x) \rangle$, 
this technique encodes the sequence of states generated 
by $f(x)$ into the ME operator $U$.  One might think  
that we have gained nothing,  since this method is 
equivalent to knowing the exact period $r$, and therefore 
Shor's algorithm would be unnecessary. However, the 
ME operators are quite forgiving, and they do not require 
knowing the full sequence of states. We can approximate 
the ME operator $U$ by a truncated version using only 
a subset of the states. This is because the continued 
fraction method does not require knowing the exact 
phase $\phi_s$, but only a sufficiently precise approximate 
phase $\tilde\phi$. This suggests that implementing 
Shor's algorithm might not be as difficult as first 
suspected. 

In closing this manuscript,  we should briefly discuss 
some practical aspects of realizing Shor's algorithm.   
References~\cite{nmr, tame, qbit_recylcing, study_ibmQ} 
have already succeeded in factoring  the numbers $N=15, 
21,  35$ on a range of existing quantum computers.
However,  these 
authors did not implement complete versions of 
Shor's algorithm,  but rather so called {\em compiled} 
versions that take advantage of the specific base
$a$ to minimize the qubit count.  This is because current 
machines lack a sufficient number of qubits even for
such small numbers.  For more realistic cases,  to factor 
a number $N$ with $n = 1024$ bits,  we would need 
$m = 2 n + 1 = 2049$ control qubits,  with the total 
number of qubits being $n + m = 3073$.   For $n=4096$ 
bits,  these numbers increase to $m = 2 n + 1 = 8193$ 
control qubits and $n + m = 12289$ total qubits.  Breaking 
RSA therefore requires thousands to tens of thousands of 
high quality qubits.  Modern quantum computers are currently 
quite far from this domain,  although future machines will 
undoubtedly be able to handle these requirements.  The 
gate count for the ME operators is also problematic.  
Reference~\cite{gen} indicates that one would require 
$72 n^3 \sim 8 \times 10^{10}$ gates for $n = 1024$ 
and $72 n^3 \sim 5 \times 10^{12}$ gates for $n = 
4096$ work qubits. The technique outlined in this 
manuscript might well lower this gate count,  but the 
requisite number of ME gates would still be quite large.  
Clearly,  automation would be required for such a large 
number of gates.  Even for the cases considered in this 
manuscript,  we employed a python script to write the 
Qiskit gates.  Finally,  implementing the 
QFT might seem to be the real challenge,  as one requires
astonishingly small phase angles for large numbers
of qubits.  Recall that the phase rotation angle is given 
by $\theta_k = 2\pi/2^k$,  and for $k=1000$,  this 
gives an angle of order $2\pi/2^{1000} \sim 10^{-301}$!
However,  this problem has already been addressed
in Refs.~\cite{coppersmith, rot_limit,nam_blumel}.
These authors show that for very small phase angles,
one can simply ignore the corresponding phase 
rotation.  In other words,  we only need to consider
phase angles larger than $\theta_{\rm min} = 
2\pi/2^{k_{\rm max}}$ for some cutoff $k_{\rm max} 
\sim 20$\,\cite{coppersmith}.  This is not surprising,
as we do not require the {\em exact} QFT,  but only 
an approximate form that captures sufficient phase
accuracy so that the method of continued fractions 
may be applied during the post-quantum processing.
We see that there are indeed very large obstacles to 
overcome in breaking RSA with Shor's algorithm,  
but none of them seem insurmountable in the long 
run.  For the immediate future,  however,  it seems that 
RSA will remain secure.

\clearpage
%%%%%%%%%%%%%%%%%%%%%%
%%
\begin{acknowledgments}
 I would like to thank David Ostby,  Mike Rodgers,
 and Max Singleton of the SavantX Quantum Division.
 This work was funded by the SavantX Research Center. 
\end{acknowledgments}
%%

%%%%%%
%\clearpage
%\pagebreak
\appendix

\section{Modular Exponentiation Operators}
\label{sec_meo}

\subsection{Composite ME operators for N = 143, a = 5, r = 20, m = 8, u\_ver=1}
\label{sec_N143_uver1}

The composite operators $U^p_{5, 143}$ from (\ref{eq_Up_N143a5})
for $p= 2, 4, \cdots, 128$. 

\vskip-1.6cm
\begin{figure}[h!] 
\includegraphics[scale=0.43, center]{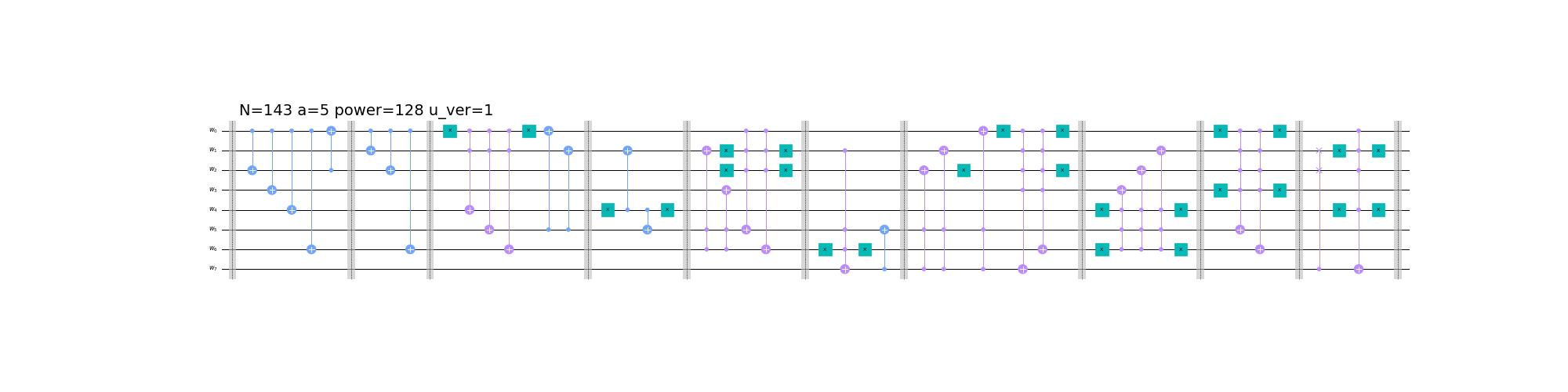} 
\vskip-1.3cm
\includegraphics[scale=0.50, center]{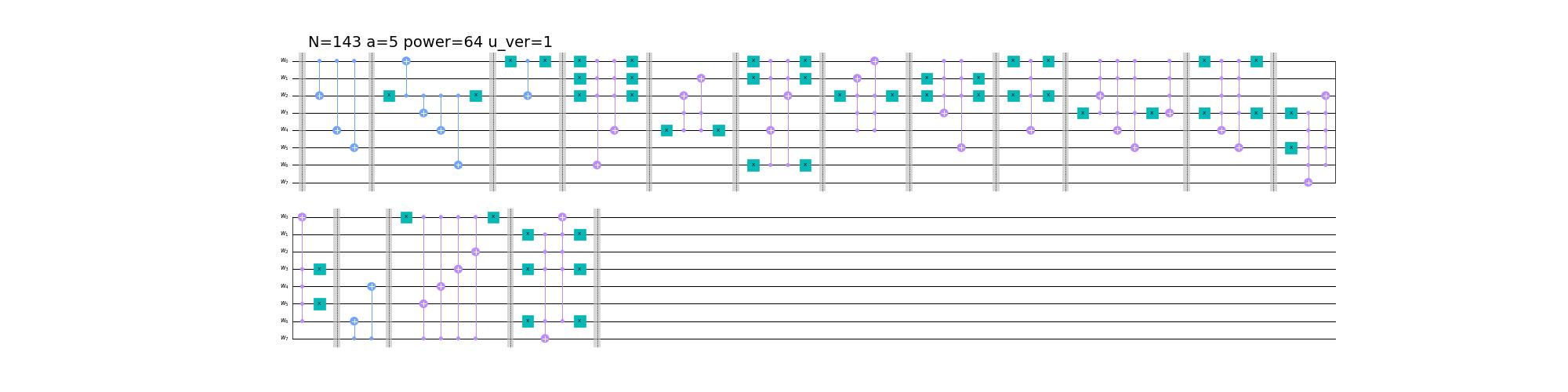} 
\vskip-0.8cm
\includegraphics[scale=0.50, center]{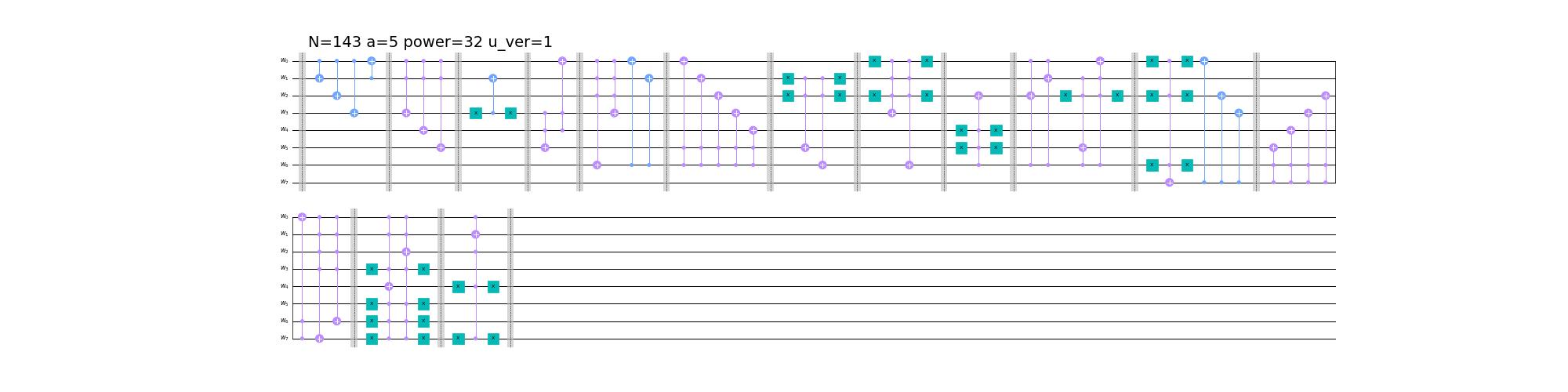} 
\vskip-0.5cm 
\caption{\footnoteskip  
$N=143$,  $a=5$,  $r=20$, $\tt{u\_ver}=1$: 
 $U_{5, 143}^{32}, U_{5, 143}^{64}, U_{5, 143}^{128}$. 
}
\label{fig_fxN143a5_Up_a}
\end{figure}
\begin{figure}[h!] 
\includegraphics[scale=0.45, center]{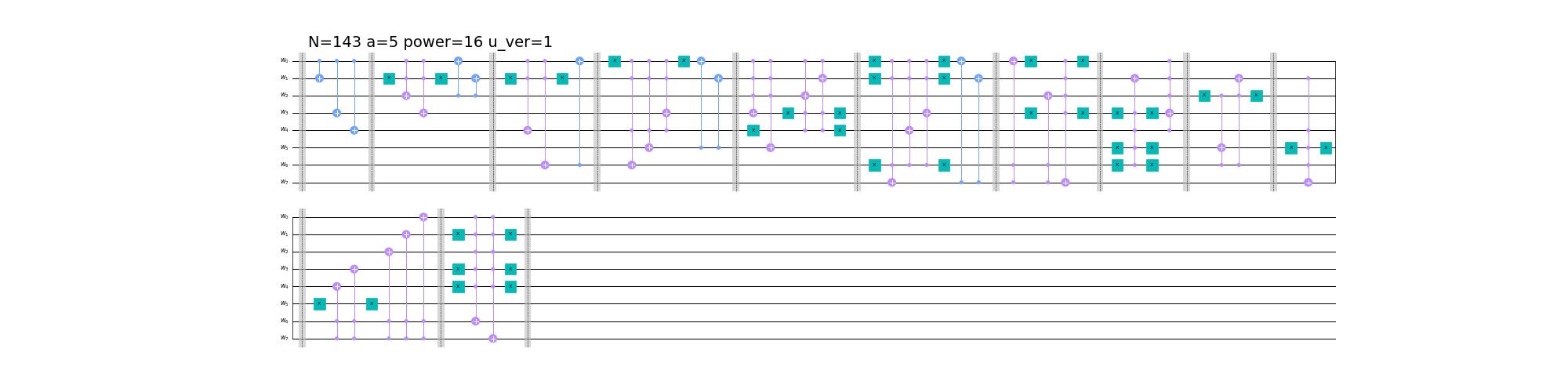} 
\vskip-1.0cm
\includegraphics[scale=0.40, center]{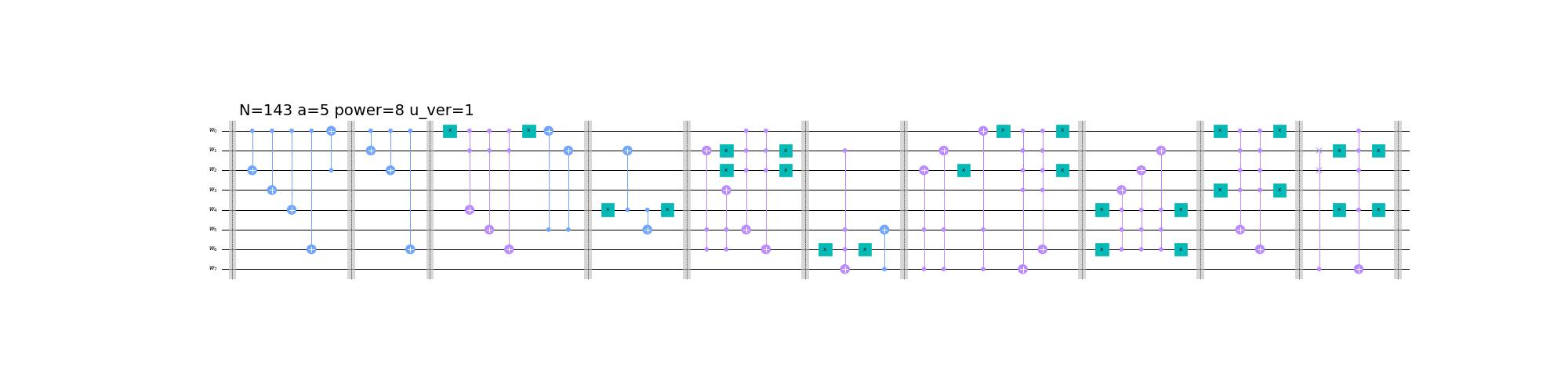} 
\vskip-1.0cm
\includegraphics[scale=0.45, center]{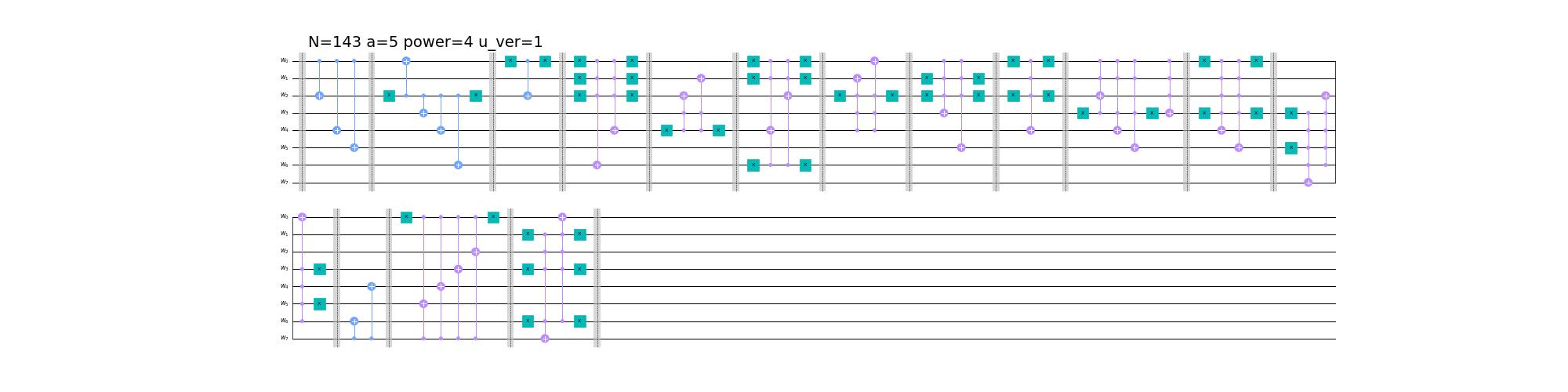} 
\vskip-0.5cm
\includegraphics[scale=0.68, center]{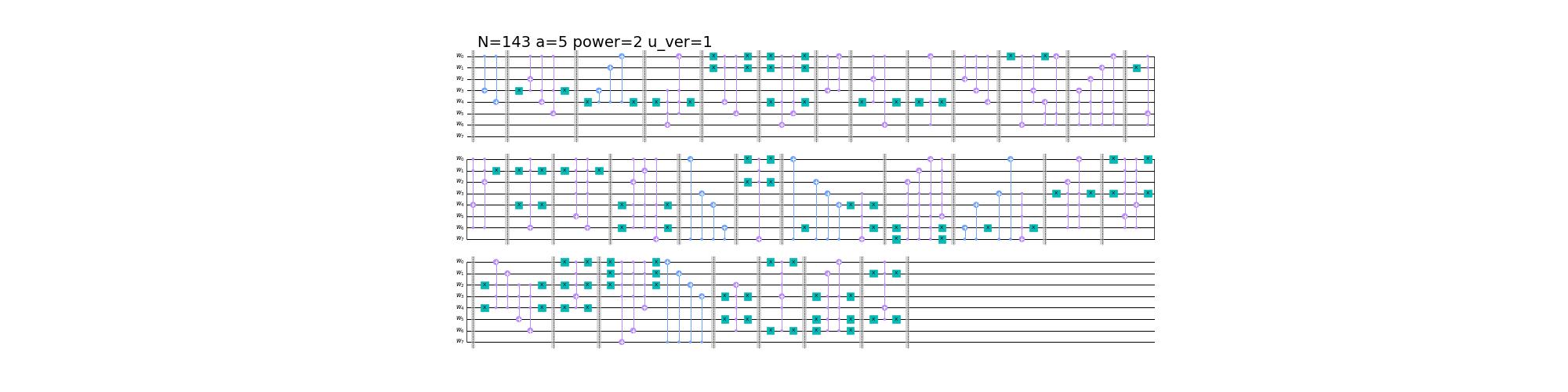} 
\vskip-1.0cm
\caption{\footnoteskip  
 $N=143$,  $a=5$,  $r=20$, $\tt{u\_ver}=1$: 
 $U_{5, 143}^2, U_{5, 143}^4,  U_{5, 143}^8, U_{5, 143}^{16}$. 
}
\label{fig_fxN143a5_Up_b}
\end{figure}
%%
%%%

\clearpage
\subsection{Composite ME operators for N = 247, a = 2, r = 36, m = 9, u\_ver=1}
\label{sec_N247_uver1}

The composite operators $U^p_{2, 247}$ from (\ref{eq_Up1_N247a2})
for $p=2, 4, \cdots, 256$.

\phantom{ff}
\vskip-1.0cm 
\begin{figure}[h!]
\includegraphics[scale=0.45, center]{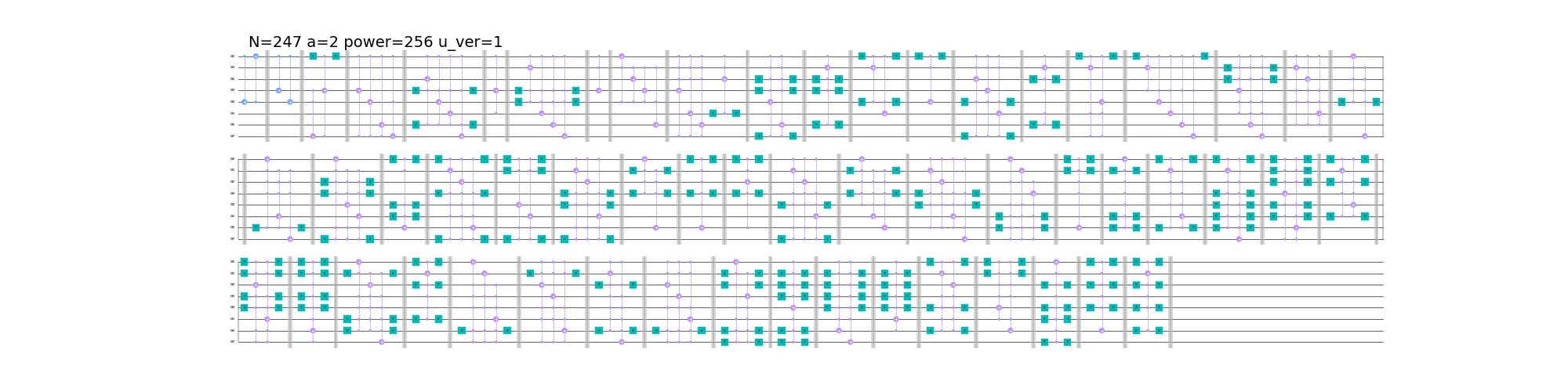} 
\vskip-0.6cm
\includegraphics[scale=0.60, center]{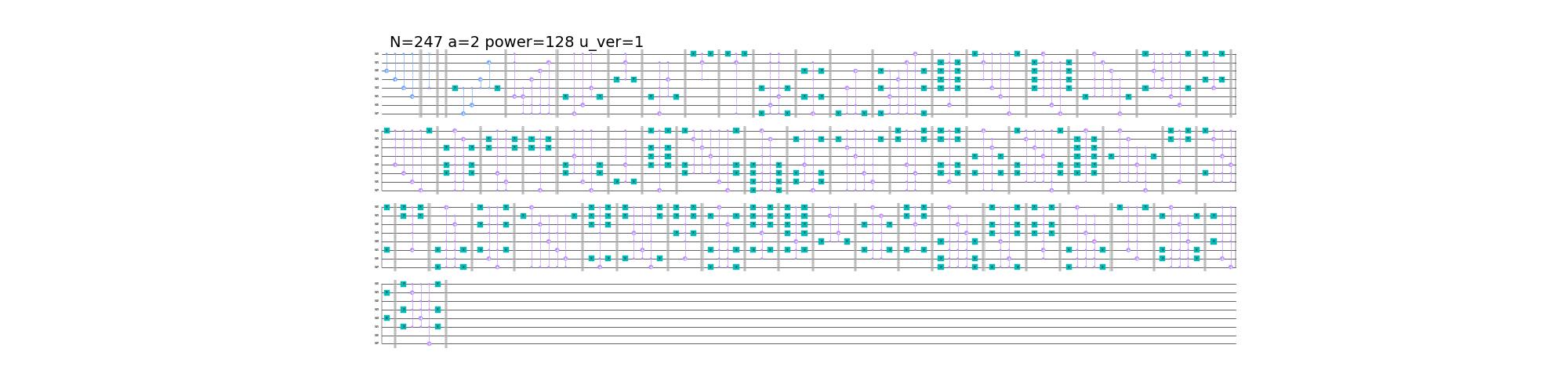} 
\vskip-0.6cm
\includegraphics[scale=0.60, center]{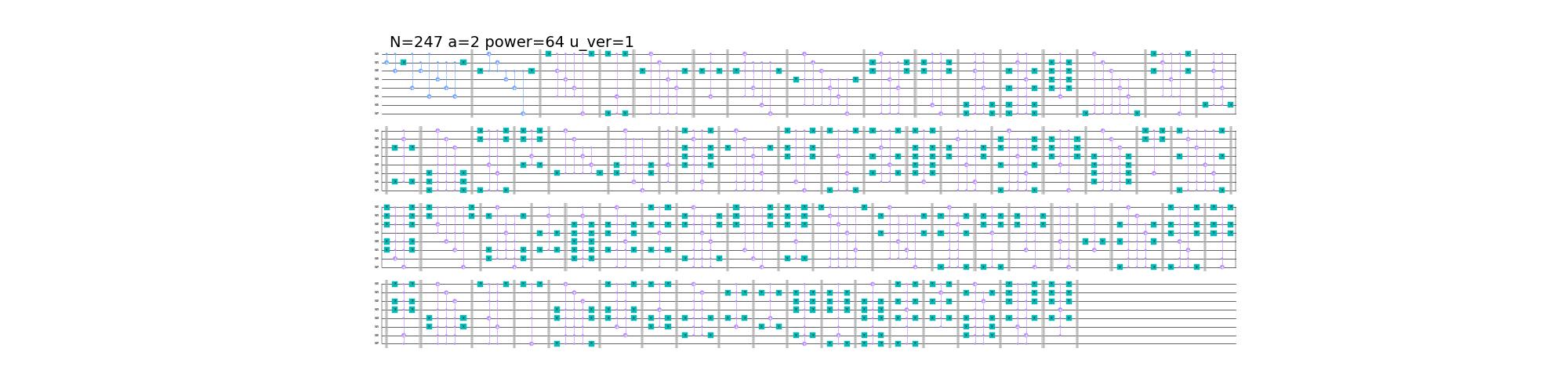} 
\vskip-1.0cm 
\caption{\footnoteskip  
$N=247$,  $a=2$,  $r=36$, $\tt{u\_ver} = 1$:  
 $U_{2, 247}^{64}$,  $U_{2, 247}^{128}$,  $U_{2, 247}^{256}$
}
\label{fig_fxN247a2_Up1_a}
\end{figure}
\begin{figure}[h!]
\vskip-0.6cm
\includegraphics[scale=0.65, center]{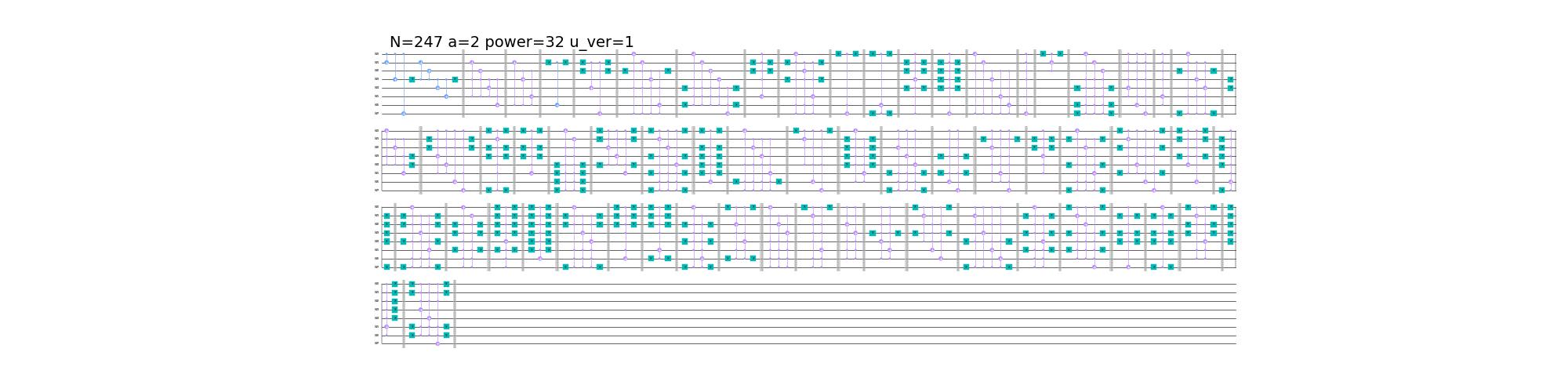} 
\includegraphics[scale=0.50, center]{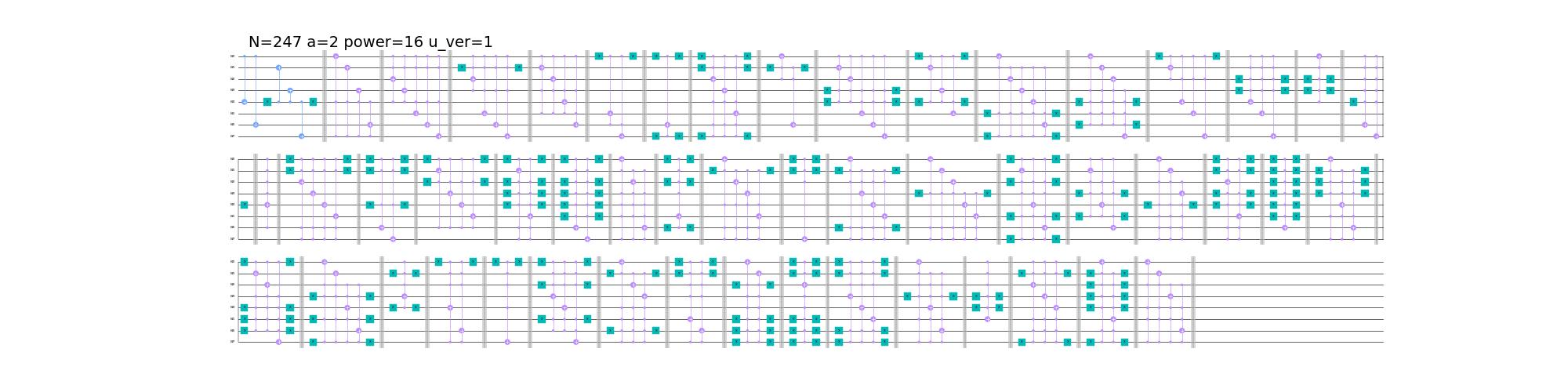} 
\includegraphics[scale=0.50, center]{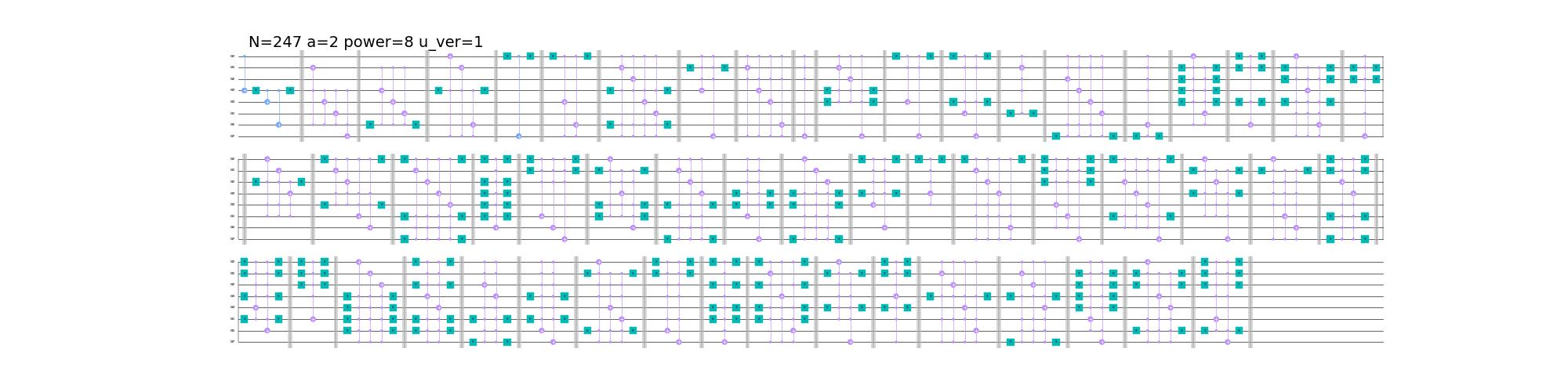} 
\vskip-0.5cm 
\caption{\footnoteskip  
$N=247$,  $a=2$,  $r=36$, $\tt{u\_ver} = 1$:  
 $U_{2, 247}^{8}$,  $U_{2, 247}^{16}$,  $U_{2, 247}^{32}$.
}
\label{fig_fxN247a2_Up1_b}
\end{figure}
\begin{figure}[h!]
\begin{centering}
\includegraphics[scale=0.67, center]{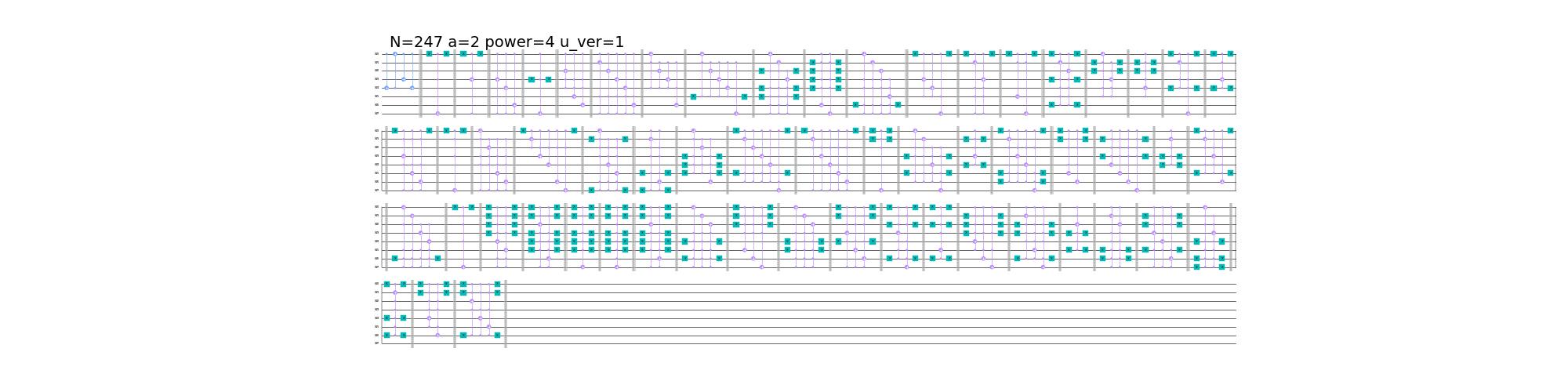} 
\par\end{centering}
\begin{centering}
\includegraphics[scale=0.50, center]{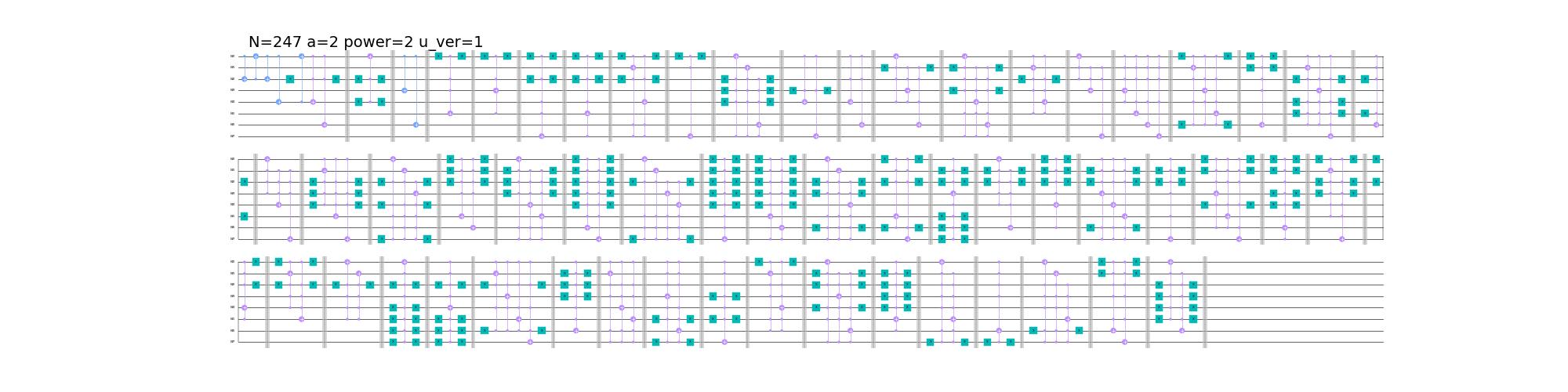} 
\par\end{centering}
\vskip-0.5cm 
\caption{\footnoteskip  
$N=247$,  $a=2$,  $r=36$,  $\tt{u\_ver} = 1$:
 $U_{2, 247}^{2}$,  $U_{2, 247}^{4}$. 
}
\label{fig_fxN247a2_Up1_c}
\end{figure}

\vfill
\pagebreak

\end{document}